%% file: PhD_thesis.tex
\documentclass[a4paper]{book}

\usepackage[latin1]{inputenc}
\usepackage[T1]{fontenc}
\usepackage{amsmath,amsfonts,amssymb,array}
\usepackage{graphics,graphicx,float,mfpic}
\usepackage{epsfig,color}
\usepackage{subfigure}
\usepackage{chapterbib}
\usepackage{lettrine}
\usepackage{geometry}
\definecolor{gris}{rgb}{.25,.25,.25}

\newcommand{\be}{\begin{equation}}
\newcommand{\ee}{\end{equation}}
\newcommand{\bea}{\begin{eqnarray}}
\newcommand{\eea}{\end{eqnarray}}
\newcommand{\ben}{\begin{enumerate}}
\newcommand{\een}{\end{enumerate}}
\newcommand{\bi}{\begin{itemize}}
\newcommand{\ei}{\end{itemize}}

\renewcommand{\th}{$^{th}$}
\newcommand{\Ord}[2]{\mathcal O \left(#1\right)^{#2}}

\newcommand{\ddf}[2]{\frac{\partial #1}{\partial #2}}
\newcommand{\dddf}[2]{\frac{\partial^2 #1}{\partial #2^2}}
\newcommand{\ddx}[1]{\frac{\partial}{\partial #1}}

\begin{document}

\title{New features of black strings and branes in higher dimensional gravity due
to a cosmological constant}
\author{T\'erence DELSATE}
\date{April 2010}

\pagestyle{empty}

\begin{titlepage}
\begin{center}
    \includegraphics{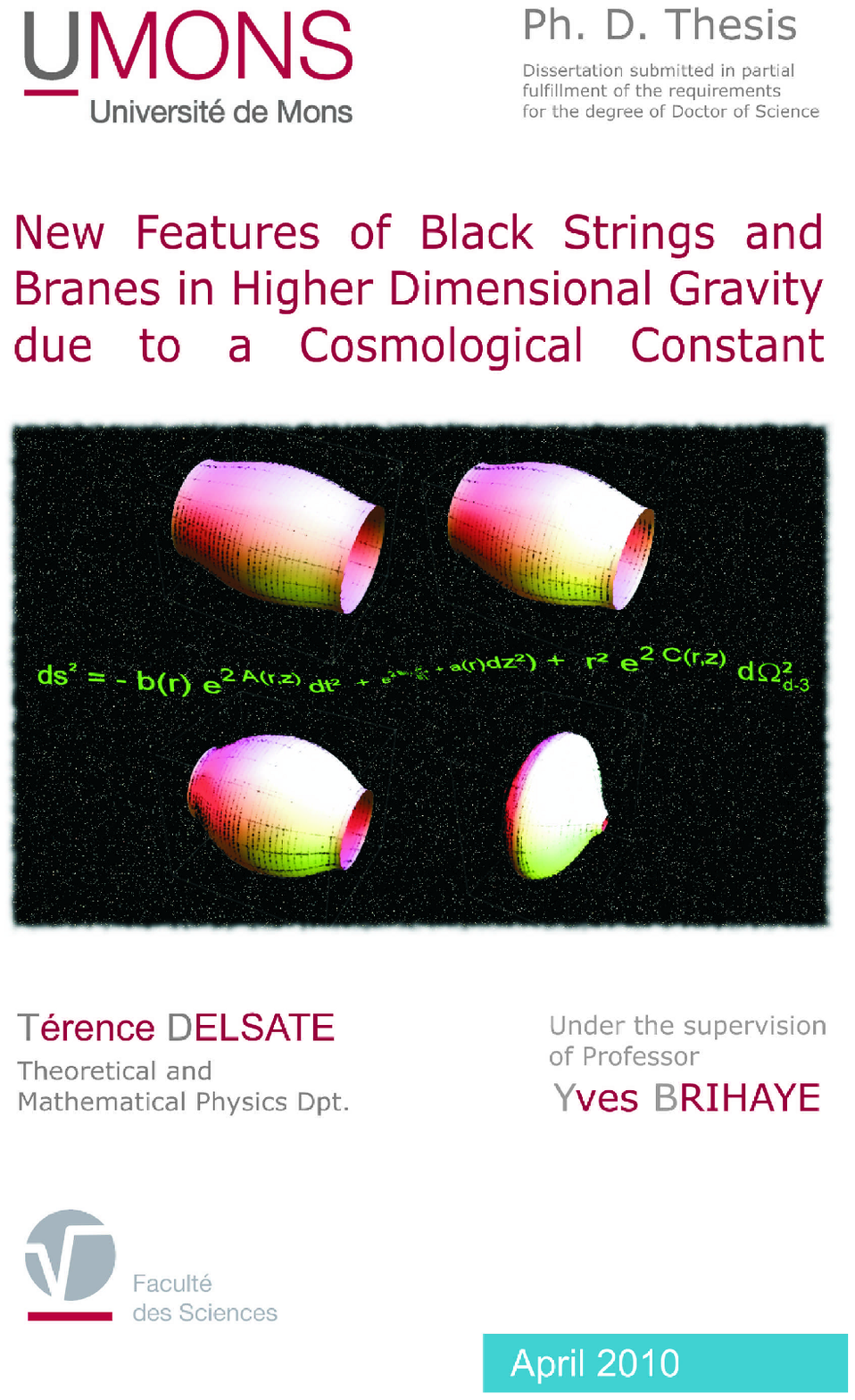}
\end{center}

%\begin{center}
%{\huge New features of black strings and branes in higher dimensional gravity due
%to a cosmological constant}\\

%\end{center}

%\vspace{7cm}
%\begin{center}
%{\huge T\'erence {\sc Delsate}}\\
%{\large Theoretical and Mathematical Physics Department\\
%Ph. D. Advisor: Prof. Dr. Yves BRIHAYE}\\
%\end{center}
%\begin{center}
%{\large March 2010}\\
%{Th\`ese pr\'esent\'ee en vue de l'obtention du grade l\'egal de}\\
%{Docteur en Sciences}
%\end{center}
%\begin{figure}[H]
%\begin{center}
    %\includegraphics*[0,0][3.72cm,2cm]{fig/logo/umacad.png}
    %\includegraphics*[scale=.4]{fig/logo/umons.eps}
%\end{center}
%\end{figure}
\end{titlepage}

\tableofcontents

\frontmatter
\chapter{Acknowledgement}
First of all, I thank my wife Natacha. She has always been a great support and made many concessions regarding my work; more than I can ask her! I also thank my father Dominique and my parents in law, Francoise and Patrice, who always pushed me to follow the way I wanted.

I would like to express my biggest gratitude to Professor Yves Brihaye without whom this thesis would never have been written. It has always been a big pleasure to work with him. Actually, quickly, he made me feel like working \emph{with} him and not \emph{for} him.
All my gratitude goes also to Professor Pierre Gillis who accepted me as an assistant for the first year students physics laboratories. Professor Gillis and his team taught me a lot regarding the pedagogical requirement in order to be as efficient as possible during trainings and labs. 

I also would like to gratefully acknowledge Dr Eugen Radu with whom I collaborated. It has been and it is a big pleasure to work with him. I also gratefully acknowledge Dr. Toby Wiseman for enlightening and stimulating discussions.

My friends have also been of great support as well as my colleagues of the Pentagone and of the building IV and VI; I kindly acknowledge Georges, Evelyne, Biscuit and Joseph, Nikita, Thierry for the pleasant breaks made with them, Olivier, Yves, Aline, Lam, Jean-Pierre and Michel, M\'elanie for having contributed to my pedagogical formation. I thank also Prof. Grard, Prof. Semay, Prof. Spindel, Prof. Nuyts, Dr Boulanger, Dr Brau, Dr Buisseret and Dr Mathieu for the interesting discussions during the coffee breaks.

May I have forgotten someone; if so I apologies and thank all the people not mentioned here that would recognise themselves.

%This is the preface and it is created using a TeX field in a
%paragraph by itself containing \verb|\chapter*{Preface}|. When the
%document is loaded, this appears if it were a normal chapter, but
%it is actually an unnumbered chapter. The \verb|markboth| TeX
%field at the beginning of this paragraph sets the correct page
%heading for the Preface portion of the document. The preface does
%not appear in the table of contents.

\pagestyle{headings}

\chapter{Introduction}
\input{intro2.tex}

\mainmatter
\chapter{Braneworld model}
\label{ch:branes}
\input{brane}

\chapter[Extended branes]{Topological braneworld models with extended branes}
\label{chextbranes}
\input{extbrane}

\chapter[Trapping realistic matter fields]{Trapping realistic matter fields: a toy model}
\label{chferm}
\input{fermions}

%\part{Higher dimensional black holes}
\chapter[Higher dimensional black objects]{Higher dimensional black holes and strings}
\label{chbhbs}
\input{bh}

\chapter[Phases of $AdS$ black strings]{Phases and thermodynamics of AdS black strings and AdS non uniform black strings}
\label{chadsbs}
\input{bsphase}

\chapter{Conclusions and perspectives}
\input{ccl}

\appendix
\chapter{Some topics in general relativity}
\label{app:reminder}
\input{reminder}

\chapter{AdS spacetime}
\label{app:ads}
\input{ads.tex}

%\chapter{Topological defects}
%\label{app:topo}
%\input{topo}

\chapter{The Poschl-Teller equation} 
\label{ptequation}
The Poschl-Teller equation is known as the following one-dimensional eigenvalue Schr\"odinger equation for the corresponding Poschl-Tellel potential.
\be
  -\frac{d^2}{dx^2} \eta - \frac{N(N+1)}{\cosh^2 x} \eta = \omega_p^2 \eta
\ee 
It is considered on an appropriate domain of the Hilbert space of square integrable functions on the real line.
It is a standard result that, for integer values of $N$, the above equation admits $N+1$ eigenvalues and eigenvectors which can be computed algebraically.

For the first few values of $N$ the eigenvalues are given by
\be
N=1: \omega_p^2 = -1,0,\  N=2: \omega_p^2 = -4,-1,0,\ N=3: \omega_p^2 = -9,-4,-1,0.
\ee 
The corresponding eigenfunctions are given by
\be
\eta_{-1} = \frac{1}{\cosh x},\ \eta_0 = \tanh x,
\ee
for the case $N=1$,
\be
\eta_{-4} = \tanh^2x-1,\ \eta_{-1}=\frac{\sinh x}{\cosh^2x},\ \eta_0 = 3 \tanh^2x-1,
\ee
for the case $N=2$ and so on.

\chapter{Supersymmetric quantum mechanics}
\label{app:susyqm}

\input{susyqm}

\chapter{Numerical method}
\label{app:num}
\input{num}

\backmatter

\addcontentsline{toc}{chapter}{List of Figures} 
\listoffigures

\addcontentsline{toc}{chapter}{Bibliography} 
\bibliography{biblio}{}
\bibliographystyle{unsrt}

\end{document}

%% file: intro2.tex
%%%%%%%%%%
%  INTRO  
%%%%%%%%%%

From an etymological point of view, physics means \emph{Science of Nature}. However, over the course of time the word 'science' had his definition changed. In early times, such as Ancient Greece, 'science' meant the knowledge one has about one subject. Later, in the early XIX\th century, 'science' meant the collection of 'mathematical science', physics and 'natural science', where 'science' in its old meaning is used to define the modern concept.

On the other hand, the definition of physics itself has changed all over the centuries. For instance, physics as the science of nature was one of the three parts of philosophy, according to Plato (about 400 B.C.), together with ethics and logic. Note that Aristotle (about 350 B.C.) also included physics, or the science of nature in his definition of philosophy, which distinguished theoretical, practical and poetic philosophy.

It is indeed interesting to note that in the XII\th century, physics had a double meaning; namely 'medicine' and 'what is related to nature'. Note that the word which designates someone who practices medicine is 'physician'. It is obvious that the words physicist and physician have the same etymologic origin. In the XV\th century, 'physics' was defined as the knowledge of natural phenomena and was also called natural philosophy. The title of the celebrated book by Newton (around 1700) is indeed \emph{Philosophi\ae Naturalis Principia Mathematica}, Latin for 'Mathematical Principles of \emph{Natural Philosophy}'.

It is only from the period of Galileo and Descartes (around 1600) on that physics took its modern meaning of classical Newtonian physics. As an illustration, the first edition of the French Academy Dictionary (1694) defines physics as \emph{Science which goal is the knowledge of natural things}. In its eighth edition (1935), the Dictionary defines 'physics' as \emph{Science which observes and classify material world's phenomena in order to extract the underlying laws}; which is closer to the modern definition of physics: \emph{The scientific study of forces such as heat, light, sound, etc., of relationships between them, and how they affect objects.}, according to the Oxford Dictionary.

Anyhow, modern physics is composed by various domains, such as condensed matter physics, astrophysics, particle physics, among many others. Moreover, there has always been an intrinsic link between theory, which formalizes abstract mathematical concepts and leads to observable predictions, and experiments, which try to measure phenomenon to which theory gives a conceptual representation. The experiments are guided by theories, while theories are developed when experiments reveal new phenomena, this is a constant interplay. As an example, in the '60s, Gell-Mann found a symmetric structure in hadronic particles, which led him to postulate the existence of quarks. These have been observed from late '60s to mid '90s. This is when theory influences experiments, but the converse is true as well: experiments on black body radiation led Planck to postulate the quantization of the energy of light.

On another hand, since the beginning of the XX\th century, there has been a distinction between theoretical physics and experimental physics, unlike in chemistry where many brilliant theorists are also brilliant experimentalists. This can be a consequence of the deep mathematisation of physics, which relies on advanced concepts of mathematics such as group representation theory, differential geometry, algebraic geometry, topology, etc.
Such a mathematisation has opened the area of theoretical and mathematical physics, at the boundary of mathematics and physics. This is an inter-dependent link between the two disciplines.

As an example, Newton and Leibniz invented differential calculus in order to formalize the laws describing motion of bodies, and Einstein relied on works by Riemann shedding the foundation of differential geometry around 1850 in order to develop his theory of General Relativity (1915). 

General Relativity is precisely the framework in which this thesis has been realized. But what is General Relativity? There are entire books dealing with this question, but we will try to give a very brief idea of the main concept of General Relativity.

The theory of General Relativity states that there is a intrinsic connection between the spacetime and its content; the energy and matter content of the spacetime curves it while the spacetime tells the matter how to move.

This interconnection is of geometric nature: the energy/matter content of the universe is closely related to the geometry of the spacetime. This is encoded in the Einstein equations:
\be
G_{ab} = \frac{8\pi G}{c^4} T_{ab},\ a,b=0,1,2,3,
\label{einsteineq}
\ee
where $G_{ab}$ is the Einstein tensor, a geometrical object related to the spacetime curvature while $T_{ab}$ is the stress tensor, representing the matter/energy content of the spacetime; $G$ Newton's constant and $c$ is the speed of light.
In particular, this confers a dynamical aspect to the spacetime, in opposition to the Newtonian theory where the spacetime is fixed and thus not dynamic.

Einstein found that his equations predicted an expanding or contracting universe; but as he believed in a static universe anyway, he refused this option and added a term in his equations: the cosmological constant term. By properly tuning the value of this cosmological constant, he could recover a static universe. But a decade later, Hubble pointed out from the observations of the relative motions of stars that our universe should be expanding. Einstein finally abandoned the assumption of the cosmological constant. Note that nowadays, cosmological observations predict a small and positive cosmological constant. We will come back to this question later.

The theory of General Relativity works quite well at the scale of the solar system and has passed numerous experimental tests, such as the displacement of Mercury's perihelion or the deviation of light rays. At larger scales however, the observation of angular velocities of stellar objects in our galaxy is not compatible with the predictions of General relativity. Physicists generally postulate the existence of some dark matter, i.e. matter that we don't observe but that has an effect on the spacetime curvature. 

Black holes, which are also one of the main aspects of this thesis, are probably among the most fundamental solutions to the Einstein field equations. The first example of black hole has been provided by Karl Schwarzschild \cite{schwarzschild} in 1916 on the Russian front during world war I. The Schwarzschild solution describes a vacuum spacetime geometry with a spherical symmetry. The Schwarzschild solution depends on an arbitrary parameter $M$, asymptotically, the gravitational field behaves like the Newtonian gravitational field of a punctual object of mass $M$ located at the origin. Note that the Schwarzschild solution also describes the region outside a neutral non-punctual static spherical massive source.

The Schwarzschild solution is further characterized by a particular radius, the Schwarzschild radius. If a spherical massive object has a radius smaller than the Schwarzschild radius, it forms a black hole, i.e. an object so massive that massive or massless particles cannot escape from the gravitational field of this object, once they are too close. 

Black holes are characterized by an event horizon coinciding with the Schwarzschild radius, it is the boundary of the region where the gravitational field is so strong that nothing can escape from it. In the coordinate system used by Schwarzschild, the event horizon corresponds to an apparent singularity in the solution; it is however possible to formulate the solution in a different coordinate system where the apparent singularity is removed. Black holes have a region of very strong gravitational attraction, which can be inside the horizon. They are therefore interesting from a theoretical point of view since they can reach the validity limit of General Relativity. Black holes are indeed sometimes called theoretical laboratories for high energy physics. We will come back on the question of black holes with more details in the end of the introduction.

From a mathematical point of view, the number of dimensions of spacetime is just a parameter of the theory, although physically, we observe four dimensions. In the twenties, Kaluza \cite{kaluza} and Klein \cite{klein} noticed that there was an elegant way of formulating gravity and electromagnetism, assuming the existence of a fifth dimension. But since we don't see five dimensions, they assumed that this fifth dimension was very small, of the order of the Planck length (of the order of $10^{-35}$m). Unfortunately, their theory suffered from numerous problems, such as the existence of a bosonic fundamental field which did not correspond to anything physical at those times.

A few years after Kaluza and Klein's theory, quantum mechanics was developed and the idea of extradimensions was not developed further. The theory of Kaluza and Klein was then completely forgotten until recently when String Theories started to develop.

\section*{What's in this thesis?}

This thesis is organized in five chapters (and an additional concluding chapter). The first three chapters are independent of the last two. Here, we briefly summarize the content of the chapters and present some theoretical aspects of the models addressed.

The {\bf first chapter} is devoted to braneworld models, we review the main ideas of Kaluza and Klein regarding extradimensions. In particular, we introduce the Kaluza-Klein model in $5$ dimensions. Next, we present the Kaluza-Klein reduction mechanism and turn to the Arkani-Hamed - Antoniadis - Dvali - Dimopoulos \cite{add1,add2} model. Then, we introduce the first warped extradimension model, namely the Randall-Sundrum model \cite{rs1,rs2}. Finally, we present other types of branes available in String Theories, namely $p$-branes and $D$-branes.

\subsection*{Someone said extradimensions?}

\leftskip=20pt As mentioned earlier, the idea of extradimensions is not new. It was postulated by Kaluza and Klein in the '20s and forgotten until the early '80s, when String theories started to develop. Originally, String theory was designed to describe hadrons and strong interaction, but it turned out that the theory contains a spin-$2$ particle, resembling the graviton. It was therefore realized that String theory can provide a good candidate for a theory of quantum gravity.
The main idea of string theory is to consider string-like instead of point-like fundamental objects. Whereas the formulation of the classical theories considering point-like objects minimizes the proper length of the worldlines of particles, String theory minimizes the area of world sheets, i.e. of the surface defined by the motion of a string along the proper time direction.

However, the quantization of String theory is consistent only if the number of spacetime dimension is $26$ (this is actually true for the bosonic String theory); or $10$ for supersymmetric String theories \cite{thooftstring}. Anyway, $26$ (and $10$) is much larger than $4$; so the idea of extradimension is nowadays necessary.

As we just said, the spectrum of String theories contains a spin-$2$ object, reminding of the graviton. Moreover, the low energy description of String theories includes gravitation. It then makes sense to study higher dimensional gravity, inspired by String theories. On the other hand, we also mentioned the fact that the theory of General Relativity can be formulated in more than four dimensions. Mathematically speaking, the number of dimensions is then just a free parameter. However, at our energy scales, we do actually observe only four dimensions. So if one deals with extradimensional theories, one has to ensure that the low energy description of his theory implies the observation of four dimensions.

This is indeed the case for braneworld models, where our four dimensional universe is embedded in a higher dimensional spacetime. Either the extradimensions are small, so that they are not observable at our energy scale, either there exists a confining mechanism that sticks our observable world on the four dimensional slices of the entire spacetime.\\

\leftskip=0pt

The {\bf second chapter} is devoted to braneworlds with an extension along the extradimensions. The brane models considered in chapter two are described by the Einstein gravity lagrangian (with a cosmological constant term) extended by matter lagrangian describing bosonic fields. The matter lagrangian are chosen as usual classical fields theory known to admit 'soliton' type solutions. Generally speaking, a soliton refers to classical localized solution to nonlinear equations. It is well known that some field theories admit soliton solutions (cosmic strings in the case of the Maxwell-Higgs model, monopoles in the case of Yang-Mills-Higgs models). 

The philosophy of the approach of chapter two is to make use of these solitons by imposing the direction of localization of the solution in the extradimensions of the full model.

This chapter is based on original results \cite{bdh,bd,sadelbri}; first we will present brane solutions to the Einstein-Maxwell-Higgs model with inflating four dimensional slices and a bulk cosmological constant (i.e. a cosmological constant for the entire spacetime) \cite{bd}. Then, we present solutions to the Einstein non abelian Higgs and Einstein-Yang-Mills-Higgs theory, again with inflating branes \cite{bd}. Finally, we consider the six dimensional inflating baby-Skyrme model \cite{sadelbri}. The baby-Skyrme is a toy model for the Skyrme model somehow describing nucleic matter. The baby Skyrme brane model can therefore be viewed as a toy model mimicking a brane composed by nucleic matter.

\subsection*{What's the cosmological constant?}

\leftskip=20pt

Roughly speaking, the cosmological constant is an extra term in the Einstein theory of gravity, which leaves the equations consistent. An important issue of this thesis is the study of the effect of a cosmological constant on some classes of classical solution to Einstein equations.  The Einstein equations with a cosmological constant read
\be
G_{ab} = \frac{8\pi G}{c^4} T_{ab} - \Lambda g_{ab},
\label{Einsteq}
\ee
where $G_{ab}$ and $T_{ab}$ are the same objects than in equation \eqref{einsteineq}, $\Lambda$ is the cosmological constant and $g_{ab}$ is the metric. The metric is the essential unknown quantity of equation \eqref{Einsteq}. It encodes the mathematical tool allowing to compute the distance between objects in spacetime.

The cosmological constant has witnessed a strange history. It was first introduced by Einstein when he realized that his theory predicted a non static universe; either expanding, either contracting. The cosmological constant term was required in order to have a quasi-static universe, compatible with the apparent staticity of the stars \cite{standingon}. However, in the late '20s it was pointed out by Hubble that stars were actually not static, instead, they were getting further and further away from each other \cite{hubble}. When Einstein heard about Hubble's discovery, he qualified the introduction of the cosmological constant as \emph{'the biggest blunder of [his] life'}.

It should be mentioned that the cosmological constant is a strange object by nature. The cosmological term can be viewed as part of the stress tensor, it leads to a constant energy density all over the universe and a constant pressure of opposite sign. In particular, the energy density $\rho$ and the pressure $P$ associated to the cosmological term are $\rho=\Lambda$ and $P=-\Lambda$, leading to the equation of state $\rho = -P$.

As already stated, current astrophysical observations point to a positive cosmological constant. It is then interesting to see how the inclusion of a cosmological constant deforms solutions to Einstein equations obtained in the absence of a cosmological constant. On the other hand, String theories predict a correspondence between gravity solutions with a negative cosmological constant ($AdS$, see appendix \ref{app:ads}) and (conformal) field theory (CFT), namely the $AdS/CFT$ correspondence \cite{adscft}. There are then motivations to study the effect of both signs of the cosmological constant.\\

\leftskip=0pt

The {\bf third chapter} presents a simplified model for the coupling of fermions to topological defects (see \cite{cosmicstrings}). This can be seen as a toy-model for the localization of fermionic fields on branes. This chapter is based on original results published in \cite{bdbellshape}.

{\bf Chapter four} is a (non-exhaustive) overview of various black holes available in higher dimensional general relativity. In higher dimensions, it turns out that the allowed horizon topology is much reacher than in four dimensions. In particular, there exists black holes with horizon topology $S_{d-2}$ as well as black strings with topology $S_{d-3}\times S_1$ (here $S_p$ refers to a $p$-dimensional sphere) among many others.
It should be mentioned that apart from their geometrical meaning, black holes can further be characterized by physical quantities which governed by laws strikingly resembling the laws of thermodynamics. A short overview of this is presented in the next section entitled black holes. As a consequence, thermodynamical properties of black objects can be emphasized and provide numerous informations about the physics of theses objects.

In this thesis, we focus on the two simpler cases: black holes and black strings. First we present known solutions generalizing the electrically charged or rotating black hole solution in four dimensions to higher dimensions. It is in particular pointed out that ($(A)dS$) electrically charged \emph{and} rotating black hole solutions are not available in higher dimensions, at least in an analytical form. We then present the numerical construction of such solutions, following the original results found in \cite{bdbh1,bdbh2}. Afterwards, we study black string solutions. Black strings are solutions to the higher dimensional Einstein equations having the particularity that the horizon topology is not spherical but cylindrical. In the case of black strings, one of the spatial dimension of the spacetimes, say $z$, plays a particular role: it is assumed to be compact (or periodic); the axis of the axial symmetry coincides with the particular direction. As far as the other dimensions are concerned, hyperspherical symmetry is assumed.

We review the basic asymptotically locally flat black string solution and its properties, in particular the fact that these objects are unstable \cite{gl}. Then we study the influence of a positive cosmological constant on the equations of black strings. In particular, we provide numerical evidences for the non-existence of an asymptotically de Sitter black string (following the original results in \cite{bdbh1}), but instead of the existence of an asymptotically singular geometry.

Next, {\bf chapter five} is devoted to the study of asymptotically locally $AdS$ black strings. In opposition to the case of a positive cosmological constant, an asymptotically locally $AdS$ black string solution is available numerically and has been constructed by \cite{rms}, extending the work of \cite{cophor} (for $d=5$). We first review the solution of \cite{rms} and its properties. In particular, we reconsider some of the thermodynamical aspects of the $AdS$ black string and present a new phase of the latter, characterized by a negative tension \cite{nubsd}. Black strings can indeed be characterized by a tension, in addition to the mass. The black string solutions mentioned here are so-called uniform, in the sense that they don't depend on $z$.

As mentioned above, asymptotically locally flat black strings are unstable. Since the $AdS$ black strings investigated in this chapter depend naturally on an supplementary parameter, namely the cosmological constant $\Lambda$, it is natural to investigate the stability properties of the latter as a function of $\Lambda$. The stability of these solutions can be studied both through the thermodynamical aspect as well as through the dynamical aspect, in the spirit of \cite{gl}.
For the families of solutions considered, we were able to show that they have a dynamically stable phase and a dynamically unstable phase, agreeing with the thermodynamically stable and unstable phases, along with a conjecture due to Gubser and Mitra \cite{gmbh}. 

Then, we perform in detail the construction of the perturbative $AdS$ non-uniform black strings and predict a new phase of thermodynamically stable non uniform black strings in $AdS$. By non uniform, we refer here to solution non trivially depending on the coordinate $z$.

All the results accumulated in the perturbative approach are useful but nevertheless are approximations of the real non uniform solution, if exists. In the last section of chapter five, we finally consider the construction of the full (non-perturbative) solution. We provide strong numerical evidences that the non uniform $AdS$ black string solution indeed exist and we foreseen that this family of solutions ends-up in an $AdS$ localized black hole, whose construction was out of the scope of this thesis. In the asymptotically locally flat case, the order of the phase transition between uniform and non uniform black strings depends on the number of spacetime dimensions. We argue that in the presence of a cosmological constant, the counterpart of this critical dimension should depend on the cosmological constant. This chapter presents original results published in \cite{pnubsads,pnubsd,nubsd}.

\subsection*{Black holes}
\leftskip=20pt
Black holes are probably among the most puzzling objects in general relativity. They are solutions to Einstein equations and they are governed by laws strikingly resembling the laws of thermodynamics. We will try to sketch the portrait of these amazing objects here.

First, what is a black hole? Formally speaking, it is a solution to Einstein's field equations admitting a trapped spacelike region, i.e. a region of spacetime such that the future light cone of this region does not extend to infinity \cite{wald}. It formally  describes the vacuum spacetime with a point-like massive source. In that sense, it is the general relativistic counterpart of the electric field of a point charged source in Maxwell theory. The solution is however valid outside a static spherical body of given mass $M$; for example, the Schwarzschild solution accurately describes the spacetime outside a (nearly) spherical object, such as the sun.

More intuitively, a black hole is a massive object, so dense that the gravitational attraction prevents any objects which are too close to escape to infinity. They can be imagined considering Newton's law of gravity; the velocity required to escape the gravitational field of a body of mass $M$ at a distance $R$ from this object is given by
\be
v_{escape} = \sqrt{\frac{2GM}{R^2}},
\label{vescape}
\ee
$G$ being the Newton constant. It can be easily seen that if the distance between the observer and the body is smaller than $2GM/c^2$, $c$ being the speed of light, the escape velocity is greater than $c$; i.e. the light itself cannot escape the gravitational field of the massive body.

In the theory of general relativity, the simplest black hole of mass $M$ is described by the Schwarzschild metric:
\be
ds^2 = -\left( 1 - \frac{2GM}{c^2}\frac{1}{r} \right) c^2dt^2+ \frac{dr^2}{\left( 1 - \frac{2GM}{c^2}\frac{1}{r} \right)} + r^2 d\Omega_2^2,
\label{schsol}
\ee
where $d\Omega_2^2$ is the square line element on a unit $2$-sphere. 

There is clearly something special around the radial coordinate $r=R_S = \frac{2GM}{c^2}$: the sign of the time and radial metric coefficients  change. Note that the surface of equation $r=R_S$ is called the event horizon of the black hole. Roughly speaking, the radial coordinate becomes a time coordinate and conversely, for radial coordinates smaller than the Schwarzschild radius $R_S$. It follows that the future of an observer crossing the Schwarzschild radius is located at $r=0$, since just before the crossing, the radial velocity is negative, $dr/dt<0$, but just after the crossing, time and radius interchange and the new time, i.e. the radial coordinate $r$,  flows towards $r=0$.
Note that a massive object is not infinitely dense. It follows that there is a critical density in order to form a black hole. From \eqref{vescape} and \eqref{schsol}, it is obvious that if the radius of the massive object is larger than $R_S$, it is not a black hole, since \eqref{schsol} is a vacuum solution - valid only outside the massive body.

Another question is: how do black holes form? There are essentially three types of black holes: stellar black holes, primordial black holes and micro black holes.

Stellar black holes result from the collapse of massive stars. A star is formed by the accretion of dust and hydrogen under their own gravitational field. As the hydrogen collapses, the heat increases, until thermonuclear processes start consuming the hydrogen into helium. During this reaction, the radiation pressure of the thermonuclear processes counterbalances the self-gravitation of the star, preventing it from further collapse. But at some stage, all the hydrogen is burnt; the star then encounters a second stage of collapse, until thermonuclear processes start again, this time using helium as a fuel for the reaction.

These stages of radiation-collapse can continue until a stable element is reached, namely Iron. Then, there are essentially three possibilities; if the star is not massive enough, nothing special occurs any more and we are left with a white dwarf. The second possibility occurs if the star is more massive: the star collapses until the degeneracy pressure of the iron nucleus, due to the exclusion principle applied on the constituent of the star, stops the process. This is a neutron star. Finally, if the star was very massive, the collapse continues until the density reaches the critical density required to form a black hole. Stars leading to the formation of black hole are expected to have masses of the order $1.5-3$ solar masses and greater, at the end of their life. This is known as the Tolman-Oppenheimer-Volkoff limit \cite{tolman,oppenvolk}.

The second type of black holes, primordial black holes, are believed to be formed in very early stage of the universe. Shortly after the Big-Bang, the density in the spacetime was extremely high. A small perturbation in the density might have resulted in the formation of a black hole, after, the medium collapsed under its own weight. The important point here is the presence of density variation; there are many models, many of them predict primordial black holes.

Finally, micro-black hole are believed to be formed during high energy collisions. Consider two particles colliding, if the collision is such that these two particles become closer to each other than the value of the Schwarzschild radius for the total mass of the particles, a micro-black hole is formed. Note that this would require ultra high energy. These processes are expected to occur for energies in the center of mass larger than the Planck mass where quantum effect are expected to break completely General Relativity down. The Planck mass is about $10^{19}\ GeV$, micro black hole formation is then unlikely to appear, except in the case of extradimensional models where the \emph{fundamental} Planck mass can be of the order of $1\ TeV$ (see chapter \ref{ch:branes}).

Note that black holes are not just theorist's fantasy: there are strong observational evidences for a black hole to reside in the center of our galaxy \cite{bhxp}.

Another amazing aspect of black holes is their link to thermodynamics. The laws governing black holes can be formulated according to three laws \cite{hawkrad,hbentropy1,hbentropy2} (we present them in units where $G=c=\hbar=1$):
\begin{itemize}
\item[0\th: ] The horizon has constant surface gravity for a stationary black hole, the surface gravity being the gravitational attraction applied by the black hole on an hypothetical observer located at the horizon.
\item[1$^{st}$: ] $dM = \frac{\kappa}{8\pi}dA+\Omega dJ+\Phi dQ$, where $M$ is the mass of the black hole, $\kappa$ the surface gravity, $A$ the horizon area, $\Omega$ the angular velocity, $J$ the angular momentum, $\Phi$ the electrostatic potential and $Q$ the total electric charge of the black hole.
\item[2$^{nd}$: ] $dA\geq0$.
\end{itemize}

The zeroth law reminds the zeroth law of thermodynamics stating that temperature is constant inside a body in thermal equilibrium, suggesting a correspondence between surface gravity and temperature.

The first law is similar to the first law of thermodynamics, expressing the conservation of energy. In General Relativity, mass and energy are the same thing. The first law then states that the energy variation of a black hole comes from the variation of rotational energy, electrostatic energy and variation of the horizon area. Usual thermodynamics include the variation of entropy, $TdS$ in the first law. It is then tempting to relate the horizon area and the entropy of a black hole. 
This is comforted by the second law, analogue to the second law of thermodynamics stating that the entropy can only increase during physical processes.

These ideas have actually been derived more formally by Hawking, Bekenstein and Carter, leading to $T_H = \kappa/(2\pi)$, $S = A/4$, $T_H$ being the black hole temperature, $S$ being its entropy.\\

\leftskip=0pt

The {\bf sixth chapter} is the final chapter of this thesis where we review the main results presented in the text and give some possible outlook and perspectives of our work.

Note that natural units will be used throughout this thesis: $\hbar=c=1$.

%% file: brane.tex
%%%%%%%%%%%%%%%%%%%%%%%
%% Brane models
%%%%%%%%%%%%%%%%%%%%%%%

In this chapter, we will present some brane models appearing in theoretical physics. Braneworld find their existence in extradimensional theories (say with $d$ dimensions); what is called the brane is then a four dimensional subspace of the $d$ dimensional spacetime, describing the four dimensional spacetime where we live. The idea of extradimension is not new and was first introduced by Kaluza and Klein in the beginning of the last century \cite{kaluza,klein}. However, the idea of Kaluza and Klein was motivated by unifying electromagnetism and gravity as we will see later, but has been forgotten for decades, until String theories started developing.

String theories are consistent in more than four spacetime dimensions, depending on the model and furthermore include general relativity as a low energy description. This is part of the motivation for considering general relativity in more than four dimensions. However, in order to explain the fact that we indeed observe four dimensions, one has to find a mechanism that leads ordinary particles to evolve in a four dimensional spacetime, just like an ink drop would live in the two dimensions of a sheet of paper, although the sheet itself lives in three spatial dimensions.

Among other, one of the first viable brane scenario was introduced by Antoniadis, Arkani-Hamed, Dimopoulos and Dvali in 1998 \cite{add1,add2} (\emph{AADD}), where the authors consider a $(n+4)$-dimensional spacetime with Ricci flat extradimensions. They were able to give an explanation to the hierarchy problem, i.e. the huge difference between the order of magnitude of the Planck mass and the electroweak unification mass scale. This is indeed one of the features of brane models. Their model excluded $n=1$ and put strong constraints on the size of the extradimensions.

Later, in 1999, another mechanism has been proposed by L. Randall and R. Sundrum \cite{rs1,rs2}, where the authors consider a five dimensional spacetime and a thin brane, described by a Dirac delta. The Randall-Sundrum model includes two branes: one is the brane on which we are living, the other is a 'mirror', phantom brane. The length of the extradimension can be infinitely large thanks to some warp factor.

This chapter is organized as follows: first, we will review the Kaluza-Klein model and the so-called Kaluza-Klein reduction mechanism in the first section. Then, we briefly review the main features of the AADD brane model before turning to the Randall-Sundrum models. Finally, we will briefly present $p$-branes, $D$-branes and black $p$-branes in (low energy) String theory for completeness, but also for the fact that they somehow link branes and black objects presented in chapters \ref{chbhbs} and \ref{chadsbs} of this thesis.

\section{Kaluza Klein model}
\label{secKK}
The original motivation of Kaluza \cite{kaluza} and Klein \cite{klein} was to give a unified formulation of gravity and electromagnetism. The construction is the following (see \cite{kkgrav} for a review): consider a five dimensional metric 
\be
ds^2 = g_{ab}dx^a dx^b,\ a,b=0,1,2,3,4.
\ee
This five dimensional metric can be written in a factorized form, without loss of generality:
\bea
&&ds^2 = e^{\phi/\sqrt{3}}\left(g_{\mu\nu}^{(4)} + \kappa^2 e^{-\sqrt{3}\phi} A_\mu A_\nu  dx^\mu dx^\nu + 2\kappa e^{-\sqrt{3}\phi} A_\mu dx^\mu dy + e^{-\sqrt{3}\phi} dy^2\right),\nonumber\\
&&\mu,\nu=0,1,2,3;
\eea
where $y$ denotes the fifth dimension, $y\in \mathcal I\subset\mathbb R$. The metric component $g^{(4)}_{\mu\nu}$ is a tensor, $A_\mu$ is a vector and $\phi$ is a scalar from the four dimensional point of view, $\kappa$ is a real constant.

Assuming that the fields $g^{(4)}_{\mu\nu}, A_\mu,$ do not depend on the extra coordinate $y$, the five dimensional Einstein-Hilbert action reduces to \cite{kkgrav2}
\bea
S &=&\frac{1}{16\pi G}\int_{\mathcal M\times \mathcal I}\sqrt{-g^{(5)}} R^{(5)}d^5x\\
  &=&\frac{1}{16\pi G_{eff}}\int_{\mathcal M} \sqrt{-g^{(4)}}\left( R^{(4)} - \frac{\kappa^2e^{-\sqrt{3}\phi}}{4} F_{\mu\nu} F^{\mu\nu} - \frac{1}{2}(\partial_\mu\phi) (\partial^\mu\phi) \right) d^4x,\nonumber
\eea
where $\mathcal M$ is the four dimensional submanifold of the $5$-dimensional spacetime, $F_{\mu\nu} = \partial_\mu A_\nu-\partial_\nu A_\mu$ is the field strength of $A_\mu$ and $G_{eff} = G/\int_{\mathcal I} dy$ is the effective Newton constant, $G$ being the $5$-dimensional Newton constant. $R^{(4)}$ is the Ricci scalar computed with $g^{(4)}_{\mu\nu}$.

Setting $\phi = \mbox{Const.}$, the action reduces to the Einstein-Maxwell model, providing a unified model of gravity and electromagnetism.
Unfortunately, this attempt of unification has been forgotten because of the success of the quantum electrodynamics theory. The idea regained interest only in the '80s when String Theories brought the question of extradimensions up to date. Anyway, it is remarkable that compactifying one dimension, reducing the model from five dimensions to four leads to the occurrence of a gauge invariant theory, the gauge group being $U(1)$. 
It should be stressed however that $\phi = \mbox{Const.}$ is not a solution of the $5$ dimensional model because of the non trivial coupling with the Maxwell field. It is anyway remarkable that $5$ dimensional gravity somehow contains Einstein-Maxwell (-dilaton) theory 

This procedure is known the Kaluza-Klein compactification; note that compactification from more than five dimensions to four dimensions induce other gauge groups, usually non abelian, depending on which manifold the extradimensions are compactified.

\subsection{Kaluza-Klein reduction}
Another mechanism enters the game once one deals with compact extradimensions: the Kaluza Klein reduction. Consider a scalar field $F(x^\mu,y)$ in five dimensions, where the extradimension is compact and Ricci-flat:
\be
ds^2 = g_{\mu\nu}^{(4)}dx^\mu dx^\nu + dy^2,\ y\in[0,L],
\ee
for some real $L$.

The five dimensional massless Klein-Gordon equation leads to
\be
\Box_5 F(x^\mu,y) = \Box_4 F(x^\mu,y) + \partial^2_y F(x^\mu,y)=0.
\label{KG5}
\ee
where $\Box_5$ is the five dimensional d'Alembertian while $\Box_4$ is the four dimensional d'Alembertian.
In order to derive a four dimensional equation, we can perform a Fourier series decomposition in the $y$ direction according to $F(x^\mu,y)=\sum_n f_n(x^\mu)e^{2\pi n i y/L}$, $n\in\mathbb Z$. Equation \eqref{KG5} then reads
\be
(\Box_4 - M_n^2)f_n(x^\mu)=0,
\ee
which is the equation of a four dimensional scalar field with mass $M_n = 2\pi |n|/L$. In other words, a massless scalar field in five dimensions appears as a tower of massive scalar fields in four dimensions.

This procedure can also be applied to vector fields or fermionic fields but is more involved, so we won't present it here (see for example \cite{nugra4}).

The Kaluza-Klein reduction furthermore generalizes to higher number of dimensions, the result being again that massless fields in higher dimensions appear as a tower of massive fields from the lower dimensional point of view. Note that this is also the case for massive fields in higher dimensions, except that the mass tower is shifted by the higher dimensional mass of the fields.

\section{AADD braneworld}

The Einstein-Hilbert action in $n+4$ dimensions has the form
\be
S = M^{n+2}_p\int \sqrt{-g}R_{n+4}d^4x d^ny,
\label{aadsS}
\ee
where $R_{n+4}$ is the $n+4$ dimensional Ricci scalar, $M_p$ the $d$-dimensional Planck mass and $g$ is the determinant of the metric.

Now, consider a $n+4$ dimensional spacetime of the form
\be
ds^2 = \eta_{\mu\nu}dx^\mu dx^\nu + \delta_{ij} dy^i dy^j,\ \eta = \mbox{ diag}(-1,1,1,1),
\label{aaddans}
\ee
where $\mu,\nu=0,1,2,3$, $i,j=1,\ldots,n$, $\delta_{ij}$ is the Kronecker symbol and where we assume the extra coordinates $y^i$ to be in $[0,L_i]$. The metric \eqref{aaddans} is a vacuum solution to the $n+4$ dimensional Einstein equations.

The action $S$ can be reduced to a four dimensional effective action. The intuitive procedure is to integrate \eqref{aadsS} over the $n$ extra coordinates, assuming that the Ricci scalar does not depend on these extra coordinates and that the geometry of the extradimensions is given by \eqref{aaddans}, leading to
\be
S_{eff} = M^{n+2}_p V_n\int \sqrt{-g_4}R_4d^4x,\ V_n = \prod_{i=1}^nL_i,
\label{aaddred}
\ee
$R_4$ being the four dimensional Ricci scalar, $g_4$ the determinant of the metric on the four dimensional slices and $V_n$ the volume of the extradimensions. This procedure implicitly assume that the four dimensional part of the metric \eqref{aaddans} can fluctuate, but not the extradimensional part.

Identifying the coefficient of the integral \eqref{aaddred} as the square of the four dimensional Planck mass $\mathcal M_p$, we have the following relation between the $n+4$ dimensional and four dimensional Planck mass:
\be
 \mathcal M_p^2 =  M^{n+2}_p V_n.
\label{pleff}
\ee

In other words, the four dimensional Planck mass appears as the product of a power of the $n+4$ dimensional Planck mass and the volume of the extradimensions.

As a consequence, it is possible to obtain a very large $4$ dimensional Planck mass with a fundamental Planck mass of the order of the $TeV$ (which is the electroweak unification energy scale), provided that the volume of the extradimensions is large enough.

A straightforward consequence of this class of models is the deviation from Newton's gravitational law. Assuming the length $L_i$ to be all of the same order of magnitude $R$, Gauss' law in $4+n$ dimensions implies that the gravitational potential felt by two test masses $m_1,m_2$ and separated by a distance $r$ is
\be
V(r) \propto \left\{\begin{array}{r}
                      \frac{m_1 m_2}{M^{n+2}_{pl}} \frac{1}{r^{n+1}},\mbox{ for } r\ll R,\\
                      \frac{m_1 m_2}{M^{n+2}_{pl}R^{n}} \frac{1}{r},\mbox{ for } r\gg R.
                    \end{array}\right.
\ee
Note that this is consistent with \eqref{pleff}: the effective Planck mass is the fundamental one times the volume of the extradimensions.

As a consequence, one can assume the fundamental $4+n$ dimensional Planck mass to be of the order of the $TeV$, along with the electroweak unification scale $m_{EW}$. The large discrepancy between the Planck scale and the electroweak scale is then a geometrical effect, due to the volume of the extradimensions. Assuming so puts constraints on the radius $R$ \cite{add1}:
\be
R\propto 10^{\frac{30}{n} - 17} cm \left( \frac{1 TeV}{m_{EW}} \right)^{1+\frac{2}{n}}.
\ee
Note that $n=1$ is directly ruled out, implying deviation from the Newton's law at astrophysical scale (for $n=1$, $R\approx 10^{11} m$ ). However, from $n=2$, the deviations would appear at the millimetric scale and beyond, compatible with experimental tests of Newton's gravity law \cite{gravxp,torsion}.

This class of models provides an elegant solution to the hierarchy problem. However, the geometry of the extradimensions is completely separated from the geometry of the $4$ dimensional branes. Moreover, these models rely on compact extradimensions; there is no natural reason that only three spatial dimensions are infinite and all others are compact. 
These geometries, where the four dimensional spacetime is completely separated by the extradimensional part of the spacetime are called factorisable geometry. Releasing the assumption of factorisability of the spacetime leads to other classes of models where the size of the extradimensions can be much less constrained. It is indeed the case of the Randall-Sundrum model \cite{rs1}.

\section{Randall-Sundrum braneworld}
Although the AADD scenario sketched above might solve the hierarchy problem, another mechanism has been introduced in 1999 by Randall and Sundrum in order to solve the hierarchy problem \cite{rs1,rs2}, without requiring compactification. Their approach is based on the use of warped spacetimes; the advantage of such spacetimes is that the warp factor can allow a large volume of the extradimensions without constraining too much the range of the extradimensional coordinates.

We will briefly review the solution and proposal. The model considered in \cite{rs1,rs2} is described by the following action:
\bea
S       &=& S_{gr} + S_{vis} + S_{hid},\nonumber\\
S_{gr}  &=& \int d^4x \int d\phi \sqrt{-g}\left( 2M^3 R - \Lambda \right),\nonumber\\
S_{vis} &=& \int d^4x \sqrt{-g_{vis}}\left( L_{vis} - V_{vis} \right), \nonumber\\
S_{hid} &=& \int d^4x \sqrt{-g_{hid}}\left( L_{hid} - V_{hid} \right).
\eea
where $g$ is the determinant of a five dimensional metric while $g_{vis},g_{hid}$ are the determinant of the induced metric on a visible (resp. hidden) brane, i.e. four dimensional subspaces where matter fields live, embedded in the five dimensional space (the bulk). $M$ is the five dimensional Planck mass, $L_{vis}$ (resp. $L_{hid}$) is the lagrangian describing the matter fields on the visible (resp. hidden) brane, $V_{vis}$ (resp. $V_{hid}$) is the vacuum energy in the visible (resp. hidden) brane.

The metric ansatz is given by
\be
ds^2 = e^{\sigma(\phi)}\eta_{\mu\nu} dx^\mu dx^\nu + r_c^2 d\phi^2,\ \phi\in[-\pi,\pi],
\ee
supplemented by a $\mathbb Z_2$ symmetry. This is typically a warped geometry and $r_c$ denotes some compactification radius, which can be infinite. The visible brane is placed in $\phi=0$ while the hidden one resides at $\phi=\pi$.

In such a setup, the solution to the Einstein equations is given by
\be
\sigma(\phi) =- r_c |\phi|\sqrt{\frac{-\Lambda}{24M^3}}.
\ee
Note that although this function is continuous, this is not the case for its derivative. This is due to the fact that the branes don't have an extension. We will see in chapter \ref{chextbranes} that considering branes with an extension regularizes the metric functions. Furthermore, the bulk cosmological constant $\Lambda$ has to be negative.

The result of Randall and Sundrum is such that the parameters of the model, namely the vacuum energy of the visible and hidden brane $V_{vis}, V_{hid}$ and the $5$-dimensional cosmological constant have to obey the following relation 
\be
V_{vis} = -V_{hid} = 24M^3 k,\ \Lambda = -24M^3k^2,
\ee
for a given value of $k\in \mathbb R$.

The effective four dimensional Planck scale in this model is given by
\be
M_p^2 = \frac{M^3}{k}\left( 1 - e^{-2kr_c\pi} \right),
\label{rspm}
\ee
allowing $r_c$ to be large, the control on the four dimensional Planck mass being essentially provided by the value of $k$.

Let us review how Randall and Sundrum found this result: consider a massless fluctuation of the metric:
\be
 ds^2 = e^{-2kT(x)|\phi|}\bar g_{\mu\nu}(x) dx^\mu dx^\nu + T(x^\mu) d\phi^2,\ \bar g_{\mu\nu}= \eta_{\mu\nu} + \bar h_{\mu\nu}(x),
\ee
where $\bar h_{\mu\nu}$ is a massless tensor fluctuation (there is no dependence in the extradimension, see section \ref{secKK}) while $T(x)$ is a massless scalar fluctuation. The authors argued that there shouldn't be off-diagonal vector fluctuation in the low energy models, since these vector modes would be massive \cite{rs1,rs2}. The effective four dimensional action resulting from this model contains a term of the form
\be
\int_{-\pi}^{\pi}  \int 2M^3 r_c e^{-2kr_c|\phi|}\sqrt{-\bar g}\bar R d^4x d\phi,
\label{preeff}
\ee
where $\bar R$ is the Ricci scalar constructed with $\bar g_{\mu\nu}$ and doesn't depend on $\phi$.

This leads to the effective four dimensional Planck mass presented in equation \eqref{rspm}, since the only term depending on the extra-coordinate in \eqref{preeff} is $e^{-2kr_c|\phi|}$, which can be explicitly integrated. This provides a possible solution to the hierarchy problem without putting constraint on the length of the extradimension: the effective Planck mass is still well defined in the limit where the hidden brane is pushed to infinity.
Note that the construction of the effective action is formally equivalent to integrating out the extradimensional dependence of the $5$-dimensional action.

This model is quite elegant and opened a new research area since it was the first model to consider warped branes. However, it contains some criticizable points: first, as already mentioned, the derivative of the metric function is not continuous, second, there is a fine tuning between the parameters of the model ($V_{vis}, V_{hid}$ and $\Lambda$). Third, it requires the existence of a mirror hidden brane. Note that the third point has been however discussed in \cite{rs2}.

\section{Branes in String theory}

For completeness, we introduce the notion of $p$-branes solutions and $D$-branes in this section. These two objects are different in their nature, but are conjectured to be actually equivalent, leading to the basis of the so-called $AdS/CFT$ duality, which relates gravity in $AdS$ and conformal field theory defined in the background of the conformal boundary of the $AdS$ spacetime. We will not enter the details of this duality, neither detail the construction of the effective models where $p$-branes live. We will actually not even give much details about how the various object entering the discussion appear theoretically; instead we will present the model and $p$-brane solution and explain briefly what are the $D$-branes. We refer the reader to \cite{polch,polch2,ictp1} between many others for more details on the construction of the theory.

In superstring theory, the fields describing the strings are bosonic and fermionic. The bosonic strings as well as fermionic strings can be closed or opened; accordingly suitable boundary conditions are imposed; in type II string theories (which is relevant for our purpose), the strings are indeed closed.

Nevertheless, fermionic strings admit two different kinds of boundary conditions: periodic (Ramond: R) or anti-periodic (Neveu-Schwarz: NS) (see for example \cite{thooftstring}).

Depending on the type of boundary conditions imposed the spectrum of the string is slightly different. Type $II$ string theories consider closed string; for example, the $NS-NS$ sector provides a fundamental 2-form $B$ while the $R-R$ sector provides $(p+1)$-forms $C^{p+1}$, with $p$ odd or even, depending if the string theory considered is of type IIA or type IIB (this is related to the chirality of the fermionic sector of the strings).

The idea here is to consider the bosonic part of low energy type II superstring theory.

So finally, the bosonic part of low energy type II string theories contains general relativity, dilaton and p-forms action:
\bea
S_{II} &=& \int d^{10}x \sqrt{-g_s} e^{-2\phi}\left\{\left[ R(g_s) + 4\partial_a\phi \partial^a \phi -\frac{1}{12} (dB)^2 \right] \right.\nonumber\\
&& \left.- \frac{1}{2}\sum_{p=0}^8 \frac{1}{(p+2)!} (dC^{p+1})^2 \right\},\ a=0,1,\ldots,9;
\label{typeIIsf}
\eea
where $g_s$ is the determinant of the metric $g_{s,ab}$, $\phi$ is the dilaton, $B$ is the $NS-NS$ $2$-form, $C^{p+1}$ are the $R-R$ $(p+1)$-forms; $d$ is the exterior derivative and the squares are to be understood as contractions on the indices of the form components. This action is the bosonic part of the $10$-dimensional supergravity \cite{nastase}, which is a good approximation of type II string theories at moderate energies. 

Note that $g_{s,ab}$ is the metric in the so-called 'string frame'; it is possible to re-express this action in the 'Einstein frame', defining $g_{e,ab} = e^{-\phi/2}g_{s,ab}$, leading to
\bea
S_{II} &=& \int d^{10}x \sqrt{-g_e} \left\{\left[ R(g_e) - 4\partial_a\phi \partial^a \phi -\frac{e^{-\phi}}{12} (dB)^2 \right] \right.\nonumber\\
&& \left.-\frac{1}{2}\sum_{p=0}^8\frac{1}{(p+2)!} e^{\frac{(3-p)}{2}\phi}(dC^{p+1})^2 \right\}.
\eea
In the Einstein frame, \eqref{typeIIsf} reduces to the Einstein-Hilbert with the Klein-Gordon action for the dilaton, a 'Maxwell-like' actions for the forms and coupling between the dilaton and the various forms.

\subsection{$p$-branes}

In order to construct the $p$-brane solution, we will consider a truncated action, taking into account only one $(p+1)$-form and setting $B$ to zero:
\bea
S_{II,p} &=&\int d^{10}x \sqrt{-g_e} \left\{\left[ R(g_e) - 4\partial_a\phi \partial^a \phi\right] \right.\nonumber\\
&& \left.-\frac{1}{2}\frac{1}{(p+2)!} e^{a_p\phi}(dC^{p+1})^2 \right\},
\label{sIIp}
\eea
where $a_p=(3-p)/2$. The equations of motions resulting from the action \eqref{sIIp} are given by
\bea
R_{ab} &=& \frac{1}{2}\partial_a\phi\partial_b\phi + S_{ab},\nonumber\\
S_{ab} &=& \frac{1}{2(p+1)!}e^{a_p\phi}\left( F_{a m_2\ldots m_{p+2}}F_b^{\ m_2\ldots m_{p+2}} - \frac{p+1}{8(p+2)}g_{e,ab} F^2 \right),\nonumber\\
0      &=& \nabla_a\left( e^{a_p\phi}F^{am_2\ldots m_{p+2}} \right),\nonumber\\
\Box\phi &=& \frac{a_p}{2(p+2)!}e^{a_p\phi}F^2,
\label{peom}
\eea
where $F = dC^{p+1}$.

There exists solutions to the equations \eqref{peom} having the properties that they extend in $p$ directions. These solutions are known as $p$-branes. For convenience, we will split the $10$ coordinate $x^a$ in two parts: $p+1$ longitudinal coordinates $x^\mu,\ \mu=0,\ldots,p$ and $9-p$ transverse coordinates $y^i,\ i=p+1,\ldots,9$. Then, the solution is given by
\bea
ds^2 &=& H_p^{(p-7)/8}\eta_{\mu\nu}dx^\mu dx^\nu + H_p^{(p+1)/8}\delta_{ij}dy^i dy^j,\nonumber\\
e^{2\phi} &=& H_p^{(3-p)/2},\nonumber\\
C^{(p+1)}_{01\ldots p} &=& H_p^{-1}-1,
\label{extpbrane}
\eea
The function $H_p$ is a function of $r=\sqrt{\vec y^2}$. and is given by 
\be
H_p = 1 + \frac{Q_p}{r^{7-p}},
\ee
where $Q_p$ is the charge associated to the $(p+1)$-form (see \cite{nastase} and references therein).

Note that in the string frame, the line element takes the simpler form
\be
ds^2 = H_p^{-1/2}\eta_{\mu\nu}dx^\mu dx^\nu + H_p^{1/2}\delta_{ij}dy^i idy^j.
\ee

It is interesting to note that low energy description of String theory provides extended solutions. In other words, these theories predict brane solutions.

\subsection{$D$-branes}
On another hand, String Theory predicts also another type of branes, namely $D$-branes. $D$-branes consist actually on hypersurfaces where opened strings end. These open strings are subject to Dirichlet boundary conditions, giving the $D$ to $D$-branes. These objects have many properties, and can be put in relation with the $p$-brane solution presented above, but this is much too far from the object of this thesis. 

\subsection{Black $p$-branes}
Let us finally remark that there exists extended objects in higher dimensional gravity which are also solutions to supergravity described by \eqref{sIIp}, referred to as black $p$-branes \cite{horostrom}. These black $p$-branes can carry some charge associated to the $p$-forms present in \eqref{sIIp}. However, we will exhibit simpler solutions, uncharged with respect to the $p$-forms:
\bea
ds^2 &=& -f(r) dt^2 + \frac{dr^2}{f(r)} + r^2 d\Omega_{8-p}^2 + \delta_{ij} dy^i dy^j,\ i,j=10-p,\ldots 10,\nonumber\\
\phi &=& \mbox{Const.},\ C^{p+1} = 0.
\label{pbgrav}
\eea
where $f(r)=1 - \left(\frac{r_h}{r}\right)^{7-p}$, $r_h\in\mathbb R^+$. This is simply a $(10-p)$-dimensional black hole with $p$ transverse flat direction. The construction of the solution with non trivial $p$-form and dilaton can be found in \cite{horostrom}.

\section{Concluding remarks}

There are many ways to approach branes; for instance, we can consider pure gravity with extradimensions and add some branes to the model, we can consider low energy String theories, full String theory, etc. Actually, once one wants to deal with extradimensions, one has to introduce the concept of branes in order to have a chance to explain why we actually don't see the extradimensions. It seems clear that branes play a very important role in actual theoretical physics. Note that extended objects also play an important role in other fields of physics, such as surface physics in condensed matter \cite{matthin} as an only example.

Another important remark is that in the Randall Sundrum model, the brane doesn't have an extension in the transverse space while $p$-branes are somehow extended in the direction transverse to the brane. Recall that the metric functions were not regular, in the sense that their derivative was not continuous, but the metric functions somehow regularize in the $p$-brane model, the extension of the $p$-brane removing the discontinuous behaviour of the metric functions.

Note also that the $p$-brane solutions \eqref{pbgrav} are actually black objects. In that sense, they are a nice junction between the considerations of chapter \ref{chextbranes} and \ref{chferm} of this thesis, dealing with extended brane models and the rest of this thesis, dealing with higher dimensional black holes and strings.

%% file: extbrane.tex
In this chapter, we will present four brane models where the brane has an extension in the extradimensions. The idea for providing an extension to the branes in the extradimensions is to use localized soliton solutions available in usual field theory and to interpret the dimensions where the soliton lives as the extradimensions of the spacetime. We will consider models containing two cosmological constant: one is a cosmological constant for the entire spacetime (the bulk cosmological constant) while the second is a cosmological constant in the four dimensional subspace of the entire spacetime, i.e. on the brane. The second cosmological constant will be modelled by an inflating four dimensional subspace, i.e. an inflating brane.

The aim of this chapter is to study the influence of these two cosmological constant on the pattern of some solitonic brane models considered before, without cosmological constant and inflation (for instance \cite{olavil, torra, sawa}).

The first model we will consider is the Einstein Abelian Higgs model \cite{bdh} in $4+2$ dimensions, the second model is the Einstein non abelian Higgs model \cite{bd} in an arbitrary number of dimensions, the third model is the Einstein-Yang-Mills non abelian Higgs model in $4+3$ dimensions and the fourth model is the gravitating baby Skyrme model \cite{babyskyrme} in $4+2$ extradimensions \cite{sadelbri}. In all these models, the brane is seen as a topological soliton extending in $n=d-4$ extradimensions, $d$ being the total number of dimensions.

In all cases, the matter fields are localized in the extradimensions, defining the region where the brane is located. The four models we consider all have solutions with non trivial topological properties. This is the reason why we call these brane models topological brane models with an extension.

In addition, as mentioned before, the four dimensional slices are chosen to be inflating, modelling a four dimensional positive cosmological constant. This is motivated by the fact that the actual observations points to a four dimensional positive cosmological constant \cite{posccxp1, posccxp2}. Note that this is also relevant for inflationary models with extradimensions.

The general action for all the models we will consider has the following form
\begin{equation}
S=S_{gravity}+S_{brane},
\label{action_genbrane}
\end{equation}
where $S_{grav}$ is the Einstein-Hilbert action:
\begin{equation}
S_{gravity}=\frac{1}{16\pi G_{d}}\int d^d x \sqrt{-g} \left(R-2\Lambda_d\right),
\label{action_gravity}
\end{equation}
$\Lambda_d$ being the bulk cosmological constant, $G_d$ is the $d$-dimensional Newton constant, related to the $d$-dimensional Planck mass $M_d$ by $G_d=1/M^{d-2}_{d}$ and $g$ the determinant of the $d$-dimensional metric.

The term $S_{brane}$ in \eqref{action_genbrane} is the action of the matter fields where the brane resides and will depend on the model under consideration; we furthermore define $L_{brane}$ as the lagrangian density such that $S_{brane} = \int\sqrt{-g} L_{brane}d^d x$.

\section{The metric ansatz and Einstein equations}
We consider a $d$-dimensional spacetime, with $4$ special dimensions describing our universe. There are then $n=d-4$ extradimensions (or codimension).

The ansatz for the $d$-dimensional metric reads:
\begin{equation}
ds^2= M^2(\rho)\left[-dt^2+e^{2Ht}\delta_{ij}d\tilde x^id\tilde x^j\right]+d\rho^2 + l^2(\rho)d\Omega_{n-1}^2,\ i,j=1,2,3,
\label{ggenans}
\end{equation}
where $\rho$ and $\theta_a,\ a=1,\ldots,n-1$ are the coordinates associated with the extra dimensions, $d\Omega_{d-5}^2 = \sum_{a=1}^{n-1}\prod_{b<a}\sin^2\theta_b d\theta_a^2$ is the square line element on the $d-5$ sphere, $\theta_1\in[0,\pi], \theta_{a\neq1}\in[0,2\pi]$, $i,j=1,2,3$ and $\tilde x^1,\tilde x^2$ and $\tilde x^3$ are the 3-dimensional spatial coordinates. $H\geq 0$ is the Hubble parameter related to the (positive) 4-dimensional cosmological constant.

In the general case, the Einstein equations read
\bea
G_\mu^\mu &=& -\frac{1}{4}\frac{R^{(4)}}{M^2} + 3\frac{M''}{M}+ 3\frac{M'^2}{M^2} + 3(n-1)\frac{M'l'}{Ml}  \nonumber\\
          &+& (n-1)\frac{l''}{l} + \frac{(n-2)(n-1)}{2}\left(\frac{l'^2}{l^2}- \frac{1}{l^2}\right)\nonumber\\
          &=& 8\pi G_dT_\mu^\mu-\Lambda_d ,\ \mu = 0, \dots , 3\nonumber\\
G_\rho^\rho     &=& -\frac{1}{2}\frac{R^{(4)}}{M^2} + 6\frac{M'^2}{M^2}+ 4(n-1)\frac{M'l'}{Ml} + \frac{(n-2)(n-1)}{2}\left(\frac{l'^2}{l^2}- \frac{1}{l^2}\right)\nonumber\\
          &=& 8\pi G_dT_\rho^\rho-\Lambda_d\nonumber\\
G_{\theta_i}^{\theta_i} &=& -\frac{1}{2}\frac{R^{(4)}}{M^2} + 4\frac{M''}{M} + 6\frac{M'^2}{M^2} + 4(n-2)\frac{M'l'}{Ml} + (n-2)\frac{l''}{l} \nonumber \\
          &+& \frac{(n-2)(n-3)}{2}\left(\frac{l'^2}{l^2} - \frac{1}{l^2}\right)\nonumber\\
          &=& 8\pi G_dT_{\theta_i}^{\theta_i}- \Lambda_d,\ i = 1, \dots , n-1
\label{geneqs}
\eea
where $n=d-4$ is the number of extradimensions, $G_{ab}$ is the Einstein tensor and $T_{ab}$ is the stress tensor (see Appendix \ref{app:reminder}) of the matter fields described by $S_{brane}$ and $R^{(4)}$ is the scalar curvature of the four dimensional slices; in our case, it is $12 H^2$.

Note that the brane appears only thought $R^{(4)}$ in the equations; it follows that we can replace the four dimensional subspace by any spacetime with a constant positive curvature.

\section{Vacuum solution}
\label{secbranevaccum}
In order to consider vacuum solutions, we set $T_{ab}=0$ in \eqref{geneqs}. According to the sign of the $4+n$-dimensional cosmological constant, vacuum solutions have constant positive, null or negative scalar curvature. The result is very similar to the 4-dimensional case where the geometry of the spacetime can be opened, closed or flat according to the sign of the cosmological constant. In the present case we emphasize a more general situation where 4-dimensional slices of the space have a de Sitter geometry, which can induce angular deficits in the $n$-dimensional subspace, as we will see later. Because the equations do not explicitly depend on the radial variable $\rho$ the solutions given below can be arbitrarily translated in $\rho$.

In this section we will consider solutions for $n>2$ for reason that will become clearer later. We will come back on special solutions for $n=2$ in section \ref{6dspecialsolutions}.

%%%%%%%%%%%%%%%%%%%%%%%%%%%%%%%%%%
%\subsubsection{$\Lambda_d  > 0$}
%%%%%%%%%%%%%%%%%%%%%%%%%%%%%%%%%%%
In the case where $H\neq0$, the Einstein equations above possess explicit solutions, depending on the sign of the bulk cosmological constant $\Lambda_d$:
\bea
M(\rho)&=&\frac{H}{\omega}\sqrt{\frac{3}{n+2}}\sin\omega\rho,\ l(\rho)=\frac{1}{\omega}\sqrt{\frac{n-2}{n+2}}\sin\omega\rho,\mbox{ for }\Lambda_d>0,\nonumber\\ 
M(\rho)&=&\sqrt{\frac{3H^2}{n+2}}\rho,\ l(\rho)=\sqrt{\frac{n-2}{n+2}}\rho,\mbox{ for }\Lambda_d=0,\nonumber\\ 
M(\rho)&=&\frac{H}{\omega}\sqrt{\frac{3}{n+2}}\sinh\omega\rho,\ l(\rho)=\frac{1}{\omega}\sqrt{\frac{n-2}{n+2}}\sinh\omega\rho,\mbox{ for }\Lambda_d<0,\nonumber\\
\omega^2 &=&  \frac{2|\Lambda_d|}{12+(n-1)(n+6)},
\label{trigo}
\eea

Note that these solution depends crucially on the two cosmological constants and from the form of the $l$ function, it is clear that $n=2$ is a special case. Indeed, in the $6$-dimensional case where $n=2$, the equation for $M$ decouples \cite{bdh} and leads to $l \propto M'$, which is not compatible with the solution above. The case $n=1$ is also special since for 5 dimensions, the function $l(r)$ is not defined.

Note also that the solution \eqref{trigo} are not regular since the Kretschmann invariant $K = R_{abcd}R^{abcd}$
\bea
K&=&\frac{24 H^4}{M^4}-\frac{48 H^2 M'^2}{M^4}+\frac{4 (n-1) l''^2}{l^2}+\frac{16 (n-1) l'^2 M'^2}{l^2 M^2}+\frac{2 (n-2) (n-1) l'^4}{l(r)^4}\nonumber\\
 &-&\frac{4 (n-2)(n-1) l'^2}{l^4}+\frac{2 (n-2) (n-1)}{l^4}+\frac{16 M''^2}{M^2}+\frac{24 M'^4}{M^4},
\eea
evaluated with the above solution gives
\be
K=\frac{2}{3} \omega^4 \left(\frac{4 (n-1) (n+1) (n+2) }{(n-2)\mbox{sin(h)}^4(\omega \rho)} + 3 (n+3) (n+4)\right),
\label{ksol}
\ee
where we define $\mbox{sin(h)}(x)=\sin(x)$ for $\Lambda_d>0$, $-\sinh(x)$ for $\Lambda_d>0$ and $x$ for $\Lambda_d=0$. The invariant \eqref{ksol} is obviously singular at the origin. Moreover, in the case $\Lambda_d>0$, it is also singular for $r= k \pi/\omega$ ($k$ integer).

The Ricci scalar, given by
\bea
R &=& 12\left(\frac{H^2}{M^2} - \frac{M'^2}{M^2}\right) - 8\left( \frac{M''}{M} +(n-1) \frac{M'l'}{Ml}\right)\nonumber\\
  &-& (n-1)(n-2)\left( 2 \frac{l''}{(n-2)l} +\frac{l'^2}{l^2}-\frac{1}{l^2} \right),
\label{ricci}
\eea
is constant for these solutions (it is proportional to $\Lambda_d$, using the equations of motion).
Let us finally emphasize that the warp factor $M$ is directly proportional to the parameter $H$, both are related to the four dimensional inflating subspace...

The solution \eqref{trigo} has a natural geometric interpretation: it describes the surface of an $n+1$ dimensional manifold of constant curvature (sphere, hyperboloid or plane) in the extra dimensions and presents an angular deficit relative to the angles $\theta_i$.

The angular deficit is computed in the following way: the extradimensional part of the line element with the solutions \eqref{trigo} reads
\be
ds_n^2 = d\rho^2 + \frac{n-2}{n+2}\frac{1}{\omega^2}\sin(h)^2(\omega\rho)d\Omega_{n-1}^2.
\ee
Setting  $\Theta = \omega \rho$, we find
\be
ds_n = \frac{1}{\omega^2}\left(d\Theta^2 +\frac{n-2}{n+2}\sin(h)^2\Theta d\Omega_{n-1}^2\right).
\ee
From this expression, it is clear that the factor $\frac{n-2}{n+2}$ induces an angular deficit $1-\frac{n-2}{n+2}$ in the angular direction. Note that the angular deficit vanishes in the limit where $n\rightarrow\infty$. 

In addition, the radius of the constant curvature manifold is found to be the parameter $1/ \omega$ defined in (\ref{trigo}). In the case of $\Lambda_d>0$, where the extradimensions present a closed geometry, it defines naturally a compactification radius for the extradimensions.

%%%%%%%%%%%%%%%%%%%%%%%%%%%%%%%%%%%%%%%%%%%%%%%%
\subsection{Melvin-type universe in $d$ dimensions}
%%%%%%%%%%%%%%%%%%%%%%%%%%%%%%%%%%%%%%%%%%%%%%%%
\label{secmelvin}
In this section, we look for solutions of the form:
\be
M(\rho) = M_0(\rho-\rho_0)^\mu,\ l(\rho) = l_0(\rho-\rho_0)^\alpha,
\label{mell}
\ee
where $M_0$, $l_0$, $\rho_0$, $\mu$, $\alpha$ are constants to be determined.

In the following, we will distinguish the cases where the solutions develop the above behaviour for $\rho \to \infty$  (asymptotic solution) and the case where the solutions develop a singularity in the neighbourhood of $\rho=\rho_0$ for some real $\rho_0$ (see section \ref{localmonopolen=3} and reference \cite{cv1} for the $n=3$ case), respectively.
%%%%%%%%%%%%%%%%%%%%%%%%%%%%%%%%%
\subsubsection{Asymptotic solution}
%%%%%%%%%%%%%%%%%%%%%%%%%%%%%%%%%
Because of the occurrence of non-homogeneous terms (e.g. $1/l^2$ and $H^2/M^2$) in the Einstein equations, power-like solutions of the form above cannot be exact for generic values of $H$, $\Lambda_d$, $n$.

However, we will see that solution of the form \eqref{mell} as asymptotic solutions and  appear as 'critical' solutions when matter fields are supplemented in the form of global and local monopoles (see Section \ref{globalmononpole} and \ref{localmonopolen=3}).

Let us for a moment neglect the non-homogeneous terms in the Einstein equations. Inserting the power law above in the Einstein equations leads to the following conditions for the exponents $\mu$, $\alpha$:
\be
3\mu(\mu-1) + 3\mu^2 + 3(n-1)\mu\alpha + (n-1)\alpha(\alpha-1)+ \frac{(n-1)(n-2)}{2}\alpha^2 = 0
\ee
\be
6\mu^2 + 4(n-1)\mu\alpha + \frac{(n-1)(n-2)}{2}\alpha^2 = 0
\ee
\be
4\mu(\mu-1) + 6\mu^2 + 4(n-2)\mu\alpha + (n-2)\alpha(\alpha-1)+ \frac{(n-2)(n-3)}{2}\alpha^2 = 0  \ .
\ee
The solutions then read:
\be
\mu = \frac{2\pm\sqrt{(n+2)(n-1)}}{2(n+3)},\ \alpha = \frac{(n-1)\mp 2 \sqrt{(n+2)(n-1)}}{(n-1)(n+3)}
\label{melvin}
\ee
Note, however, that with these exponents, it is not justified to neglect the inhomogeneous terms $\propto 1/l^2$ and $\propto 1/M^2$ except in the particular case of the critical solutions (see Section \ref{globalmononpole}), where such terms vanish.

%%%%%%%%%%%%%%%%%%%%%%%%%%%%%%%%
\subsubsection{Singular solutions}
%%%%%%%%%%%%%%%%%%%%%%%%%%%%%%%%
If we want to interpret the functions (\ref{mell}),(\ref{melvin}) as the dominant terms of a solution of the vacuum Einstein equations which is 
singular in the limit $\rho \to \rho_0$, the exponents should be such that $\mu < 1$, $\alpha < 1$ (along with the assumption that the inhomogeneous terms are sub-dominant). It turns out that these conditions are fulfilled for {\it both} values of the sign $\pm$ in (\ref{melvin}).

Note the relation between $\mu$ and $ \alpha$: $4\mu + (n-1)\alpha = 1$. This reminds the Kasner conditions \cite{gravitation}:
\be
4\mu^2 + (n-1)\alpha^2 =  1,\ 4\mu + (n-1)\alpha = 1,
\ee
except that only the \emph{linear} relation is fulfilled. We will refer to this type of solution as Kasner type.

\section{Einstein abelian Higgs Model}
In this section, we consider a six dimensional spacetime with matter fields described by the Abelian-Higgs model.

The action $S_{brane}$ for the Einstein-Abelian-Higgs (EAH) string is given in analogy to the 4-dimensional case \cite{no,gstring} by:
\begin{equation}
S_{brane}=\int d^6 x \sqrt{-g} \left(-\frac{1}{4} F_{MN}F^{MN}-\frac{1}{2}\left(D_M\phi\right) \left(D^M\phi\right)^*-\frac{\lambda}{4}(\phi^*\phi-v^2)^2  \right)
\end{equation}
where $D_M=\nabla_M-ieA_M$ is the gauge covariant derivative, $F_{MN}=\partial_M A_N-\partial_N A_M,\ M,N=0,\ldots,5$ the field strength of the U(1) gauge potential $A_M$, $e$ is the gauge coupling, $v$ the vacuum expectation value of the complex valued Higgs field $\phi$ and $\lambda$ the self-coupling constant of the Higgs field.

%%%%%%%%%%%%%%%%%%%%%%%%%%
\subsection{The ansatz}
%%%%%%%%%%%%%%%%%%%%%%%%%%
The ansatz for the $6$-dimensional metric is given by \eqref{ggenans} with $n=2$. Note that here, $\theta_1\in [0,2\pi]$. We will denote the angular coordinate $\theta$ since there is no possible confusion with other angular variables here.
%and that the 4-dimensional metric satisfies the 4-dimensional Einstein equations:
%\begin{equation}
%G_{\mu\nu}^{(4)}=3H^2 g_{\mu\nu}^{(4)}  \ \ , \ \ \mu,\nu=0,1,2,3 \ .
%\end{equation}

The ansatz for the non vanishing gauge and Higgs field reads \cite{no}:
\begin{equation}
\phi(\rho, \theta)=v f(\rho) e^{i N\theta},\ A_{\theta}(\rho,\theta)=\frac{1}{e}(N-P(\rho)),
\end{equation}
along with the ansatz of the Nielsen-Olesen cosmic string \cite{no}, and where $N$ is the vorticity of the string, which throughout this section will be set to $N=1$.

Note that this ansatz is symmetric under rotations in the two extradimensions.

%%%%%%%%%%%%%%%%%%%%%%%%%%%%%%%%%%
\subsection{Equations of motion}
%%%%%%%%%%%%%%%%%%%%%%%%%%%%%%%%%%
Introducing the following dimensionless coordinate $r$ and the dimensionless function $L$:
\begin{equation}
r=\sqrt{\lambda}v \rho \ , \ \ \ L(r)=\sqrt{\lambda}v l(\rho),
\end{equation}
the set of equations depends only on the following dimensionless coupling constants:
\begin{equation}
\alpha=\frac{e^2}{\lambda} \ ,\ \gamma^2=8\pi G_6 v^2 \ , \ \Lambda=\frac{\Lambda_6}{\lambda v^2} \ , \ \kappa=\frac{3H^2}{\lambda v^2}.
\end{equation}
The gravitational equations then read
\bea
\label{eqbgr}
&&3 \frac{M''}{M} + \frac{L''}{L}+ 3 \frac{L'}{L} \frac{M'}{M} + 3 \frac{M'^2}{M^2}+\Lambda-\frac{\kappa}{M^2}  =-\gamma^2 \left(\frac{f'^2}{2}+\frac{(1-f^2)^2}{4}\right.\nonumber\\
&&\left.+\frac{f^2 P^2}{2L^2}+\frac{P'^2}{2\alpha L^2}   \right)\\
&&6 \frac{M'^2}{M^2} + 4 \frac{L'}{L} \frac{M'}{M} +\Lambda-2\frac{\kappa}{M^2}= -\gamma^2 \left(-\frac{f'^2}{2}+ \frac{(1-f^2)^2}{4}+\frac{f^2 P^2}{2L^2} - \frac{P'^2}{2\alpha L^2} \right),\nonumber\\
&&6 \frac{M'^2}{M^2}+4 \frac{M''}{M} +\Lambda -2\frac{\kappa}{M^2} =-\gamma^2\left(\frac{f'^2}{2}+\frac{(1-f^2)^2}{4}-\frac{f^2P^2}{2L^2}
-\frac{P'^2}{2\alpha L^2} \right),\nonumber
\eea

the equations for the matter fields read
\bea
\label{eqaehm}
\frac{(M^4 L f')'}{M^4L}+(1-f^2)f-\frac{P^2}{L^2}f&=&0,\\
\frac{L}{M^4}\left(\frac{M^4 P'}{L}\right)'-\alpha f^2 P&=&0,\nonumber
\eea
the prime denoting the derivative with respect to $r$. For later use, we note that these equations are solved by $f=1,\ P= 0$. We will refer to this configuration as the vacuum solution.

The equations \eqref{eqbgr} can be combined to obtain the following two differential equations for the two unknown metric functions:
\bea
\label{eqbgr2}
\frac{(M^4 L')'}{M^4 L}+\frac{\Lambda}{2}&=&\frac{\gamma^2}{2}\left(\frac{P'^2}{2\alpha L^2} - \frac{1}{4}(1-f^2)^2      \right),\\
\frac{(LM^3 M')'}{M^4 L}+\frac{\Lambda}{2}-\frac{\kappa}{M^2}&=&-\frac{\gamma^2}{2}\left(\frac{2 P^2 f^2}{L^2} + \frac{1}{4}(1-f^2)^2
+\frac{1}{2} \frac{P'^2}{\alpha L^2}\right).\nonumber
\eea

%The value of the parameter $\tilde c$ can be approximated by its value in absence of matter fields: $\tilde c = 4/\vert 2\kappa - \Lambda \vert$.

%%%%%%%%%%%%%%%%%%%%%%%%%%
\subsection{Six dimensional vacuum and asymptotic solutions}
%%%%%%%%%%%%%%%%%%%%%%%%%%%
\label{6dspecialsolutions}
Explicit solutions to equations \eqref{eqbgr2} can be constructed for $f(r)=1$ and $P(r)=0$. These are by themselves of interest, of course, but are also interesting for the generic solutions since we would expect that the metric fields take the form presented below far away from the core of the string.
A similar analysis has been done for global defects in \cite{olo}.

The solution for $M$, $L$ can be written in terms of quadratures as follows:
\begin{equation}
r - \tilde{r}_0 =  \int dM \sqrt{\frac {M^3}{{\frac{\kappa}{3}M^3- \frac{\Lambda}{10}M^5+C}}},\ L = \tilde{L}_0 \frac{dM}{dx}  = \tilde{L}_0 \sqrt{ \frac{  \frac{\kappa}{3}M^3 - \frac{\Lambda}{10} M^5+ C} {M^3}}
\label{quadrature}
\end{equation}
where $\tilde{r}_0$, $C$ are integration constants and $\tilde{L}_0$ is arbitrary.

Unfortunately, we could not find a general solution to the integral \eqref{quadrature}. We will now discuss some particular cases where it is indeed possible to perform the integral explicitly. 

%%%%%%%%%%%%%%%%%%%%%%%%%%%%%%%
\subsubsection{Static branes ($\kappa=0$)}
%%%%%%%%%%%%%%%%%%%%%%%%%%%%%%%%%

For $\Lambda=0$ the system admits two different types of solutions \cite{bh1,bh2}:
\begin{equation}
M_S = 1 \ \ , \ \ L_S = r - r_0
\label{sol1}
\end{equation}
and
\begin{equation}
M_M = \tilde{C_1}(r-\tilde{r_0})^{2/5},\ L_M = \tilde{C_2}(r-\tilde{r_0})^{-3/5}
\label{sol2}
\end{equation}
where $\tilde{C_1}$, $\tilde{C_2}$ are constants related to $C$ and $\tilde L_0$.

By analogy to the case of section \ref{secmelvin}, we refer to the second type of solution as the 'Melvin' (M) branches. The first type of solution is the flat space solution, if we add localised matter fields in the $2$ extradimensions, it will look like a string, the geometry of the two extradimensions being similar to a plane in polar coordinates. This is the reason why we will refer to the first type of solution as the 'String' (S) branch. The Melvin solution can be found by setting $\kappa=\Lambda=0,\ C\neq0$ in \eqref{quadrature}. The string solution is the Minkowski space and corresponds to $\Lambda=\kappa=C=0$ in the integral.%Note that the only difference to the 4-dimensional case are the powers in (\ref{sol2}).

For $\Lambda > 0$ the explicit solution reads:
\begin{equation}
M = C_1 \sin\left(\lambda (r- \tilde r_0)\right)^{2/5},\ L = C_2 \frac{\cos \left(\lambda (r-\tilde r_0)\right)}{\sin \left(\lambda (r-\tilde r_0)\right)^{3/5}}
\label{vacuumtri}
\end{equation}
where $C_1$, $C_2$ are parameters again related to $C, \tilde L_0$ and $\lambda^2 \equiv  5 \Lambda/8$.

This solution is periodic in the metric functions. In the 4-dimensional analogue, i.e. Nielsen-Olesen strings in de Sitter space, periodic
solutions also appear for trivial matter fields \cite{bbh,linet}.

It is easy to see that the Melvin solutions \eqref{sol2} can be obtained from this solution for special choices of the free parameters and specific limits (for $\Lambda \to 0$) of the trigonometric solution in \eqref{vacuumtri}.

The solutions for $\Lambda < 0$ are given by
\begin{equation}
M = A_{\pm} \exp (\pm \sigma r),\ L = B_{\pm} \exp (\pm \sigma r),\ \sigma^2 = -\frac{\Lambda}{10}
\end{equation}
where $A_{\pm}$, $B_{\pm}$ are constants related to $\tilde L_0$ in the case where $C=0$ and
\begin{equation}
M = \hat C_1 \sinh\left(\lambda (r- \tilde r_0)\right)^{2/5},\ L = \hat C_2 \frac{\cosh \left(\lambda (r-\tilde r_0)\right)}{\sinh \left(\lambda (r-\tilde r_0)\right)^{3/5}}
\end{equation}
where $\hat C_1$, $\hat C_2$ are parameters again related to $C, \tilde L_0$ and $\lambda^2 \equiv  -5 \Lambda/8$.

%%%%%%%%%%%%%%%%%%%%%%%%%%%%%%%%%
\subsubsection{Inflating branes ($\kappa > 0$)}%%%%
%%%%%%%%%%%%%%%%%%%%%%%%%%%%%%%%%%%%%
\label{secinflbranekappaneq0}
The general explicit form of $M$ is involved; it depends on elliptic functions. If $\Lambda = 0$, in the case $\kappa=0$, the integration can be done by an elementary change of variable, as we have just seen. However, in some particular cases, it is possible to integrate \eqref{quadrature} explicitly for $\kappa\neq0$.

In the particular case $\Lambda=0$, $C=0$ we find
\begin{equation}
M = \sqrt{\frac{\kappa}{3}}(r-\hat{r}_0),\ L = L_0 \equiv \tilde{L}_0 \sqrt{\frac{\kappa}{3}}= {\rm constant}.
\label{kappasol}
\end{equation}
This latter solution corresponds to a cigar-type solution (in the extra dimensions; 'Cigar-type' refers to the fact that $g_{\theta\theta}$ is constant) and we find that for $C\neq 0$ the solutions are also of this type, however they cannot be given in an explicit form. 

Consequently, the circumference of a circle in the two extra dimensions becomes independent of the bulk radius $\rho$. An analogue solution in $7$ dimensions with monopoles residing in the three extra dimensions has been found previously in \cite{cv1,cv2}.

In the case $\Lambda > 0$, $\kappa > 0$ and $C=0$, we find
\begin{equation}
M = \sqrt{\frac{10 \kappa}{3 \Lambda}}\sin\left(\sqrt{\frac{\Lambda}{10}}(r-\hat{r}_0)\right),\ L = L_0 \cos \left( \sqrt{ \frac{\Lambda}{10}}(r-\hat{r}_0)\right).
\label{trigonometric}
\end{equation}
Again, the metric functions are periodic. The periodicity of string-like solutions seems to be a generic feature - independent of the number of space-time dimensions or of the type of brane present in the case of a positive cosmological constant.

Note that by analytic continuation we find 
\begin{equation}
M = \sqrt{-\frac{10 \kappa}{3 \Lambda}} \sinh\left(\sqrt{-\frac{\Lambda}{10}}(r-\hat{r}_0)\right),\ L = L_0\cosh \left( \sqrt{ -\frac{\Lambda}{10}}(r-\hat{r}_0)\right)
\end{equation}
for $\Lambda < 0$, $\kappa > 0$ and $C=0$.

%%%%%%%%%%%%%%%%%%%%%%%%%%%%%%%%%
\subsection{Boundary conditions}
%%%%%%%%%%%%%%%%%%%%%%%%%%%%%%%%%
We will now consider the system \eqref{eqaehm}, \eqref{eqbgr2} for general $f,\ P$. Before solving the equations, we need suitable boundary conditions.
We require regularity at the origin $r=0$ which leads to the following boundary conditions:
\begin{equation}
\label{bcx0}
f(0)=0,\ P(0)=1,\ M(0)=1,\ M^{'}|_{r=0}=0,\ L(0)=0,\  L^{'}|_{r=0}=1.
\end{equation}
Along with \cite{shapo1,shapo3,cv1,cv2}, we assume the matter fields to approach the vacuum configuration far from the string core (i.e. for $r \gg  1$):
\begin{equation}
f(r \to \infty)=1,\ P(r \to \infty)=0.
\end{equation}

%%%%%%%%%%%%%%%%%%%%%%%%%%%%%%%%%%%%%%%%%%%
\subsection{Behaviour around the origin}
%%%%%%%%%%%%%%%%%%%%%%%%%%%%%%%%%%%%%%%%%%%
Close to the origin $r=0$, the functions have the following behaviour:
\begin{eqnarray}
f(r\ll 1)&\simeq & f_0r+f_0\left(\frac{1}{12}f_0^2 \gamma +\frac{\gamma}{6\alpha} p_0+ \frac{1}{4}p_0+\frac{1}{48} \gamma + \frac{1}{12} \Lambda\right.\nonumber\\
&-&\left.\frac{1}{8} - \frac{1}{16} \kappa\right)r^3 + \Ord{r^5}, \nonumber\\
P(r\ll 1)&\simeq & 1+p_0 r^2 + \Ord{r^4}, \nonumber\\
M(r\ll 1)&\simeq & 1+ \left(\frac{1}{4} p_0^2 \frac{\gamma}{\alpha}-\frac{1}{32}\gamma -\frac{1}{8}\Lambda+ \frac{3}{16}\kappa  \right) r^2 + \Ord{r^4},\\
L(r\ll 1)&\simeq &  r+ \left(-\frac{1}{6} f_0^2\gamma -\frac{5}{6} p_0^2 \frac{\gamma}{\alpha}+\frac{1}{48} \gamma +\frac{1}{12}\Lambda -\frac{1}{4}\kappa   \right) r^3 + \Ord{r^5}, \nonumber
\end{eqnarray}
where $p_0,\ f_0$ are real constants.

Note that the behaviour found in \cite{shapo1} is recovered for $\kappa=0$.

For large values of $r$, the matter functions reach their asymptotic values $P\to 0$, $f\to 1$. As a consequence, the metric functions asymptotically approach the special solutions described in section \ref{6dspecialsolutions}.

%%%%%%%%%%%%%%%%%%%%%%%%%%%%%%%%%%%%%%%%%%%
\subsection{Four dimensional effective action}
%%%%%%%%%%%%%%%%%%%%%%%%%%%%%%%%%%%%%%%%%%%%%
In this section, we will apply the ideas presented in the first chapter regarding the effective lower dimensional theory. To this end, we will consider the vacuum solution, assuming that the radial extradimension ranges in the interval $[r_1,r_2]$, $r_1<r_2$. Note that we don't exclude the possibility $r_1=0,\ r_2\rightarrow\infty$. In this context, the length of the extradimension in the radial direction is $r_2-r_1$.

In the vacuum configuration, where $f=1,\ P=0$, it follows from the last two equations in \eqref{eqbgr} that $L(x)\propto M'(x)$ unless $M(x)$ is constant. This will allow to write explicitly the effective action in four dimensions by formally integrating the extradimensional dependence of the fields in the Einstein-Hilbert action.

Due to the ansatz we use for the metric, a dimensional reduction of the gravity action (\ref{action_gravity}) can be performed. For this purpose notice that we can write the 6-dimensional Ricci scalar $R$ according to:
\begin{equation}
R=\frac{R^{(4)}}{M^2} - 8\frac{M''}{M} - 12\frac{M'^2}{M^2} - 8\frac{L' M'}{LM}
-2 \frac{L''}{L}
\end{equation} 
where $R^{(4)}$ is the 4-dimensional Ricci scalar associated with the metric on the brane.

Using the fact that $L(r)=\tilde{c}M'(r)$, $\tilde{c}$ constant, we can integrate the extradimensional dependence, along with the remark following equation \eqref{preeff} of chapter \ref{ch:branes}. Doing so, we obtain the following 4-dimensional effective action:
\begin{equation}
S_{gravity}=\int d^4 x \sqrt{-g^{(4)}} 
\frac{1}{16\pi G_{eff}}\left(R^{(4)}-2\Lambda^{(4)}_{eff}\right),
\label{action_gravity_effective}
\end{equation}
where 
\begin{equation}
\Lambda^{(4)}_{eff}=\left.\frac{4\tilde{c}\pi G_{eff}}{G_6}\left(\frac{2}{5} \Lambda M^5 - 
2 M'' M^4 -4 M^3 M'^2\right)
\right\vert_{r=r_1}^{r=r_2}.
\end{equation}
and where
\begin{equation}
\frac{1}{G_{eff}}=M_P^2=\left.\frac{2\tilde{c}\pi}{3G_6} M^3\right
\vert_{r=r_1}^{r= r_2},
\label{plmas}
\end{equation}
where $M_P$ is the four dimensional Planck mass.

Furthermore, we can evaluate the value of $\tilde c$ for the vacuum solution, using the boundary conditions: $L'(0)=\tilde c M''(0)=1$. Using the equation of motion and the boundary conditions on $M,M'$, we can compute the value of $M''(0)$, leading to $\tilde c= \frac{4}{\kappa-\Lambda}$.

It should be noted that due to the solitonic nature of the matter fields, we expect this low energy effective action to be a good approximation of the case where the matter fields are not in a vacuum configuration; the four dimensional Planck mass should be well approximates by \eqref{plmas} with $\tilde c=\frac{4}{\kappa-\Lambda}$.

\subsection{Numerical results}
In absence of explicit solutions for non trivial $f,P$, we have solved the system \eqref{eqaehm}, \eqref{eqbgr2} numerically and present the solutions in the next few sections.
Following the investigations in \cite{cv1,cv2}, here we mainly aim at a classification of the generic solutions available in the system. The results are organised according to the domain of $\kappa$ and $\Lambda$.
%%%%%%%%%%%%%%%%%%%%%%%%%%%%%%%%%%%%%%%%%%%%%
\subsection{Static branes ($\kappa=0$)}
%%%%%%%%%%%%%%%%%%%%%%%%%%%%%%%%%%%%%%%%%%%%%%%%
We solve the system of ordinary differential equations subject to the above boundary conditions numerically. The system depends on three independent coupling constants $\alpha$, $\gamma$, $\Lambda$. We here fix $\alpha=2$, corresponding to the self dual case in flat space, and we will in the following describe the pattern of solutions in the $\gamma$-$\Lambda$ plane. Results for the 4-dimensional gravitating string \cite{gstring} make us believe that the pattern of solutions for $\alpha\neq 2$ is qualitatively similar. 

For vorticity $n >1$, the situation might change (see \cite{gstring}), however, we don't discuss this case here. As will become evident, the presence of two cosmological constants leads to a rather complicated pattern of solutions. The interconnection of the different type of solution available is illustrated in Fig.\ref{fig11}.

%%%%%%%%%%%%%%%%%%%%%%%%%%%%%%%%%%%%%%%%%%%%%%%
\subsubsection{Zero or negative bulk cosmological constant}
%%($\Lambda \neq 0$)}
%%%%%%%%%%%%%%%%%%%%%%%%%%%%%%%%%%%%%%%%%%%%%%%%

\begin{figure}[H]
\center
\includegraphics[scale=.6]{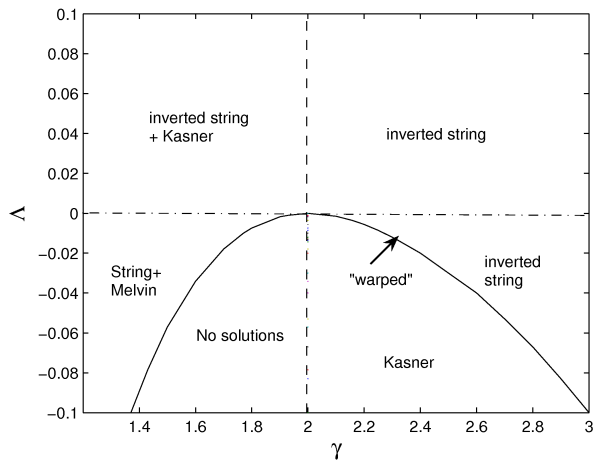}
\caption{\label{fig1}The pattern of brane world solutions in the $\Lambda$-$\gamma$ plane for $\alpha=2$.}
\end{figure}

The case $\Lambda=0$ was studied in detail in \cite{bh1,bh2}. We review the main results here to fit it into the overall pattern of solutions. The
pattern of solution in the $\gamma$-$\Lambda$ plane is given in Fig.\ref{fig1}.

For $\Lambda=0$ and $\gamma < 2$ two branches of solutions exist, with an asymptotic behaviour of the metric functions given by (\ref{sol1}) and  (\ref{sol2}). Referring to their counterparts in a four-dimensional space-time \cite{gstring} we denote these two families of solutions as the 'string' and 'Melvin' branches, respectively. The terminology used e.g. in \cite{gstring} will be used throughout the rest of the chapter.

Specifically for $\alpha=2$, we have $M(r)\equiv 1$ and $L(r\gg 1) \sim ar+b$, $a>0$. For $\gamma = 2$ the two solutions coincide and $L(r\gg 1)=1$.
When the parameter $\gamma$ is increased to values larger than $2$ the string and Melvin solutions get progressively deformed into closed solutions with zeros of the metric functions. For $\gamma > 2$ indeed, the string branch smoothly evolves into the so called inverted string branch (again using the terminology of \cite{gstring}). The inverted string solutions are characterized by the fact that the slope of the function $L(r)$ is constant and negative, $L(r)$ therefore crosses zero at some finite value of $r$, say $r=r_{IS}$.

\begin{figure}[H]
\center
\includegraphics[scale=.5]{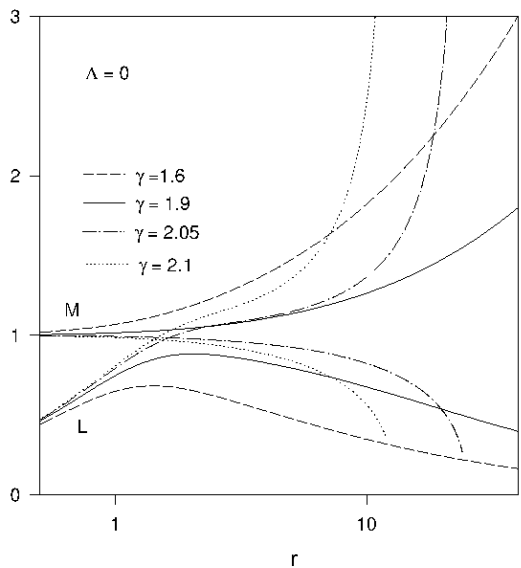}
\caption{\label{fig2}The pattern of the transition from the Melvin solution ($\gamma < 2$) to the Kasner solution ($\gamma > 2$) for $\Lambda = 0$ if given for the metric functions $M$ and $L$.}
\end{figure}

On the other hand, the Melvin branch evolves to the Kasner branch, a configuration for which the function $M(r)$ develops a zero at some finite value of $r=r_K$ while $L(r)$ becomes infinite for $r\to r_K$. More precisely, for $0 \ll r < r_K$  these solutions have the behaviour $M \propto (r_K-r)^{2/5}$, $L \propto (r_K-r)^{-3/5}$. The transition between Melvin solutions (for $\gamma < 2$) and Kasner solutions (for $\gamma > 2$) is illustrated in Fig. \ref{fig2}.

For negative $\Lambda$ and $\gamma < 2$, the string and Melvin solutions are still present and merge into a single solution at some critical value of $\Lambda$.  For $\Lambda <0$ and $\gamma > 2$, we have the zero $r_{IS}$ of the inverted string solution increasing with decreasing $\Lambda$. It reaches infinity for the exponentially decreasing solution, the so-called ``warped'' solution that localizes gravity on the brane \cite{bh1,bh2}. 
If the cosmological constant is further decreased the solution becomes of Kasner-type.

%%%%%%%%%%%%%%%%%%%%%%%%%%%%%%%%%%%%%%%%%%%%%%%%%%%%%%%%%%%%%
\subsubsection{Positive bulk cosmological constant ($\Lambda >  0$)}
%%%%%%%%%%%%%%%%%%%%%%%%%%%%%%%%%%%%%%%%%%%%%%%%%%%%%%%%%%%%%%%

Up to now static branes have only been discussed $\Lambda=0$ or $\Lambda<0$. Here, we also consider the case of static branes
with $\Lambda>0$. First, we examine the evolution of the string and Melvin solutions for $\Lambda > 0$. For all solutions constructed with $\Lambda > 0$, we were able to recover the behaviour \eqref{vacuumtri} asymptotically.

\begin{figure}[H]
\center
\subfigure[\label{fig3}The profiles of the metric functions $M$, $L$  of the string solution (for $\Lambda=0$, $\Lambda=-0.005$), respectively of
the Kasner solution (for $\Lambda=0.005$) for $\gamma=1.6$. ]{\includegraphics[scale=.38]{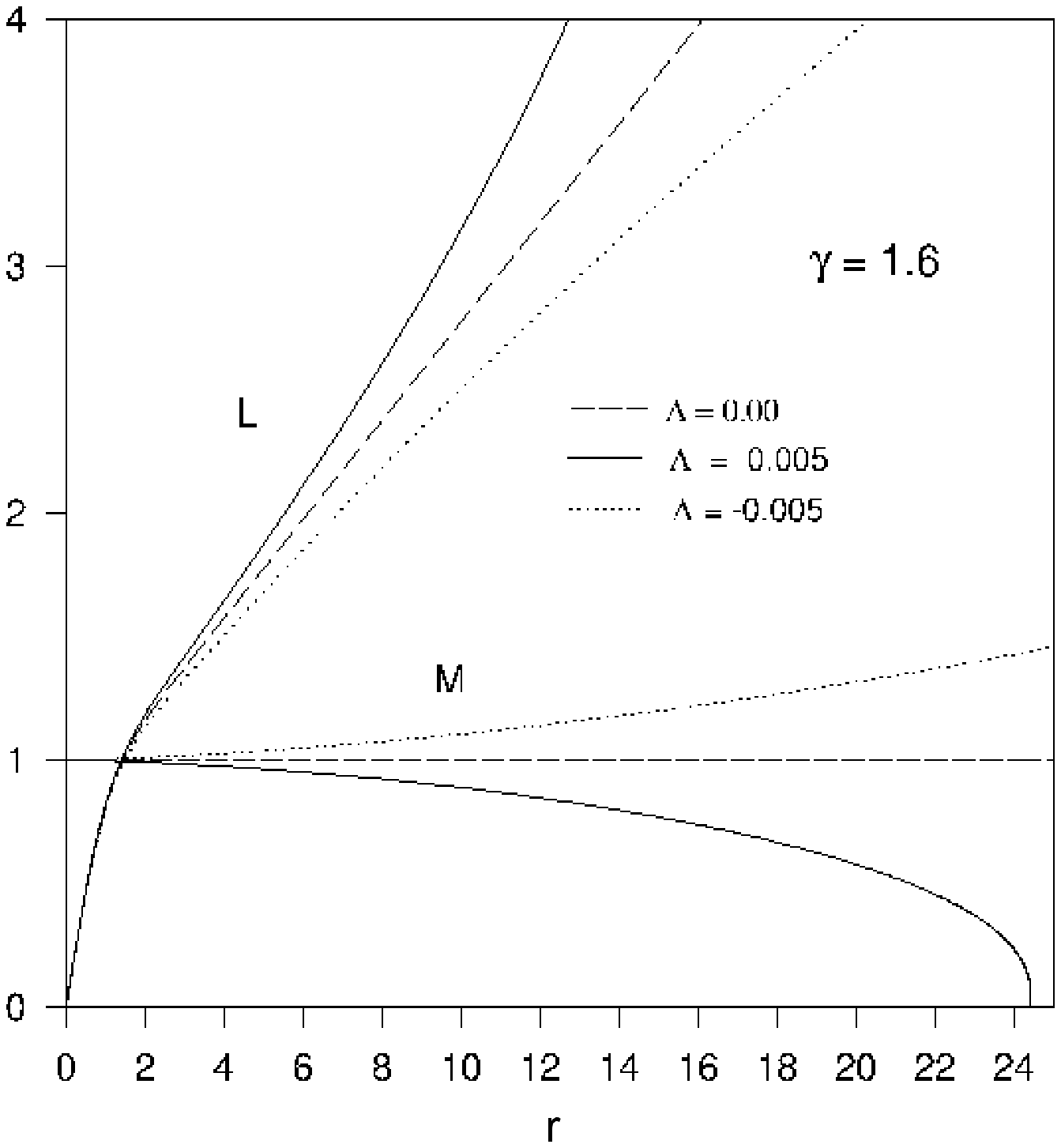}}
\hspace{.1cm}
\subfigure[\label{fig4}The profiles of the metric functions $M$, $L$  of the Melvin solution (for $\Lambda=0$, $\Lambda=-0.005$), respectively
of the inverted string solution (for $\Lambda=0.005$) $\gamma=1.6$.]{\includegraphics[scale=.38]{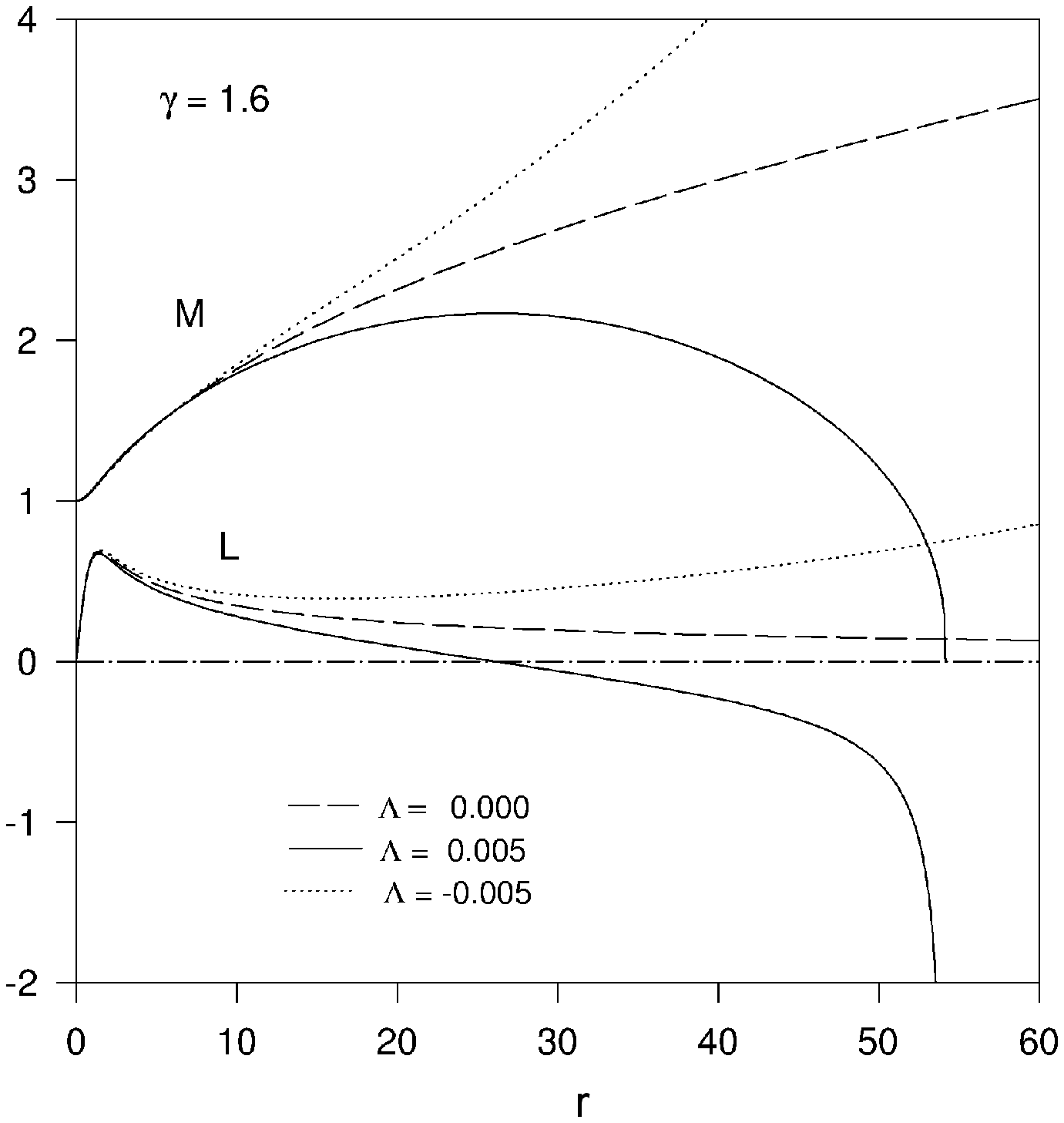}}
\caption{Profiles of the metric functions $L$ and $M$ for the string, Kasner and Melvin solution.}
\end{figure}

This evolution is illustrated in Fig.\ref{fig3} and  Fig.\ref{fig4} for $\gamma = 1.6$ respectively for the string and Melvin solutions.
Here, we show the metric functions $M$, $L$ for  $\Lambda = 0$ and $\Lambda = \pm 0.005$. For $\Lambda > 0$ we find a solution with the  metric function $M(r)$ possessing a zero at some finite $r$, say $r=r_1$. At the same time $L(r\to r_1)\to + \infty$.  These solutions tend to the
string solutions in the limit $\Lambda \to 0$. Following the convention used in the $\Lambda=0$ case, we refer to these solutions
as of 'Kasner' type.  

The second type of solutions that we find has metric functions behaving for $0< r_2 < r_3$ as
\begin{equation}
              L(r_2)= 0 \ \ , \ \ M'(r_2)=0 \ \ , \ \
              M(r_3)= 0 \ \ , \ \ \lim_{r \to r_3} L(r) = - \infty  \ .
\end{equation}
These tend to the Melvin solutions in the limit $\Lambda\to 0$ and we will refer to them as of ``inverted string''-type.

\begin{figure}[H]
\center
\includegraphics[scale=.5]{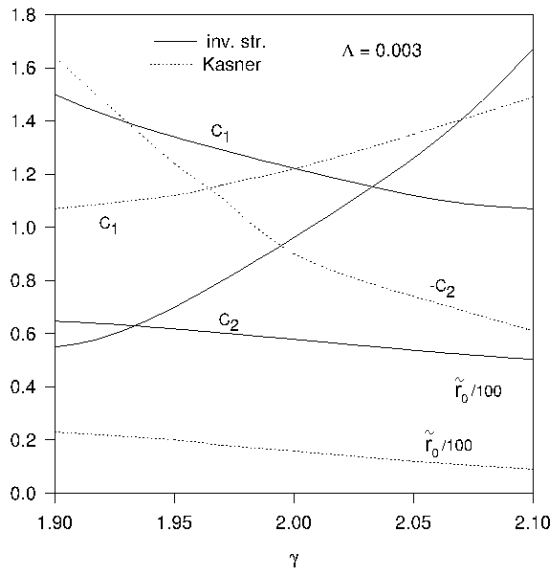}
\caption{\label{fig5}The values of the parameters $C_1,C_2,r_0$ defined in \eqref{sol2} for the inverted string and Kasner-like solutions as functions
of $\gamma$ for $\Lambda = 0.003$}
\end{figure}

For fixed $\Lambda > 0$ and increasing $\gamma$ we find that both Kasner and inverted string solutions exist for all values of $\gamma$.
This is demonstrated in Fig.\ref{fig5} where the values of the parameters $C_1$, $C_2$, $\tilde r_0$  (defined in Eq.\eqref{vacuumtri}) are plotted as functions of $\gamma$.

%%%%%%%%%%%%%%%%%%%%%%%%%%%%%%%%%%%%%%%%%%%%%%%%%%%%
 \subsection{Inflating branes $\kappa > 0$}
%%%%%%%%%%%%%%%%%%%%%%%%%%%%%%%%%%%%%%%%%%%%%%%%%%%%%
Here, we discuss inflating branes ($\kappa > 0$). Again, we fix $\alpha=2$. The pattern of solutions can largely be characterized by the integration constant $C$ appearing in \eqref{quadrature}.

%%%%%%%%%%%%%%%%%%%%%%%%%%%%%%%%%%%%%%%%%%%%%%%%%%%%%%%%%%%%%%%%%%%%%
\subsubsection{Zero bulk cosmological constant}
%%%%%%%%%%%%%%%%%%%%%%%%%%%%%%%%%%%%%%%%%%%%%%%%%%%%%%%%%%%%%%%%%%%%%

\begin{figure}[H]
\center
\includegraphics[scale=.5]{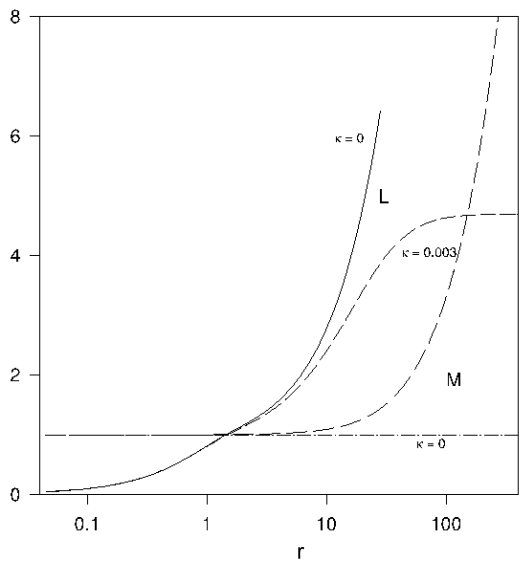}
\caption{\label{fig6}The profiles of the metric functions $M$, $L$ for $\Lambda= 0$ and for $\kappa=0$ and $\kappa = 0.003$, respectively.}
\end{figure}

First, we have constructed solutions corresponding to deformations of the string solutions residing in the extra dimensions. We present the profiles in Fig. \ref{fig6} for $\kappa = 0.003$ and for comparison for $\kappa=0$. Obviously, the presence of an inflating brane ($\kappa>0$) changes the asymptotic behaviour of $M$, $L$ drastically in comparison to a Minkowski brane. The function $M(r)$ now behaves linearly far from the core of the string, while $L(r)$ tends to a constant. The solutions approach asymptotically \eqref{kappasol}. The space-time is then cigar-like. 

This can be explained as follows: In the case $\alpha = 2$ the equations are self dual, in particular the equation determining the function $M$ on the string branch is $M M'' + (3/2) (M')^2 - \kappa/2 = 0$ (the combination of the energy momentum tensor vanishes identically for $\alpha=2$ and for the string like solution).

The value of $C$ compatible with the boundary condition turns out to be $C = - \kappa/4$, since we are interested is small values of $\kappa$, the integral \eqref{quadrature} can be reasonably approximated by \eqref{kappasol}, in complete agreement with our numerical results.

\begin{figure}[H]
\center
\subfigure[\label{fig7}The parameter $L_0$.]{\includegraphics[scale=.43]{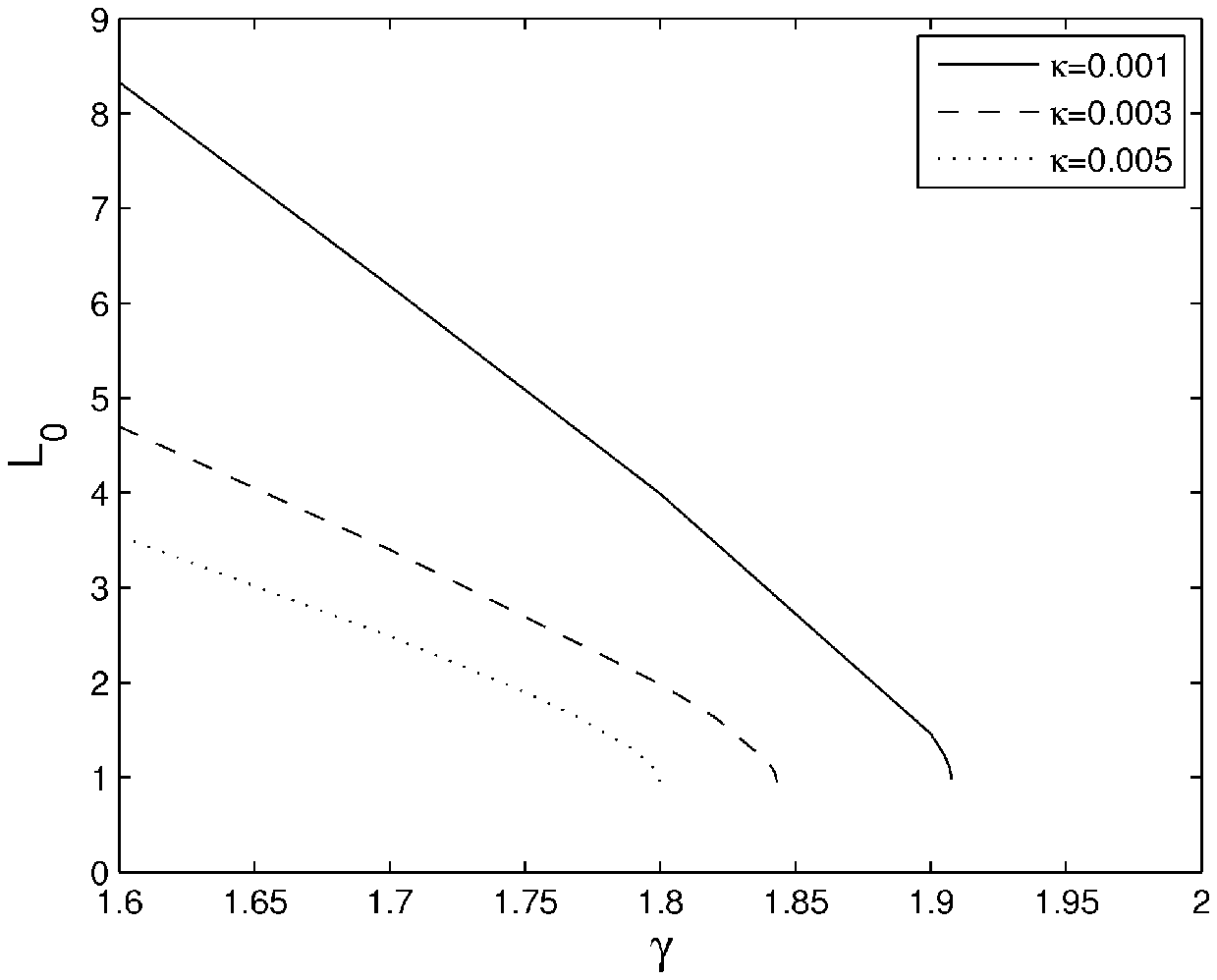}}
\hspace{.1cm}
\subfigure[\label{fig8}The parameter $C$.]{\includegraphics[scale=.43]{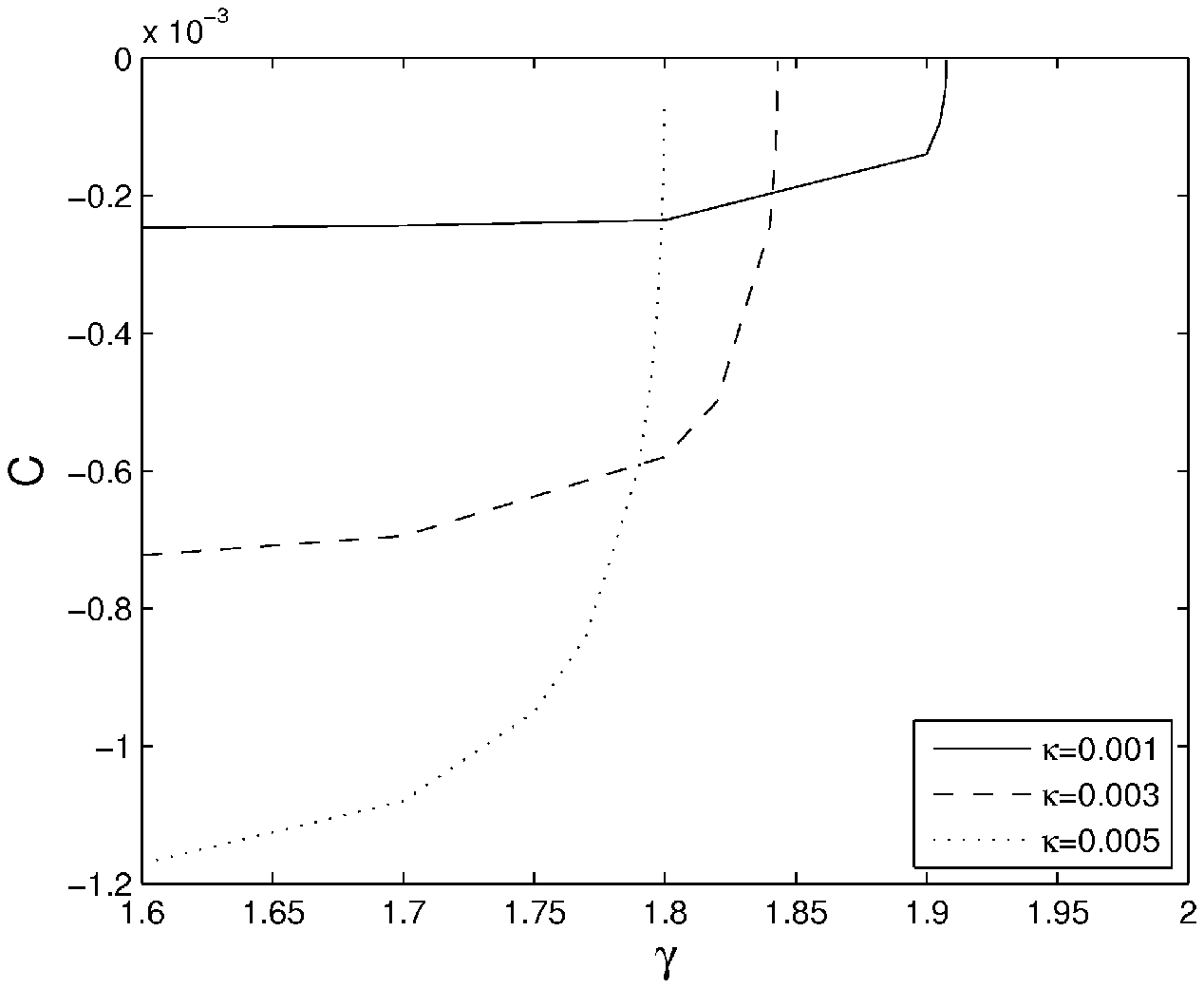}}
\caption{The parameters $C$ and $L_0$ for the cigar-like solution as functions of $\gamma$ for different values of $\kappa$ and $\Lambda=0$.}
\end{figure}

The parameters $L_0$ and $C$ appearing in \eqref{quadrature} can be determined numerically. It turns out that for the cigar-like solutions we always 
have $C < 0$, while $L_0$ is positive. For a fixed value of $\kappa$ and varying $\gamma$, we find that $C\to 0$ for a critical, $\kappa$-dependent  value of $\gamma$, $\gamma_{cr}(\kappa)$. At the same time $L_0 \to 0$ for $\gamma \to \gamma_{cr}$. This is shown for three different values of $\kappa$ in figure \ref{fig7} and \ref{fig8}. In the limit $\gamma \to \gamma_{cr}$, the diameter
of the cigar tends to zero and the cigar-like solutions cease to exist.

We have also studied solutions corresponding to deformations of the Melvin solutions. It turns out that the $\kappa = 0$ Melvin solution is smoothly
deformed for $\kappa > 0$.  $C$ is always positive and tends exponentially to zero as function of $\gamma$. For $\gamma > 2$ we find that the solutions have a zero of the metric function $L$ at some $\gamma$ and $\kappa$ dependent value of $r$. The solutions are thus of inverted string-type.

\begin{figure}[H]
\center
\includegraphics[scale=.55]{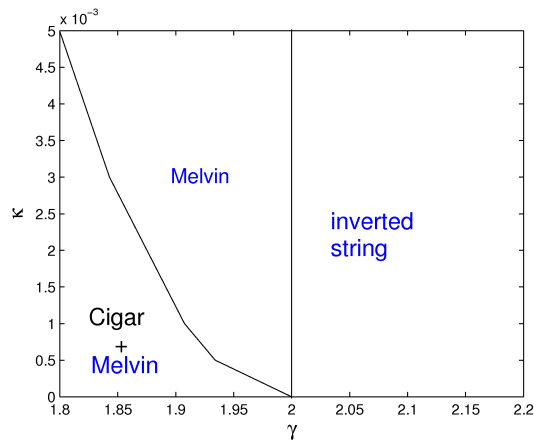}
\caption{\label{fig9}The pattern of inflating brane solutions in the $\gamma$-$\kappa$ plane.}
\end{figure}

We present the pattern of solutions in the $\gamma$-$\kappa$ plane in Fig.\ref{fig9}.

%%%%%%%%%%%%%%%%%%%%%%%%%%%%%%%%%%%%%%%%%%%%%%%%%%%%%%%%%%%%%%%%%%%%%%%%%%%%%%%%%%%%%%%%%%%%%
\subsubsection{Positive/negative bulk cosmological constant}
%%%%%%%%%%%%%%%%%%%%%%%%%%%%%%%%%%%%%%%%%%%%%%%%%%%%%%%%%%%%%%%%%%%%%%%%%%%%%%%%%%%%%%%%%%%%%
We have limited our analysis here to $\gamma \leq 1.8$. For $\gamma > 1.8$, the numerical analysis becomes very unreliable, this is why we don't report our results here.

Fixing $\gamma$, e.g. to $\gamma=1.8$, the inverted string solutions (available for $\Lambda > 0$, $\kappa=0$) gets smoothly deformed for $\kappa > 0$. The constant $C$ is positive for all solutions. For fixed $\kappa$ and $\Lambda \to 0$ the Melvin solution is approached.

\begin{figure}[H]
\center
\includegraphics[scale=.45]{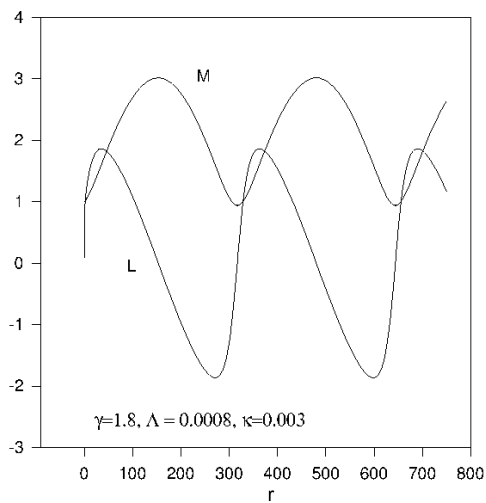}
\caption{\label{fig10}The profiles of the metric functions $M$, $L$ for $\Lambda= 0.0008$ and for $\kappa=0.003$ and $\gamma = 1.8$.}
\end{figure}

The Kasner solutions (for $\Lambda >0$) get deformed and the functions $M$, $L$ become periodic in $r$ asymptotically. These are deformations of the periodic solution \eqref{trigonometric} with $C\neq 0$. This is illustrated in Fig.\ref{fig10}.

The value of $C$ for these solutions is negative. The function $M$ oscillates around a mean value given by $M_{mv}= (\frac{-3 C}{2 \Lambda})^{1/5}$
and stays strictly positive, the solution is therefore regular. The period of the solution depends weakly on $\kappa$. In the limit, $\Lambda \to 0$ the periodic solutions tend  to the cigar-like solutions. 

We notice that the values of the constants $\kappa$ and $\Lambda$ can be chosen in such a way that $2\kappa - \Lambda$ becomes arbitrarily small. This seems compatible with the domain of parameters leading to periodic solutions. The corresponding four dimensional Planck mass determined through \eqref{plmas} can therefore be made arbitrarily large, compatible with observations. It is remarkable that the region allowing a realistic Planck mass is naturally compact.

\begin{figure}[H]
\center
\includegraphics[scale=.5]{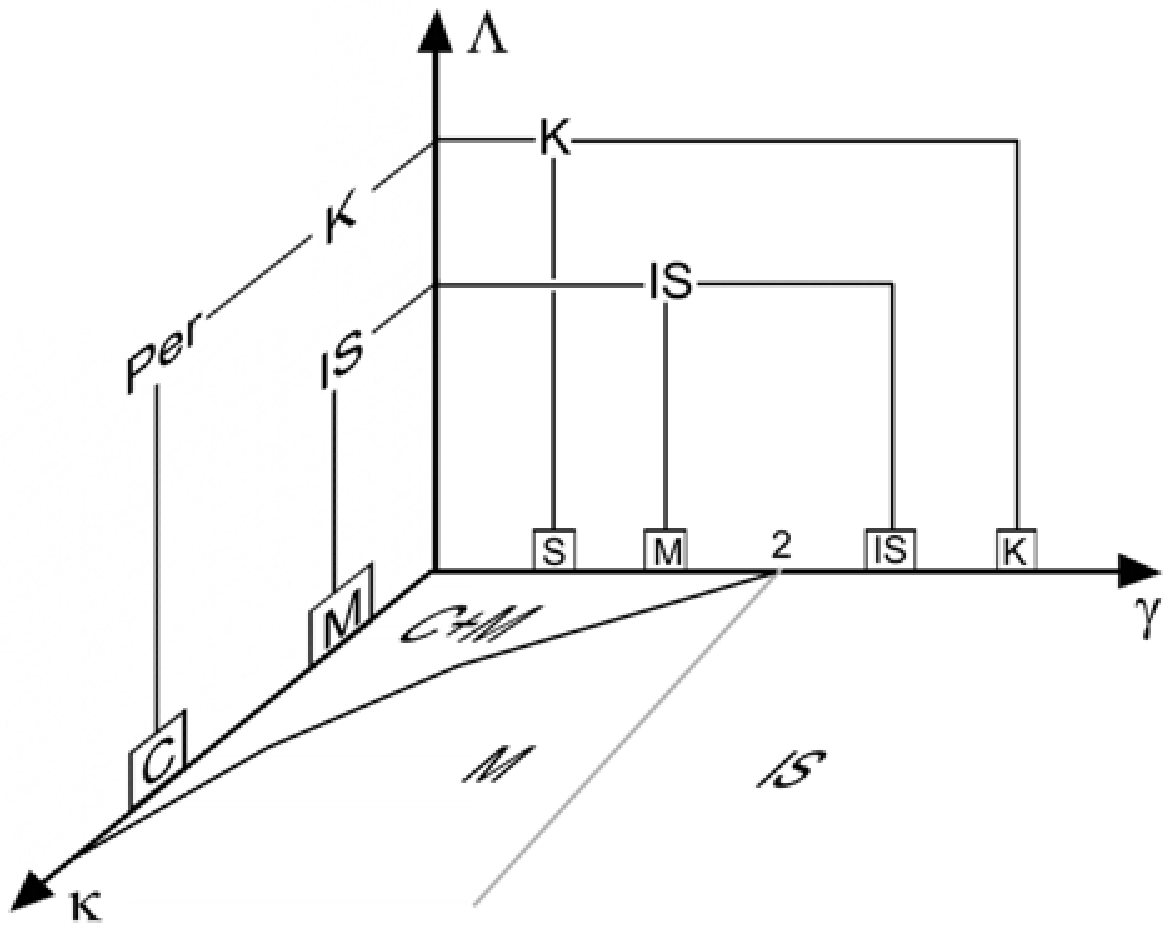}
\caption{\label{fig11}The pattern of brane solutions in the $\kappa$-$\gamma$-$\Lambda$ domain. The origin of the coordinate system here corresponds to $\kappa=0$, $\gamma=1.8$, $\Lambda=0$. M, S, IS, K, Per, C denote Melvin, string, inverted string, Kasner, periodic and cigar-type solutions, respectively.}
\end{figure}

A sketch of the pattern of the solutions is proposed in figure \ref{fig11}. The fact that we obtain periodic solutions constitutes a natural framework to define finite volume brane world models. With view to the discussion on the effective 4-dimensional action of section \ref{action_gravity_effective}, we could imagine to put branes with 'large' effective gravitational coupling at $r=r_k$, $k=1,2,3,...$, where $r_k$ corresponds to the position of the maxima of $M(r)$.
Correspondingly, branes with 'small' effective gravitational coupling could be put at $r=\tilde{r}_k$, $k=1,2,3,..$, where $\tilde{r}_k$ are the positions of the minima of $M(r)$. Though there is no 'warping' in the sense of Randall and Sundrum \cite{rs1,rs2}, nevertheless gravity looks stronger on the branes positioned at the maxima of $M(r)$ in contrast to gravity on the branes positioned at the minima. Note that at the positions of these branes, the metric function $L(r)$ vanishes, i.e. $L(r_k)=L(\tilde{r}_k)=0$. Thus the cylindrical geometry of the extra dimensional space 
shrinks to a point and the whole space-time geometry becomes effectively 5-dimensional.

In order to complete this diagram, we summarize the various type of solutions with the corresponding domain of parameters for $\Lambda\geq0$ in the following table:
\begin{center}
\begin{tabular}{lllccc}
\hline
 Type	&	Form of the asymptotic 						&	Range 		&$\kappa$&	$\gamma$   &  $\Lambda$\\
\hline
\hline
String		&   $M\propto \mbox{Cste}, L\propto L_0(r-r_0)$			&	$[0,\infty]$   	& $\geq0$	&	$<2$	     &  $0$\\
Melvin		&   $M\propto (r-r_0)^\mu, L\propto (r-r_0)^\nu$ 		&	$[0,\infty]$	& $0$		&	$<2$	     &  $0$\\
Cigar		&   $M\propto (r-r_0), L\propto \mbox{Cste}$	 		&	$[0,\infty]$	& $>0$		&	$<\gamma_c<2$&  $0$\\
Inv. String	&   $M\propto \mbox{Cste}, L\propto -(r-r_0)$	 		&	$[0,\infty]$	& $>0$		&	$>2$	     &  $0$\\
or		&   $M\rightarrow0, L\rightarrow -\infty, \mbox{ for }r\rightarrow r_c)$& $[0,r_c]$	& $\geq0$	&	-            & $>0$\\
Kasner		&   $M\propto (r-r_c)^\mu, L\propto (r-r_c)^\nu$ 		&	$[0,r_c]$	& $\geq0$	&	-	   &$\geq0$\\
Periodic	&   $M\propto \cos , L\propto \sin $ 				&	$[0,\infty]$	& $>0$		&	-	     & $>0$\\
\hline
\end{tabular}
\end{center}

\section{Global $O(n)$ monopoles}
\label{globalmononpole}
In this section, we will consider global $O(n)$ monopole models in $n+4$ spacetime dimensions, $n$ being the number of extradimensions. 
In this model, we put the the $O(n)$ symmetry of the extradimensions in relation with the internal symmetry group $O(n)$.
%The $O(n)$ symmetry will be related to the $O(n)$ symmetry of the extradimensions. 

Global monopole solutions occurs in the Goldstone model describing $n$ scalar fields in a field theory globally invariant under the $O(n)$ transformations. The  symmetry is spontaneously broken by a potential. In the present context, the Goldstone model and the corresponding scalar fields are formulated with respect to the $n$ extradimensions:
\be
S_{brane}=\int \left((\partial_A\Phi)^\dagger(\partial^A\Phi) -\frac{\lambda}{4}(\Phi^\dagger\Phi - v^2)^2\right)d^{n+4}x
\ee
where the $n$ scalar fields $\Phi = (\phi^1,\ldots,\phi^n)$ form a fundamental representation of the group $O(n)$. $\lambda$ is the
self-coupling of the potential, $v$ the vacuum expectation value of the scalar field.
%For notational consistency, 
We define the following quantities, which should not be confused with the rescaled variables of the previous section:
\be
 \gamma=\Lambda_d,\   \beta= 8\pi G_d.
\ee

The ansatz for the scalar $n$-uplet reads
\be
\phi^i = \phi(\rho)\xi^i/\rho  \ ,
\ee
where the $\xi^i$ denote the Cartesian coordinates representing the extra dimensions. 

Correspondingly, the energy momentum tensor has only diagonal components given by \cite{japanese}:
\be
T_\mu^\mu = \phi'^2 + \frac{(n-1)\phi^2}{2l^2} + \frac{\lambda}{4}(\phi^2 - v^2)^2
\ee
\be
T_\rho^\rho = -\phi'^2 + \frac{(n-1)\phi^2}{2l^2}
+ \frac{\lambda}{4}(\phi^2 - v^2)^2
\ee
\be
T_{\theta_i}^{\theta_i} = \phi'^2
 + \frac{(n-3)\phi^2}{2l^2} + \frac{\lambda}{4}(\phi^2 - v^2)^2
\ee
where $\phi' \equiv d \phi / dr$. The equation corresponding to the Goldstone field reads:
\be
\phi'' + (n-1)(\frac{l'}{l} + \frac{4}{n-1}\frac{M'}{M})\phi' - \frac{1}{l^2} \phi = \lambda \phi(\phi^2-v^2)
\ee

The appropriate boundary conditions read:
\be
M(0)=1 \ \ , \ \ M'(0)=0 \ \ , \ \ l(0) = 0 \ \ , \ \ l'(0)=1
\ee
for the metric functions. In the case when the radial variable
can be extended to $\rho=\infty$, the usual boundary conditions 
for the scalar field function are
\be
\phi(0)=0 \ \ , \ \  \phi(\infty) = v,
\ee
for regularity and finiteness of the energy.

Note that the presence of a cosmological constant can lead to a cosmological horizon at $\rho=\rho_c$. In such case, an appropriate boundary condition for $\phi(\rho_c)$ has to be imposed; this will be discussed in due course.

The expressions of the energy momentum tensor contain terms of the form $1/l^2$ which also appear in the Einstein tensor.  If the gravitational constant $\beta$ is chosen such that
\be
\sqrt{\beta} v = \sqrt{n-2}  \ \ , \ \
\ee
the two inhomogeneous terms cancel. This value determines the so-called 'critical monopole'.  In this case, (and assuming in addition $H=\Lambda_d=0$) the asymptotic Melvin solutions of section \ref{secmelvin} are acceptable as asymptotic solutions \cite{cv1,japanese}.

%%%%%%%%%%%%%%%%%%%%%%%%%%%%%%%%%%%%%
\subsection{Sub-critical monopoles}
%%%%%%%%%%%%%%%%%%%%%%%%%%%%%%%%%%%%%%%
The case  $\sqrt{\beta} v < \sqrt{n-2}$ corresponds to the case of sub-critical monopoles \cite{japanese}. The vacuum solutions for which the functions $M$, $l$ go asymptotically to infinity (i.e. corresponding to $\Lambda_{n+4} \leq 0$) are such that the term $\phi^2/l^2 \to 0$.

The asymptotic behaviour of the metric function $M(\rho)$ remains the same irrespectively of the presence of a global monopole while the function  $l(\rho)$ must be renormalized according to
\be
l(\rho)_m = \frac{l(\rho)_v}{\sqrt{1-\frac{(\sqrt{\beta} v)^2}{n-2}}}
\ee
Here $l_v$ corresponds to the function of the vacuum solution while $l_m$ corresponds to the solution in the presence of the monopole.

For $\gamma > 0$ the arguments above do not apply because the terms $\phi / l^2$ cannot be neglected, since in this case $l\rightarrow 0$ for $r\rightarrow \infty$. However, the profiles of the metric functions $M$, $l$ remain very close to those of the vacuum solution. For larger values of $\rho$, the metric  become singular at some finite value of $\rho$. The singularity is of Melvin type and is of the same nature as in the case of local monopoles (see discussion below).

%%%%%%%%%%%%%%%%%%%%%%%%%%%%%%%%%%%%%%%%
\subsection{Mirror symmetric solutions}
%%%%%%%%%%%%%%%%%%%%%%%%%%%%%%%%%%%%%%%%
The coupled system of equations possesses several symmetries, namely invariance under translations of the radial variable $\rho$ and invariance under the reflections $\rho \to -\rho$ and $\phi \to -\phi$. These symmetries suggest that solutions which are invariant under suitable combinations of the symmetries could exist.

In the case of vacuum solutions, the solutions corresponding to $\gamma > 0$  possess such a symmetry. The most natural combination of the symmetries suggests to look in particular for solutions with the following properties
\be
      l(\rho_0-\rho) = l(\rho_0+\rho),\ M(\rho_0-\rho) = M(\rho_0+\rho),\ \phi(\rho_0-\rho) = \epsilon \phi(\rho_0+\rho),\ \epsilon = \pm 1
\ee
where the reflection point $\rho_0$ depends on the various coupling constants.
The existence of solutions presenting one of the above symmetries can be analysed by solving the equations supplemented by the following
boundary conditions at $\rho=\rho_0$:
\be
 l'(\rho_0)=0 \ , \ M'(\rho_0)=0 \ , \ \phi(\rho_0) = 0 \ \ {\rm if} \ \epsilon=-1
 \ \ , \ \ \phi'(\rho_0) = 0 \ \ {\rm if} \ \epsilon=1  \ .
\ee
These conditions complete the ones given above  for $\rho=0$ and allow for a numerical study of the equations (no explicit solution was found for
$\phi \neq 0$). Our numerical analysis suggests that 
\begin{enumerate}
\item solutions corresponding to $\phi(\rho_0-\rho) = \phi(\rho_0+\rho)$ do not seem to exist. In fact we were able to construct numerically such configurations but they do not persist when increasing the accuracy.
\item Solutions obeying $\phi(\rho_0-\rho) =-\phi(\rho_0+\rho)$ do indeed exist for peculiar values of the coupling constants.
\end{enumerate}

The existence of these 'odd' solutions can be related to the fact that, in the neighbourhood of the symmetric point $\rho_0$, the Goldstone field equation can be put into the form
\be
\phi ''- \lambda \phi^3 + (\lambda v^2 - \frac{1}{l(\rho_0)^2})\phi  \sim 0
\ee
where we used the fact that $M'(\rho_0)=l'(\rho_0)=0$.

This simplified version of the Goldstone equation is identical to the kink equation provided $\sqrt{\lambda} l(\rho_0) v > 1$ which turns out to hold true in the region of parameters that we have explored. Kink-like solutions can therefore be expected.

\begin{figure}[H]
\begin{center}
\subfigure[The profiles of the metric and scalar functions for the mirror
symmetric solution corresponding to $\rho_c = 40$. This is a solution
on the first branch. ]{\includegraphics*[scale=.38]{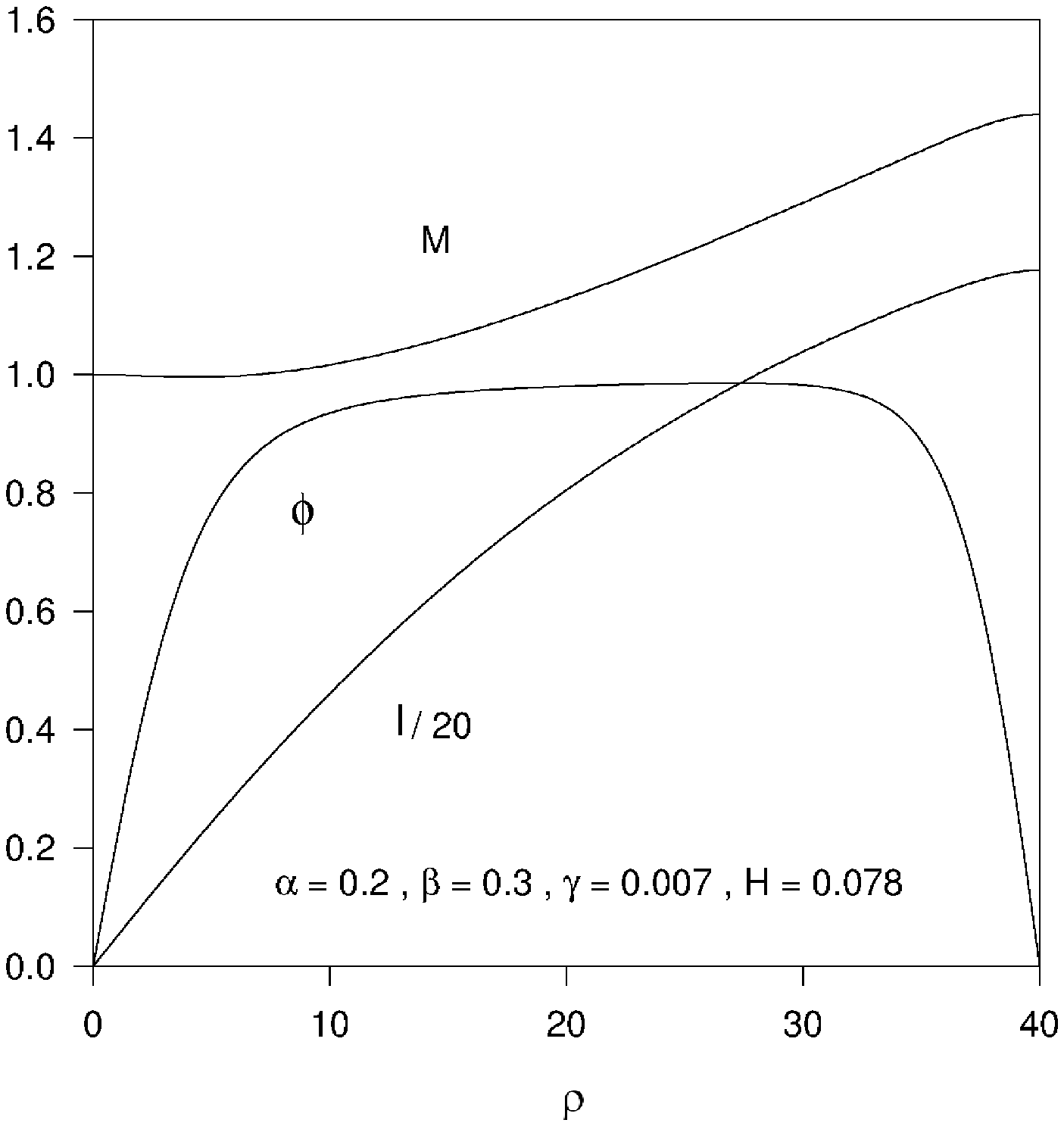}
\label{fig1p2}
}
\hspace{.1cm}
\subfigure[The profiles of the metric and scalar functions for the mirror
symmetric solution corresponding to $\rho_c = 40$. This is a solution
on the second branch.]{\includegraphics[scale=.38]{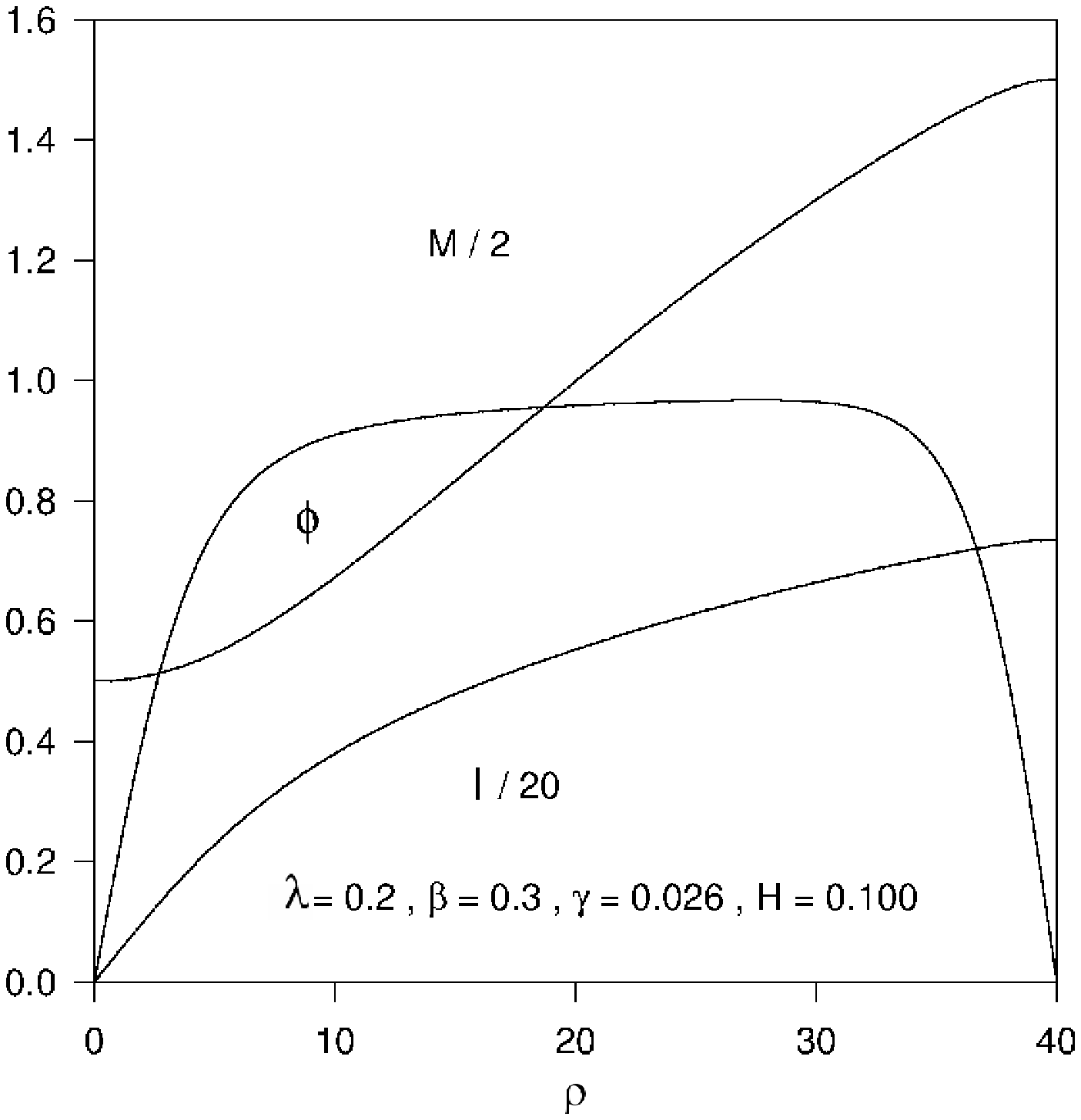}
\label{fig2p2}
}
\caption{Typical profiles of the two mirror symmetric branches.}
\label{fig:fig1-2p2}
\end{center}
\end{figure}

Two solutions of this type are presented in figure \ref{fig:fig1-2p2}. These solutions are similar to the so called 'Bag of Gold' solutions discovered in the 4-dimensional Einstein-Yang-Mills equations with a positive cosmological constant \cite{lavre}. Similar phenomena in 4-dimensional space-time were observed in \cite{bfm} and more recently in \cite{fr}. 
However, to our knowledge, it is the first time that compact solutions relative to the spatial extra dimensions are constructed.

\begin{figure}[H]
 \centering
\includegraphics[scale=.5]{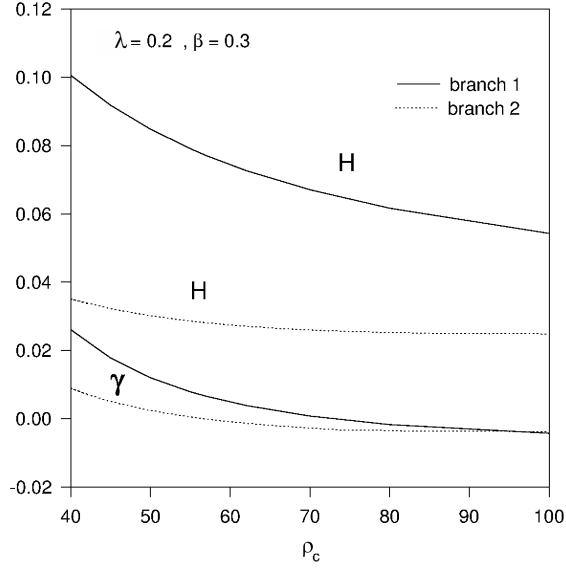}
\caption{The relations between the parameters $\gamma$, $H$, $\rho_0$ allowing mirror
symmetric solutions.}
\label{fig:fig3p2}
\end{figure}

It has to be stressed that this type of solutions exist only for peculiar values of the cosmological constants  $\gamma$, $H$ once the coupling constants  $\lambda$, $\beta$ are fixed. Setting $\lambda=0.2$, $\beta=0.3$, the relations between $\gamma$, $H$ and $\rho_c$ allowing for mirror symmetric solutions are given in figure \ref{fig:fig3p2}. Our results suggest that several branches of solutions could exist. We were able to find two non trivially different branches, see figure \ref{fig:fig3p2}. The construction of the second branch is however involved numerically.

It is clear that these solutions can be extended symmetrically for  $\rho \in [\rho_c,2 \rho_c]$ and further for $\rho \in [2 \rho_c, 4 \rho_c]$.
This leads naturally to periodic solutions on $[0, 4\rho_c]$. The corresponding spacetimes are deformed hyper-spheres with angular deficit.

%%%%%%%%%%%%%%%%%%%%%%%%%%%%%%%%%%%%%%%%%%%%%%%%%%%%%%%%%%%%%%%%%%%
%%%%%%%%%%%%%%%%%%%%%%%%%%%%%%%%%%%%%%%%%%%%%%%%%%%%%%%%%%%%%%%%%%%
%%%%%%%%%%%%%%%%%%%%%%%%%%%%%%%%%%%%%%%%%%%%%%%%%%%%%%%%%%%%%%%%%%%
\section{Einstein Yang-Mills Higgs $SO(3)$ monopole}
\label{localmonopolen=3}
%%%%%%%%%%%%%%%%%%%%%%%%%%%%%%%%%%%%%%%%%
It is well know that monopoles exist in some spontaneously broken non-abelian gauge theories. The simplest case is the Georgi-Glashow model with gauge group $SO(3)$ and a Higgs triplet. In this context, the matter lagrangian reads
\be
L_m= -(D_A\Phi^a)(D^A\Phi^a) - \frac{1}{4} F_{AB}^a F^{a,AB}-\frac{\alpha}{4 e^2}(\Phi^a \Phi^a - v^2)^2,
\ee
where $A,B=0,1,\dots,6$, $a=1,2,3$ and where the usual definitions for the covariant derivative and field strengths are used:
\be
D_M \Phi^a = \partial_M  \Phi^a + e \epsilon^{abc} W^b_M \Phi^c,\ F^a_{MN} = \partial_M W^a_N - \partial_N W^a_M + e \epsilon^{abc} W^b_M W^c_N
\ee
Along with \cite{chovil}, we use a spherically symmetric ansatz for the gauge and Higgs fields, where the $SO(3)$ symmetry is imposed in the three-dimensional space defined by the three extradimensions:
\be
W_{\mu}^a = 0,\ W_i^a = (1-w(\rho)) \epsilon_{aij}\frac{\xi^j}{e \rho^2},\ a,i,j=1,2,3,
\ee
\be
\Phi^a =\phi(\rho) \frac{\xi^{a}}{\rho},\ a = 1,2,3,
\ee
$\xi^1,\xi^2,\xi^3$ being the Cartesian coordinates in the extradimensions such that $\rho^2 = \xi^2_1 + \xi^2_2+ \xi^2_3$. The metric ansatz is given by \eqref{ggenans} with $n=3$.

The resulting energy momentum tensor has the following non-vanishing components:
\bea
T_{\mu}^{\mu} &=& \frac{(\phi')^2}{2} + \frac{\phi^2 w^2}{l^2}  + \frac{1}{e^2 l^2}( \frac{(1-w^2)^2}{2 l^2} + w'^2) + \frac{\alpha}{4}(\phi^2-v^2)^2 \nonumber \\
T_{r}^{r}    &=&  - \frac{(\phi')^2}{2} + \frac{\phi^2 w^2}{l^2} + \frac{1}{e^2 l^2}( \frac{(1-w^2)^2}{2 l^2} - w'^2) +\frac{\alpha}{4}(\phi^2-1)^2\nonumber  \\
T_{\theta_i}^{\theta_i} &=&  \frac{(\phi')^2}{2}- \frac{(1-w^2)^2}{2 e^2 l^4}+  \frac{\alpha }{4}(\phi^2-v^2)^2.
\eea

In the absence of a bulk cosmological constant, this model was studied in \cite{chovil}. Here, the equations are solved with the usual boundary
conditions for the functions $w(\rho),\phi(\rho)$:
\be
\phi(0)= 0,\ w(0) = 1
\ee
for regularity at the origin and
\be
\phi(\rho\to \infty)= v,\ w(\rho \to \infty) = 0
\ee
outside the monopole core, for finiteness of the energy.

We will analyse the influence of the bulk cosmological constant on the solutions obtained in \cite{chovil}.  Again, mirror symmetric solutions will be available, specific boundary conditions being imposed at some intermediate value $\rho=\rho_c$. In the discussion and the figures, we use the following rescaling of the parameters: 
\be
\beta = 8\pi e^2 v^2 G_d,\ \gamma = \Lambda_d/e^2 v^2, H/ev \to H.
\ee

\begin{figure}
\centering
\includegraphics[scale=.5]{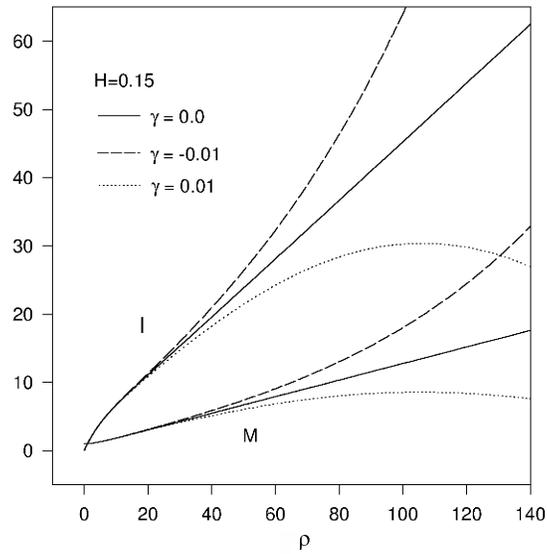}
\caption{\label{fig:fig4p2}
The profiles of the metric functions $l$, $M$ corresponding to an inflating brane with $H=0.15$ regularized by a monopole are given for
$\gamma = 0, \pm 0.01$.}
\end{figure}

In general, the three types of vacuum solutions presented in section \ref{secbranevaccum} are deformed by the presence of the local monopole on the brane.
The deformation of the metric fields for $\gamma=0,\pm 0.01$ are sketched in figure \ref{fig:fig4p2} for $H= 0.15$.
We will now discuss the different solutions more qualitatively.

\subsection{$\gamma = 0$}
In this case, the solutions have been analysed in details in \cite{chovil}. Far from the monopole core, the metric functions behave linearly:
\be
     M(\rho) = \sqrt{\frac{3}{5}} H \rho + m_0 + \frac{m_1}{\rho},\ l(\rho) = \sqrt{\frac{1}{5}} \rho + l_0 + \frac{l_1}{\rho}.
\ee
These approach the corresponding vacuum solutions of section \ref{secbranevaccum} for $\rho\to \infty$.

\subsection{$\gamma < 0$}
For $\gamma < 0$ our numerical analysis reveals that there are regular solutions which obey asymptotically the form
\be
   M(\rho) = M_0 \sinh{\mu \rho} \ \ , \ \ l(\rho) = l_0 \sinh{\mu \rho},\  \mu = \sqrt{\frac{-\beta \gamma}{15}}.
\ee

\begin{figure}
\centering
\includegraphics[scale=.5]{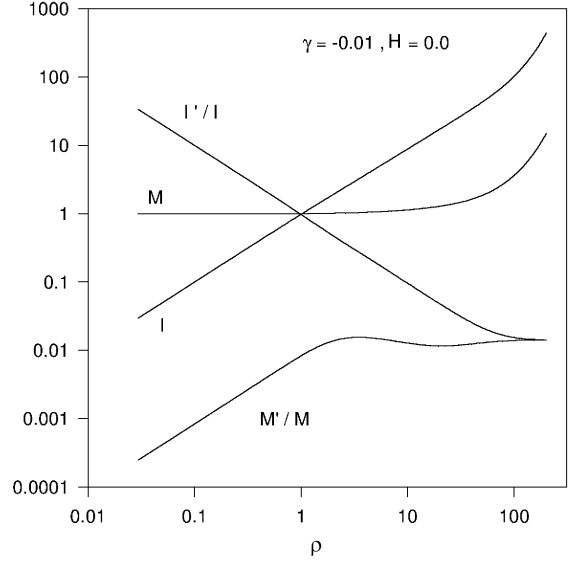}
\caption{\label{fig:fig5p2}
The profiles of the metric functions $l$, $M$ and the ratio $l'/l, M'/M$ 
corresponding to a static brane are shown for $\gamma = -0.01$}
\end{figure}

The profile of such a solution is presented in figure \ref{fig:fig5p2} for $H=0$ and $\gamma=-0.01$. In particular, we can see the deviation of the functions $M$, $l$  from their asymptotic behaviour.

\subsection{$\gamma > 0$}
\begin{figure}
\centering
\subfigure[\label{fig:fig6p2}
The profiles of the metric functions of an inflating brane solution with $H=0.07$ and positive cosmological constant $\gamma = 0.005$ are shown.]{\includegraphics[scale=.4]{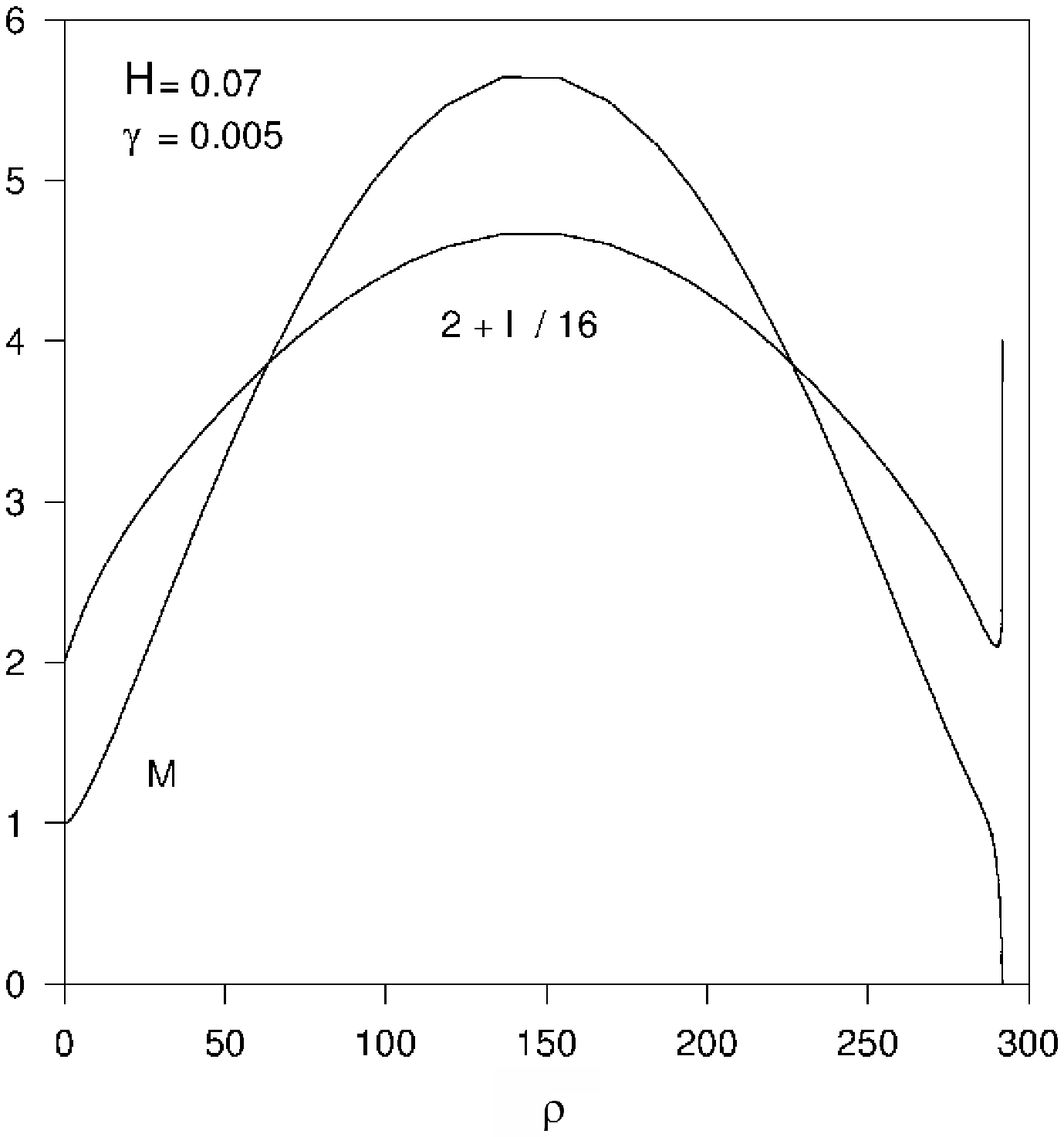}}
\hspace{.1cm}
\subfigure[\label{fig:fig7p2}
The details of figure \ref{fig:fig6p2} in the region of the singular point $\rho\approx 291.7$.]{\includegraphics[scale=.4]{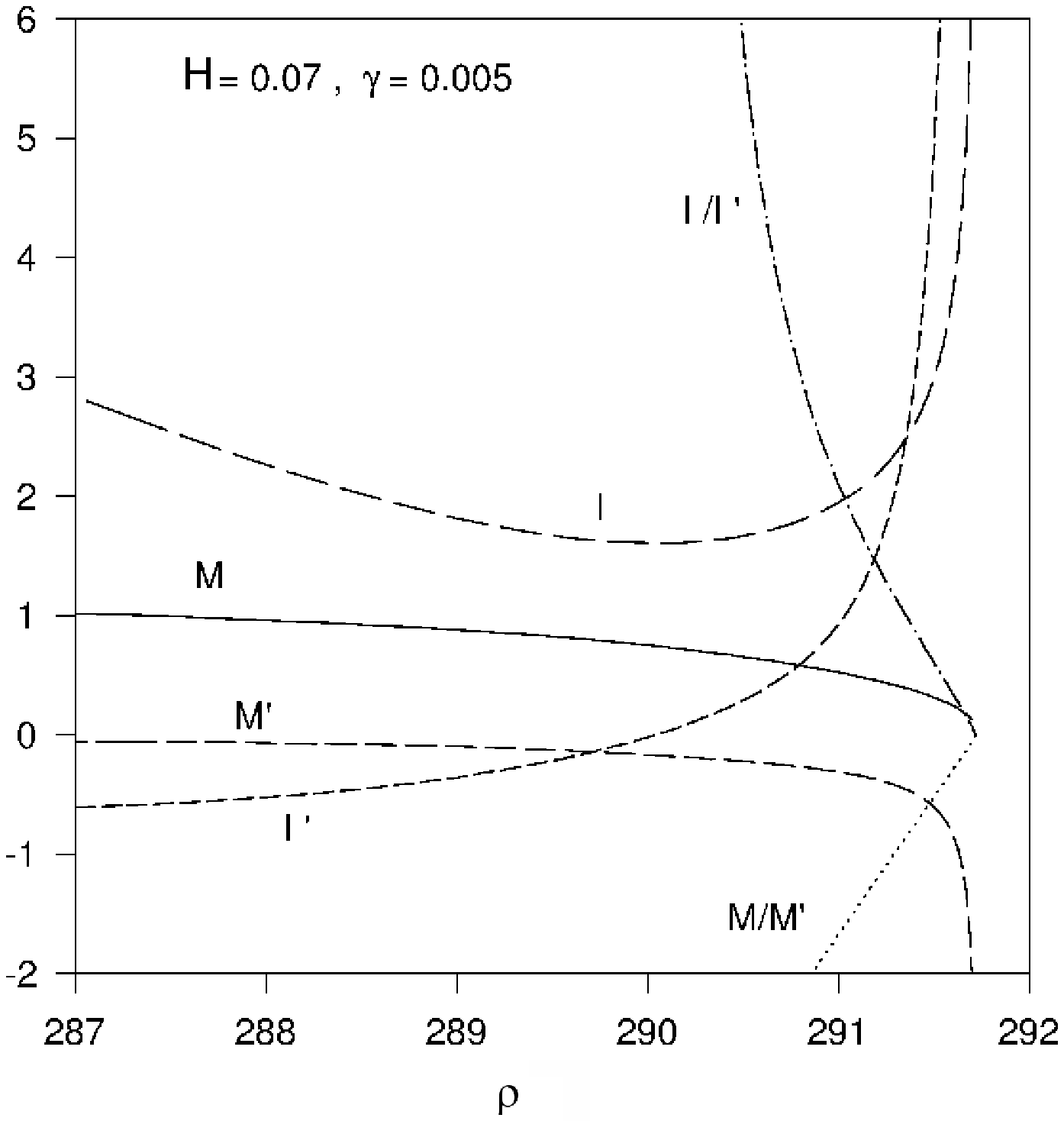}}
\caption{Profile of metric functions for a singular solution.}
\end{figure}

We studied the solution for $\gamma >0$ and found qualitatively how the vacuum trigonometric solutions are deformed by the matter fields. This is illustrated in figure \ref{fig:fig6p2}. As expected, we observe that the monopole regularizes the singular vacuum solutions such that the spacetime become regular on the brane $\rho=0$. 
The functions $M$, $l$ reach their maximum value when the matter fields have already reached their asymptotic values.
The mirror symmetry of the vacuum solutions is broken, in particular the numerical analysis reveals that the two functions $M$, $l$ do not reach their maximum exactly at the same value of $\rho$. It is likely that mirror symmetric solution exist in this case as well for tuned values of the constants $H$, $\gamma$. Mirror symmetric solutions will be discussed in the next section.

The challenging question is to understand how these  solutions (regularized at the origin) behave in the neighbourhood of the first period of the 
associated periodic solution of the vacuum equation. 

This is illustrated in figure \ref{fig:fig7p2}, which clearly indicates that the solution develops a singularity at some $\rho=\rho_c$. For $H=0.07,\gamma=0.005$ we have $\rho_c\approx 291.7$.
A detailed analysis of the numerical solution for $\rho \approx \rho_c$ reveals that the singularity is of Melvin type discussed in section \ref{secmelvin}. For generic values of the inflating parameter and positive bulk cosmological constant, the local monopole solutions present a singularity at some finite value of the radial variable relative to the extra dimensions.

%%%%%%%%%%%%%%%%%%%%%%%%%%%%%%%%%%%%%%%
\subsection{Mirror symmetric solutions}
%%%%%%%%%%%%%%%%%%%%%%%%%%%%%%%%%%%%%%%
In the previous sections we reported solutions obtained for generic values of the different coupling constants. The counterpart of the vacuum periodic solutions turned out to be irregular. However it could be that regular solutions exist for specific values of $H$, $\gamma$. Mirror symmetric solutions, like the ones obtained in the context of global monopoles are good candidates. It turns out that also in the case of local monopoles, it is possible to tune $H$ and $\gamma$ in such a way that a mirror symmetric solutions exist.
In order to construct such solution, we imposed the constraints  $\phi=0$, $w=0$ $M'=0$, $l'=0$ as supplementary boundary conditions at some finite value $\rho=\rho_c$; then we tried to determine numerically if the values of $H, \gamma$ can be adjusted for all constraints to be fulfilled. It turns out to be possible.

For fine tuned  values of the two cosmological constants, indeed, our numerical integration of the equations exhibits solutions such that the function $\phi(\rho)$ bends backwards and reaches the value $\phi=0$ at $\rho=\rho_c$ after developing a plateau where $\phi \sim 1$ for $0 < \rho < \rho_c$. At the same time  the function $w(\rho)$ and the derivatives of the metric functions $l$ and $M$, $l'$, $M'$ approach zero for $\rho \to \rho_0$.

\begin{figure}
\centering
\subfigure[\label{fig:fig8p2}
The profiles of the metric functions $M,l$ and their derivatives
for a periodic monopole solution. Note that the slope of the functions $M,l$ vanish for $\rho$ close to $100$ on the figure.]{\includegraphics[scale=.39]{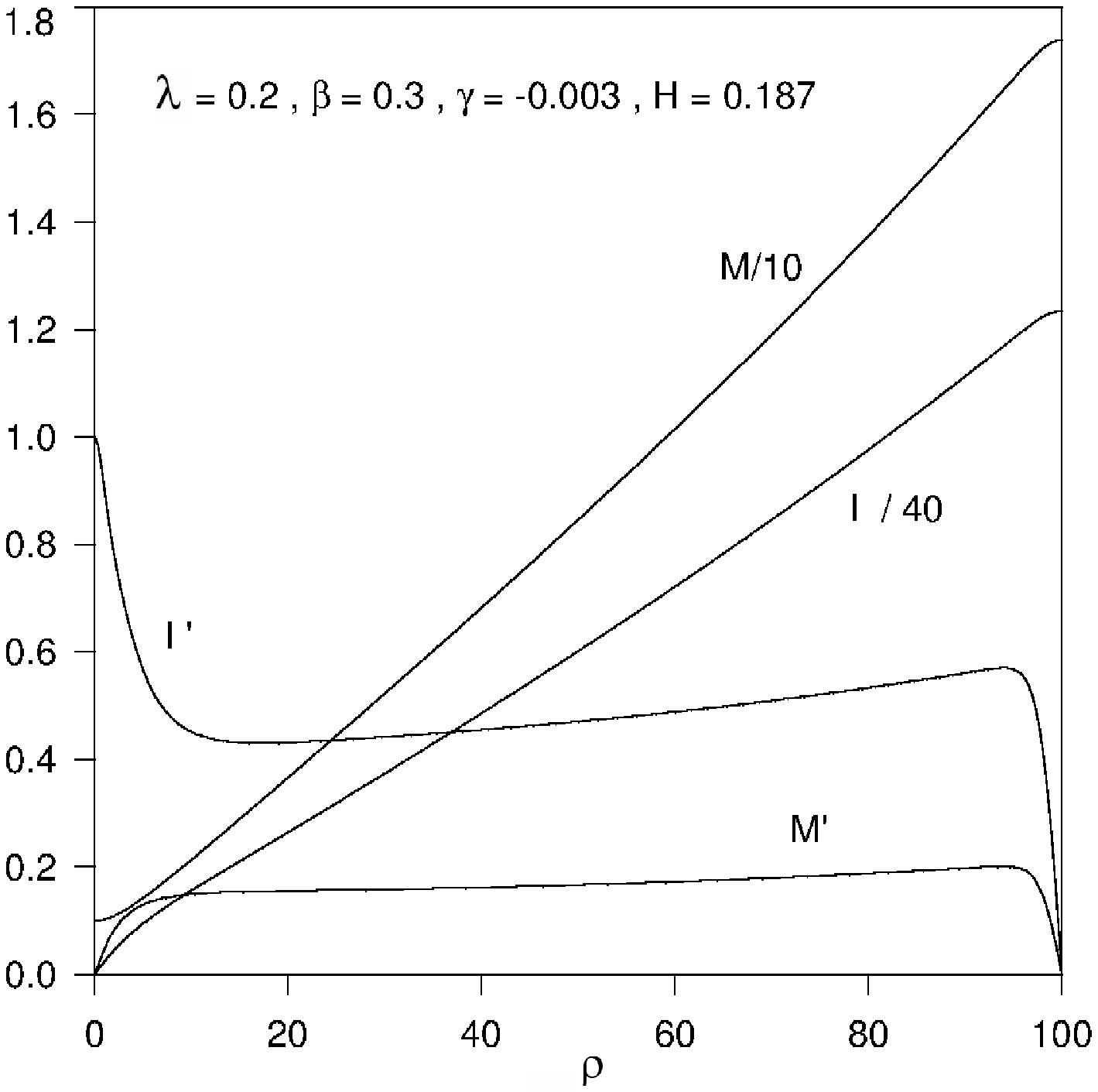}}
\hspace{.1cm}
\subfigure[\label{fig:fig9p2}
The profiles of the matter functions $w,\phi,\phi'$
and the derivatives $M',l'$ for a periodic monopole solution.]{\includegraphics[scale=.39]{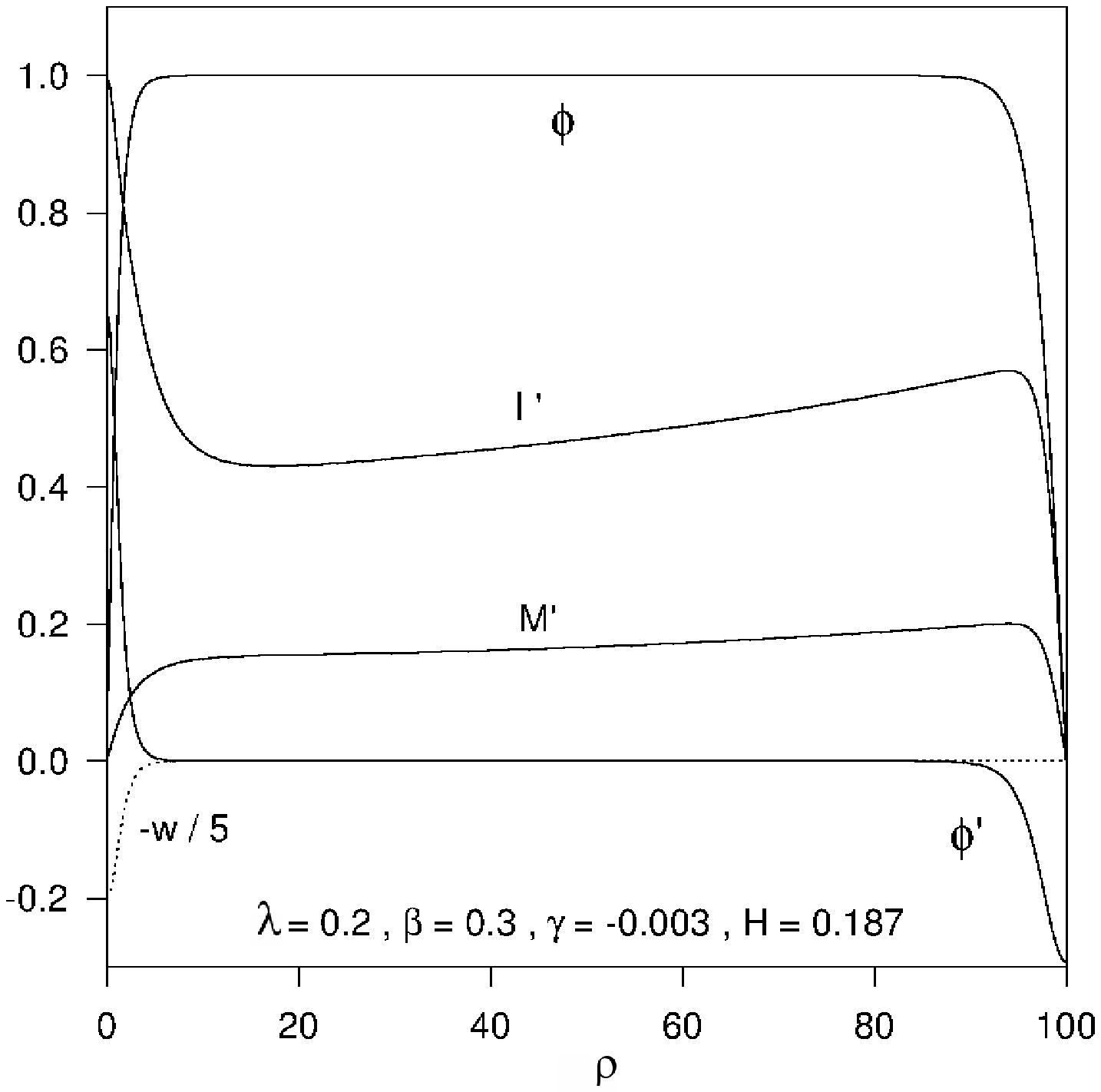}}
\caption{Profiles of metric functions for the periodic monopole solutions.}
\label{fig:fig8-9p2}
\end{figure}

An example of such a solution is given in figure \ref{fig:fig8-9p2}. It corresponds to $\rho_c = 100$ which fixes $\gamma \approx -0.003$,
$H \approx 0.187$. We constructed a branch of solutions for lower values of $\rho_c$ and obtained , e.g. for $\rho_c=50$ the values $\gamma \approx 0.018$, $H \approx 0.217$.
The sign of the bulk cosmological constant $\gamma$ leading to these kind of solutions seems to be negative for $\rho_c > 82$ and positive for $\rho_c < 82$. The value $H$ is positive and  does not vary significantly with $\rho_c$.

The properly speaking mirror symmetric solutions can then be obtained by extending the numerical solution discussed above by mirror symmetry for $\rho \in [0,2 \rho_c]$ ($M,l,w$ are symmetric under the reflection $\rho \to 2 \rho_c-\rho$ while $\phi$ is antisymmetric). This is possible by using the discrete symmetries of the equations as discussed above. These mirror symmetric solutions can further be extended periodically on the interval $\rho\in [0, 2p \rho_c]$ for $p$ integer. The periodicity of the function $l(\rho)$ suggest that the geometry of these solutions in the extra dimension consists of the surface of a deformed 3-sphere. We can then interpret the radial coordinate $\rho$ as the "colatitude" and the solution looks like a magnet (with non-abelian fields) whose positive and negative poles lay respectively on the north ($\rho=0$) and south ($\rho=2\rho_c$) poles of the sphere.

\begin{figure}
\centering
\includegraphics[scale=.4]{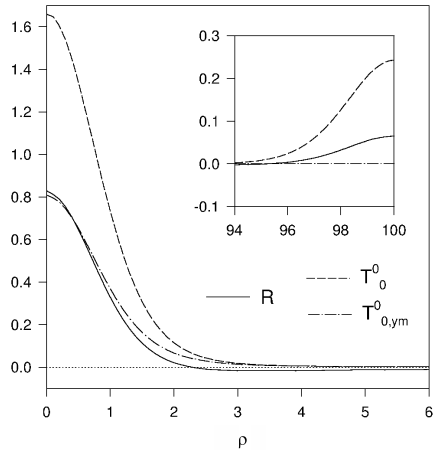}
\caption{\label{fig:fig10p2}
The Ricci scalar $R$, the energy momentum tensor $T^0_0$ and the Yang-Mills contribution to the energy $T^0_{0,ym}$ corresponding to the periodic monopole of figure \ref{fig:fig8p2}, \ref{fig:fig9p2}.}
\end{figure}

In the region of the equator ($\rho=\rho_c$) the Higgs field form a domain wall which separates these two poles. This is illustrated by figure \ref{fig:fig10p2} where the Ricci scalar $R$ (see Eq. (\ref{ricci})), the full energy density $T^0_0$ and the contribution to the energy of the Yang-Mills fields $T^0_{0,ym}$ of the solution of figure \ref{fig:fig8-9p2} are represented in the north-pole region and in the equator region. In particular, we see that only the Yang-Mills energy is zero in the equator region, so that the energy content in this region is due to the Goldstone boson.

%%%%%%%%%%%%%%%%%%%%%%%%%%%%%%%%%%%%%%%%%%%%%%%%%%%%%%%%%%%%%%%%%%%%%%%%%%%%%%%%%%

%%%%%%%%%%%%%%%%%%%%%%%%%%%%%%%%%%%%%%%%%%%%%%%%%%%%%%%%%%%%%%%%%%%%%%%%%%%%%%%%%%

\section{Gravitating baby Skyrme model}
In this final section, we consider another type of solitonic matter fields, namely the baby-Skyrme model. The standard baby-Skyrme model is formulated in $2$ spatial dimensions and admits a solution: the baby-Skyrmion.

The original Skyrme model was proposed in the 60's \cite{skyrme} as an effective model to describe the pion-nucleon interaction. It is based on a mapping between the $3+1$ dimensional spacetime and the Lie group $SU(2)$, which is known to be equivalent to the manifold $S_3$. The model admits a fundamental soliton, the Skyrmion which is interpreted as a nucleon. The Skyrme model and the corresponding classical equations are notoriously difficult. For instance, the description of nuclei of high baryonic number is possible but technically difficult to construct since it involves several interacting Skyrmions.

The baby-Skyrme model was proposed in the 90's \cite{babyskyrme} in order to provide a toy model where the theoretical ideas of the grand brother Skyrme model could be tested. The baby-Skyrme model is constructed for a $2+1$ dimensional spacetime and involves fields taking values on the manifold $S_2$. Skyrme and baby-Skyrme models are non linear sigma models. 

In our framework, the spatial dimensions of the various models we considered were interpreted as the extradimensions. It is therefore natural to consider a spacetime in $6$ dimensions and to interpret the baby Skyrmion as the localized brane. The $2$ extradimensions are the spatial dimensions where the baby Skyrme resides, following the philosophy of the various extended brane models considered in this chapter. The case of the static baby-Skyrme branes has already been considered in \cite{sawa,sawa2}. Here we will focus on the case where the brane are inflating.

In this section, we used a modified version of the baby-Skyrme model: 
\begin{itemize}
 \item we consider the static version of the model
 \item we replace the usual $\mathbb R^2$ space by the extradimensional part of the six dimensional metric \eqref{ggenans} with $n=2$.
\end{itemize}
Modified in this way, the relevant action of the baby-Skyrme model reads
\be
S_m= \int\sqrt{-g}\left[-\frac{\kappa_2}{2}(\partial_i \vec n)\cdot(\partial^i\vec n) - \frac{\kappa_4}{4}(\partial_i\vec n\times\partial_j\vec n)(\partial^i\vec n\times\partial^j\vec n) - \kappa_0 V(\vec n)\right]d^6x,
\label{sgbs}
\ee
where $\kappa_0,\kappa_4,\kappa_2$ are coupling constants, $\vec n$ is a three dimensional vector field taking values on the unit sphere (i.e. $\vec n^2=1$), $V$ is the potential, which will be detailed later, $g_{ij}$ is the six dimensional spacetime metric \eqref{ggenans} with $n=2$.

Note that following the modification of the model, $\vec n$ depends only on the extradimensional coordinates, i.e. $r, \theta$. It follows that the action can be written in the form
\bea
S_m&=& \int\int\sqrt{-g_4}\sqrt{h}h^{ij}\left[-\frac{\kappa_2}{2}(\partial_i \vec n)\cdot(\partial_j\vec n) - h^{ik}h^{jl}\frac{\kappa_4}{4}(\partial_i\vec n\times\partial_k\vec n)(\partial_j\vec n\times\partial_l\vec n)\right.\nonumber\\
&-&\left.  \kappa_0 V(\vec n)\right]drd\theta d^4x,
\label{sbsk}
\eea
where $g_4$ is the determinant of the $4$ dimensional part of the metric and $h$ is the determinant of the extradimensional part of the metric, given by
\be
ds^2_{ED} = dr^2 + L(r)^2 d\theta^2.
\ee
It should be noted that the four dimensional part of the metric somehow plays a spectator role, since the vector field $\vec n$ doesn't depend on the four dimensional coordinates. However, the warp factor $M$ will contribute non trivially.

The aim of this final section is to investigate the counterparts of the three sorts of extradimensional geometries due to the two cosmological constants, in a different context. The first three section considered models where Higgs fields were present; here it is not the case. As we will see, the geometry of the extradimensions will again be opened, closed or flat.

\subsection{The ansatz and boundary conditions}
The baby Skyrmion is usually constructed by using the hedgehog ansatz for the vector $\vec n$:
\be
\vec n = \left( \sin(f(r)) \cos(n\theta),\sin(f(r))\sin(n\theta), \cos f(r)  \right),
\ee
where $n$ is an integer and $f(r)$ an arbitrary function. We supplement the ansatz with the following boundary conditions:
\be
f(0)= -(m-1)\pi,\ f(\infty) = \pi,
\ee
where $m$ is also an integer. The condition on $f(0)$ is necessary for the mapping to be well defined at the origin and the condition on $f(\infty)$ ensures that all points at infinity in the extradimensional manifold are mapped to the south pole of the sphere.

Note that this is a parametrisation of $S_2$, where $f(r)$ plays the role of an colatitude angle and $\theta$ an azimuthal angle; it is clear from the ansatz that $\vec n^2=1$ for all values of $f(r)$ and $\theta$. However, it is important to stress that the presence of the integer $m$ and $n$ have the consequence that $\vec n$ covers many times the sphere: $m$ times in the colatitude direction and $n$ times azimuthally.

The winding number $N$ of the mapping $\vec n$ is a topological invariant, defined by
\be
N = \frac{1}{4\pi}\int \vec n\cdot \left( \partial_r\vec n\times\partial_\theta\vec n \right)drd\theta = \frac{n}{2}\left( 1- (-1)^m \right)\in \mathbb Z.
\ee
This is nothing else than the flux of the vector $\vec n$ thought the sphere.

We choose the potential term to be \cite{sadelbri}
\be
V(\vec n) = 1 - \vec n\cdot\vec n^{(\infty)} = 1+\cos(f(r)),
\ee
where $\vec n^{(\infty)} = \lim_{r\rightarrow\infty} \vec n=(0,0,-1)$ is the vacuum configuration of the baby Skyrme.

As usual, the boundary conditions for the metric fields are given by
\be
M(0)=1,\ M'(0)=0,\ L(0)=0,\ L'(0)=1.
\ee

\subsection{Equations of the baby-Skyrme brane}
First we introduce the following dimensionless coordinates and quantities
\bea
r    &=& \sqrt{\frac{\kappa_{2}}{\kappa_{4}}}\rho,\ l(r) = \sqrt{\frac{\kappa_{2}}{\kappa_{4}}}\,L(r),\\
\alpha &=& 8\pi G_6\kappa_{2},\ \beta = \frac{\Lambda_6\kappa_{4}}{\kappa_{2}^{2}},\ \gamma =\frac{3H^2\kappa_{4}}{\kappa_{2}},\ \mu =\frac{\kappa_{0}\kappa_{4}}{\kappa_{2}^{2}}
\eea
and introduce the functions $u(r), v(r)$ such that
\be
u(r)= 1 + \frac{n^{2}}{L^{2}(r)}\sin^{2}f(r),\ v(r)= 1 - \frac{n^{2}}{L^{2}(r)}\sin^{2}f(r).
\ee

Then, the equations of motions are given by:
\begin{align}
& uf'' + \left( 4\frac{M'}{M} + \frac{L'}{L} + \frac{u'}{u} \right)uf'- (1 + f'^{2})\frac{n^{2}}{L^{2}}\sin f\cos f + \mu\sin f = 0,\nonumber\\
& 3\frac{M'^{2}}{M^{2}} + \frac{L''}{L} + 3\frac{M'L'}{ML} + 3\frac{M'}{M} - \frac{\gamma}{M^{2}}	= \alpha( \tau_{0} - \beta ),\nonumber\\
& 6\frac{M'^{2}}{M^{2}} + 4\frac{M'L'}{ML} - 2\frac{\gamma}{M^{2}}= \alpha( \tau_{r} - \beta ),\nonumber\\
& 4\frac{M''}{M} + 6\frac{M'^{2}}{M^{2}} - 2\frac{\gamma}{M^{2}} = \alpha( \tau_{\theta} - \beta ).
\end{align}
where
\bea
\tau_{0}&=& -\frac{1}{2}f'^{2} - \frac{1}{2}\frac{n^{2}}{L^{2}}\sin^{2}f- \frac{1}{2}\frac{n^{2}}{L^{2}}f'^{2}\sin^{2}f - \mu(1 + \cos f),\nonumber\\
\tau_{r}&=& \frac{1}{2}f'^{2} - \frac{1}{2}\frac{n^{2}}{L^{2}}\sin^{2}f+ \frac{1}{2}\frac{n^{2}}{L^{2}}f'^{2}\sin^{2}f - \mu(1 + \cos f),\nonumber\\
\tau_{\theta}&=& - \frac{1}{2}f'^{2} + \frac{1}{2}\frac{n^{2}}{L^{2}}\sin^{2}f+ \frac{1}{2}\frac{n^{2}}{L^{2}}f'^{2}\sin^{2}f - \mu(1 + \cos f)
\eea
are components of the dimensionless energy-momentum tensor.

\subsection{Asymptotic expansion}
\subsubsection{Large $r$ limit}
In the limit where $r$ becomes very large, the function $f$ tends to the vacuum configuration, $f=\pi$. The asymptotic solutions for the functions $M$ and $L$ are then the one described in section \ref{6dspecialsolutions}.

The decay of the functions $f,M,L$ can be computed easily for $\beta<0$ and $\beta=0$ by considering the vacuum configuration plus a correction and keeping the dominant terms. Doing so, we find
\bea
\delta f&\approx& f^-_1 e^{ -\frac{r}{4}\left(\sqrt{-10\alpha\beta} +\sqrt{-10\alpha\beta+16\mu}\right)} +f^-_2 e^{ -\frac{r}{4}\left(\sqrt{-10\alpha\beta} -\sqrt{-10\alpha\beta+16\mu}\right)}\nonumber\\
\delta M&\approx& M^-_1 e^{\sqrt{-\alpha\beta/10}r} + M^-_2 e^{-\sqrt{-8\alpha\beta/5}r},\\
\delta L&\approx& L^-_1 e^{\sqrt{-\alpha\beta/10}r} + L^-_2 e^{-\sqrt{-8\alpha\beta/5}r},\nonumber
\eea
for $\beta>0$ and where $\delta M,\ \delta L$ and $\delta f$ denote the correction around $M(r),\ L(r)$ and $f(r)$, i.e. 
\bea
M(r) &=& \sqrt{-10\gamma/(3\alpha\beta)}\sinh(\sqrt{-\alpha\beta/10}r) + \delta M(r),\nonumber\\
L(r) &=& L_0 M'(r) + \delta L(r),\\
f(r) &=& \pi + \delta f(r).\nonumber
\eea

The parameters $f^-_1,f^-_2,M^-_1,M^-_2,L^-_1,L^-_2$ are arbitrary constants. Of course, we are looking for the modes with $M^-_1= f^-_2 =L^-_1=0 $ such that $\delta f,\ \delta M$ and $\delta L$ vanish asymptotically and are indeed fluctuations.

For $\beta=0$, however, it is possible to solve for the corrections without approximations:
\bea
\delta f &=& f^0_1 x^{-\frac{3}{2}}K_{\frac{3}{2}}(x),\nonumber\\
\delta M &=& M^0_1 + \frac{M^0_2}{r^2},\\
\delta L &=& L^0_1 + \frac{L^0_2}{r^2},\nonumber
\eea
where $x = r\sqrt{\frac{n^2}{L_0^2\gamma} + \mu}$, $f^0_1, M^0_1, M^0_2, L^0_1,L^0_2$ are arbitrary constants and $K_{n}(x)$ is the modified Bessel function of second kind.

Note that for large $r$, the function $f$ decays as $\pi+f_1\frac{e^{-r\sqrt{\frac{n^2}{L_0^2\gamma} + \mu}}}{r^2}$ for a real constant $f_1$.

\subsubsection{Expansion around the origin}
Close to the origin, the expansion of the functions $f, L, M$ depends on $m$ and $n$. It is possible to give the leading terms in a generic form, but the case $n=0$ has to be considered separately:
\bea
f(r)&=& -(m-1)\pi + f^{(1)} r + \Ord{r}{2},\\
L(r)&=&r+\frac{\alpha  \left(-5 (f^{(1)})^4-4 (f^{(1)})^2-2 (-1)^m \mu +2 \beta +2 \mu \right)-8 \gamma}{4}\frac{r^3}{3!} +\Ord{r}{4},\nonumber\\
M(r)&=&1+\frac{2 \alpha  \left(2 (f^{(1)})^4+2 \left((-1)^m-1\right) \mu -2 \beta \right)+8 \gamma}{4} \frac{r^2}{2!}+\Ord{r}{3}.\nonumber
\eea

In the case where $n>0$, we can write the first terms in a general form:
\bea
f(r)&=& -(m-1)\pi + f^{(n)} r^n + \Ord{r}{n+1},\nonumber\\
L(r)&=&r+\frac{-2\gamma-\alpha\left(\beta+\mu-{\left(-1\right)}^m\mu\right)}{4}\frac{r^3}{3!}+\Ord{r}{4},\\
M(r)&=&1+\frac{2\gamma-\alpha\left(\beta+\mu-{\left(-1\right)}^m\mu\right)}{4}\frac{r^2}{2!}+\Ord{r}{3},\nonumber
\eea
where $f^{(1)}, f^{(n)}$ are real constants.

\subsection{Numerical solutions}
We solved the system of ordinary differential equations numerically with the solver Colsys \cite{colsys} for many values of the parameters. Due to the large number of parameters, we decided to adopt the following approach: we keep fixed the value of the gravitational coupling $\alpha$, the bulk cosmological constant $\beta$ and $m,n$. Then we vary the Hubble factor $\gamma$ for different values of the Skyrme coupling $\mu$; the way we treat the problem allows to have a direct visualisation of the influence of $\gamma$ on the solutions.

In the analysis, we focused on the energy density of the Skyrmion $E$ and its mean square radius $MSR$ \cite{skft}:
\be
E = 2\pi \int_0^\infty T_0^0 L(r)dr\ \ ,\ \  MSR = \int_0^\infty r^2 f'(r)\sin^2f(r) dr.
\ee
The mean square radius allows to characterise the extension of the brane in the transverse direction; the more $MSR$ is small, the more the brane is localized.

\begin{figure}
\centering
\includegraphics[scale=.34]{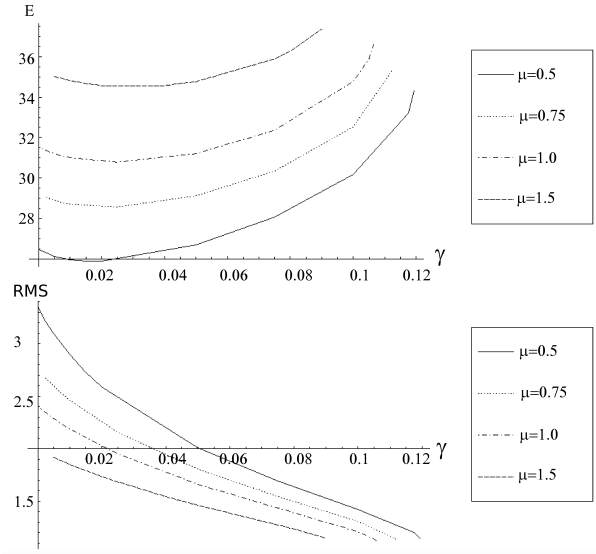}
\caption{The value of the energy and of the mean square radius as a function of $\gamma$ for various values of $\mu$ and for $\alpha = 0.1,\ \beta=0.1$.}
\label{fig:0101}
\end{figure}

\begin{figure}
\centering
\includegraphics[scale=.38]{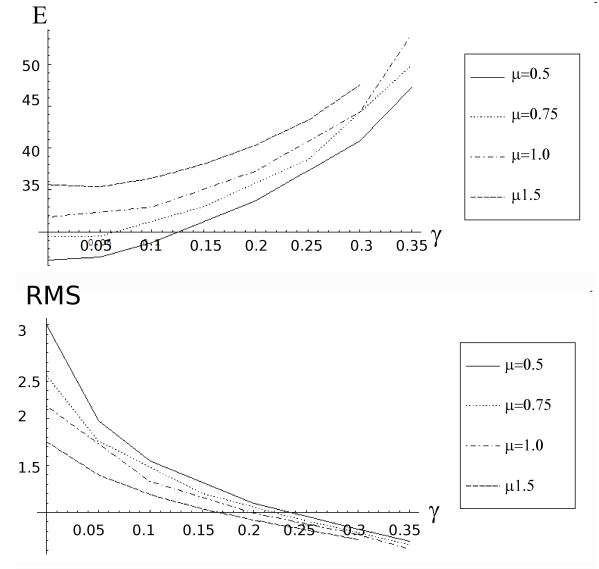}
\caption{The value of the energy and of the mean square radius as a function of $\gamma$ for various values of $\mu$ and for $\alpha = 0.05,\ \beta=0.1$.}
\label{fig:00501}
\end{figure}

Our results are presented in figure \ref{fig:0101} and \ref{fig:00501} for $\alpha = 0.1,\ \beta=0.1$ and $\alpha = 0.05,\ \beta=0.1$ respectively and $m=n=1$ for both cases. Note that these are preliminary results, a more detailed analysis, including stability is in preparation \cite{sadelbri}.

For fixed $\alpha,\ \beta,\ \mu $, the mean square radius decreases when the value of $\gamma$ increases, while the energy decreases for small values of $\gamma$ and increases for larger values. In some intermediate values of $\gamma$, the energy of the baby-Skyrmion passes thought a minimum. The minimum occurs at smaller values of $\gamma$ when the value of $\mu$ is smaller. Note also that increasing values of $\mu$ leads to decreasing values of the mean square radius and decreasing values of the energy.

\begin{figure}[H]
\centering
\includegraphics[scale=.45]{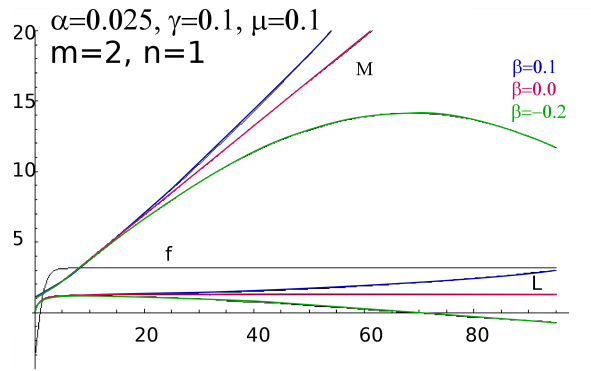}
\caption{The typical profiles of the solutions for different signs of $\beta$ for $m=2$. The function $f$ tends quickly to its asymptotic value, so it is not possible to distinguish the different profiles for the function $f$.}
\label{fig:prof m=2}
\end{figure}

\begin{figure}[H]
\centering
\includegraphics[scale=.45]{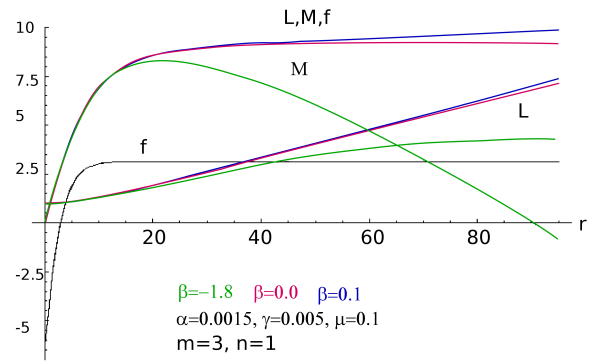}
\caption{The typical profiles of the solutions for different signs of $\beta$ for $m=3$. Here again, it is not possible to distinguish the different profiles for the function $f$.}
\label{fig:prof m=3}
\end{figure}

We present typical profiles of the solutions for $m=2,3$ in figure \ref{fig:prof m=2} and \ref{fig:prof m=3} respectively for non vanishing values of the parameters. These figures show that there are three possible geometries depending on the sign of the cosmological constant, as in section \ref{secinflbranekappaneq0}; namely opened ($\beta>0$), flat ($\beta=0$) and closed ($\beta<0$). Again, all three geometries have an angular deficit.

\section{Concluding remarks}
In this chapter, we investigated the effect of both a bulk cosmological constant and a brane cosmological constant on essentially four classes of warped topological braneworld models: the gauged Einstein-abelian-Higgs model, the gauged and global Einstein-non abelian-Higgs models and the Einstein-baby-Skyrme model.

We reviewed solutions available in a wide region of parameters, giving a classification for the solutions to the $6$-dimensional abelian Higgs brane model, and we presented new solutions where the branes are inflating, completing the classification. We also considered models with more than $6$ dimensions, since the case $d=6$ is particular. 

The four dimensional cosmological constant have been chosen to be positive in accordance with the current observational status and was modeled by inflating branes. Note that the four-dimensional slices of the generic metric \eqref{ggenans} can be replaced by any Einstein four dimensional spacetime (i.e. a spacetime satisfying $R_{ab}=C g_{ab}$, $C\in\mathbb R$) with positive curvature, since only the $4$ dimensional scalar curvature appears in the equations of motions.

It should be stressed that the inclusion of such a cosmological constant has heavy consequences on the solutions available in the models. It seems to be a generic feature that the geometry of the extradimensional spacetime with positive cosmological constant on the brane is either opened, closed or flat, with deficit angles. This is comforted by the vacuum solutions presented in section \ref{secbranevaccum}, which are precisely of opened, closed or flat extradimensional geometry. Far from the brane core, the matter fields should asymptote their vacuum value, so the metric functions should approach the vacuum solutions asymptotically. 

This idea has been checked on different models, three of them (Einstein-Maxwell-Higgs, Einstein non abelian Higgs, Einstein Yang-Mills Higgs) contained a Higgs field, the fourth is somehow different in its nature (baby-Skyrme model). In every cases, we found the three sorts of extradimensional geometries.

Note also that the effect of considering branes with an extension in the extradimensions somehow regularizes the metric functions, as it was already foreseen in the previous chapter.

%% file: fermions.tex
In this chapter, we consider a model describing an interaction between a boson and a fermion field in $1+1$ dimensions. The procedure will actually consist in considering solutions available in the bosonic sector of the theory and then coupling the fermionic fields to these \emph{background} solutions. The bosonic sector is chosen in such a way that it admits localized 'bell-shaped' classical solutions.

This kind of models can be seen as toy models where the scalar fields mimics a brane (and/or an antibrane) on which we investigate bound (localized) states of fermions. In the first section, we will discuss the fermionic modes while in the second section, we will study the stability of three particular bell shaped scalar field solutions and show that the stability equation reduces to the Poschl-Teller equation (see appendix \ref{ptequation}), which admits analytic solutions. These two points constitute our original results, published in \cite{bdbellshape}.

\section[Fermions on kink, kink anti-kink]{Fermionic modes on kink and kink-anti-kink systems}
Up to now, we considered brane models described by gravitating topological defects such as cosmic strings or monopoles. However, realistic matter is of fermionic nature. The goal of this section is to consider a toy model of a topological defect seen as a brane and to couple fermionic fields to it. The simplest way to do so is to consider something which resemble a domain wall or a system of wall/anti-wall (see \cite{cosmicstrings}) in $1+1$ dimensions. Among other, one possibility is to consider the Sine-Gordon model \cite{bdbellshape}. It is well known that the Sine-Gordon model admits kink and anti-kink solutions, as well as superposition of these solutions. The solution of interest for our purpose is the kink-anti-kink system.

Previous work has been done in this direction, considering a $\phi^4$ model, where kink or anti-kink are available \cite{vachaspati}. The author of \cite{vachaspati} studied coupling of fermions to a single kink and to a superposition of the kink and anti-kink solution. However, the superposition is valid as long as the kink and anti-kink are distant enough. We will briefly review their result before turning to the Sine-Gordon model. But first, we present the formalism used throughout this section.

\subsection{Fermion - Boson coupling in $1+1$ dimensions}
The model we consider describes a coupling between a bosonic field to a fermionic field and an effective mass for the fermion generated by a Yukawa interaction of the fermion with the scalar field $\phi$. Using the notations of Chu-Vachaspati \cite{vachaspati}, the lagrangian is given by (with the metric signature $(+,-)$)
\be
{\cal L}=\frac{1}{2}\partial_{\mu}\phi\partial^{\mu}\phi-V(\phi) + i \overline \psi \gamma^{\mu} \partial_{\mu} \psi- g (\phi-C) \overline \psi 
\psi 
\label{lagrangian_full}
\ee
where $\psi$ is a two component spinor and where the two dimensional Dirac matrices can be chosen in terms of the Pauli matrices $\gamma^0 = \sigma_3$, $\gamma^2=i \sigma_1$, $g$ is the Yukawa coupling constant and the shift constant $C$ has to be chosen appropriately according to the vacuum expectation value of the scalar field. $V$ is a given potential, defining the model considered for the bosonic part of the theory.

Let $\phi_c$ be a classical solution of the bosonic equation derived from \eqref{lagrangian_full}, i.e. a solution of
\be
\Box \phi - \ddf{V}{\phi}=0.
\ee

In order to solve the Dirac equation in the background of a classical solution $\phi_c$ available in the bosonic sector of the model, it is convenient to parameterize \cite{vachaspati} the Dirac spinor according to 
\be
  \psi = \left(\begin{array}{c}
 \psi_1\\
  \psi_2
\end{array}\right)\ , \ \psi_1 = e^{-iEt} (\beta_+ - \beta_-) \ , \ \psi_2 = e^{-iEt} (\beta_+ + \beta_-),
\ee
where $\beta_{\pm}$ are functions of $x$. 

Then the system of first order Dirac equations is transformed into two decoupled second order equations for $\beta_-$ and $\beta_+$~:
\be
(-\partial_x^2 + V_{\pm}(\phi_c)) \beta_{\pm} = E^2 \beta_{\pm} \ , \  V_{\pm} (\phi_c) = g^2 \phi_c^2 \mp g \partial_x \phi_c.
\label{dirac}
\ee

\subsection{Coupling to a $\phi^4$ kink}

In the case where the potential $V$ is given by
\be
V(\phi)=\frac{1}{2}\left( \phi^2 - 1 \right)^2,
\ee
the bosonic part of the lagrangian \eqref{lagrangian_full} admits static localised solution given by
\be
\phi(x) = \pm \tanh x.
\label{kinksol}
\ee
The solution corresponding to the plus sign is called a kink while the solution with the minus sign is an anti-kink. The solutions, kink or anti-kink, extrapolate between the two vacua of the potential $\phi_\pm = \pm 1$, between $x\rightarrow-\infty$ and $x\rightarrow+\infty$.

The Dirac equation in the background of one kink corresponding to this model was studied in detail in \cite{vachaspati}. It was found in particular that the equations \eqref{dirac} with the solution \eqref{kinksol} are Poschl-Teller equations respectively with $N=g$ and $N=g-1$ and $ \omega_p^2 = E^2- g^2$. 
For any value of $g$ such that $n-1 < g \leq n$ there exist $n$ normalisable solutions with eigen energy
\be
 E_j = \sqrt{j(2g-j)} \ \ , \ \ j = 0,1, \dots , n-1
\ee
At each integer value of $g$, a supplementary bound state emerges from the continuum $E=g$ and exist for  $g > n$.

Next, the Dirac equation was studied in the background of a kink-anti-kink configuration say $\phi_{K\overline K}$ approximated by a superposition of a Kink centred at $x = -L$ and an anti-Kink centred at $x=L$ with $L \gg 1$:
\be
\label{superposition}
      \phi_{K\overline K}(x) = \tanh(x+L) - \tanh(x-L) - 1.
\ee
Note that this is indeed a good approximation of a kink/anti-kink solution, due to the localized properties of the solution. However, if the separation $L$ between the kink and anti-kink becomes too small, then the approximation does not hold any more.

Configurations like \eqref{superposition} are "bell-shaped", in the sense that the graph of the solutions looks like a bell. 

The main result is that the spectrum of the Dirac equation in the background of $\phi_{K\overline K}$ deviates only a little from the spectrum available in the background of a single Kink (or an anti-Kink) for $L \gg 1$. The analysis of the spectrum in the region $L\sim 1$ is unreliable since the linear superposition \eqref{superposition} is an approximate solution of the classical equation only for large values of $L$.

\subsection{Coupling to the Sine Gordon kink and kink-anti-kink}
Since the considerations of the previous section cannot be extended in the region where the kink and anti-kink are too close, we reconsidered the spectral problem in a different theory, namely the Sine-Gordon model where explicit solution are known for the kink, anti-kink as well as for a kink-anti-kink.

The potential in the Sine-Gordon model is given by
\be
V(\phi) = \frac{1 - \cos \phi}{2},
\ee
and the kink and anti-kink solutions are given by
\be
\phi = \pm 4 {\rm Atan}(e^x),
\label{kinksinegordon}
\ee
the $+$ (resp. $-$) sign describing the kink (resp. the anti-kink).

\begin{figure}[H]
\centering
\includegraphics[scale=.5]{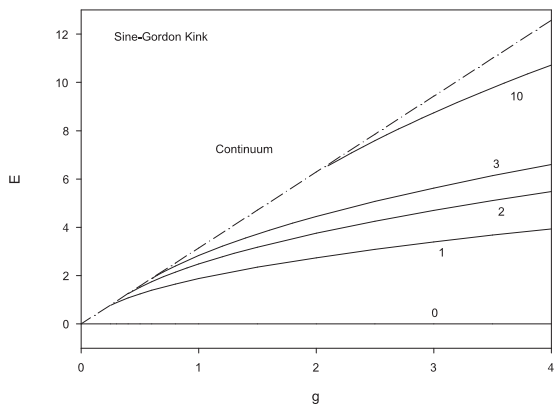}
\caption{The fermionic bound state corresponding to the single Sine-Gordon kink as function
of the constant $g$. The level $n=0,1,2,3$ and $10$ are represented.}
\label{fig:fig1}
\end{figure}

We first solved equation \eqref{dirac} in the background of a single kink \eqref{kinksinegordon}. On figure \ref{fig:fig1}, a few fermionic bound states are represented as function of the parameter $g$ (for clarity, we omitted the modes $n=4,\dots 9$ on the graphic). The figure present exactly the same pattern than in the case of the $\phi^4$ kink. In particular, new bound states emerge regularly from the continuum at critical values of $g$, say $g=g_n$. This is in the case of the single kink solution.

%%%%

%Let us mention that the normal modes about the sphaleron and bisphaleron solutions \cite{brihaye_kunz1} of the standard model of electroweak interactions also lead to a similar pattern. 

Because these qualitative properties of the solutions are similar to the one of the $\phi^4$ kink we expect the features discovered in \cite{vachaspati} to hold in the case of well separated sine-Gordon kink and anti-kink. In the sine-Gordon model, we can take advantage of the fact that an exact form of the kink-anti-kink solution is available:
\be
\phi_{K \overline K} = 4 {\rm Atan} \frac{\sinh(ut/\beta)}{u\cosh ( x/\beta)} \ , \ \beta = \sqrt{1-u^2},
\label{kak}
\ee
where $u$ is a constant related to the relative velocity of the two lumps. We used this solution to study the Dirac equation in the background of a moving wall-anti-wall system (approaching or spreading each other, according to the sign of $t$) and were able to study the spectrum of the bound states in the domain of the parameter $t$ become small, i.e. when the kink and the anti-kink interact. 

\begin{figure}[H]
\centering
\includegraphics[scale=.5]{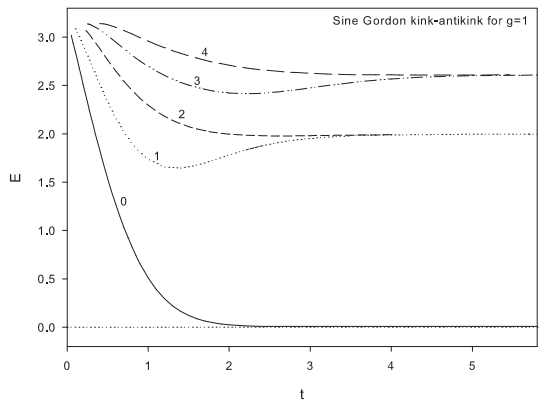}
\caption{The fermionic bound state corresponding to the sine Gordon kink-anti-kink solution
 as function of time for $g=1$ and $u=0.5$. The level $n=0,1,2,3,4$ are represented.}
\label{fig:fig2}
\end{figure}

Solving \eqref{dirac} in the background \eqref{kak}, we implicitly assume that the motion of the kink and of the anti-kink is treated adiabatically.
The results are summarized on figure \ref{fig:fig2} where the fermionic eigenvalue $E$ for the ground state $n=0$ and the first few excited states are represented as function of $t$.

%%%%%%%  

Our numerical integration of the spectral equations \eqref{dirac} for several values of $t$ reveals that only the ground state subsists
in the $t\to 0$ limit where it enters in the continuum (this could be expected since the classical background solution vanishes in this limit) and that the excited states join the continuum at finite values of $t$, depending  on the level of the excitation.

It should be pointed out that the fermionic eigenvalue of the fundamental solution (line $n=0$) approaches $E=0$ for $t \to \infty$ although staying positive. It decays exponentially with time.

The fact that the excited energy levels become pairwise degenerate in the large $t$-limit can be understood from the form of the potential. Indeed, for $t \gg 1$, the potential possesses two well separate valleys centred about the points $x \sim \pm ut$. In the neighbourhood of $x=\pm ut$, the form of the potential is given by $V_{\pm} = g^2 \phi^2 \mp g \phi'$. Far away from the two regions of the $x$-line situated around $x=\pm ut$ (i.e in the region of the origin and in the asymptotic regions), the potential is exponentially small. In fact the effective potential under investigation results from a superposition of two (suitably shifted) potentials which are supersymmetric partners of each other;  accordingly they have the same spectrum apart from the ground state of $V_+$. It is one of the striking property of supersymmetric quantum mechanics (see appendix \ref{app:susyqm} or \cite{khare} for a review) that if $V_{\pm}(x)$ are supersymmetric partner potentials, the $k^{th}$ energy level of $V_-$ coincides with the $(k+1)^{th}$ energy level of $V_+$. 

\begin{figure}[H]
\centering
\includegraphics[scale=.5]{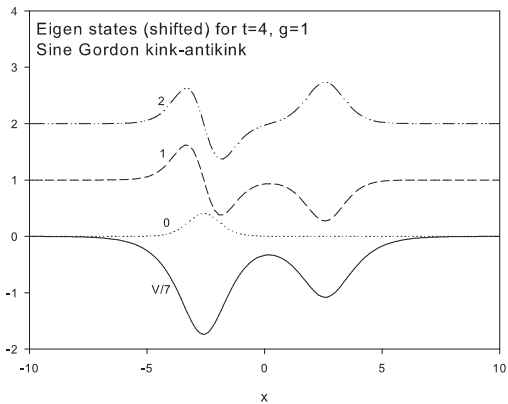}
\caption{The profile of the potential and of the first three fermionic bound states 
corresponding to the single sine Gordon kink-anti-kink solution for $t=4$, $g=1$ and $u=0.5$ are represented.}
\label{fig:fig3}
\end{figure}

\begin{figure}[H]
\centering
\includegraphics[scale=.5]{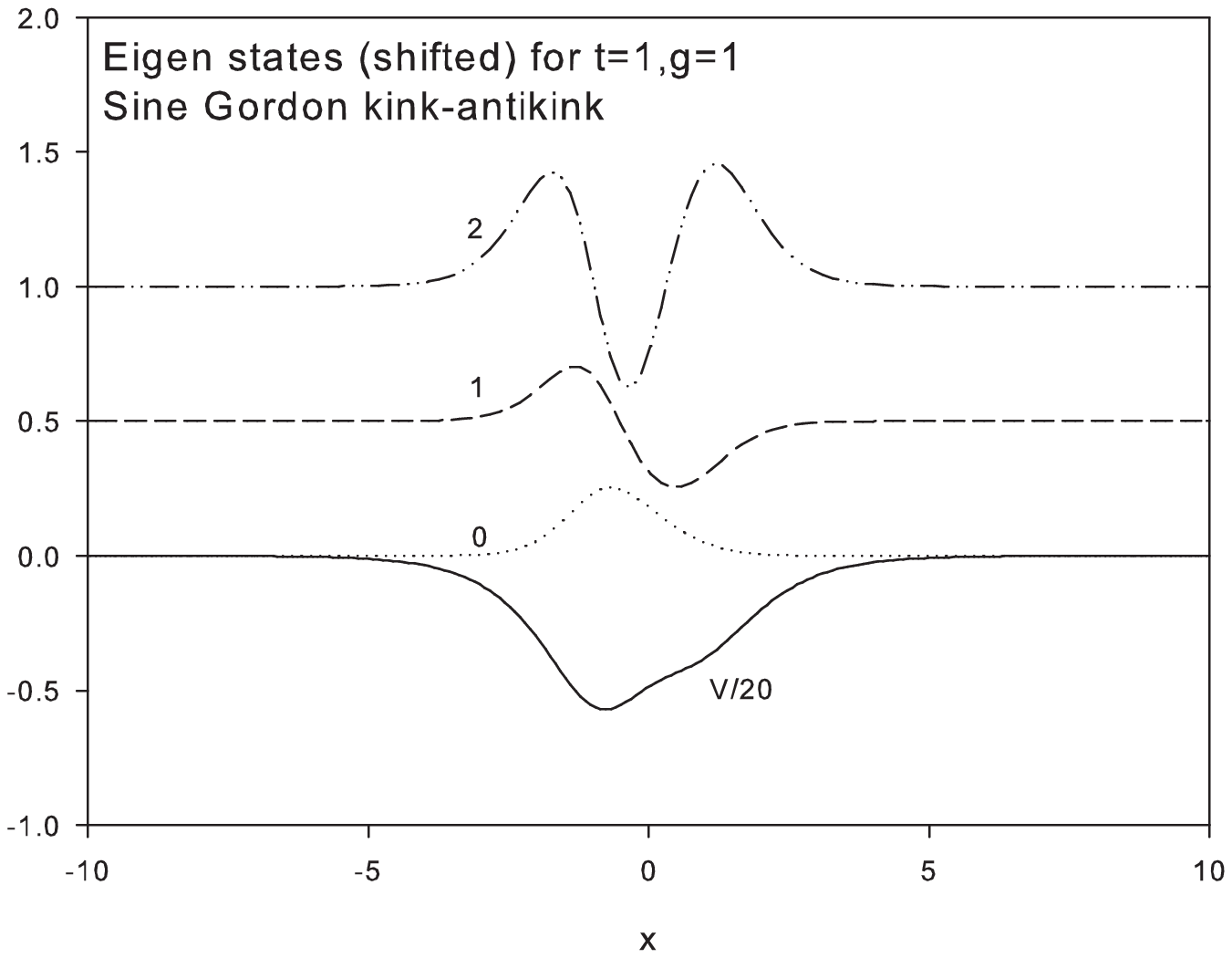}
\caption{The profile of the potential and of the first three fermionic bound states corresponding to the single sine Gordon kink-anti-kink solution
for $t=1$, $g=1$ and $u=0.5$ are represented.}
\label{fig:fig4}
\end{figure}

As a consequence, in our case, two eigenvectors exists with roughly the same eigenvalue when the wall and the anti-wall are well separated, like e.g. in figure \ref{fig:fig3}. While $t$ decreases, the two valleys of the potential have tendency to merge in the region of the origin, see figure \ref{fig:fig4}, and the degeneracy of the two energy levels corresponding to each other by the supersymmetry is lifted, as shown on figure \ref{fig:fig2}. It is tempting to say that the supersymmetry of the spectrum occurring for $|t| \gg 1$ is broken in the $|t| \to 0$ limit. 

%%%%%%%%%%%%%%%%%%%%%%%%%%%%%%

Said in other words, it turns out that when the wall and anti-wall approach each other, the system cannot support fermion bound states and their spectrum merge into the continuum of the Dirac equation. Seen oppositely, fermion bound states (ground state and a number of excited modes, the number of them depending on the coupling constant $g$) can emerge from the continuum when the wall and anti-wall separate from each other.

\section{Bell-shaped lumps: stability}
In the previous section, several lumps which have the shape of a bell were used, mimicking a kink-anti-kink non topological solution. The stability of the underlying fundamental solution is well known and can even be formulated in terms of a Poschl-Teller potential (see appendix \ref{ptequation})  where some relevant stationary solutions can be computed algebraically.
However, a more general question can be addressed: which potential $V(\phi)$ admit kink like configuration as static solution and for which of them is it possible to formulate the stability equation as an algebraic equation? We won't answer fully this question, because classifying all the possibilities can be a bit ambitious, but we will focus on three models, namely the $\phi^4$, inverted $\phi^4$ and $\phi^3$ models. New lumps solutions to these models were presented recently in \cite{bazeia}. 

These models are interesting by themselves but once again, they can be seen as toy brane models.
The models considered are 1+1 dimensional scalar field theory of the form 
\be
    {\cal L} = \frac{1}{2} \partial_{\mu}\phi \partial^{\mu}\phi - V(\phi) 
\ee  
with suitable potential $V(\phi)$. 

For static fields, the classical equation has the form $\partial_x^2 \phi = \frac{d V}{d \phi}$.
The linear stability of a solution, say $\phi_c(x)$, of this equation, can be studied by diagonalizing the quadratic part of the lagrangian in a perturbation around $\phi_c$.

Concretely, one has to compute the eigenvalues allowing for normalisable eigenfunctions of the spectral equation
\be
     ( -\frac{d^2}{dx^2} + \left. \frac{d^2 V}{d \phi^2} \right|_{\phi=\phi_c})\eta = \omega^2 \eta \ \ \ , \ \ \ 
     \phi(t,x) = \phi_c(x) + e^{i \omega t} \eta(x)
\ee
Negative eigenvectors $\omega^2 < 0$, reveal the existence of instabilities of the classical solution $\phi_c$.

\subsection{Case 1: $\phi^4$ potential}
The case of a quartic potential is well known but we mention it for completeness and for the purpose of comparison of the eigenmodes with the cases mentioned earlier.

The potential and the corresponding kink-solution read
\be
V(\phi) = \frac{1}{2} (\phi^2 - 1)^2,\ \phi_c(x) = \pm \tanh(x),
\label{kink}
\ee
and the stability equation corresponds to the Poschl-Teller equation (see appendix \ref{ptequation}) with $N=2$ and $\omega^2 = \omega_p^2 +4$. The eigenvalues are then  $\omega^2= 0,3,4$. It is well know that the $\lambda \phi^4$ kink  has no negative mode and is stable. The corresponding eigenfunctions are
\be
    \eta_0 = \frac{1}{\cosh(x)^2} = - \frac{d \phi_c}{dx},\ \eta_3 = \frac{\sinh(x)}{\cosh^2(x)},\ \eta_4 = \tanh^2(x) - \frac{1}{3}. 
\ee
The zero mode corresponds to infinitesimal translations of the classical solution $\phi_c$ in the space variable.

\subsection{Case 2: Inverted $\phi^4$ potential}
This case corresponds to a quartic potential with inverted sign. The potential and the corresponding lump are given by
\be
V(\phi) = \frac{1}{2} \phi^2 (1- \phi^2), \ \phi_c(x) = \pm \frac{1}{\cosh(x)}.
\ee
In this case also, the stability equation corresponds to the Poschl-Teller equation with $N=2$ but with a different shift of the eigenvalue, for instance $\omega^2 = \omega_p^2 +1$; leading to $\omega^2= -3,0,1$. 

As a consequence the lump is characterized by one negative mode and one zero mode. Up to a normalisation they are given by
\be
\tilde\eta_{-3} = \frac{1}{\cosh(x)^2},\ \tilde \eta_0 = \frac{\sinh(x)}{\cosh^2(x)} = -\frac{d \phi_c}{dx},\ \tilde \eta_1 =\tanh^2(x)-\frac{1}{3}
\ee
as usual, the zero mode is associated with the translation invariance of the underlying field theory.

\subsection{Case 3: Cubic $\phi^3$ potential}
The last case investigated corresponds to a  potential of third power in the field $\phi$. The potential and the corresponding lump are given by
\be
V(\phi) = 2 \phi^2 (1- \phi),\ \phi_c(x) = \pm \frac{1}{\cosh^2(x)}.
\ee
The corresponding  normal mode equation corresponds to the Poschl-Teller equation with $N=3$ with an appropriate shift of the spectrum:
$\omega^2 = \omega_p^2 +4$; leading to $\omega^2=-5,0,3,4$. 

The lump here is then characterized by one negative mode, one zero mode and two positive modes. Up to a normalisation
they are given by
\bea
\tilde\eta_{-5} &=& \frac{1}{\cosh^3(x)},\ \eta_0 = \frac{\sinh(x)}{\cosh^3(x)} = -\frac{1}{2}\frac{d \phi_c}{dx},\\
 \eta_3 &=& \frac{1}{\cosh^2(x)}( \tanh^2(x) - \frac{1}{5}),\ \eta_4 = \tanh(x) ( \tanh^2(x) - \frac{3}{5}).\nonumber
\eea

In \cite{bazeia}, deformations  of the three potentials given above are studied as well (e.g. is cases 1 and 3, deformations breaking the reflection symmetry $\phi \to -\phi$). The corresponding stability equation is not of the Poschl-Teller type but their spectrum could be studied perturbatively from the ones obtained above. 
It should be interesting to see in particular how the unstable mode (or the zero mode in the case 1) evolve in terms of the  coupling constant parameterizing the deformation.

\section{Concluding remarks}
In this chapter, we studied fermionic localization on kink and kink anti-kink solutions available in the Sine-Gordon model in $1+1$ dimensions. This kink/anti-kink system can be viewed as a toy model for a brane - anti-braneworld solution. Note however that the solution is time dependent, but we treated $t$ as an external parameter, fixing the distance between the kinks. As mentioned in the text, this consists in treating the system adiabatically. The interesting result is that the fermionic field is indeed localized on the kinks. Bound states emanate from the continuum as the coupling between the kink or kink-anti-kink and the fermionic field is increased. 

In the case of the kink/anti-kink system, fermionic bound states emanating from the continuum appear when the kink and anti-kink start spreading apart. Moreover, when the kink and anti-kink are well separated, consecutive fermionic levels degenerate, i.e. level $1$ and $2$, level $3$ and $4$, etc. 

We also considered the stability of three types of potential, namely $\phi^4$, inverted $\phi^4$ and cubic $\phi^3$ potential.
The stability equation for these three models are of the form of the Poschl-Teller equation.

As a result, lump solutions with the $\phi^4$ potential are stable, but solutions with the two other potential have a negative mode and are then unstable. This seems reasonable since the first potential is positive while the two last have a local minima and $\lim_{\phi\rightarrow\infty}=-\infty$. Intuitively, such potentials allow a decay of localized solutions to the unbounded minima via quantum tunnelling.

%% file: bh.tex
%%%%%%%%%%%%%%%%%%%%%%%%%%%%%%%%%%%%%%%%%%%%%%%
%%% Black Hole solutions in higher dimension
%%%%%%%%%%%%%%%%%%%%%%%%%%%%%%%%%%%%%%%%%%%%%%%

%The last few decades has witnessed a growing interest for extradimension, mainly motivated by String Theory. On the other hand, from a mathematical point of view, general relativity is consistent in more than four dimensions; moreover, higher dimensional gravity is also a consequence of String Theory. 

Black holes solutions have always been of considerable interest in general relativity. Such objects are solutions to the Einstein equations describing spacetimes having the property that there exists a spatial region whose causal future is trapped \cite{wald}. The boundary of this spatial region is called the black hole horizon. Black holes have been intensively studied in four dimensions. In particular, the only possible horizon topology in four dimensions is $S_2$ - the surface of a $2$-sphere - and stationary (charged) black hole solutions are classified in a three parameters family, namely the Kerr-Neumann family \cite{kn}. These three parameters are the charge, mass and angular momentum of the black hole. Black hole solutions are of considerable interest from an observational point of view, since they can result from a collapse of massive stars, but they are also fundamental objects from a theoretical point of view. They are actually theoretical labs for general relativity. Black holes of the Kerr-Neumann family are characterised by a curvature singularity inside the horizon; close to the singularity, the curvature blows up and the gravitational interaction becomes very strong.

On the other hand, higher dimensional gravity has been considered long time ago, as discussed in chapter \ref{ch:branes}, and have regained interest the last three decades, mainly motivated by String theory.

It seems then natural to investigate the higher dimensional counterparts of the black hole solutions available in four dimensional gravity. However, in more than four dimensions, there are several drastic changes with respect to what is known in four dimensions. First, in higher dimensions, say in $d$ dimensions, there are different possible horizon topology, such as $S_{d-2}$ (spherical), $S_{d-3}\times S_1$ (cylindrical or toroidal), $S_{d-2-p}\times R^p$ (black branes), etc. There is also the possibility of having several asymptotical spacetime topology, such as $\mathcal M_4\times S_{d-4}$, $\mathcal M_d$,... while in four dimensions, the asymptotic spacetime is $\mathcal M_4$, $\mathcal M_d$ denoting the $d$-dimensional Minkowski spacetime. In four dimensions, a black hole is uniquely specified once the mass, angular momentum and charge is given \cite{israel,uniq0,uniq1,uniq2}. It is not the case any more in higher dimensions (see for instance the case of the black ring and of the Myers Perry solution, \cite{revempreall,solubr1}).

Note that the three parameter family of solutions describing four dimensional black holes has a counterpart in the presence of a cosmological constant, namely the $(A)dS$- Kerr-Neumann solutions, where the asymptotical spacetime is the $(A)dS_4$ spacetime. 

In this chapter, we will first focus on higher dimensional black objects presenting a spherical horizon topology ($S_{d-2}$). We will present such known solutions in pure Einstein gravity (vacuum) and for the Einstein-Maxwell equations (i.e. in the presence of an electromagnetic field). Furthermore, we will present the construction of new solutions, namely the de Sitter charged, rotating black hole \cite{bdbh1, bdbh2}. 

The second part of this chapter is devoted to black strings, black objects presenting a $S_1\times S_{d-3}$ horizon topology, in asymptotically locally flat spacetime, which have been intensively studied in the literature (stability \cite{gl}, non uniform black strings \cite{gubser, wiseman}, etc.) and with a positive cosmological constant. Black strings with a negative cosmological constant have been studied in \cite{rms} and will be discussed in more details in chapter \ref{chadsbs}; black strings with a positive cosmological constant have not been discussed in the literature yet.

In asymptotically locally flat spacetime, black strings and more generally black branes are known to suffer from a long wavelength dynamical instability, the so-called Gregory-Laflamme instability \cite{gl}. We will briefly review this instability in the end of the chapter.\\

To sum up, this chapter is organised as follows: in section \ref{hdbh}, we give a brief overview of the analytically known higher dimensional black hole solutions with a spherical horizon topology and their properties. In section \ref{newbh} we present new black hole solutions, namely the charged rotating de Sitter black holes and their properties. Then we review the black string solution and its instabilities in section \ref{flatgl}; finally we discuss the effect of a positive cosmological constant on the black string solution in section \ref{bss}. In particular, we give evidences that such black strings do not approach an asymptotically locally de Sitter spacetime but indeed lead to a singular spacetime. We briefly conclude this chapter in section \ref{cclbh}.

\section{The general framework}
The various solutions presented here are all extrema of the Einstein-Hilbert action with boundary term (see appendix \ref{app:reminder}), minimally coupled matter field and a cosmological constant:
\be
S = \frac{1}{16\pi G}\int_{\mathcal M}\sqrt{-g}\left(R-2\Lambda + L_m\right)d^dx + \frac{1}{8\pi G}\int_{\partial\mathcal M}\sqrt{-h}Kd^{d-1}x,
\label{sgen}
\ee
where $G$ is the $d$-dimensional Newton constant, $\mathcal M$ is a $d$-dimensional manifold, $\partial\mathcal M$ the boundary manifold, $g$ the determinant of the metric, $R$ the scalar curvature, $\Lambda$ the cosmological constant, $h$ the determinant of the induced metric on the boundary manifold, $K$ the trace of the extrinsic curvature $K_{ab}$ (see appendix \ref{app:reminder}) of $\partial\mathcal M$ and $L_m$ is the matter lagrangian, depending on the model we will be interested in.

We will not discuss in detail all the thermodynamical properties of the various known solutions apart from the simplest case, namely the Tangherlini black hole, but we will present the various solutions instead. The interested reader can find more details on the thermodynamical properties in the literature \cite{tangh,solubh,solubh1,solubh2,solubh3}, between many others (see also \cite{revempreall} for a review).

\section{Known higher dimensional black holes}
\label{hdbh}
\subsection{Tangherlini black hole}
The Tangherlini solution is the generalization of the four dimensional Schwarzschild solution to any number of dimensions. It extremizes the action \eqref{sgen} with $\Lambda=0$, $L_m=0$ and is given by
\be
ds^2 = -f(r)dt^2 + \frac{dr^2}{f(r)} + r^2d\Omega_{d-2}^2,\ f(r)=1-\left(\frac{r_h}{r}\right)^{d-3}.
\label{tang}
\ee

The event horizon is located at $r=r_h$. The solution \eqref{tang} generalizes the spatial spherical symmetry of the Schwarzschild black hole and is time translationally invariant (it is a static solution). The vector $\partial_t$ is then a Killing vector and the associated conserved quantity (the mass, $M$) can be computed using the standard Komar integral (see Appendix \ref{app:reminder}):
\be
M = \frac{(d-2) V_{d-2} r_h^{d-3}}{16 \pi  G},
\ee
where $V_{d-2}$ is the surface of the unit $(d-2)$-sphere.

The entropy $S$ and temperature $T_H$ are computed in the standard way; using the Hawking-Bekenstein formula \cite{hbentropy1,hbentropy2} and by demanding regularity in the euclidean section, i.e. by avoiding conical singularity in the near horizon Euclideanised geometry of the black hole:
\be
S= \frac{1}{4G}V_{d-2}r_h^{d-2},\ T_H= \frac{d-3}{4\pi r_h}.
\ee

Note that $S\propto 1/T_H^{d-2}$; this implies that the specific heat $C_p=T_H\ddf{S}{T_H}$ of the Tangherlini black hole is negative. This is the typical signature of a thermodynamical instability.

The solution obeys the first law of black hole thermodynamics 
\be
dM = T_HdS,
\ee 
and the thermodynamical quantities are related to each other by the Smarr relation, 
\be
(d-3)M = (d-2)T_HS.
\ee
These two relations can be checked explicitly for the particular case of the Tangherlini solution but it is possible to prove such relation more formally (see for example \cite{wald}).

The Tangherlini solution generalizes to asymptotically $(A)dS$ spacetimes with 
\be
f(r)= 1-\epsilon\frac{r^2}{\ell^2} - \left(\frac{r_h}{r}\right)^{d-2},
\ee
where $\ell=\frac{(d-1)(d-2)}{2\left| \Lambda \right|}$ is the $(A)dS$ curvature radius and $\epsilon$ is the sign of the cosmological constant.

The mass is still given by $M=\frac{(d-2) V_{d-2} r_h^{d-3}}{16 \pi  G}$, but the horizon radius now depends on the cosmological constant. As a consequence, the entropy and temperature depend on the cosmological constant as well. The presence of the cosmological constant has serious implication on the thermodynamical properties of the black hole, for instance, in asymptotically $AdS$ spacetime, the black hole develops two phases: small black holes ($r_+/\ell<<1$) and big black holes $r_+/\ell\approx 1$, $r_+$ being the largest root of $f(r)$. Small black holes are thermodynamically unstable while big black holes are thermodynamically stable; this is purely an effect of the $AdS$ curvature \cite{hptr}.

\subsection{Myers-Perry black hole}
The Myers-Perry black hole is the higher dimensional generalization of the Kerr solution. However, in more than four dimensions, there are multiple independent planes of rotation. It follows that the Myers-Perry solution is characterized by multiple angular momenta, associated with the rotation in these various planes.

In $d$ dimensions, the number of independent planes $N$ is given by $\lfloor(d-1)/2\rfloor$, $\lfloor X\rfloor$ denoting the floor integer value of $X$.
For simplicity, we will consider only odd number of dimensions and equal angular momenta. More general case can be found in \cite{revempreall}.

In order to describe rotating black holes, we consider the following ansatz:
\bea
\label{rotansatz}
ds^2 &=& -b(r)dt^2 +  \frac{ dr^2}{f(r)} +g(r)\sum_{i=1}^{N-1}  \left(\prod_{j=0}^{i-1} \cos^2\theta_j \right) d\theta_i^2  \nonumber \\
&+& h(r) \sum_{k=1}^N \left( \prod_{l=0}^{k-1} \cos^2 \theta_l  \right) \sin^2\theta_k \left( d\varphi_k - w(r)  dt\right)^2  \\
&+&p(r) \left\{ \sum_{k=1}^N \left( \prod_{l=0}^{k-1} \cos^2  \theta_l \right) \sin^2\theta_k  d\varphi_k^2 \right. \nonumber\\
&-&\left. \left[\sum_{k=1}^N \left( \prod_{l=0}^{k-1} \cos^2  \theta_l \right) \sin^2\theta_k   d\varphi_k\right]^2 \right\}.\nonumber
\eea
In the above formula  $\theta_0 := 0$ and  $\theta_N := \pi/2$ are assumed; the  non trivial angles have $\theta_i \in [0,\pi/2]$ for $i=1, \dots, N-1$, while $\varphi_k \in [0,2\pi]$ for $k=1,\dots , N$. Note that the ansatz \eqref{rotansatz} imposes $p(r)=g(r)-h(r)$ for consistency.

The Myers-Perry solution \cite{mpbh} with equal angular momenta is then given by \eqref{rotansatz} with
\begin{eqnarray}
 f(r)=1-\frac{2M}{r^{d-3}}+\frac{2Ma^2}{r^{d-1}},~ h(r)=r^2(1+\frac{2Ma^2}{r^{d-1}}),~ \nonumber  \\
w(r)=\frac{2Ma}{r^{d-3}h(r)},~~g(r)=r^2,~~ b(r)=\frac{r^2f(r)}{h(r)},
\label{mpsol}
\end{eqnarray}
where $M$ and $a$ are two constants related to the solution's mass and angular momentum.

The symmetries of \eqref{rotansatz} are given by the following (commuting) Killing vectors: $\partial_t, \partial_{\varphi_i},\ i\in[1,\lfloor(d-1)/2\rfloor]$ and once again, conserved quantities can be computed using standard Komar integrals leading to the Mass $\mathcal M$ and the angular momenta $J_i$ associated to the $N$ independent rotation planes (here, we assume $J_i = J,\ i=1,\ldots,N$:
\be
\mathcal M = \frac{V_{d-2}}{4 \pi G} \frac{d-2}{2}M,\ J = \frac{V_{d-2}}{4 \pi G} \frac{d-1}{2} M a,
\ee
where $V_{d-2}$ is the surface of the unit $(d-2)$-sphere.
The entropy and Hawking temperature are computed in the standard way, leading to
\be
T_H =\frac{\sqrt{b'(r_+)f'(r_+)}}{4\pi},\ S = \frac{V_{d-2}r_+^{d+2}}{4},
\ee
where $r_+$ is the event horizon, defined as the largest root of $f(r)$.
The first law becomes $d\mathcal M=T_HdS + \Omega^{(i)}_H dJ_i$, where $\Omega^{(i)}_H$ is the angular velocity at the horizon defined by the requirement that $k = \ddx{t} + \Omega^{(i)}_H \ddx{\varphi_i}$ is null on the event horizon.
In the case under interest, the angular velocity of the horizon reduces to $\Omega_H^{(i)} = \omega(r_h)$. In the following, we will omit the superscript $(i)$ and the subscript $H$ since all angular velocities at the horizon are equal and are always to be understood as evaluated at the event horizon, except if stated explicitly.

The counterpart of the Myers-Perry solution in the presence of a cosmological constant has also an analytic solution \cite{solubh2},
\begin{eqnarray}
f(r)&=&1-\epsilon\frac{r^2}{\ell^2}-\frac{2M\Xi}{r^{d-3}}+\frac{2Ma^2}{r^{d-1}},~ h(r)=r^2(1+\frac{2Ma^2}{r^{d-1}}),~ \nonumber  \\
w(r)&=&\frac{2Ma}{r^{d-3}h(r)},~~g(r)=r^2,~~ b(r)=\frac{r^2f(r)}{h(r)},
\end{eqnarray}
where $M$ and $a$ are still related to the solution's mass and angular momentum:
\be
\mathcal M = \frac{V_{d-2}}{4 \pi G} M (\frac{d-2}{2} - \epsilon \frac{a^2}{2 \ell^2}),\ J = \frac{V_{d-2}}{4 \pi G} \frac{d-1}{2} M a,
\ee
while $\Xi=1+\epsilon a^2/\ell^2$.
The parameter $\epsilon$ is again the sign of the cosmological constant.

\subsection{Charged higher dimensional black hole}
The generalization of the four dimensional Reissner-Nordstr\"om black hole with (or without) cosmological constant is a solution of the equations of motions resulting from \eqref{sgen} with a minimally coupled Maxwell field \cite{solubh6}:
\be
L_m=-\frac{1}{4}F_{AB}F^{AB},\ F_{AB} = \partial_A A_B - \partial_B A_A,
\label{maxlag}
\ee
where $A,B$ run from $0$ to $d-1$ and $A_A$ is the Maxwell field.

The equations resulting from the variation of \eqref{sgen} with respect to the gravitational field and the Maxwell field are then the Einstein-Maxwell equations, eventually supplemented by a cosmological constant.

The $d$-dimensional $(A)dS$ charged black hole solution is given by
\bea
ds^2&=&-f(r)dt^2 + \frac{dr^2}{f(r)} + r^2d\Omega_{d-2}^2,\nonumber\\
f(r)&=&-\epsilon\frac{r^2}{\ell^2}+1-\frac{2M}{r^{d-3}} + \frac{Q^2}{2(d-2)(d-3) r^{2(d-3)}},\\
A_Adx^A &=& V(r)dt,\ V(r) = \frac{Q}{(d-3)r^{d-3}},\nonumber
\eea
where $M$ and $Q$ are the mass and charge of the solution, $\ell = \sqrt{\epsilon\frac{(d-1)(d-2)}{2\Lambda}}$ is the $(A)dS$ radius and $\epsilon=|\Lambda|/\Lambda$ if $\Lambda\neq0$ and $\epsilon=0$ if the cosmological constant vanishes.

\section[$(A)dS$ charged rotating black hole]{$(A)dS$ charged rotating higher dimensional black hole}
\label{newbh}
As we have seen in the previous sections, known solution are either rotating, either charged (with or without a cosmological constant). The missing point is the charged \emph{and} rotating solution. Unfortunately it seems that finding analytic solutions is hopeless, so we have to rely on numerical computations.

Here we consider the $dS$ charged rotating black hole in odd dimensions. The charged rotating solution have been constructed in asymptotically flat spacetime and asymptotically $AdS$ spacetime \cite{Kunz:2006eh,knlr}. The difficulty for the construction of the $dS$ charged rotating black hole is essentially due to the presence of a cosmological horizon. The number of independent angular momenta is given by $N=(d-1)/2$. We will use the ansatz \eqref{rotansatz} for the metric field, the matter content will be described by the minimally coupled Maxwell action, as in the previous section. The ansatz for the Maxwell field reads:
\be
A_Adx^A =  V(r)dt+a_\varphi(r) \sum_{k=1}^N\left( \prod_{l=0}^{k-1} \cos^2\theta_l \right) \sin^2\theta_k d\varphi_k,
\label{adscrmaxwell}
\ee
where $V(r)$ and $a_\varphi(r)$ are the electric and magnetic potentials.

Note that we include a magnetic part in the vector potential. This is related to the fact that the solutions are not static (but they are stationary). Obtaining charged rotating black holes indeed requires electric and magnetic potentials.

\subsection{Equations and boundary conditions}
The action \eqref{sgen} with minimally coupled Maxwell fields \eqref{maxlag} supplemented by the ansatz \eqref{rotansatz}, \eqref{adscrmaxwell} leads to a system of non-linear coupled differential equations:
\bea
&& f'+\frac{f}{(d-2)}\bigg(-\frac{rh}{2b}w'^2+\frac{2r}{ b}V'^2+\frac{4rw}{b}a_\varphi 'V'+\frac{h'}{h}(1-\frac{rb'}{2b}) \nonumber\\
&&-2r(\frac{1}{h}-\frac{w^2}{b})a_\varphi'^2+\frac{b'}{b}+\frac{(d-1)(d-4)}{r}\bigg)+\frac{1}{(d-2)r^3}((3d-5)h\nonumber\\
&&+4(d+1)a_\varphi^2-(d-1)^2r^2)+\frac{(d-1)r}{\ell^2}=0,\nonumber\\
\label{ec1}
\eea
\bea
&& b''+ \frac{1}{d-2}\bigg( 4(5-2d)w a_\varphi'V' +\frac{(d-3)}{2h}b'h' -\frac{2(d-3)b}{2rh}h' \nonumber\\
&&-2(2d-5)V'^2+\frac{1}{2}(3-2d)hw'^2-2(\frac{b}{h}+(2d-5)w^2)a_\varphi'^2+\nonumber\\
&&(d-2)\left(\frac{b'f'}{2f}-\frac{b'^2}{2b}\right)+\frac{(d-3)^2}{r}b'-\frac{(d-3)b}{r^4f}(12 a_\varphi^2+h)\nonumber\\
&&+\frac{(d-3)b}{r^2}\left(\frac{d-1}{f}+4-d\right)+\frac{(d-1)(d-2)b}{\ell^2f}\bigg)=0,\nonumber\\
\label{ec2}
\eea
\bea
&& h''+\frac{1}{(d-2)}\bigg(\frac{(2d-5)h^2}{2b}w'^2 +\frac{2h}{b}V'^2+\frac{4hw}{b}a_\varphi'V' \\
&&-\frac{(d-2)h'}{2}(\frac{h'}{h} -\frac{f'}{f})+\frac{(d-3)}{2b}b'h' +\frac{(d-3)^2}{r}h'\nonumber\\
&&+2(\frac{hw^2}{b} +2d-5)a_\varphi'^2-\frac{(d-3)h}{rb}b'-\frac{(d-3)(2d-3)h^2}{r^4f}\nonumber\\
&&-\frac{12(d-3)a_\varphi^2h}{r^4f}-\frac{(d-3)(d-4)h}{r^2}+\frac{(d-1)h}{f}(\frac{d-2}{\ell^2}-\frac{d-3}{r^2})\bigg)=0,\nonumber\\
\label{ec3}
\eea
\bea
&&w''-\frac{4w }{h}a_\varphi'^2-\frac{4a_\varphi'V'}{h}+\frac{(d-3)w'}{r}+\frac{1}{2}\left(-\frac{b'}{b}+\frac{f'}{f}+\frac{3h'}{h}\right)w'=0,\nonumber
\label{ec4}
\eea
for the metric components and 
\bea
&&V''-\frac{w}{b}b'a_\varphi'+\frac{w}{h}a_\varphi'h'+\frac{1}{2}(\frac{2(d-3)}{r}-\frac{b'}{b}+\frac{f'}{f}+\frac{h'}{h})V' \nonumber \\
&&+(1+\frac{hw^2}{b})a_\varphi'w'+\frac{hw}{b}V'w'+\frac{2(d-3)a_\varphi hw}{r^4f}=0,
\label{ec5}
\eea
\be
a_\varphi''+\frac{1}{2}(\frac{2(d-3)}{r}+\frac{b'}{b}+\frac{f'}{f}-\frac{h'}{h})a_\varphi'-\frac{h}{b}(wa_\varphi'+V')w'-\frac{2(d-3)a_\varphi h}{r^4f}=0,
\label{ec6}
\ee
for the matter fields. Here, the gauge choice $g=r^2$ is used and prime denotes derivative with respect to $r$.

It can easily be seen that the equations of motion present the first integral
\begin{eqnarray}
 r^{(d-3)}\sqrt{\frac{fh}{b}}(wa_\varphi'+V')=(d-3)q,
\label{fiwbh}
\end{eqnarray} 
where $q$ is an integration constant related to the charge of the solution.

Thus, similarly to the asymptotically flat case  \cite{Kunz:2006eh} case, the electric potential can be eliminated from the equations \eqref{ec1}-\eqref{ec6} by making use of the first integral \eqref{fiwbh}. 

Note also that the cosmological constant parameter can be arbitrarily rescaled by a rescaling of the radial variable $r$ and of the fields $h$ and $w$. In the following, we use this arbitrariness to choose $r_h=1$ without loosing generality.

\subsubsection{Constraint of regularity about the horizon}
We are interested in black hole solutions, with an horizon located at $r=r_h$.
The solutions can be expanded in the neighbourhood of the horizon leading to
\begin{eqnarray}
b(r)&=&b_1(r-r_h)+O(r-r_h)^2,~~f(r)=f_1(r-r_h)+O(r-r_h)^2, \nonumber
\\
h(r)&=&h_0+h_1(r-r_h)+O(r-r_h)^2,w(r)=w_h+w_1(r-r_h)+O(r-r_h)^2, \nonumber \\
\nonumber a_\varphi(r)&=&a_0+a_1(r-r_h)+O(r-r_h)^2,~~V(r)=V_0+V_1(r-r_h) + O(r-r_h)^2
\end{eqnarray}
where $a_0,w_h,V_0,h_0,a_1,b_1,w_1,V_1,h_1$ are real constants.
%\begin{eqnarray}
%f_1&=&\frac{b_1(d-2)(d-1)r_h^4+b_1(-4a_0^2(d-1)+(d-2)(-2h_0+(d-1)r_h^2))\ell^2}{r_h^3\ell^2(b_1(d-2)+2r_h(v_1+a_1w_h)^2)},\nonumber\\
%a_1&=&\frac{((d-3) h_0r_h^{-d-4}}{\sqrt{b_1f_1^3h_0}}(2a_0\sqrt{b_1f_1h_0}r_h^d+f_1qr_h^7w_1),\\
%v_1&=&\frac{(d-3)h_0r_h^{-d-4}}{\sqrt{b_1f_1^3h_0}}(-2a_0\sqrt{b_1f_1h_0^3}r_h^d w_h+f_1 q r_h^7(b_1-h_0w_1w_h)),\nonumber
%\end{eqnarray}
For the solutions to be regular at the horizon $r_h$ (or at the cosmological horizon $r_c$), 
the equation for $h$ leads to the  condition $\Gamma_1(r=r_h) = 0$ with
\bea
  \Gamma_1(r) &\equiv & 8 b' h^2 (12 a^2 + 7 h) + 4 x b' h h' (12 a^2 + 5 h) \nonumber \\
              &-& 32 b' h^2 x^2 + 8 x^3 b' h (f' h - 4 h')          \\
              &+& 2 h x^4 ( \frac{24}{\ell^2} b' h^2 - f' (4 (a')^2 h w^2 + 8 a' h w V' + b' h' + 5 h^2 (w')^2 + 4 h (V')^2  )) \nonumber \\
              &+&  h' x^5 ( - \frac{24}{\ell^2} b' h^2 - f' (4 (a')^2 h w^2 + 8 a' h w V' - b' h' -  h^2 (w')^2 + 4 h (V')^2  ))\nonumber
\eea
where we posed $a_{\varphi} \equiv a$.
The value $f'(r_h)$ can be extracted from the equation for $f$, giving
\be
      f'(r_h) = \frac{4 b_1 h_0}{r_h^3} 
      \frac{6 r_h^4/ \ell^2 + 8 r_h^2 - 12 a_0^2 - 5 h_0}
      {8 b_1 h_0 + r_h(4h_0(V_1+a_1 w_0)^2 - b_1 h_1 - h_0^2 w_1^2)}
\ee
In the same way, the two Maxwell equations \eqref{ec5},\eqref{ec6} lead to a single condition $\Gamma_2(r = r_h) = 0$ with
\be
\Gamma_2 (r) \equiv 4 a b' h + r^4 f'(h w' V' + a' h w w' - a' b')
\ee

\subsection{Thermodynamical quantities}
\subsubsection{Horizon quantities}
Although it is not clear if the thermodynamical properties are well defined in the presence of the cosmological horizon, one may still define them in the standard way. The Hawking temperature and the event horizon area of these solutions are given by
\bea
T_H=\frac{\sqrt{ f_1b_1}}{4\pi},\ A_H=V_{d-2}r_h^{d-2}.
\eea
Along with \cite{knlr}, we also write the mass and the angular velocity at the horizon. They are obtained by using the appropriate Komar integrals:
\be
   M_H = \left.\frac{V_{d-2}}{8 \pi G} \sqrt{\frac{f h g^2}{b}} (b' - h w w')\right|_{r=r_h} \ \ , \ \ 
   J_H =\left. \frac{V_{d-2}}{8 \pi G} 2\sqrt{\frac{f g^2 h^3}{b}} \  w' \right|_{r=r_h}
\label{komar}
\ee
where $V_{d-2}$ denotes the area of the $(d-2)$-sphere.
Similar quantities, say $M_C, J_C$ can be associated with the cosmological horizon. In fact, the masses and angular velocities at the two horizons can be related to each other by Smarr-like formulae, as we now demonstrate.

We consider a Killing vector $K_{\mu}$ and integrate the two sides of the identity
\begin{equation}
              \nabla^{\mu} \nabla_{\mu} K_{\nu} = - R^{\mu} _{\ \nu} K_{\mu} = 
              - 8 \pi (T^{\mu}_{\ \nu} - \frac{1}{d-2} T \delta^{\mu}_{\nu}) K_{\mu}
\end{equation}
over the (truncated) hyper-volume $\Sigma$ covering the space-like region between the two horizons. Then, using both the Einstein equations and the Stokes theorem appropriately result into the identities we are interested in.

Choosing first $K = \partial_{\varphi}$ leads to
\be
 J_C - V_{d-2} \sqrt{\frac{fhg^2}{b}}(w a_{\varphi}+ V')\vert_{r=r_c} = 
 J_H - V_{d-2} \sqrt{\frac{fhg^2}{b}}(w a_{\varphi}+ V')\vert_{r=r_h} \ \ ,
\ee
expressing the conservation of the total angular momentum.

The identity associated with the time-translation invariance ($K = \partial_t$) is a bit more involved but leads to
\be
M_C + \Phi_C Q_C - M_H - \Phi_H Q_H = \frac{1}{\pi \ell^2} V_{d-2} \int_{r_h}^{r_c} \sqrt{\frac{fhg^2}{b}} dr,
\ee
where $Q_{H}$  (the electric charge  at the horizon \cite{maeda}) and $\Phi_H$ (the electrostatic potential)  are defined by 
\be
Q_{H} = \int_0^{\pi/2} d\theta (\sqrt{\frac{fhg^2}{b}})\vert_{r=r_h},\ \Phi_H = (V + \frac{d-1}{2} \Omega a_{\varphi})\vert_{r=r_h}.
\ee
The same relations hold at the cosmological horizon  for $Q_C$, $\Phi_C$, but with the quantities evaluated on the cosmological horizon. Similarly to \cite{Kunz:2006eh}, the mass at the horizon can further be expressed in terms of the temperature and angular momentum according to (recall that $\Omega=w(r_h)$)
\be
   \frac{d-3}{d-2} M_H = \frac{1}{4} T_H A_H + \frac{d-1}{2} \Omega J_H.
\ee
The various identities above are fulfilled by our numerical solutions, providing several nice crosschecks of our method.

%%%%%%%%%%%%%%%%%%%%%%%%%%%%%%%%%%%%%%%%%%%%%%%%%%%%%%%%%%%%%%%%%%%%
\subsection{The asymptotic and global charges}
%%%%%%%%%%%%%%%%%%%%%%%%%%%%%%%%%%%%%%%%%%%%%%%%%%%%%%%%%%%%%%%%%%%%

In this section, we present the global charges characterizing the solutions asymptotically.
%This uses a formalism developed in \cite{counter,Brown:1993} and also used in \cite{bhr,bhrs}, namely the counterterm formalism (see also chapter \ref{chadsbs}). 
The metric functions have the following asymptotic behaviour in terms of three arbitrary constants $\alpha,\beta$ and $\hat J$
\bea
b(r)&=&- \frac{r^2}{\ell^2}+1+ \frac{\alpha}{r^{d-3}} +O(1/r^{2d-6}),\ f(r)=- \frac{r^2}{\ell^2}+1+\frac{\beta}{r^{d-3}} +O(1/r^{d-1}),\nonumber\\
h(r)&=&  r^2(1+  \frac{\beta-\alpha}{r^{d-1}} +O(1/r^{2d-4})),\ w(r)=    \frac{\hat J}{r^{d-1}} +O(1/r^{2d-4 }).
\label{asdschrot}
\eea
The asymptotic expression of the gauge potential is similar to the asymptotically flat case:
\be
 V(r)=- \frac{q}{r^{d-3}} +O(1/r^{2d-4 }),
 a_\varphi(r)= \frac{\hat \mu}{r^{d-3}} +O(1/r^{2d-4}) .
\ee
 The mass-energy and angular momentum of the solutions are given by
\begin{eqnarray}
 E=\frac{V_{d-2}}{16\pi G}(\beta-(d-1) \alpha),\ J=\frac{V_{d-2}}{8\pi G}\hat J.
\label{grav-charges}
\end{eqnarray}
%%%
The above relations can be obtained by using a background subtraction approach (see appendix \ref{app:reminder}) or using a suitable counterterm technique \cite{dscft}.

The electric charge and the magnetic moment of the solutions are given by \cite{bdbh2}
\begin{eqnarray}
 Q=\frac{(d-3)V_{d-2} }{4\pi G}q,\ \mu=\frac{(d-3)V_{d-2}}{4\pi G}\hat \mu.
\label{gauge-charges}
\end{eqnarray}

The ansatz in $f,b,h$ has the advantage to present a direct connection with the closed form vacuum rotating solution. This is not the case in \cite{knlr} where an isotropic coordinate system is used.

\subsection{Numerical procedure and results}

We solved the equations \eqref{ec1}-\eqref{ec6} in the case $d=5$ and we hope that this case catches the qualitative properties of the pattern of the solutions for $d>5$. The equations have been integrated numerically with the ODE solver Colsys \cite{colsys}.

The positive cosmological constant leads to the occurrence of a cosmological horizon at $r=r_c$ (with $r_h < r_c < \infty$) where $f(r_c) = b(r_c)=0$.
This creates a difficulty since $r_c$ constitutes an apparent singular point of the equations. In order to overcome this difficulty, we have solved the equations in two steps.

In the first step, we supplemented the system by the trivial equation $d \ell^2 /dx = 0$ and imposed the conditions of two regular  horizon at $r = r_h$ and $r=r_c$, fixing $r_c,r_h$ by hand and solved the equations for $r \in [r_h,r_c]$.
The appropriate set of twelve boundary conditions at the two horizons then reads
\bea
&&f = 0,\ b = 0,\ b' = 1,\ w = w_h,\  V = 0,\ a' = a'_h, \nonumber \\
&&\Gamma_1 = 0,\ \Gamma_2 = 0 \mbox{ for } r = r_h;\\
&& f = 0,\ b = 0,\ \Gamma_1 = 0,\ \Gamma_2 = 0 \mbox{ for } r = r_c.\nonumber
\eea
where the functions $\Gamma_{1,2}$ are defined above. Here we take advantage of the arbitrary scale of the field $b(r)$ to impose $b'(r_h)=1$ and of the arbitrary additive constant of the electric potential to assume $V(r_h)=0$. The function $b(r)$ will be renormalized appropriately after the second step in order to follow the asymptotic behaviour \eqref{asdschrot}, say $b\rightarrow b/N_b$. Then, the functions $w$ and $V$ have to be normalized as well, according to $w\rightarrow w/N_b$, $V\rightarrow V/N_b$.

The values $w_h$ and $a'_h$ are, a priori, arbitrary and control the total angular momentum and the (electric and magnetic) charges of the black hole.
The numerical value of $\ell^2$ is determined by the first step, together with the values of all the fields at $r=r_c$. The value of these fields at $r=r_c$ and of $\ell^2$ can then be used as a suitable set of Cauchy data to solve the equations for $r \in [r_c,\infty]$. The disadvantage of the method is that we cannot perform a systematic analysis of the solution for a fixed value of the cosmological constant, but fortunately, the numerical value of $\ell^2$ depends only a little on $w(r_h)$ and $a(r_h)$ once $r_h,r_c$ are fixed.

In the following, we present the results corresponding to $r_h=1,r_c=3$, this case corresponds to a large value of the cosmological constant but allows one to analyse the effect of $\Lambda$ on the solution. We hope that the results for this case capture the main features of the solutions for
generic values of $\Lambda$. 

\begin{figure}[H]
 \centering
  \includegraphics[scale=.6]{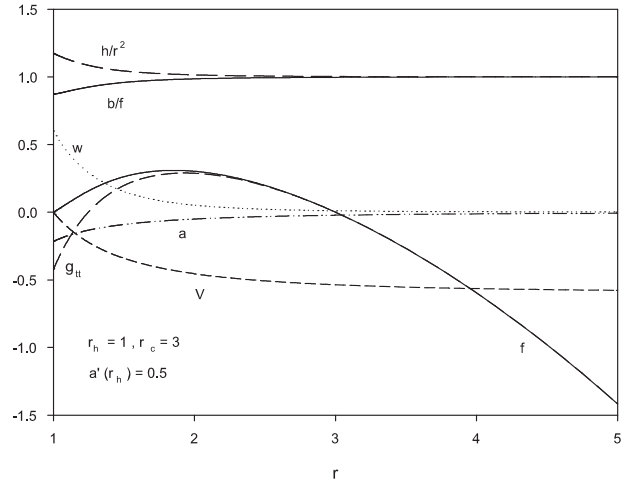}
  \caption{The profiles of a generic solution corresponding to $r_h=1,r_c=3$ , $w_h = 0.6, a'_h=0.5$}
 \label{fig:dschrot} 
\end{figure}

A generic profile of the solution corresponding to $r_h=1,r_c=3$ , $w_h = 0.6, a'_h=0.5$ is presented in figure \ref{fig:dschrot}, it corresponds to  $1/\ell^2 = 0.0945$ and $q = -0.08538$. The smoothness of the profile at $r=r_c$ can be appreciated on the plot. It is also worth to point out that the numerical solutions approach the asymptotic behaviour \eqref{asdschrot} although all boundary conditions are imposed at $r=r_c$ for the second step of our construction.
We supplemented figure \ref{fig:dschrot} with the plot of the $t-t$ component of the metric $g_{tt} = b - h w^2$, showing that there is a small region around the event horizon where $g_{tt}$ is negative, defining an ergoradius $r_e$ where $g_{tt}(r_e)=0$ (on the figure, $r_e \approx 1.17$).

\begin{figure}[H]
\centering
\includegraphics[scale=.55]{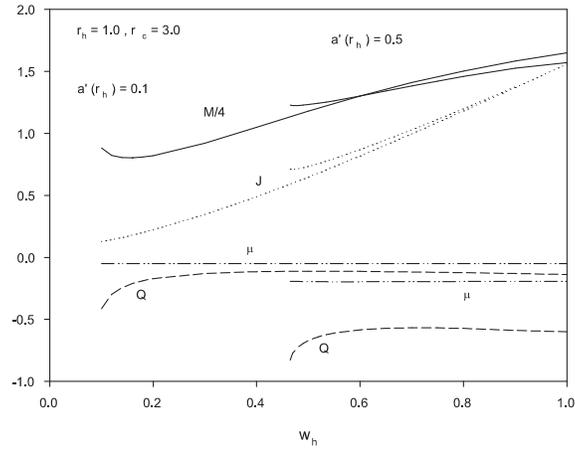}
\caption{The asymptotic charges $M,J,Q, \mu$ are presented as functions of $w_h$ for $a'_h=0.1$ and $a'_h=0.5$}
\label{fig:figasym} 
\end{figure}

We managed to construct several branches of solutions for different values of $a'_h$ and increasing the parameter $w_h$. For fixed $a'_h$, the solutions exist only for sufficiently large values of $w_h$; this can be interpreted as an effect of the balance between electric attraction and centrifugal repulsion. This is illustrated on figure \ref{fig:figasym} where we plot the asymptotic charges $M,J,Q,\mu$ as functions of $w_h$ for two values of $a'_h$; for $a'_h = 0.1$ (resp. $a'_h = 0.5$) the branch exist for $w_h > 0.1 $ (resp. $w_h > 0.465$). The figure further suggests that, if a second branch exists, its asymptotic mass will be larger than the one of the first branch that we constructed. The magnetic moment depends only a little in the parameter $w_h$.

\begin{figure}[H]
\centering
\includegraphics[scale=.55]{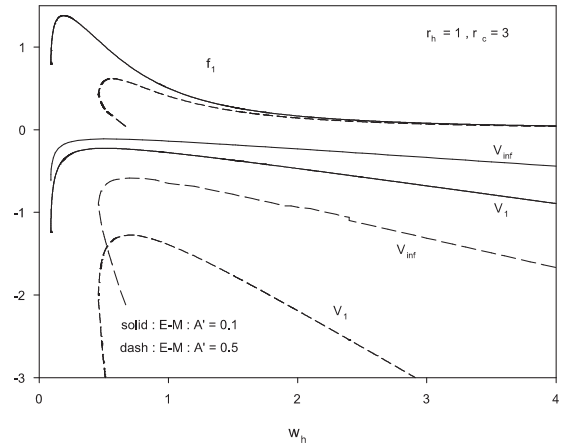}
\caption{The values $f_1$, $V_1$ and $V_{inf}\equiv V(\infty)$ are reported as functions of $w_h$ 
for $a_h'=	0.1$ and $a_h'=0.5$.}
\label{fig:2branches} 
\end{figure}

\begin{figure}[H]
\centering
\includegraphics[scale=.6]{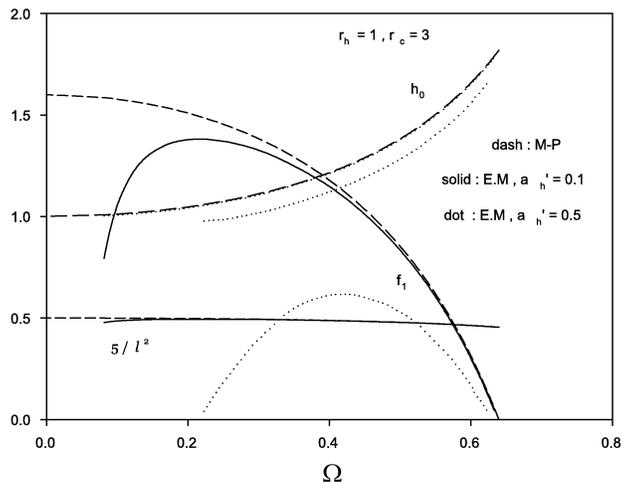}
\caption{The values $f_1$, $h_0$ and $1/\ell$ are reported as functions of $\Omega_h$ (i.e. $w(r_h)$, properly normalized) 
for $a_h'=0.1$ and $a_h'=0.5$.}
\label{fig:2branchesomega} 
\end{figure}

In fact, our analysis indicates that the second branch indeed exists: for $a'_h$ fixed, we find solutions for $w_h$ larger that some minimal value of $w_h$, say $w_{h,min}$. It turns out that a second branch of solution emerges at the minimal value $w_{h,min}$. When we increase $w_h$, we have seen that the second branch ends in an extremal solution where $f'(r_h)=0$. This is shown in figure \ref{fig:2branches} where we present a few parameters that are significantly different in the two branches as a function of $w_h$. Figure \ref{fig:2branchesomega} presents the value of $f_1$, $h_0$ and $1/\ell$ as a function of $\Omega$, suggesting that the solution exists in a finite range of $\Omega$, for fixed value of $a'_h$. When we solved the equation for larger values of $r_c$, corresponding to smaller values of $\Lambda$, we observe that the asymptotic quantities plotted on figure \ref{fig:figasym} depend weakly on $r_c$.

\section{Black strings and instabilities}
\label{flatgl}
The simplest higher dimensional black object presenting another horizon topology than $S_{d-2}$ is the black string, presenting an horizon topology $S_1\times S_{d-3}$. The black string metric reads
\be
ds^2 = -f(r)dt^2 + \frac{dr^2}{f(r)} + r^2 d\Omega_{d-3}^2 + dz^2,\ z\in{0,L}.
\label{flatbs}
\ee
for some length $L$ of the $S_1$ circle and where $f(r) = 1-\left(\frac{r_h}{r}\right)^{d-4}$.

This is simply a $(d-1)$-Tangherlini black hole \cite{tangh} with an extra Ricci-flat direction. Note that the asymptotic spacetime is \emph{locally} Minkowskian, more precisely, it is of topology $\mathcal M_{d-1}\times S_1$.

\subsection{The black string solution and its properties}
The black string solution has obviously $\ddx{t}$ and $\ddx{z}$ as killing vectors. Using the Komar formula (see appendix \ref{app:reminder}), we can compute two global charges associated to these killing vectors: the mass $M$ and the tension $\mathcal T$. 
Obers and Harmark \cite{ho} have shown that for the general black objects (e.g. black strings, localised black holes with or without matter) asymptotically approaching $\mathcal M_{d-3}\times S_1$ spacetime, the metric component $g_{tt}$ and $g_{zz}$ are of the form
\be
g_{tt} = -1 + \frac{c_t}{r^{d-4}} + \mathcal O\left(\frac{1}{r} \right)^{d-3},\ g_{zz} = 1 - \frac{c_z}{r^{d-4}} + \mathcal O\left( \frac{1}{r} \right)^{d-3}.
\ee
They then established the following formula for the mass and tension of these objects: 
\be
M= \frac{V_{d-3}L}{16\pi G} \left( (d-3)c_t - c_z \right),\ \mathcal T = \frac{V_{d-3}}{16\pi G} \left( -(d-3)c_z + c_t \right).
\ee

The black string solution \eqref{flatbs} is characterised by $c_t = r_h^{d-4}$ and $c_z= 0$, leading to
\be
M= \frac{V_{d-3}L}{16\pi G} (d-3)r_h^{d-4},\ \mathcal T = \frac{V_{d-3}}{16\pi G} r_h^{d-4}.
\ee

The entropy $S$ and the temperature $T_H$ are computed in the standard way:
\bea
S = V_{d-3}\frac{r_h^{d-3} L }{4G},\ T_H = \frac{d-4}{4\pi r_h}.
\eea

These quantities are related to each others by the Smarr relation
\be
(d-3)M - \mathcal T L  = (d-2)T_H S,
\ee
which can be checked explicitly using the definitions above. Note that the Smarr relation can be derived in a formal way, see \cite{tensdef}.
The first law of thermodynamic has an additional contribution coming from the gravitational tension:
\be
dM = T_H dS + \mathcal T dL,
\ee
which can also be checked explicitly, considering that the varying parameters are $r_h$ and $L$.

\subsection{Dynamical stability}
The dynamical stability of the black string \eqref{flatbs} has been studied in 1993 by R. Gregory and R. Laflamme \cite{gl}. This was partly motivated by the thermodynamical instability and from the fact that a black hole is more entropic than a black string of same mass when the length of the string is too large as a simple calculation shows:
\bea
&& S_{bs} \propto r_h^{d-3}L, M_{bs}\propto r_h^{d-4}L,\\
&& S_{bh}\propto r_h^{d-2},   M_{bh}\propto r_h^{d-3},\nonumber
\eea
where $S_{bs}, M_{bs}$ (resp. $S_{bh}, M_{bh}$) is the black string (resp. black hole) entropy and mass.

Eliminating the horizon radius, we can express the entropy as a function of the mass in each cases:
\be
S_{bs}\propto \left(\frac{M_{bs}^{d-3}}{L}\right)^{\frac{1}{d-4}},\  S_{bh}\propto M_{bh}^{\frac{d-2}{d-3}},
\ee

Then, comparing the entropy assuming the same value of the mass $M_{bh}=M_{bs}=M$ leads to 
\be
\frac{S_{bs}}{S_{bh}} \propto \left(\frac{M^{\frac{1}{d-3}}}{ L}\right)^{\frac{1}{d-4}}.
\ee
Then, it turns out that if the length is too large for a given mass, the entropy of the black string is less than the entropy of the black hole. Note that the mass is roughly $r_h^{d-3}$ (for the black hole) so the ratio is essentially proportional to $r_h/L$.

Note also the analogy between this thermodynamical instability of the black string and the Plateau Rayleigh instability (a fluid cylinder is unstable if the cylinder is too long) \cite{glrp}. 

In order to consider the dynamical stability of the string, one has to consider a small perturbation around the metric \eqref{flatbs}:
\be
ds^2 = \left(g_{AB} + \epsilon h_{AB}\right)dx^Adx^B,
\ee
where $\epsilon$ is a small arbitrary parameter, $g_{AB}$ are the component of the metric \eqref{flatbs}, $h_{AB}$ are the component of the fluctuation and $A,B=0,1,\ldots,d-1$.

To first order in $\epsilon$, the equation are given in the transverse-traceless gauge ($\nabla^A h_{AB} =0,\ g^{AB}h_{AB}=0$) by
\be
\delta R_{AB} = (\Delta_L)^{CD}_{AB}h_{CD}= -\frac{1}{2}\left(\Box \delta_A^C\delta_B^D + 2 R_{A\ B}^{\ C\ D}  \right)h_{CD}=0,
\label{lichbs}
\ee
where $\Delta_L$ is the Lichnerowitz operator and the d'Alembertian and Riemann tensor are evaluated using the background metric $g_{AB}$ (see Appendix \ref{app:reminder}).

We consider a perturbation of the form
\be
h_{AB} = \left(
		\begin{array}{cccccc}
		f(r) H_{tt} 	& H_{tr} 	& 0 	& 0 			& 0	& 0\\
		H_{tr}		& H_{rr}/f(r)	& 0 	& 0 			& 0 	& 0\\
		0		& 0		& r^2K	& 0			& 0	& 0\\
		0		& 0		& 0	& r^2 K\sin^2\theta 	& 0	& 0\\
		0		& 0 		& 0 	& 0 			&\ddots	& 0\\
		0		& 0		& 0	& 0			& 0	& 0
		\end{array}
	\right) e^{\Omega t-ikz},
\ee
where $H_{tt}, H_{tr}, H_{rr}, K$ are functions of $r$ only and $k=2\pi/L$. The values of $\Omega$ and $k$ are not independent as will be shown later in a dispersion diagram.

It is possible to reduce the linearised Einstein equations given by \eqref{lichbs} to a single equation, eliminating all but one field, say $H_{tr}$ following \cite{gl}, using the gauge conditions and the Lichnerowitz equations. The system then reduces to a single second order ordinary differential equation.

The reduced equation is of the form
\be
H_{tr}'' + \mathcal Q(r) H_{tr}' + \mathcal S(r) H_{tr}=0,
\label{egl}
\ee
where of course, $\mathcal Q(r),\mathcal S(r)$ depend on $k,\Omega,r,d$ and can be found in \cite{glrev,gl} between many others. 
The asymptotic solution of the equation is given by
\be
H_{tr}\propto e^{\pm r\sqrt{k^2+\Omega^2}}
\ee
while close to the horizon, the field $H_{tr}$ behaves like
\be
H_{tr}\propto (r-r_h)^{-1 \pm \frac{r_h\Omega}{d-4}}.
\label{glhor}
\ee
Gregory and Laflamme found that the conditions required in order to describe well behaved perturbation \cite{glrev,gl,gl2} leads to the requirement that close to the horizon, $H_{tr}$ behaves like \eqref{glhor} with the $+$ sign and tends to zero asymptotically. They found solutions to \eqref{egl} with $\Omega>0$ for small values of $k$, which means that black strings are unstable towards long wavelength (small wavenumber) perturbations. However, larger values of $k$ lead to $\Omega<0$, the black strings are then stable toward short wavelength perturbations. 

\begin{figure}
\centering
 \includegraphics[scale=.15]{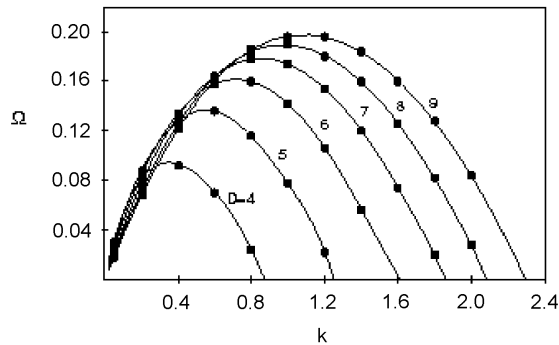}
 \caption{The dispersion relation $\Omega(k)$ for various number of dimensions. the total number of dimensions is $d=D+1$. Figure reprinted from \cite{glrev}.}
\label{disprel}
\end{figure}

This is illustrated in figure \ref{disprel} where the value of $\Omega$ as a function of $k$ is plotted. Note the presence of a zero mode; i.e. a solution with $k\neq 0,\mbox{ say }k_c,\ \Omega=0$.

\subsection{Phases of asymptotically locally flat black strings}
The existence of the zero mode implies that the particular perturbation with wavenumber $k=k_c$ is static. This is a hint for a phase transition.

It was widely believed that the string would decay to a localized black hole, i.e. a black hole in a spacetime which is asymptotically $\mathcal M_{d-1}\times S_1$ and with horizon topology $S_{d-2}$. However, it has been shown by Horowitz and Maeda that such a phase transition would take an infinite proper time at the horizon \cite{horowitz}. This led to postulate an intermediate phase, namely the non uniform black string phase. 

The non uniform phase is characterized by the fact that the solution depends non trivially on $z$; as a consequence, $\ddx{z}$ is no longer a Killing vector. However, $\ddx{z}$ is still a Killing vector in the asymptotic region of the spacetime, allowing the definition of the tension. The horizon topology of the non uniform black string is still $S_{d-3}\times S_1$, in contrast with the localised black hole characterise by an horizon topology $S_{d-2}$ (the asymptotic spacetime is $\mathcal M_{d-1}\times S_1$ in both cases).

\begin{figure}[H]
\center
\subfigure[\label{phase1}$d=5$.]{\includegraphics[scale=.27]{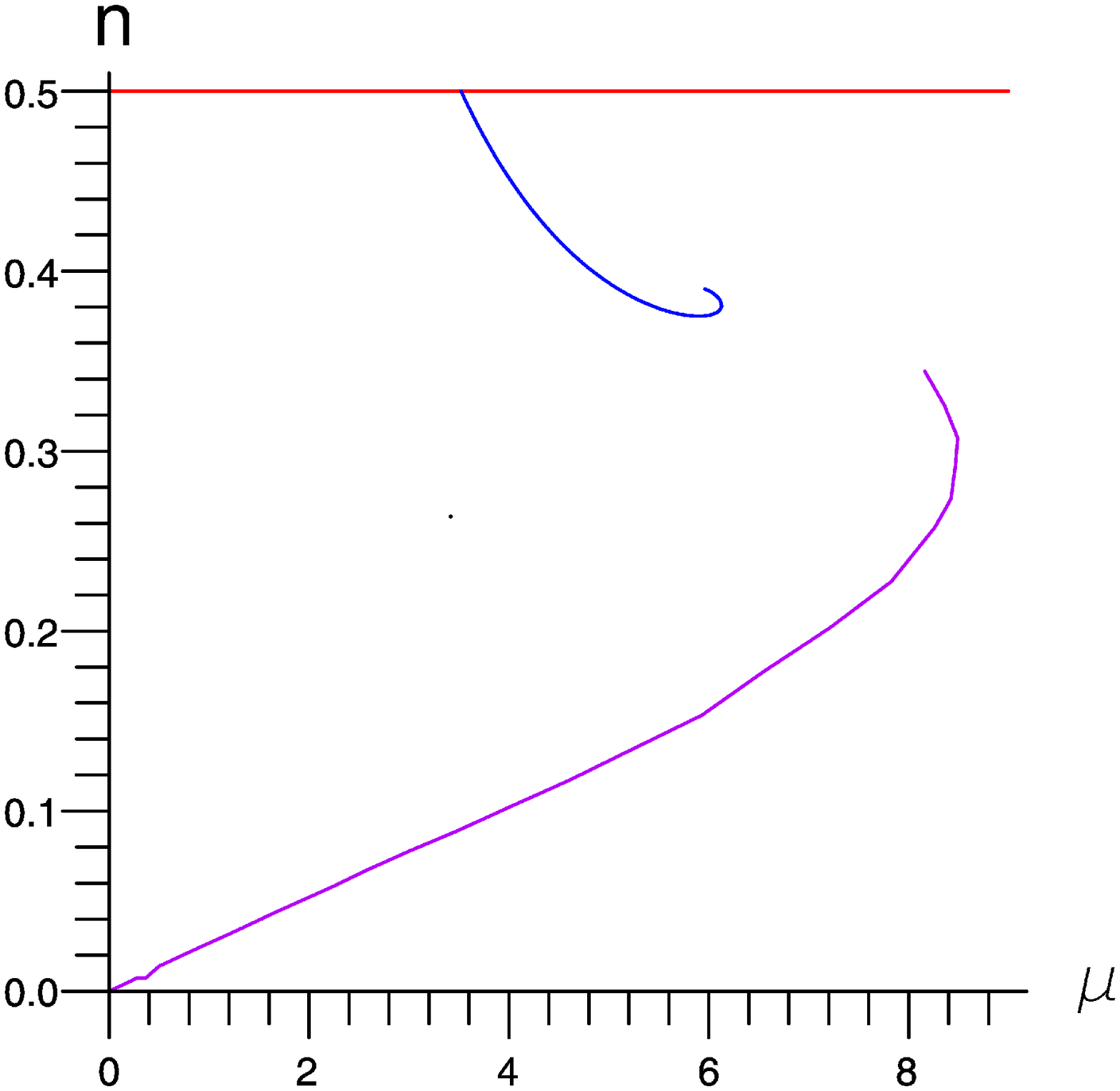}}
\hspace{.5cm}
\subfigure[\label{phase2}d=6]{\includegraphics[scale=.4]{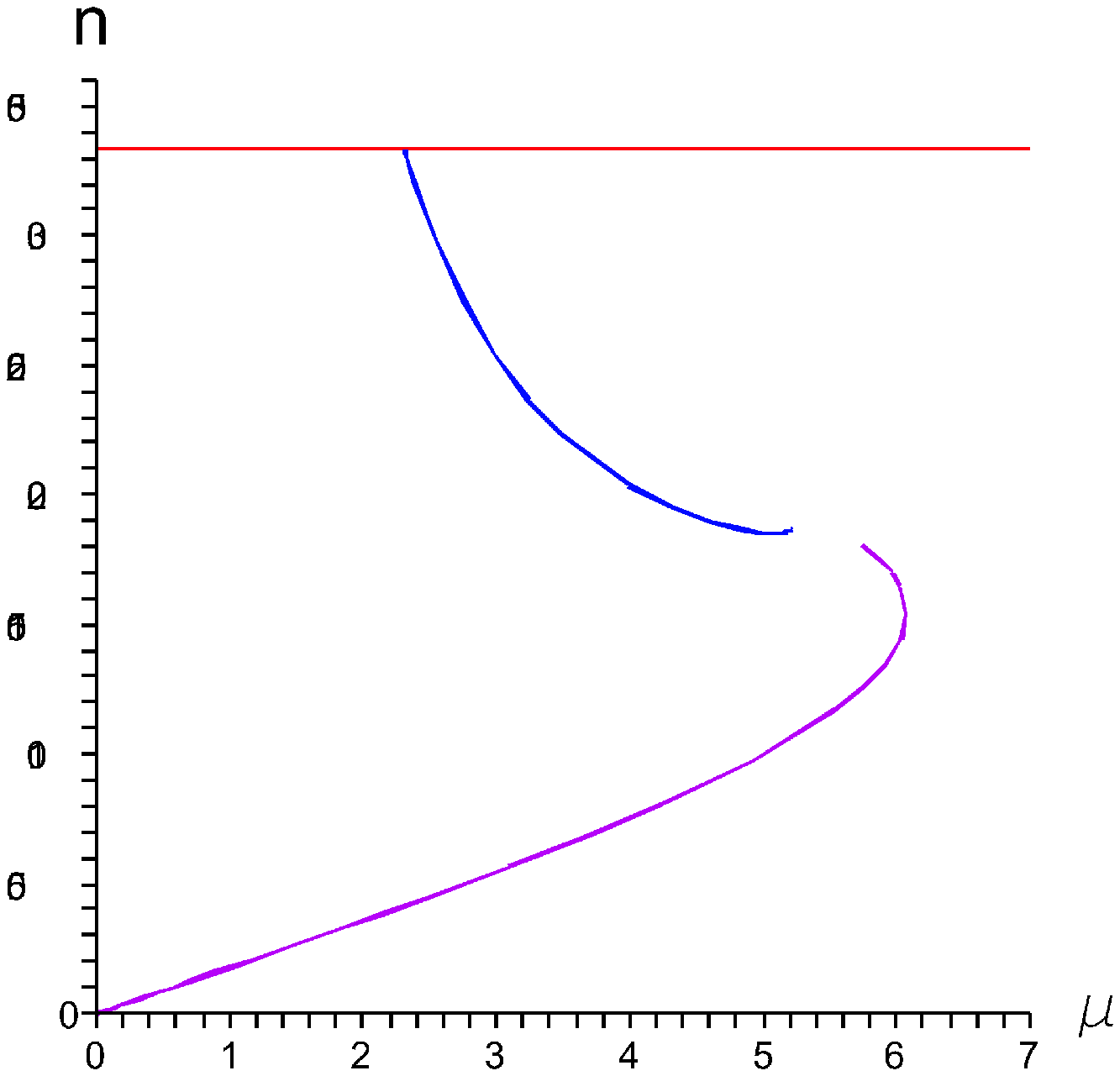}}
\caption{The $\mu-n$ phase diagram, $n$ being the relative tension $\mathcal T L/M$ and $\mu = M/L^{d-4}$ is the rescaled mass. Reprint from \cite{glrev}. The straight line is the uniform string phase, the line going down and emanating from the uniform phase is the non uniform phase while the line emanating from the origin is the localized black hole phase.}
\label{phasediag}
\end{figure}

The non uniform black string solution is a static solution to the vacuum Einstein equations and have been studied first perturbatively by Gubser \cite{gubser} and in the full non linear regime by Wiseman \cite{wiseman}. The non uniform phase is connected to a localised black hole phase; at the connection point (merger point), there is a horizon topology change. The localised black hole phase have been studied by T. Wiseman and Kudoh \cite{wiseman,kuwi} and the resulting phase diagram connecting the localised black hole, the non uniform black hole and the uniform string has been studied in \cite{kuwi}. The phase diagram is presented in figure \ref{phasediag} for $d=5,6$. The lines represent the various static solution which should be equilibrium configurations since they don't depend on time. 

It should be noted that various 'problems' appear in this context; for example, it is not clear that cosmic censorship (which roughly state that every singularities should be hidden behind an horizon) is not violated. At the merger point, the string disconnects and form a localized black hole; precisely at this point, the areal radius (square root of the coefficient of the angular part of the metric) of the horizon vanishes leaving the singularity naked.
Another issue is that the horizon topology change from $S_{d-3}\times S_1$ to $S_{d-2}$; it is indeed possible to study these solutions starting from the black hole side or by the string side, but not both in once. This of course makes the picture quite involved \cite{kuwi}.

The transition from black strings to black hole is also from considerable interest from the thermodynamical point of view. For instance, the non uniform string is less entropic than the uniform string for small deformations, up to certain number of dimensions (13). This is called the critical dimension \cite{sorkin}. We will come back to the question of critical dimension in the next chapter; we refer the interested reader to the review from Kol \cite{kolrev} for a deeper discussion on phase transition between localized black hole and black strings.

\section[$\Lambda>0$ Black Strings]{Black string solutions with a positive cosmological constant}
\label{bss}
Considering \eqref{sgen} with a positive cosmological constant and restricting to pure gravity ($L_m=0$), we can try to construct the generalization of the black string solution presented in the previous section. This was actually already done in \cite{rms} for a negative cosmological constant, but we will come back to the $AdS$ black strings in more details in the next chapter.

In order to construct a black string solution with a cosmological constant, we supplement the action \eqref{sgen} by the following ansatz
\be
ds^2 = -b(r) dt^2 + \frac{dr^2}{f(r)} + r^2d\Omega_{d-3}^2 + a(r)dz^2.
\label{ansdsubs}
\ee
where $d\Omega^2_{d-3}$ denotes the square line element on the unit $(d-3)$-sphere.

The reason why we add the new field $a$ can be intuitively understood by the fact that the cosmological constant gives a curvature to the extra-direction. From a mathematical point of view, the ansatz without the function $a$ doesn't lead to a consistent set of equations.

\subsection{Equations and boundary conditions}
The resulting equations of motion are given by
\begin{eqnarray}
f'&=&\frac{2(d-4)}{r}-\frac{2(d-1)r}{\ell^2}-\frac{2(d-4)f}{r}-f\left(\frac{a'}{a}+\frac{b'}{b} \right),\nonumber\\
b''&=&\frac{(d-3)(d-4)b}{r^2} -\frac{(d-3)(d-4)b}{r^2f} +\frac{(d-1)(d-4)b}{\ell^2f} +\frac{(d-3)ba'}{ra}\nonumber\\ &&+\frac{(d-4)b'}{r}-\frac{(d-4)b'}{rf}+\frac{(d-1)rb'}{\ell^2f}+\frac{a'b'}{2a}+\frac{b'^2}{b},\\
\frac{a'}{a}&=& 2\frac{ b\big[-\ell^2(d-3)(d-4)(1-f)+(d-1)(d-2)r^2\big]+(d-3)r\ell^2fb'}{-r\ell^2f\big[rb'+2(d-3)b\big]}\nonumber
\label{eqdsubs}
\end{eqnarray}
We supplement these equations by the boundary conditions corresponding to a black string:
\be
b(r_h) = f(r_h)=0,\ b'(r_h)=1,\ a(r_h)=1,
\label{bcdsbs}
\ee
where we set the normalisation of $a,b$ such that $a(r_h)=1,\ b'(r_h)=1$ without loss of generality since the normalisation can be absorbed in a redefinition of the time and $z$ coordinate. The conditions on $b(r_h),\ f(r_h)$ are necessary to implement a regular horizon at $r=r_h$.

It is interesting to notice that the above equations may be written in a decoupled form. The decoupled form of the equation is valid for both $\Lambda>0$ and $\Lambda<0$. In the following, we define $\epsilon = -\Lambda/|\Lambda|$.

First, let us introduce
\be
A(r) = \frac{ra'}{a}\ \ B(r) = \frac{rb'}{b}
\ee
in terms of which the equations for $f,a,b$ rewrite:
\be
rf' = 2(d-4)\left(k-f\right) - 2(d-1)\frac{r^2}{\epsilon l^2} - f(r)\left(A+B\right),
\label{eqter1}
\ee
\be
rB'f = (d-1)(2-B) \frac{r^2}{\epsilon l^2} - (d-4)kB,
\label{eqter2}
\ee
\bea
\label{eqter3}
\left(AB + 2(d-3)(A+B)\right. &+&\left. 2(d-3)(d-4)\right)f \\
		       &=& 2k(d-3)(d-4) - 2(d-1)(d-2)\frac{r^2}{\epsilon \ell^2}.\nonumber
\eea
Solving the third equation for $A = A(B,f)$, substituting $A(B,f)$ in the first equation and solving for $B=B(f,f')$ and finally evaluating the second equation with $A=A(B(f,f'),f),\ B=B(f,f')$ leads to the following decoupled equations:

\bea
r^2ff'' &=&  - rff' + \left(rf'\right)^2 + 2\left( (d-1) - (d-4)\frac{r^2}{\epsilon l^2}\right)^2 \\
&-& 2(d-4)\left( (d-1) - (d-4)\frac{r^2}{\epsilon l^2}\right)f - 3\left( k(d-1) - (d-4)\frac{r^2}{\epsilon l^2}\right)rf'\nonumber
\label{eqf}
\eea

\bea
B &=& -\tilde B \\
      &\pm & \sqrt{ \tilde B^2 - 2(d-3)\left((d-4)\left(1-\frac{1}{f} \right) - \frac{rf'}{f}\right) - \epsilon \frac{2r^3(d-1)(d-4)r^2}{\ell^2 f}} \nonumber
\eea

where $\tilde B := \left( (d-4) + \frac{1}{2}\frac{rf'}{f} + (d-4)\frac{1}{f} - (d-1)\frac{1}{f}\frac{r^2}{\epsilon l^2}\right)$ and
\be
A = \frac{-2(d-1)(d-2)\frac{r^2}{\epsilon l^2} + 2(d-3)(d-4)f + 2(d-3)fB}{2(d-3)f + fB}.
\ee
Note that the decoupling is still valid in the more general case considered in \cite{rms} where
topological black holes are investigated as well (i.e. space-times  where the spherical part $d\Omega^2_{d-3}$ in \eqref{ansdsubs} is replaced by the metric of an hyperbolic or flat manifold with the same dimensions).

\subsection{Asymptotic solution}
We have not been able to solve \eqref{eqf} explicitly, but we can study the asymptotic behaviour of $f$ with this equation. In fact, to the leading order, it appears that
\be
f(r)\approx F_0r^\phi
\ee
for a constant $F_0$ and for any $\phi \in \{2\}\cup]4,\infty[$, so we cannot conclude on the asymptotic behaviour of $f$ at this stage. In fact, it is not possible to fix analytically $\phi$, but it is possible to give a description of the asymptotic behaviour of the metric functions in terms of only one parameter.

We have to consider two cases : $\phi = 2$ and $\phi>4$. In the case $\epsilon = +1$, where the cosmological constant is negative, the asymptotic solution admits $\phi=2$ \cite{rms}. In the case $\epsilon=-1$, there are strong numerical evidences that $\phi=2$ leads to a solution where the metric functions diverge at the event horizon \cite{bd}. Since this section is precisely devoted to the case $\epsilon=-1$, we will then assume $\phi>4$.

Assuming $a(r)\propto r^\alpha,\ b(r)\propto r^\beta$ and $f=F_0 r^\phi$ for large values of $r$, it is possible to express $\alpha$ and $\phi$ as a function of $\beta$ only, i.e. to describe the asymptotic behaviour of the metric functions with only one parameter:
\bea
f&\rightarrow& F_0r^\phi\ , \ \phi = -\frac{\beta^2 + 2(d-4)\beta + 2(d-3)(d-4)}{\beta + 2(d-3)},\nonumber\\
A(r)&\rightarrow& \alpha(\beta) = -\frac{2(d-3)(d-4) + 2(d-3)\beta}{\beta + 2(d-3)},\nonumber\\
B(r) &\rightarrow& \beta,
\label{kasneras}
\eea
for large values of $r$.

Note the particular relation between the exponents:
\be
\alpha + \beta + \phi + 2(d-4) = 0
\label{link}
\ee

Since these behaviours are compatible with the equations of motion for each value of $\beta$ such that $\phi > 4$, it is not possible to fix $\beta$ without more assumptions, but there are strong numerical evidences for $a(r) = b(r)$ at the leading order in the asymptotic region. 
This means that the value of $\beta$ chosen by the system is a fixed point of $\alpha(\beta)$ and is given by
\be
\alpha =  \beta = -2(d-3) - \sqrt{2(d-2)(d-3)}.
\label{exvalue}
\ee

We have constructed the higher order corrections for the function $f(r)$ and obtained
\be
f(r) \rightarrow F_0r^\phi - \frac{d-1}{d-3+\beta} \frac{r^2}{\ell^2} + \frac{d-4}{d-4+\beta}, \mbox{ for }r\rightarrow\infty,
\label{secord}
\ee
with $\phi$ discussed previously, $\phi = -\frac{\beta^2 + 2(d-4)\beta + 2(d-3)(d-4)}{\beta + 2(d-3)}$.

Note the subleading term $-r^2/\ell^2$; it is the term we expected to the leading order to describe an asymptotically locally de Sitter spacetime. As a consequence, it is obvious that the solution does indeed not describe an asymptotically locally de Sitter spacetime. In the next sections, we will evaluate the energy and try to determine whether this solution can be physically relevant.

\subsection{Energy of the solution}
The energy of the solution is given by the following quantity \cite{hoha}:
\be
E = -\frac{1}{8\pi G}\int_{S_t^\infty} N\left( K_{d-2} - K_{0,d-2} \right)
\label{masshawk}
\ee
where $N$ is the lapse function, $S_t^\infty$ is the boundary of the fix time sliced manifold $\Sigma_t$, $K_{d-2}$ is the trace of the extrinsic curvature of $S_t^\infty$ in $\Sigma_t$ and $K_{0,d-2}$ is the trace of the extrinsic curvature of the fixed time boundary of some reference background spacetime (see appendix \ref{app:reminder}).

The trace of the extrinsic curvature of the $r=Const.$ hypersurfaces, say $K_{d-2}$ is given by:
\be
K_{(d-2)} = \sqrt{f}\left(  \frac{(d-3) + \frac{\beta}{2}}{r}\right).
\ee
This quantity, once integrated according to \eqref{masshawk}, is not divergent for $r\rightarrow\infty$ despite of the unusual asymptotic form of the solution. It follows that we don't need a reference background in order to have a finite energy. Anyway, it wouldn't have been not clear which background to use with such an asymptotic.

We can now compute the energy of the solutions which exhibits the asymptotic behaviour as described in the previous section. Using \eqref{link}, the energy is simply given by
\be
E =  \frac{1}{8\pi G}V_{d-3} L \sqrt{ 2F_0(d-2)(d-3)}
\ee
where $L$ is the length of the extradimension in the $z$ direction and $V_{d-3}$ is the surface of the unit $d-3$ sphere. This energy is finite and depend only on $F_0$ which depends essentially on the dimension of space-time and of the cosmological constant. Note that $F_0$ is determined numerically.

%%%%%%%%%%%%%%%%%%%%%%%%%%%%%%%%%%%%%%%%%%%%%%%%%%%%%
%%%% Geodesic equations and Curvature invariants  %%%
%%%%%%%%%%%%%%%%%%%%%%%%%%%%%%%%%%%%%%%%%%%%%%%%%%%%%
\subsection{Geodesic equations and curvature invariant}
We will now compute some curvature invariants in order to check whether the solution with the asymptotic behaviour described above have a chance to be regular. The Kretschmann curvature invariant is given by:
\be
K=R_{abcd}R^{abcd} = (d-3)\frac{2(d-4)-4(d-4)f +2(d-4)f^2 + r^2f'^2}{r^4}.
\label{kret}
\ee
Plugging the asymptotic behaviour of the function $f$ in this expression shows that $K$ is diverging for large values of $r$:
\be
K\rightarrow (d-3)F_0^2 \left(2(d-4)+\phi^2\right) r^{2 \phi -4}\rightarrow \infty\mbox{ for }r\rightarrow\infty,
\ee
since $\phi>4$ by assumption. This shows that the solution with the asymptotic behaviour of previous section is asymptotically singular.

Worst, the asymptotical singularity can be reached in a finite proper time for a massive or massless observer (say with mass $m$) starting from outside the horizon. This can be seen by integrating the geodesic equation at fixed $\theta_i$ (the angular sector of the metric).

Using the fact that $k_a \dot x^a$ is a constant along geodesics, provided $k$ is a Killing vector\footnote{In order to prove that, one must use the geodesic equation: $\dot x^a\nabla_a \dot x^b = 0$, dot denoting derivation with respect to an affine parameter parametrising the geodesic, and the Killing equation: $\nabla_a k_b + \nabla_b k_a =0$.}, the geodesic equation can be written in the form \cite{bdbh1}:
\be
\dot r^2 = f\left(\frac{\mathcal E^2}{b} - \frac{\mathcal Z^2}{a} - m^2\right),
\ee
where $\mathcal E, \mathcal Z$ are conserved quantities along the geodesic, related to the observer's energy and momentum along the $z$ direction and dot denotes derivative with respect to an affine parameter along the geodesic, say $\lambda$. This equation can be integrated numerically with particular solutions admitting the singular asymptotic behaviour. One can also assume that the observer is in a region well described by the asymptotic solution and integrate the geodesic equation using $a=b=b_0 r^\beta$ and $f=F_0 r^\phi$ to first order in $r$, with $\beta$ defined above and obtain:
\be
\lambda-\lambda_0\approx\int_{r_0}^{r(\lambda)} \frac{dr}{\sqrt{r^{\phi-\beta}}}\approx r^{1+\frac{\beta-\phi}{2}} - r_0^{1+\frac{\beta-\phi}{2}}.
\ee
It turns out that $1+\frac{\beta-\phi}{2}<0$ for all values of $d\ge5$. Consequently, in the limit $r\rightarrow\infty$, $\lambda$ is finite. In other words, an observer far from the horizon can reach the asymptotic singularity in a finite affine time.
A numerical computation for observers that start close to the horizon reveals that the singularity can still be reached in a finite time for a massive or massless observer.

\subsection{Numerical results}

\begin{figure}
\centering
\includegraphics[scale=.6]{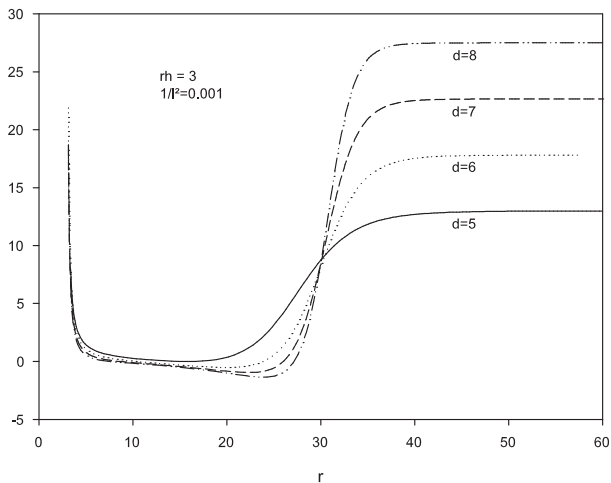}
\caption{The ratio $rf'/f$ is given for several values of $d$}
\label{fig:rfpsf1} 
\end{figure}

We solved numerically equations \eqref{eqter1}, \eqref{eqter2} and \eqref{eqter3} with the boundary conditions \eqref{bcdsbs} for many values of the horizon $r_h$.
We present the evolution of the ratio $rf'/f$ for $\Lambda>0$ black string solutions and for several values of $d$ and $r_h=0.5$ in figure \ref{fig:rfpsf1}. The figure clearly demonstrates that the power law configuration \eqref{kasneras} is approached.
The exact values of the exponents \eqref{exvalue} coincide with our numerical values within the numerical accuracy required for the numerics, i.e. typically $10^{-8}$.

We also investigated the solution in the interior region, i.e. for $r \leq r_h$, to check whether there exists a second horizon which would 'hide the asymptotic singularity', but it turned out that this was not the case. We have a naked singularities both, at $r=0$ and at $r=\infty$. We are tempted to interpret this solution as an hypercylindre with an horizon at some equator by matching the origin with the point
at infinity by means of an appropriate system of coordinates, still to be found, but that we expect to exist.

\section{Concluding remarks}
\label{cclbh}
In this chapter, we quickly reviewed black hole and black string solutions available in higher dimensional general relativity. The analytically known black hole solutions are either rotating and charged, either rotating with a cosmological constant, either charged with a cosmological constant. The missing part was the charged rotating with cosmological constant solutions. Such solutions have been constructed numerically for a negative cosmological constant in \cite{solubh3} and we constructed their counterpart for a positive cosmological constant. The main difficulty in the case of a positive cosmological constant is the fact that there exists a cosmological horizon where some metric coefficients become zero. We imposed regularity constraints at the cosmological horizon and added an equation, expressing the fact the cosmological constant is indeed a constant in order to have sufficient freedom to impose the constraints.

We also reviewed the black string solution, in flat spacetime and with a positive cosmological constant. We left the $AdS$ black string for the next chapter. We briefly introduced the Gregory-Laflamme instability and the phase diagram of the black string in flat spacetime.
We presented black string solutions with a positive cosmological constant; an interesting result is that there doesn't exist an asymptotically de Sitter black string, the spacetime becoming asymptotically singular. This can be interpreted as the fact that the compact direction cannot support the pressure induced by the cosmological constant.

%% file: bsphase.tex
%%%%%%%%%%%%%%%%%%%%%%%%%%%%%
%%% Black Strings phases

%%%%%%%%%%%%%%%%%%%%%%%%%%%%%
In this chapter, we consider the black string solution in asymptotically locally $AdS$ spacetime. First, we review the $AdS$ uniform solution and revisit some particular thermodynamical properties of the latter. These uniform black strings in $AdS$ have been considered by several authors in different contexts \cite{rms,klemmbs,cophor}. Then we investigate the stability of the $AdS$ uniform black string by looking for the existence of the counterparts of the Gregory-Laflamme zero modes. This approach was introduced by S. S. Gubser \cite{gubser} for asymptotically locally flat black strings. It consists in considering the non uniform solution as a perturbation of the uniform solution and in solving the first order equations in the perturbation. 

In a third step, we present solutions to the non uniform problem, first in a perturbative approach then in the full non linear regime. Whereas the uniform black string does not depend on the extra-coordinate, the non uniform black string is characterized by a non trivial dependence on this extra-coordinate. Up to our knowledge, this is the first time this technique is applied to asymptotically locally $AdS$ spacetimes.

The second order in the perturbative approach is already sufficient to see that the cosmological constant provides drastic changes in the phase diagram of $AdS$ black strings. In particular, it suggests the existence of a thermodynamically stable phase of non uniform black string in a regime where the corresponding uniform black string with same length is thermodynamically unstable. This is one of the unexpected features of the asymptotically locally $AdS$ spacetime.

Finally, the solution to the system of partial differential equation provides hints for the existence of a localized black hole phase and for the possibility that the phase transition between the uniform and non uniform phase can be of second order in every number of dimensions, provided the horizon radius is large enough. The $AdS$ localized black hole solution has not been constructed yet, up to our knowledge.

This chapter presents the counterpart of the black string (uniform and non uniform) which we briefly discussed in the previous chapter. Our original results have been presented in \cite{nubsd,pnubsads,pnubsd,rbd}.

\section{The generic model}
We consider the $d$-dimensional Einstein Hilbert action with the presence of a (negative) cosmological constant and the Hawking-Gibbons boundary terms,
\be
S = \frac{1}{16\pi G}\int_{\mathcal M} \sqrt{-g}\left( R - 2 \Lambda \right)d^dx + \frac{1}{8\pi G}\int_{\partial\mathcal M}\sqrt{-h}Kd^{d-1}x,
\label{snubs}
\ee
where $G$ is the $d$-dimensional Newton constant, $g$ the determinant of the metric, $R$ the scalar curvature computed with the metric $g_{AB},\ A,B=0,\ldots,d-1$ associated to the manifold $\mathcal M$, $\Lambda$ the cosmological constant, $h$ the determinant of the induced metric on the boundary $\partial\mathcal M$ of the manifold and $K$ the extrinsic curvature of the boundary manifold.

In order to investigate the various phases of asymptotically locally $AdS$ black strings, we consider the equations resulting from the variation of \eqref{snubs} with respect to the metric components. We then study the equations for metrics of the form
\be
ds^2 = -e^{2A(r,z)}b(r) dt^2 + e^{2B(r,z)}\left(\frac{dr^2}{f(r)} + a(r)dz^2\right) + e^{2C(r,z)}r^2 d\Omega_{d-3}^2,
\label{ansatzbs}
\ee
for some functions $a,b,f,A,B,C$ and where $d\Omega_{d-3}$ is the line element of the $(d-3)$-unit sphere. We impose an horizon radius at a given value of the radial coordinate, say $r_h$. The variable $r$ then ranges from $r_h$ to $\infty$ while the coordinate $z$ ranges from $0$ to $L\in\mathbb R_+$. We furthermore define the $AdS$ curvature radius $\ell = \sqrt{-\frac{(d-1)(d-2)}{2\Lambda}}$, which will be useful later.

The metric ansatz \eqref{ansatzbs} is a generalization of the uniform black string metric \eqref{ansdsubs} of chapter \ref{chbhbs}; we recover \eqref{ansdsubs} for $A=B=C=0$. The functions $A(r,z),B(r,z),C(r,z)$ parametrise the non uniformity and we will refer to these as the correction functions (although they are more than 'just' corrections in the non perturbative approach presented in the last sections of this chapter). 

In the presence of a non vanishing cosmological constant, the field equations imposes
\be
b(r)= \frac{r^2}{\ell^2}+\Ord{r}{0}\ \ ,\ \ f(r)= \frac{r^2}{\ell^2}+\Ord{r}{0}\ \ ,\ \ a(r)= \frac{r^2}{\ell^2}+\Ord{r}{0},
\label{ubsas}
\ee
to the leading order for $r\rightarrow\infty$ for any value of the $z$ coordinate. This asymptotic behaviour describes an asymptotically locally $AdS$ space (Appendix \ref{app:ads}). The correction functions vanish at infinity; if it was not the case, the asymptotic spacetime wouldn't be of the form \eqref{ubsas}. It is worth noting that the decay of the correction functions cannot be arbitrary, in particular it shouldn't spoil the asymptotic form of the metric components. This will be discussed in due course.

The ansatz \eqref{ansatzbs} is a general axisymmetric parametrisation (in a certain gauge); the functions $A,B,C$ parametrise the $z$ component dependence of the line element whereas $a,b,f$ are relevant for a uniform setup. The need of the function $a(r)$ was already discussed in the previous chapter, in the black string with positive cosmological constant; this is in contrast with the asymptotically locally flat case where the extra-coordinate $z$ was Ricci-flat.

In the next sections, we will consider the uniform string phase, its stability, the non-uniform string phase in a perturbative approach and in a non perturbative approach.

From now on, we will omit the coordinate dependence of the various metric functions.

%%%%%%%%%%%%%%%%%%%%%%%%%%%%%%%%%%%%%%%%%%%%%%%%%%%%%%%%%%%%%%%%%%%%%%%%%%%%%%%%%%%%%%%
%%%%%%%%%%%%%%%%%%%%%%%%%%%%%%%%%%%%%%%%%%%%%%%%%%%%%%%%%%%%%%%%%%%%%%%%%%%%%%%%%%%%%%%
%%%%%%%%%%%%%%%%%%%%%%%%%%%%%%%%%%%%%%%%%%%%%%%%%%%%%%%%%%%%%%%%%%%%%%%%%%%%%%%%%%%%%%%
%%%             Uniform String
%%%%%%%%%%%%%%%%%%%%%%%%%%%%%%%%%%%%%%%%%%%%%%%%%%%%%%%%%%%%%%%%%%%%%%%%%%%%%%%%%%%%%%%

\section{Uniform string phase}
The uniform phase is characterized by
\be
A=B=C=0,
\label{ABC0}
\ee
since the functions $A,B,C$ induce a non trivial dependence on the $z$ coordinate when non vanishing, making the string non uniform (along the $z$ direction). The line element of the uniform black string then reduces to 
\be
ds^2 = -b(r) dt^2 + \frac{dr^2}{f(r)} + r^2d\Omega_{d-3}^2 + a(r) dz^2.
\label{ansubs}
\ee
Unfortunately, there is no analytic solution describing $AdS$ $d$-dimensional black strings, so one has to rely on a numerical resolution of the Einstein equations, except in some specific case, where other fields are present (for example with an additional $2$-form and particular relations between the coupling constants, see \cite{klemmbs}). Solutions for the asymptotically locally $AdS$ black string have been first constructed in \cite{rms}; we will review the main properties of the uniform string in the following section.

\subsection{The equations and boundary conditions}
The variation of the action \eqref{snubs} with respect to the metric components, supplemented by the ansatz \eqref{ansubs} and \eqref{ABC0} leads to a coupled system of three non linear ordinary differential equations :
\bea
f'&=&\frac{2(d-4)}{r}+\frac{2(d-1)r}{\ell^2}-\frac{2(d-4)f}{r}-f\left(\frac{a'}{a}+\frac{b'}{b} \right),\nonumber\\
b''&=&\frac{(d-3)(d-4)b}{r^2} -\frac{(d-3)(d-4)b}{r^2f} -\frac{(d-1)(d-4)b}{\ell^2f} +\frac{(d-3)ba'}{ra}\nonumber\\
&&+\frac{(d-4)b'}{r} -\frac{(d-4)b'}{rf}-\frac{(d-1)rb'}{\ell^2f}+\frac{a'b'}{2a}+\frac{b'^2}{b},\nonumber\\
\frac{a'}{a}&=& 2\frac{ b\big[\ell^2(d-3)(d-4)(1-f)+(d-1)(d-2)r^2\big]-(d-3)r\ell^2fb'}{r\ell^2f\big[rb'+2(d-3)b\big]}
\label{equbs}
\eea

\subsubsection{Boundary conditions}
In order to describe black objects, we impose an event horizon on a given value of the radial coordinate, say $r_h$ by setting $b(r_h)=0$. The regularity of the equation then imposes $f(r_h)=0$. Since the system of differential equations has a total degree 4, we still need two boundary conditions. It is important to notice that the equations \eqref{equbs} are invariant under independent rescaling of the functions $a$ and $b$. So we choose to arbitrarily fix $a(r_h)=b'(r_h)=1$ and globally rescale the $a$ and $b$ functions \emph{a posteriori} in order to obtain the correct asymptotic $a= r^2/\ell^2+\Ord{r}{0},\ b=r^2/\ell^2+\Ord{r}{0}$. Note that the equations \eqref{equbs} are not invariant under a rescaling of the $f$ function, moreover, once $f(r_h)=0$ is fixed, the asymptotic behaviour of the $f$ function is naturally $r^2/\ell^2$ to the leading order.

To sum up, the boundary conditions read
\be
 f(r_h)=b(r_h)=0\ \ ,\ \ a(r_h)=b'(r_h)=1.
\label{bcubs}
\ee
for a given value of the horizon radius $r_h$.

Note that there exists a solution of \eqref{equbs} regular on $r\in[0,\infty]$. However, in this case, regularity conditions and thus boundary are different:
\be
f(0)=1,\ \ b(0)=b_0,\ a(0)=a_0,\ b'(0)=0,
\ee
for real values of $a_0, b_0$ chosen such that the solution is asymptotically locally $AdS$.

The black string solution obeying \eqref{bcubs} approach the regular solution for $r_h\rightarrow0$; the convergence is point-like for $r\in]0,\infty]$.
Further details on the globally regular solution can be found in \cite{rms}.

\subsection{Near horizon and asymptotic expansion}
In the region where $r\approx r_h$, the solution can be expended as a Taylor series according to
\bea
\label{nhubs}
a(r) &=& a_h + \frac{2a_h\left(d-1\right)r_h}{\left(d-4\right)\ell^2+\left(d-1\right)r_h^2}(r-r_h) \\
&+& \frac{2a_h\left(d-1\right)^2r_h^2}{\left(\left(d-4\right)\ell^2+\left(d-1\right)r_h^2 \right)^2}\frac{(r-r_h)^2}{2}+\mathcal O(r-r_h)^3,\nonumber\\ 
b(r) &=& b_h(r-r_h) + \frac{b_h\left(d-4\right)\left(\left(d-3\right)\ell^2+\left(d-1\right)r_h^2 \right)}{\left(d-4\right)\ell^2r_h + \left(d-1\right)r_h^3}\frac{(r-r_h)^2}{2}+\mathcal O(r-r_h)^3,\nonumber\\ 
f(r) &=& \left(\frac{(d-1)}{r_h} + \frac{(d-4)r_h}{\ell^2}\right)(r-r_h) - (d-4)\left(\frac{d-1}{\ell^2}+\frac{d-3}{r_h^2}\right)\frac{(r-r_h)^2}{2}\nonumber\\
&+& \mathcal O(r-r_h)^3;\nonumber
\eea
where $a_h=a(r_h),\ b_h=b'(r_h)$ are real constants.

The asymptotic behaviour of the background functions is given in \cite{rms}:
\bea
a&=& \frac{r^2}{\ell^2}  + \sum_{i=0}^{\left\lfloor \frac{d-4}{2}\right\rfloor} a_i\left(\frac{\ell}{r}\right)^{2i} + c_z\left(\frac{\ell}{r}\right)^{d-3} + \sum_k\delta_{d,2k+1}\xi\log\frac{r}{\ell}\left(\frac{\ell}{r}\right)^{d-3} + \mathcal O\left(\frac{\ell}{r}\right)^{d-1},\nonumber\\
b&=& \frac{r^2}{\ell^2}  + \sum_{i=0}^{\left\lfloor \frac{d-4}{2}\right\rfloor} a_i\left(\frac{\ell}{r}\right)^{2i} + c_t\left(\frac{\ell}{r}\right)^{d-3} + \sum_k\delta_{d,2k+1}\xi\log\frac{r}{\ell}\left(\frac{\ell}{r}\right)^{d-3} + \mathcal O\left(\frac{\ell}{r}\right)^{d-1},\nonumber\\
f&=& \frac{r^2}{\ell^2}  + \sum_{i=0}^{\left\lfloor \frac{d-4}{2}\right\rfloor} f_i\left(\frac{\ell}{r}\right)^{2i} + (c_t+c_z+c_0)\left(\frac{\ell}{r}\right)^{d-3} + \sum_k\delta_{d,2k+1}\xi\log\frac{r}{\ell}\left(\frac{\ell}{r}\right)^{d-3}\nonumber\\ 
&+& \mathcal O\left(\frac{\ell}{r}\right)^{d-1},
\label{asbckg}
\eea
where $\delta$ is the Kronecker delta, $\lfloor X \rfloor$ stands for the floor integer value of $X$, $c_t,c_z$ are parameters to be determined numerically and $a_i,f_i,c_0,\xi$ depend on $d$ and are given in \cite{rms}.

For fixed values of $r_h$ and $\ell$, the black string solution extrapolating between \eqref{nhubs} and \eqref{asbckg} has very specific values of $a_h, b_h$.

\subsection{Thermodynamical properties of the uniform black string}
\label{thbs}
The Hawking temperature of the string can be found by computing the surface gravity or by demanding regularity in the Euclidean section near the horizon. Both procedure give the same result:
\be
T_H= \frac{1}{4\pi}\sqrt{\frac{b_h}{r_h\ell^2}\left[(d-1)r_h^2+(d-4)\ell^2\right]}.
\ee

The entropy of the black string is computed using the Hawking-Bekenstein formula and is given by one quarter of the horizon area:
\be
S = \frac{1}{4G}r_h^{d-3}V_{d-3}\sqrt{a_h}L,
\ee
where $V_{d-3}$ is the surface of the unit $(d-3)$-sphere.

The quantities defined in the asymptotic region, namely the mass and tension require more attention. Standard approaches like Komar integrals (see appendix \ref{app:reminder}) are not applicable because the action is diverging asymptotically. However, it is possible to use a regularising procedure; such a procedure has been first proposed by Balasubramanian and Krauss \cite{counter} and was inspired by the $AdS/CFT$ correspondence. This procedure consists on adding to the action \eqref{snubs} appropriate boundary terms $I_{ct}$, which are functional only of curvature invariants of the induced metric on the boundary. Such terms will not interfere with the equations of motion because they are intrinsic invariants of the boundary metric. By choosing appropriate counterterms, which cancel the divergences, one can then obtain well-defined expressions for the action and the energy momentum of the spacetime. This procedure is intrinsic to the spacetime of interest and it is unambiguous once the counterterm action is specified.

Thus we have to supplement the action \eqref{snubs} with \cite{counter}:
\begin{eqnarray}
\label{Lagrangianct}
I_{\mathrm{ct}}^0 &=&\frac{1}{8\pi G}\int_{\partial\mathcal M} d^{d-1}x\sqrt{-\gamma}\left\{ -\frac{d-2}{\ell }-\frac{\ell \mathsf{\Theta }\left( d-4\right)}{2(d-3)}\mathsf{R} \right.\nonumber \\
&&\left.-\frac{\ell^3\mathsf{\Theta}(d-6)}{2(d-3)^2(d-5)} \left(\mathsf{R}_{ab}\mathsf{R}^{ab} - \frac{d-1}{4(d-2)}\mathsf{R}^{2}\right)\right. \\
&&\left.+\frac{\ell^5\mathsf{\Theta}(d-8)}{(d-3)^3(d-5)(d-7)}\left(\frac{3d-1}{4(d-2)}\mathsf{RR}^{ab}\mathsf{R}_{ab}-\frac{d^2-1}{16(d-2)^2} \mathsf{R}^3 \right.\right.\nonumber\\
&&\left.\left.-2\mathsf{R}^{ab}\mathsf{R}^{cd}\mathsf{R}_{acbd}-\frac{d-1}{4(d-2)}\nabla_a\mathsf{R}\nabla^a\mathsf{R}+\nabla^c\mathsf{R}^{ab} \nabla_c\mathsf{R}_{ab}\right) +\ldots\right\},\nonumber
\end{eqnarray}
where $\mathsf{R}$ and $\mathsf{R}^{ab}$ are the curvature and the Ricci tensor associated with the induced metric $\gamma$ on $\partial\mathcal M$ and $\mathsf{\Theta }(x)$ is the step function such that $\mathsf{\Theta }\left( x\right) =1$ provided $x\geq 0$, and vanishes otherwise.

The series truncates for any fixed dimension, with new terms entering at every new even value of $d$, as denoted by the step-function.

However, given the presence of $\log(r/\ell)$ terms in the asymptotic expansions \eqref{asbckg} (for odd $d$), the counterterms \eqref{Lagrangianct} regularise the action for even dimensions only. For odd values of $d$, we have to add the following extra terms to \eqref{snubs} \cite{counterlog}:
\begin{eqnarray}
I_{\mathrm{ct}}^{s} &=&\frac{1}{8\pi G}\int d^{d-1}x\sqrt{-\gamma }\log(\frac{r}{\ell})\left\{ \mathsf{\delta }_{d,5}\frac{\ell^3 }{8}(\frac{1}{3}\mathsf{R}^2-\mathsf{R}_{ab}\mathsf{R}^{ab})\right.\\
&&-\frac{\ell ^{5}}{128}\left(\mathsf{RR}^{ab}\mathsf{R}_{ab}-\frac{3}{25}\mathsf{R}^{3} -2\mathsf{R}^{ab}\mathsf{R}^{cd}\mathsf{R}_{acbd}\left. -\frac{1}{10}\mathsf{R}^{ab}\nabla_{a}\nabla_{b}\mathsf{R}\right.\right.\nonumber\\
&&\left.\left.+\mathsf{R}^{ab}\Box \mathsf{R}_{ab}-\frac{1}{10}\mathsf{R}\Box \mathsf{R}\right)\delta_{d,7} +\dots\right\},\nonumber
\end{eqnarray}
where $\Box$ denotes the d'Alembertian computed with the boundary metric $\gamma_{ab}$.

Using these counterterms in odd and even dimensions, one can construct a  boundary stress tensor free of divergence from the total action 
$I=S+I_{\mathrm{ct}}^0+I_{\mathrm{ct}}^s$ by defining a boundary stress-tensor: 
\be
T_{ab}=\frac{2}{\sqrt{-\gamma}}\frac{\delta I}{\delta \gamma^{ab}}.
\ee
Thus a conserved charge 
\begin{equation}
{\mathfrak Q}_{\xi }=\oint_{\Sigma }d^{d-2}S^{a}~\xi ^{b}T_{ab},
\label{Mcons}
\end{equation}
can be associated with a closed surface $\Sigma $ whose normal unit vector is noted $n^{a}$, provided the boundary geometry has an isometry generated by a Killing vector $\xi ^{a}$. If $\xi =\partial /\partial t$ then $\mathfrak{Q}$ is the conserved mass/energy $M$. 
Similar to the $\Lambda=0$ case \cite{flatphase}, there is also a second charge associated with  $\partial/\partial z$, corresponding to the solution's tension ${\mathcal T}$.

For example, in the case $d=6$, the components of the boundary stress tensor reduce to
\bea
T_{tt} &=&\frac{1}{16\pi G}\left( -4 c_t + c_z \right)\frac{\ell^2}{r^3},\\
T_{zz} &=&\frac{1}{16\pi G}\left( -c_t + 4c_z \right)\frac{\ell^2}{r^3},\nonumber\\
T_{\theta\theta} &=&-\frac{1}{16\pi G}\left( c_t + c_z \right)\frac{\ell^4}{r^3},\nonumber\\
T_{\phi_1\phi_1}&=&\sin^2\theta T_{\theta\theta},\ T_{\phi_2\phi_2}=\sin^2\phi_1 T_{\phi_1\phi_1}\nonumber,
\eea
where $\theta,\phi_1,\phi_2$ are angle of the angular sector.

After some algebra, the mass (associated with $\partial_t$ and with $\Sigma$ chosen as the asymptotic cylinder $S_{d-3}\times S_1$) and tension (associated with $\partial_z$ and the asymptotic sphere $S_{d-3}$) are given by
\bea
M &=&  M_0 +  M_c,\ M_0 =\frac{\ell^{d-4}LV_{d-3}}{16\pi G}\left(c_z - (d-2)c_t\right),\nonumber\\
\mathcal T &=& \mathcal T_0 + \mathcal T_c,\ \mathcal T_0 =\frac{\ell^{d-4}V_{d-3}}{16\pi G}\left((d-2)c_z - c_t\right),
\label{ubsmt}
\eea
and where $M_c = -L\mathcal T_c =\frac{\ell^4LV_{d-3}}{16\pi G}\left(\frac{1}{12}\delta_{d,5} - \frac{333}{3200}\delta_{d,7} + \ldots\right)$ are Casimir-like terms in the dual theory defined on the boundary spacetime \cite{rms}; however, they are not relevant for our purpose.

Note that the thermodynamical quantities defined above obey a Smarr relation \cite{rms}
\be
M + L\mathcal T = T_HS.
\label{smarrads}
\ee
This relation can be obtained by evaluating the action with suitable boundary terms, using the equations of motions: $R-2\Lambda = 2 R_t^t$ or $R-2\Lambda = 2 R_z^z$. But since $\partial_t$ and $\partial_z$ are Killing vectors, $R_t^t$ and $R_z^z$ can be expressed as divergences. For example, in $d=6$, we find:
\be
\sqrt{-g}R_t^t = -\frac{1}{2}\left( r^3b'\sqrt{\frac{a}{bf}} \right)',\ \sqrt{-g}R_z^z = -\frac{1}{2}\left( r^3a'\sqrt{\frac{b}{af}} \right)'.
\ee

Then, using the suitable counterterms in $d=6$, we find that the total action $I$ evaluated using $R_t^t$ is given by
\be
I = \frac{V_3 L}{16\pi G}\left((-4c_t + c_z)\ell^2 + r_h^3 \sqrt{a_h \frac{b_h}{r_h\ell^2}\left[(d-4)r_h^2+(d-1)\ell^2\right]}\right) = M - T_HS,
\ee
while the total action evaluated with $R_z^z$ is 
\be
I = \frac{V_3 L}{16\pi G}\left(-4c_z + c_t\right) = -\mathcal T L.
\ee
It then follows that $M + \mathcal T L = T_H S$, as stated previously. The computations are similar in higher number of dimensions.

The solutions also obey the first law of thermodynamic:
\be
dM = T_H dS + \mathcal T dL.
\label{flaw}
\ee

\subsection{Integration of the first law and the string tension}
\label{flibg}
As described above, the mass and tension are defined in terms of quantities appearing in sub-leading orders of the asymptotic solution \eqref{asbckg}. In practice, these quantities are difficult to extract with a sufficient precision, especially when the number of dimensions is large. An alternative is to integrate the first law of thermodynamic \eqref{flaw} and to use the Smarr relation \eqref{smarrads}. This procedure has been used previously in the context of non uniform black strings in asymptotically locally flat spacetime by Kudoh and Wiseman \cite{kuwi}; it is however the first time it is used in asymptotically locally $AdS$ black strings and with numerical solutions.

Since the length $L$ doesn't enter explicitly in the equations of motions, we can assume working at fixed length, without loss of generality. Moreover, the length appears as a multiplying factor in the mass and the entropy and doesn't enter the definition of the tension nor of the temperature. So we choose to work with mass and entropy per unit length.

The first law then reduces to $dM/L = T_HdS/L$. In practice, for a given value of the cosmological constant, $T_H$ and $S/L$ are thermodynamical quantities defined on the horizon and we compute them as functions of the horizon radius.
So one can construct $T_H(r_h),S(r_h)/L$ and integrate the first law in order to find the mass as a function of the horizon radius:
\be
\frac{M(r_h)}{L} = \frac{M(r_h^*)}{L} + \frac{1}{L}\int_{r_h^*}^{r_h} T_H(x)\ddf{S(x)}{x} dx.
\ee

Of course, we still need one value for the mass, $M(r_h^*)/L$, which we can be computed in a region of the parameters where the extraction of the numerical coefficient is still manageable. Moreover, the effort is to be done only once for each value of the number of dimensions.

Once the mass is computed, we use the Smarr relation \eqref{smarrads} to construct the tension:
\be
\mathcal T(r_h) = \frac{T_H(r_h)S(r_h) - M(r_h)}{L}.
\ee

From the thermodynamical quantities, one can compute whether a given uniform black string is thermodynamically stable or not. The object needed to deal with thermodynamical stability is the specific heat, defined as
\be
C_p =T_H \ddf{S}{T_H},
\ee
where the length $L$ is kept fixed.
If $C_p>0$, we have thermodynamically stable solutions while if $C_p<0$, we have thermodynamically unstable solutions. Positive specific heat means that increasing the temperature increases the entropy and conversely for negative specific heat. Since $T_H$ is always positive, it is sufficient to look at the sign of $\ddf{S}{T_H}$.

\subsection{Scale invariance}

Note that the equations \eqref{equbs} are invariant under the following scaling transformation
\be
\tilde \ell = \lambda \ell,\ \ \tilde r =  \lambda r,
\label{ubsscale}
\ee
for a real constant $\lambda$.

This transformation has an effect on the various thermodynamical quantities as follows
\bea
\tilde M &=& \lambda^{d-4} M,\ \mathcal{\tilde T} = \lambda^{d-4} \mathcal T\nonumber\\
\tilde S &=& \lambda^{d-3}S,\ \ T_H= T_H/\lambda.
\eea

We take advantage of the scaling relations \eqref{ubsscale} to vary the $AdS$ radius and work with a fixed horizon radius, since it is technically easier to keep the horizon radius fixed. We set arbitrarily $r_h=1$ and varied the $AdS$ radius $\ell$. Afterwards, we rescaled all the thermodynamical quantities in order to investigate the physics of the black strings at fixed $AdS$ curvature, which seems to make more sense physically.

\subsection{Numerical solution}
\begin{figure}[H]
 \center
  \includegraphics[scale=.7]{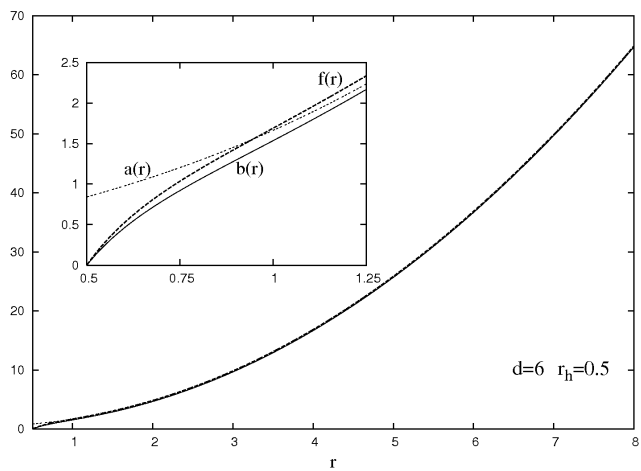}
 \caption{Profile of the metric functions $a,b,f$ for $d=6,r_h=0.5, \ell=1$ (reprinted from \cite{rms}). The profiles are qualitatively the same for every number of dimensions, difference arising at sub-leading terms in the asymptotic development.}
\label{fig:profd6}
\end{figure}

The profiles of the $a,b,f$ functions are generically the same for every number of dimensions and are presented in figure \ref{fig:profd6} for $\ell=1$.

\begin{figure}[h]
 \center
  \includegraphics[scale=.7]{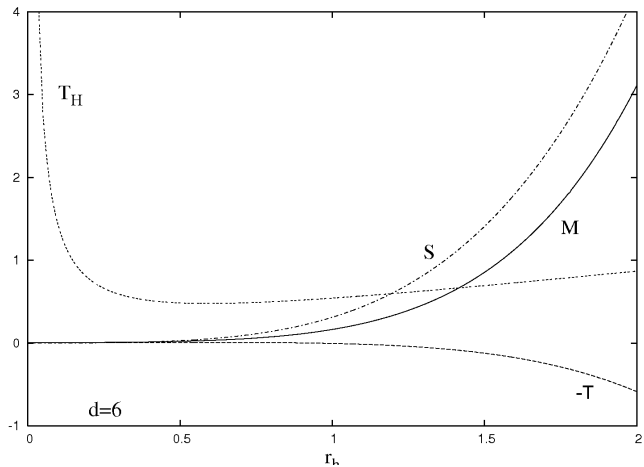}
 \caption{Temperature $T_H$, mass $M$, tension $T$ and entropy $S$ for $d=6,\ell=1$ as functions of the horizon radius. Here again, the figure is reprinted from \cite{rms}.}
\label{fig:thermorh}
\end{figure}

Thermodynamical quantities defined in the previous section are computed as functions of $r_h$ and are presented in figure \ref{fig:thermorh} for $d=6,\ell=1$. Once again, the general case with $d$ arbitrary presents the same main features than in $6$ dimensions. 

It should be noted that the entropy is an increasing function of the horizon radius while the temperature first decreases for small values of $r_h$ and start increasing about $r_h=r_h^c$ ($r_h^c\approx0.5$ for $d=6, \ell=1$). In other words, there is a phase of uniform black strings where the entropy increases when the temperature decreases, i.e. $C_p<0$ and another phase where $C_p>0$; these phases are thermodynamically unstable (resp. stable) and are referred to as small (resp. big) $AdS$ black strings. This phenomenon is similar to the $AdS$ black hole case \cite{hptr}.

Moreover, there exists a regime where the tension is negative. This is an unexpected feature of the $AdS$ uniform black string and has been pointed out in \cite{nubsd}, as we explain now.

\begin{figure}[H]
   \center 
    \includegraphics[scale=.42]{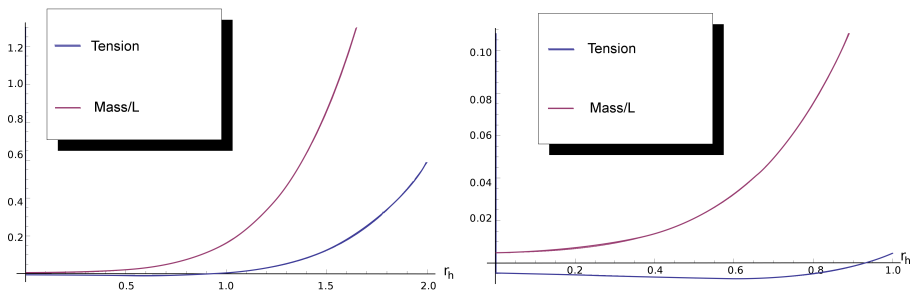}
    \caption{The mass and tension per unit length for the uniform black string in $d=6, \ell=1$. Large values of $r_h$ exhibit a positive tension while smaller values of $r_h$ have negative tension. The figure on the right is a zoom of the region where the tension is negative.}
    \label{fig:d6mt}
  \end{figure}

\begin{figure}[h]
   \center 
    \includegraphics[scale=.45]{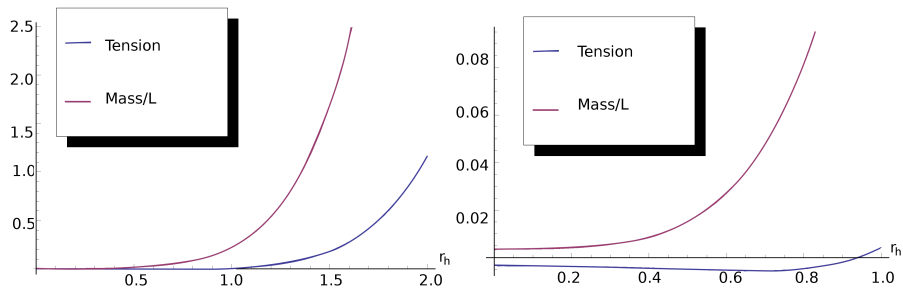}
    \caption{This figure is the same as \ref{fig:d6mt} for $d=7$.}
    \label{fig:d7mt}
  \end{figure}

A regular solution in the asymptotically locally flat spacetime would be the $(d-1)$-dimensional Minkowski spacetime times a circle, $\mathcal M_{d-1}\times S^1$. Of course, this solution has zero mass and zero tension. The $d$ dimensional uniform black string in the asymptotically locally flat case is characterise by a positive mass and tension. However, the $AdS$ counterpart of the regular solution is characterized by a positive non vanishing mass \cite{rms}, a temperature and entropy such that $T_H S = 0$. Using the Smarr relation, it follows that the tension of the regular solution is negative:
\be
M + \mathcal T L = 0\ \Rightarrow \mathcal T = - \frac{M}{L}.
\ee
It turns out that the black strings with small values of $r_h$ also have negative tension since these black strings are continuously connected to the $AdS$ regular solution. This is confirmed by a numerical analysis as shown in figure \ref{fig:d6mt} and \ref{fig:d7mt} for $d=6, \ell=1$ and $d=7, \ell=1$. These values were obtained using the definition \eqref{ubsmt} of the mass and tension as functions of $c_t$, $c_z$ (in units where the Newton constant is set to one) and are in accordance with the first law integration procedure proposed in section \ref{flibg}. Note however that the regular solution ($r_h=0$) has a vanishing total energy since $E = M + \mathcal T L = 0$, where $E$ denotes the total energy \cite{ho} (intuitively, $M$ is the energy due to the mass of the black string and $\mathcal T L$ is the binding energy of the string).

It should be stressed that uniform black string solutions in asymptotically locally $AdS$ spacetime have been constructed previously in different contexts \cite{rbd,klemmbs,cophor} (e.g. with or without gauge fields). Our solution and thermodynamical quantities are compatible with these previous results; however, the negative tension phenomenon was not noticed in these references. This phenomenon should hold for every number of dimensions, provided the mass in the $r_h\rightarrow0$ limit does not vanish.

The negative tension is an effect of the $AdS$ curvature since this phenomenon doesn't appear in the asymptotically locally flat uniform string problem. This might be interpreted as follows: on the one hand, a negative cosmological is characterized by a positive constant pressure. On the other hand, tension on strings typically acts oppositely to the  pressure. In this picture, when the gravitational tension is too small, the pressure from $AdS$ dominates, resulting in a negative tension. If the string is massive enough, the gravitational tension is higher and then dominates the $AdS$ pressure. The effect on the sign of the tension then results from a balance between the $AdS$ pressure and the gravitational tension.

We checked the sign of the tension numerically for $d=6,7$ but we expect this property to hold for every value of the number of dimension.

%%%%%%%%%%%%%%%%%%%%%%%%%%%%%%%%%%%%%%%%%%%%%%%%%%%%%%%%%%%%%%%%%%%%%%%%%%%%%%%%%%%%%%%
%%%%%%%%%%%%%%%%%%%%%%%%%%%%%%%%%%%%%%%%%%%%%%%%%%%%%%%%%%%%%%%%%%%%%%%%%%%%%%%%%%%%%%%
%%%%%%%%%%%%%%%%%%%%%%%%%%%%%%%%%%%%%%%%%%%%%%%%%%%%%%%%%%%%%%%%%%%%%%%%%%%%%%%%%%%%%%%
%%%             Stability
%%%%%%%%%%%%%%%%%%%%%%%%%%%%%%%%%%%%%%%%%%%%%%%%%%%%%%%%%%%%%%%%%%%%%%%%%%%%%%%%%%%%%%%
\section{Stability of the $AdS$ uniform string}
\label{sec:stab}
From the black string instability in asymptotically locally flat spacetime (see section \ref{flatgl}), it is expected that the $AdS$ uniform black strings are also unstable, at least for small values of the cosmological constant. In order to investigate the stability of the $AdS$ black strings, we consider a static perturbation around the string by choosing the functions $A,B,C$ in \eqref{ansatzbs} as:
\bea
\label{ansstab}
A &=& \epsilon A_1(r)\cos kz  + \mathcal O(\epsilon)^2,\nonumber\\
B &=& \epsilon B_1(r)\cos kz  + \mathcal O(\epsilon)^2,\\
C &=& \epsilon C_1(r)\cos kz  + \mathcal O(\epsilon)^2,\nonumber
\eea
where $\epsilon$ is a small real parameter and $k=\frac{2\pi}{L}$ is the critical wavelength. We expand the Einstein equation in a series in $\epsilon$ and keep the equations appearing at order $\epsilon$.

Note that all the wavelength should not lead to a static perturbation, so $k$ will have to assume a precise value depending on the parameters of the model in order to lead to a static perturbation. This precise value of $k$ will be the $AdS$ counterpart of the critical Gregory-Laflamme wavenumber.

Note also that once the length $L$ is fixed, one can say whether the black string is stable or not: for a given value of $L$ and a given perturbations with wavenumber $K$, the perturbation cannot fit in the compact extra-direction if $K<2\pi/L$; by analogy with the asymptotically locally flat case, smaller value of $K$ are unstable perturbations. As a consequence, black strings are unstable if they are longer than the length associated with the critical wavenumber.

\subsection{The equations and boundary conditions}
We consider the Einstein equations with cosmological constant supplemented by the ansatz \eqref{ansatzbs} where the correction functions are developed as a Series in $\epsilon$ according to \eqref{ansstab}. The relevant equations for the stability problem are the equations appearing at order $\epsilon$.

Following \cite{gubser}, we express $B_1$ as a function of $A_1,C_1$ and their first derivative using the equations of motions:
\be
B_1=\frac{a \left(2 b \left(r \left(A_1'+(d-3)C_1'\right)+(d-3) C_1\right)+r A_1 b'\right)-r b
   a' (A_1+(d-3) C_1)}{a \left(r b'+2 (d-3) b\right)}
\ee

The resulting equations of motion for the fields $A_1,C_1$ are given by
\be
A_1''= \alpha_1 A_1 + \alpha_2 A_1' + \alpha_3 C_1 + \alpha_4 C_1',~~
C_1'' = \varphi_1 A_1 + \varphi_2 A_1' + \varphi_3 C_1 + \varphi_4 C_1',
\label{eqadsstab}
\ee
where prime denotes the derivative with respect to the $r$ coordinate and where  
\bea
\alpha_1 &=& \frac{2b\left( \left( -3 + d \right) k^2\ell^2 + \left( r - dr \right) a' \right)  + 
    r\left( k^2 \ell^2 + 2\left( -1 + d \right) a \right) b'}{\ell^2af
    \left( 2\left( -3 + d \right) b + rb' \right) },
\nonumber
\\
\alpha_2 &=&-\frac{1}{r} - \frac{b'}{2b} - 
  \frac{1}{\ell^2
     rf}\left[\left( -1 + d \right) r^2 + \left( -4 + d \right) \ell^2  - 
     \frac{4\left( -1 + d \right) r^2b}{2\left( -3 + d \right) b + rb'}\right],
\nonumber \\    
     \alpha_3 &=&\frac{2\left( -3 + d \right) \left( -1 + d \right) b\left( 2a - ra' \right) }
  {\ell^2af\left( 2\left( -3 + d \right) b + rb' \right) },~~
 \nonumber\\
\alpha_4 &=&-\frac{\  \left( -3 + d \right) b'   }{2b} + 
  \frac{4\left( -3 + d \right) \left( -1 + d \right) rb}
   {\ell^2 f\left( 2\left( -3 + d \right) b + rb' \right) },
\nonumber
\eea
\bea
\nonumber
\varphi_1 &=&\frac{2\left( \left( -1 + d \right) r^2 + \left( -4 + d \right) \ell^2 \right) 
    \left( -  b a'  + a b' \right) }{\ell^2 r a f
    \left( 2\left( -3 + d \right) b + r b' \right) },~~~
\nonumber
\\
\varphi_2 &=&\frac{1}{r}\left[-1 + \frac{4 b\left( \left( -1 + d \right) r^2 + \left( -4 + d \right) \ell^2  \right)
         }{\ell^2 f\left( 2\left( -3 + d \right) b + rb' \right) }\right],
 \\        
         \nonumber
\varphi_3 &=&\frac{1}{\ell^2 raf\left( 2(d-3)b+rb'\right)}\left[ 2b(d-3)\left(k^2\ell^2 r+2(d-1)ra-\left( (d-1)r^2 \right.\right.\right.\nonumber\\
 &+&\left.\left.\left.(d-4)\ell^2\right) a' \right)  + \ell^2\left( k^2r^2 - 2\left( -4 + d \right) a \right) b'\right],\nonumber\\
\varphi_4 &=&\frac{1}{\ell^2 raf\left( 2(d-3)b+rb'\right)}\left[2b\left( -3 + d \right) \left( k^2 \ell^2r + 2\left( -1 + d \right) ra\right.\right. \nonumber\\
&-&\left.\left.\left( \left( -1 + d \right) r^2 + \left( -4 + d \right) \ell^2  \right) a' \right)  + \ell^2\left( k^2r^2 - 2\left( -4 + d \right) a \right) b'\right]
    .
\nonumber
\eea

This system of equations requires five boundary conditions; the appropriate boundary conditions are given by
\bea
A_1(\infty) &=& C_1(\infty) = 0,\ \ C_1(r_h)= 1,\nonumber\\
A_1'(r_h)   &=& A_{11}\ \ ,\ \ C_1'(r_h)=C_{11}.
\label{bco1}
\eea
The first two conditions are required in order to describe asymptotically vanishing perturbations, the third condition is imposed using the linearity of the equations and the last two conditions are required for regularity of the equations at the horizon, with $A_{11}, C_{11}$ given later by \eqref{nhstab}.

Note that we have $5$ boundary conditions for a system of total order $4$; $k^2$ will have to assume very specific values in order that all the boundary conditions are fulfilled. The system of equations is in fact similar to an eigenvalue problem given by a system of two second order coupled differential equations, the eigenvalue being the square critical wavenumber $k^2$. The technical way to tackle this problem is to supplement the system of equations with the trivial equation
\be
\frac{d k^2}{dr} = 0
\ee
and to hope that the numerical algorithm will be able to find the appropriate solution and the corresponding value of $k^2$. It turns out to be the case.

Let us also stress that the system of equations still admits the same scaling relations that the uniform string one \eqref{ubsscale}, supplemented by the scaling of the critical wavenumber:
\be
\tilde k = k/\lambda.
\ee

\subsection{Near horizon and asymptotic expansion}
In the limit where $r$ is close to $r_h$, one can perform a Taylor series expansion of the solution according to
\bea
A_1(r) &=& A_{10} + A_{11}(r-r_h) + \mathcal O (r-r_h)^2\ ,\nonumber\\
C_1(r) &=& C_{10} + C_{11}(r-r_h) + \mathcal O (r-r_h)^2,
\eea
where
\bea
A_{11} &=& -\frac{2a_h\left(A_{10}-C_{10}\right)\left(d-4\right)\left(d-3\right)\ell^2}{3a_hr_h\left(\left(d-4\right)\ell^2+\left(d-1\right)r_h^2 \right) }\\
&&+\frac{\left( 2a_hA_{10}\left(d-5\right)\left(d-1\right)+\left(-2A_{10}+C_{10}\left(d-3\right)  \right)k^2\ell^2\right)r_h^2}{3a_hr_h\left(\left(d-4\right)\ell^2+\left(d-1\right)r_h^2 \right) }\nonumber\\
C_{11} &=& \frac{2\left(A_{10}-C_{10}\right)}{r_h} + 
  \frac{C_{10}\left(2a_h\left(d-1\right)+k^2\ell^2\right)r_h}{a_h\left(d-4\right)\ell^2+a_h\left(d-1\right)r_h^2},
\label{nhstab}
\eea
and $a_h$ was defined previously, $A_{10}, C_{10}$ are real constants. Note that due to the linearity of the first order equation, either $A_{10}$ or $C_{10}$ can be fixed arbitrarily, only the ratio $A_{10}/C_{10}$ is non arbitrary. We used this freedom to impose $C_1(r_h) = C_{10}=1$ in the boundary conditions

In the asymptotic region, where $r\rightarrow\infty$, the fields $A_1(r)$ and $C_1(r)$ decay according to
\bea
A_1(r) &=& -(d-3)\gamma_1 \left(\frac{\ell}{r}\right)^{d-1}+ \mathcal O\left(\frac{\ell}{r}\right)^{d+2},\nonumber\\
C_1(r) &=& \gamma_1 \left(\frac{\ell}{r}\right)^{d-1} + \mathcal O\left(\frac{\ell}{r}\right)^{d+2},
\label{asstab}
\eea
where $\gamma_1$ is a real constant to be determined numerically. The influence of the logarithmic terms appearing in the background expansion for odd values of $d$ shows up in higher order terms in \eqref{asstab}. In fact, this is not the most general asymptotic expansion for the first order fields. The most general expansion is of the form
\bea
A_1(r) &=& \gamma_0\left(1 + \frac{C_1}{r^{d-4}}\right) -(d-3)\gamma_1 \left(\frac{\ell}{r}\right)^{d-1}+ \mathcal O\left(\frac{\ell}{r}\right)^{d+2},\nonumber\\
C_1(r) &=& \gamma_2\left(1 + \frac{C_2}{r^{d-4}}\right)+\gamma_1 \left(\frac{\ell}{r}\right)^{d-1} + \mathcal O\left(\frac{\ell}{r}\right)^{d+2},
\eea
for real values of $\gamma_0, C_1,C_2$. These additional terms all cancel once the boundary conditions \eqref{bco1} are imposed.

\subsection{Numerical results and the Gubser-Mitra conjecture}
We used the ODE solver Colsys in order to integrate the equations \eqref{eqadsstab}. Once given a starting profile sufficiently close to the solution, the solver was able to produce an accurate solution for both the functions $A_1,C_1$ and the corresponding eigenvalue $k^2$ for a given value of the number of dimension, the horizon radius and of the cosmological constant. 

A positive value of $k^2$ indicates the presence of an unstable mode of the uniform black string, as in the asymptotically locally flat case. A negative value of $k^2$ leads to the absence of an instability, since in this case, the critical wavelength is imaginary.

It should be stressed that the thermodynamical quantities are not affected by the perturbations. Indeed the perturbations are linear in $\cos kz$ and the thermodynamical quantities involve an integration over $z$ from $0$ to $L=2\pi/k$.

As mentioned before, we worked at fixed horizon radius and rescaled the solutions (in particular the critical wavenumber) in order to bring the stability analysis in a framework where the $AdS$ radius is fixed. Our results suggest that the Gregory-Laflamme instability persists in asymptotically locally $AdS$ spacetimes, at least when the ratio $r_h/\ell$ is small (compared to one). The dynamically unstable uniform black strings (with $r_h/\ell\ll1$) are referred to as small black strings while the dynamically stable black strings are referred to big black strings. The same phenomenon occurs for $AdS$ black holes, small $AdS$ black holes are unstable while big $AdS$ black holes are stable. As stated in \cite{hptr}, the $AdS$ radius acts like a confining box; if the horizon radius is small, the black objects doesn't 'feel' the curvature, while for larger values of the horizon radius, the black objects feel the 'box' and are stabilized.

\begin{figure}[H]
 \center
\includegraphics[scale=.72]{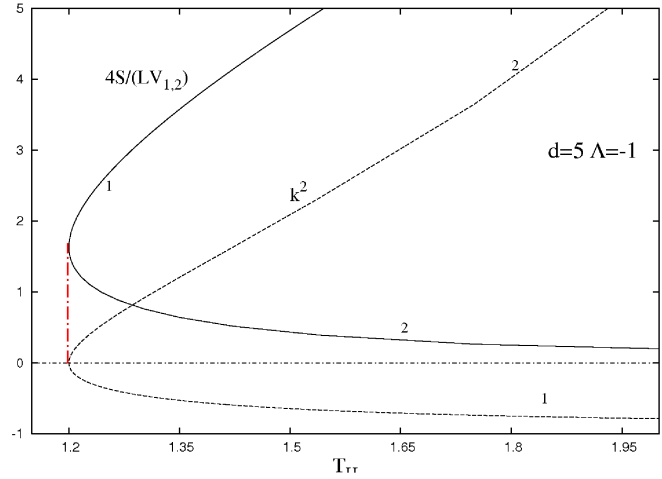}
\caption{Plots of the entropy and $k^2$ as functions of $T_H$ for $d=5$ and $\Lambda=-1$. The red line corresponds to the temperature where the thermodynamically stable phase ($\ddf{S}{T_H}>0$) meets the thermodynamically unstable phase (resp. $\ddf{S}{T_H}<0)$. This temperature is precisely the temperature separating the dynamically stable ($k^2<0$) and unstable ($k^2>0$) phase. This confirms the Gubser-Mitra conjecture in the case of $AdS$ black strings.}
\label{fig:GM58}
\end{figure}

Note that the dynamical instability matches the thermodynamical instability, in accordance with the Gubser-Mitra conjecture \cite{gmbh}, stating essentially that there is a correlation between thermodynamically stability and dynamical stability. This is shown in figure \ref{fig:GM58}. Note that the Gubser-Mitra conjecture is not always satisfied \cite{gmcounter}, but seems to hold for simpler models where matter fields are absent.

\begin{figure}[H]
 \center
\includegraphics[scale=.72]{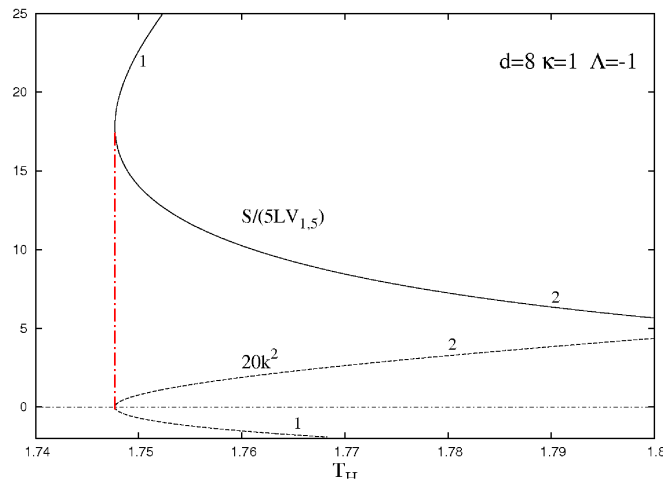}
\caption{The same figure as \ref{fig:GM58} for $d=8$.}
\label{fig:GM8}
\end{figure}

\begin{figure}[H]
 \center
\includegraphics[scale=.8]{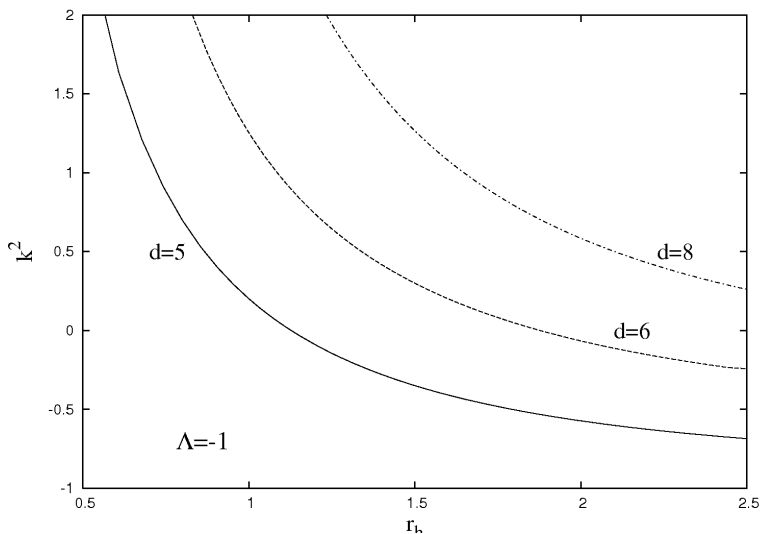}
\caption{The critical wavenumber as a function of the horizon radius for different values of $d$ and for $\Lambda=-1$.}
\label{fig:k2rh}
\end{figure}

In asymptotically locally flat spacetimes, we can consider that there is only one unstable black string, since the only dimensionless parameter is $r_h/L$. So one can fix arbitrary $r_h$ and compute the critical wavenumber, fixing the critical length $L_C=2\pi/k_c$. However, in asymptotically locally $AdS$ spacetimes, we have a new dimensionful parameter, namely the $AdS$ radius $\ell$. So now, we can construct two dimensionless quantities: the rescaled horizon radius $r_h/\ell$ and the rescaled length $L/\ell$. In other words, for each values of $r_h$ under consideration, we will find a critical length $L_c$ for a given value of $\ell$ (figure \ref{fig:k2rh}). This of course complicates the picture since now we find a continuum of unstable uniform black strings characterized by $r_h \in [0,r_h^*]$, $r_h^*$ depending on the number of dimension. This critical value defines the boundary between small uniform black strings and large uniform black strings.
Note also that for every number of dimensions we considered, the value of $k^2$ increases when $r_h\rightarrow0$. This implies that the globally regular solution also presents a Gregory-Laflamme instability.

In analogy with the flat case, we expect a non uniform string phase to emerge from the unstable black strings. In the next sections, we will present numerical arguments for the existence of non uniform $AdS$ black strings. We use first a perturbative approach, following the lines of Gubser \cite{gubser}, then we perform a non-perturbative approach, solving the appropriate system of coupled non linear partial differential equations.

%%%%%%%%%%%%%%%%%%%%%%%%%%%%%%%%%%%%%%%%%%%%%%%%%%%%%%%%%%%%%%%%%%%%%%%%%%%%%%%%%%%%%%%
%%%%%%%%%%%%%%%%%%%%%%%%%%%%%%%%%%%%%%%%%%%%%%%%%%%%%%%%%%%%%%%%%%%%%%%%%%%%%%%%%%%%%%%
%%%%%%%%%%%%%%%%%%%%%%%%%%%%%%%%%%%%%%%%%%%%%%%%%%%%%%%%%%%%%%%%%%%%%%%%%%%%%%%%%%%%%%%
%%%             pNUBS
%%%%%%%%%%%%%%%%%%%%%%%%%%%%%%%%%%%%%%%%%%%%%%%%%%%%%%%%%%%%%%%%%%%%%%%%%%%%%%%%%%%%%%%
%\section{Non uniform string phase}
The non uniform string phase has $A,B,C \neq0$ and emerges from the unstable black strings. There are two approaches in order to construct the non uniform branch. The first involves a perturbative approach, following the lines introduced by Gubser \cite{gubser} for asymptotically flat black strings and leads to a system of ordinary differential equations. In this case, however, only an approximated solution can be produced.

The second approach consists in solving the non linear partial differential equation problem (see \cite{wiseman} for the asymptotically locally flat non uniform black strings). We will present the perturbative analysis in the next section and the solution to the system of non linear partial differential equations, using results based on the perturbative results in section \ref{sec:nubs}.

\section{Perturbative non uniform strings}
\label{pertu}
In the perturbative approach, we develop the functions $A,B,C$ as a Fourier series and in term of a small parameter $\epsilon$ \cite{gubser}:
\bea
A &=& \epsilon A_1(r)\cos kz + \epsilon^2\left( A_0(r) + A_2(r) \cos 2kz \right) + \mathcal O(\epsilon)^3,\nonumber\\
B &=& \epsilon B_1(r)\cos kz + \epsilon^2\left( B_0(r) + B_2(r) \cos 2kz \right) + \mathcal O(\epsilon)^3,\\
C &=& \epsilon C_1(r)\cos kz + \epsilon^2\left( C_0(r) + C_2(r) \cos 2kz \right) + \mathcal O(\epsilon)^3,\nonumber
\label{anspert}
\eea
where $k$ is the critical wavenumber. The perturbative equations are given by the Einstein equations developed as a series in $\epsilon$.

As already noted, order $\epsilon^0$ is the uniform black string problem, while order $\epsilon^1$ is the stability problem treated in section \ref{sec:stab}. Order $\epsilon^2$ involves the backreacting modes $X_0$ of the metric, induced by the first order perturbation and the massive modes $X_2$, $X$ denoting generically $A,B,C$.

The backreacting and massive modes are decoupled to the order $\epsilon^2$ since they appear linearly to this order and with linearly independent Fourier modes. In this thesis, we will consider only the backreacting modes since they are the only modes that contributes to the thermodynamical corrections at order $\epsilon^2$, as we will see later. 

\subsection{Equations of the backreacting modes}
Like for the first order equations, it is still possible to solve algebraically $B_0$ and to reduce the problem to a system of two second order coupled differential equations. However, the expression for $B_0$ is quite long but is straightforward to compute. We will not give the expression for $B_0$ but only mention that it is of the form
\be
B_0= \mathsf{L}(A_0,C_0) + \mathsf{Q}(A_1,C_1),
\label{eqb0}
\ee
where $\mathsf L $ is a linear differential operator and $\mathsf Q$ is a quadratic expression involving $A_1,C_1$ and their derivatives, with coefficients depending on the background functions, on the number of dimensions and on the $AdS$ radius.

The resulting equations are of the form
\bea
A_0'' &=& \alpha_0 A_0' + \alpha_1 C_0' + \mbox{ terms quadratic in } A_1,C_1,A_1',C_1',\nonumber\\
C_0'' &=& \gamma_0 C_0' + \gamma_1 A_0' + \mbox{ terms quadratic in } A_1,C_1,A_1',C_1',
\label{breq}
\eea
where
\begin{small}
\bea
\alpha_0 &=&-\left\{2 r b b' \left((d-2) (2 d-5) f+\frac{(d-2) (d-1) r^2}{\ell^2}+(d-4) (d-2)\right)\right.\nonumber\\
&&+(d-2) r^2 f b'^2+2 b^2 \Big[(d-3)(d-2)^2 f+(d-4) \Big[\frac{(d-2) (d-1) r^2}{\ell^2}\nonumber\\
&&\left.+(d-3) (d-2)\Big]\Big]\right\}/\left\{2 (d-2) r b f \left(r b'+2 (d-3) b\right)\right\}\nonumber\\
\alpha_1&=&\left[(d-3) \left(-4 (d-3) r b f b'-r^2 f b'^2+2 b^2 \left(-(d-4) (d-3) f\right.\right.\right.\nonumber\\
&&\left.\left.\left.+\frac{(d-2) (d-1) r^2}{\ell^2}+(d-4)   (d-3)\right)\right)\right]/\left[2 r b f \left(r b'+2 (d-3) b\right)\right]\nonumber\\
\gamma_0&=&\Big\{-r b b' \left(-(d-2) (2 d-9) f+\frac{(d-2) (d-1) r^2}{\ell^2}+(d-4) (d-2)\right)\nonumber\\
&&+(d-2) r^2 f b'^2+b^2 \left((d-8) (d-3)(d-2) f-(d-4) \Big(\frac{(d-2) (d-1) r^2}{\ell^2}\right.\nonumber\\
&&\left.+(d-3) (d-2)\Big)\right)\Big\}/\Big\{(d-2) r b f \left(r b'+2 (d-3) b\right)\Big\}\nonumber\\
\gamma_1&=&\frac{b \left((d-3) (d-2) f+\frac{(d-2) (d-1) r^2}{\ell^2}+(d-4) (d-3)\right)}{(d-3) r f \left(r b'+2 (d-3) b\right)}.
\eea
\end{small}
The quadratic terms in \eqref{breq} are really long and we prefer not writing them explicitly. They are anyway straightforward to compute.

We note that only the derivatives of $A_0,C_0$ appear in equations \eqref{breq}. As a consequence, the equations are invariant under
\be
A_0\rightarrow A_0 + \mbox{ Const.}\ ,\ C_0\rightarrow C_0+\mbox{ Const.}
\label{shiftinv}
\ee
This will be useful to integrate the equations since it allows to shift the solutions down to zero asymptotically and thus not to care whether the functions $A_0, C_0$ go to zero at infinity as long as they go to constants.

\subsection{Near horizon and asymptotic solution}

The second order fields have the following near horizon expansion:
\bea
A_0(r) &=& A_{00} + A_{01}(r-r_h) + \mathcal O(r-r_h)^2 ,\nonumber\\
C_0(r) &=& C_{00} + C_{01}(r-r_h) + \mathcal O(r-r_h)^2,
\eea
where $A_{00},C_{00}$ are to be fixed using the invariance \eqref{shiftinv} such that the fields $A_0,C_0$ decay to $0$ at infinity, $C_{01}$ is an arbitrary real constant and 
\bea
A_{01} &=& -\frac{a_hC_{01}\left(d-4\right)\left(d-3\right)\ell^2 + \left(A_{10}^2+ 2A_{10}C_{10}\left(d-3\right)- C_{10}^2\left(d-3\right)\right)k^2\ell^2r_h}{3a_h\left(d-4\right)\ell^2+3a_h\left(d-1\right)r_h^2}\nonumber\\
&&+\frac{a_hC_{01}\left(d-3\right)\left(d-1\right)r_h^2}{3a_h\left(d-4\right)\ell^2+ 
      3a_h\left(d-1\right)r_h^2}.
\eea
This expression for $A_{01}$ is obtained by imposing regularity at the horizon; if $A_{01}$ is not chosen to be the above expression, there will exist terms proportional to $(r-r_h)^{-1}$ in the near the horizon development. Such terms are clearly diverging near the horizon.

The backreacting fields $A_0,C_0$ follow the same asymptotic form as the first order fields to the leading order:
\bea
A_0(r) &=& -(d-3)\gamma_0 \left(\frac{\ell}{r}\right)^{d-1}+ \mathcal O\left(\frac{\ell}{r}\right)^{d+2}, \nonumber\\ 
C_0(r) &=& \gamma_0 \left(\frac{\ell}{r}\right)^{d-1} + \mathcal O\left(\frac{\ell}{r}\right)^{d+2},
\label{bras}
\eea
where $\gamma_0$ is a real constant, also to be determined numerically. Once again, the most general expansion involves lower order terms which would give infinite contribution to the mass and thus are non-physical \cite{pnubsads}. We will impose boundary conditions in such a way that the non physical terms vanish.

\subsection{Boundary conditions and numerical technique}
The numerical technique is the following: we first integrate the background fields then we integrate the first order fields with the solver Colsys. The backreacting fields are then integrated with the following initial conditions:
\bea
A_0(r_h) &=& \alpha_0,\ \ C(r_h) = \chi_0,\ \ C'(r_h)=\chi_1\nonumber\\
A_0'(r_h)&=&-\Big\{a_h\chi_1\left(d-4\right)\left(d-3\right)\ell^2 + \left(A_{10}^2+ 2A_{10}C_{10}\left(d-3\right)- \right.\nonumber\\
&&\left.C_{10}^2\left(d-3\right)\right)k^2\ell^2r_h+a_h\chi_1\left(d-3\right)\left(d-1\right)r_h^2\Big\}/\Big\{3a_h\left(d-4\right)\ell^2\nonumber\\
&&+ 3a_h\left(d-1\right)r_h^2\Big\},
\eea
where $\alpha_0,\chi_0$ are arbitrary constant to be fixed a posteriori using the shift invariance \eqref{shiftinv}, while $\chi_1$ is a constant which is tuned such that the fields $A_0,C_0$ follow the decay \eqref{bras}.

In practice, we integrate the second order equations using a Runge-Kutta algorithm at order 4. The integration is carried out from the horizon radius, $r_h$ to some $R\gg r_h$. The fields $A_0,C_0$ follow the asymptotic \eqref{bras} if $R C_0'(R) + (d-1) C_0(R) = 0$. The problem here is that $C_0$ can always be shifted by an arbitrary constant so we impose the decay \eqref{bras} transposed to the derivative of $C_0$:
\be
RC_0''(R) + dC_0'(R) = 0,
\label{brbc}
\ee
the value of $C_0''(R)$ being obtained using the field equations.

The value of $RC_0''(R) + dC_0'(R)$ is a function of $\chi_1$, say $\mathcal C(\chi_1)$. The backreacting fields will follow the asymptotic decay \eqref{bras} if $\chi_1$ is chosen to be a root of $\mathcal C(\chi_1)$. We used a Newton algorithm in order to find the value of $\chi_1$ such that $\mathcal C(\chi_1) = 0$.

\subsection{Thermodynamical corrections}

As mentioned before, the first corrections to the thermodynamical quantities arise from the backreacting modes, at order $2$ in $\epsilon$. This is due to the fact that there is an integration over $z\in[0,L]$ which discards contributions from massive fields at this order in $\epsilon$. Of course, the corrections at order $\epsilon^2$ depends on the first order fields since they come squared in the development of the metric component.

The correction at order $\epsilon^2$ on the entropy is given by \cite{pnubsd}
\bea
\delta S/S_0 &=& \frac{a_h\left(-A_{10}^2+2A_{10}C_{10}(d-3)+\left( 4C_{00}+C_{10}^2(d-3)  \right)(d-3)  \right)(d-1)}{4a_h(d-1) }\nonumber\\
&&+\frac{C_{10}(A_{10}+C_{10})(d-3)k^2\ell^2}{4a_h(d-1) },
\eea
where the background quantities are evaluated with the length $L = 2\pi/k$ and $S_0$ is the background entropy. The total entropy is given by $S_0 + \epsilon^2\delta S$.

The correction to the temperature at the horizon is obtained by demanding regularity in the Euclidean sections and developing the condition as a series in $\epsilon$. Here again, the first correction arise at order $\epsilon^2$ and is given by:
\be
\delta T_H/T_H =\frac{2a_h(d-1)\left( 2A_{00} + A_{10}^2\right) + (d-3)(A_{10}-C_{10})C_{10}k^2\ell^2}{4a_h(d-1)},
\ee
where $T_H$ is the background temperature and $\delta T_H$ the correction from the backreaction, the temperature of the non uniform phase being denoted by $T_H + \epsilon^2\delta T_H$.

The mass and tension can be computed using the counterterm procedure \cite{counter} defined in section \ref{thbs}. Note that an application of the counterterm procedure to perturbative non uniform black strings in $6$ dimensions can be found in \cite{pnubsads}.

However, the thermodynamical quantities involved in these procedures are difficult to extract from the numerical solution in the general $d$-dimensional case because of the backreacting fields asymptotical decay \eqref{bras}. Instead, it is in principle possible to compute the mass by integrating the first law with fixed asymptotical length ($L$ fixed) and to use the Smarr formula \cite{rms} to extract the tension. However, we will postpone discussion of the mass and tension to the next section where the full non perturbative solution will be presented. Instead, we will investigate the thermodynamical stability properties of the perturbative non uniform black string.

In order to investigate the thermodynamical stability of the non uniform phase, we consider the specific heat, defined as the derivative of the entropy with respect to the temperature. In the case of perturbative non uniform black strings at order $\epsilon^2$, the specific heat reduces to
\be
C_p^{NU} = T_H\frac{\delta S}{\delta T_H},
\ee
with $L$ fixed, $\delta S$ and $\delta T_H$ as defined above.

This quantity gives the variation of entropy with respect to the Hawking temperature in the non uniform phase at the emanating point (where the non uniform phase emerges from the uniform phase); non uniform solutions with $C_p<0$ are thermodynamically unstable, while solutions with $C_p >0$ are thermodynamically stable. Since the temperature $T_H$ of the background is always positive, it is sufficient to compare the sign of $\delta S$ and $\delta T_H$ in order to investigate the existence of a stable phase.

%%%%%%%%%%%%%%%%%%%%%%%%%%%%%%%%%%%%%%%%%%%%%%%%%%%%%%%%%%%
\subsection{Numerical results and discussion}

\begin{figure}[h]
\begin{center}
\includegraphics[scale=.7]{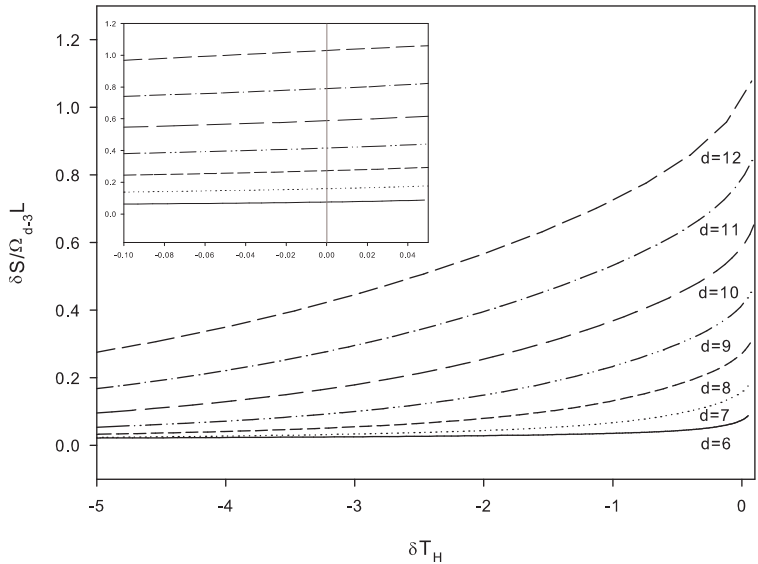}	
\caption{The correction on the entropy $\delta S$ as a function of the correction on the Hawking temperature $\delta T_H$ for various number of dimensions. There are points on this figure with $\delta T_H >0, \delta S>0$ corresponding to a thermodynamically stable phase. The box in the upper left corner is a zoom of the region with $\delta T_H>0$.}
\label{dSdT}
\end{center}
\end{figure}

We integrated the field equations for values of $d$ from $5$ to $15$. We considered perturbative non uniform black strings emanating from the uniform phase with length fixed by the critical length $L=2\pi/k$ found in the stability analysis. Our results show that for every number of dimensions considered, there exists a phase of non uniform black string presenting a positive specific heat (figure \ref{dSdT}).

This stable phase exists for small values of the critical wave number, i.e. large value of the length (with $\ell$ fixed). This is the analogue of the small/big black strings or black hole in $AdS$ where the relevant parameter for the thermodynamical stability was $r_h/\ell$. Here, the relevant parameter is $\mu_1 = L/\ell$; short non uniform black strings solutions with $\mu_1\ll 1$ are thermodynamically unstable while long non uniform black strings with $\mu_1\approx1$ are thermodynamically stable (see table \ref{tab:muc} where we present the value of $\mu_1$ separating the stable phase from the unstable phase, say $\mu_1^s$).

\begin{table}[H]
\centering
$\begin{array}{ccccccccc}
\hline
\hline
d   &\vline & 6      & 7      &  8  & 9  & 10  & 11 & 12 \\
\hline
\mu_1^{s}& \vline & 8.66409 & 8.76464 & 9.02511 & 9.3052 & 9.56981 & 9.80617 & 10.1825\\
\hline
\hline
\end{array}$
\caption{The value of the ratio $\mu_1^{s}=L/\ell$ where the stable phase occurs for various $d$.}
\label{tab:muc}
\end{table}

Let us stress the fact that these short/long black strings are present in the small $AdS$ phase, since it is the phase we are dealing with.
We expect this property to be a generic feature of non uniform black strings in arbitrary number of dimensions.

\begin{figure}[H]
\begin{center}
\includegraphics[scale=.7]{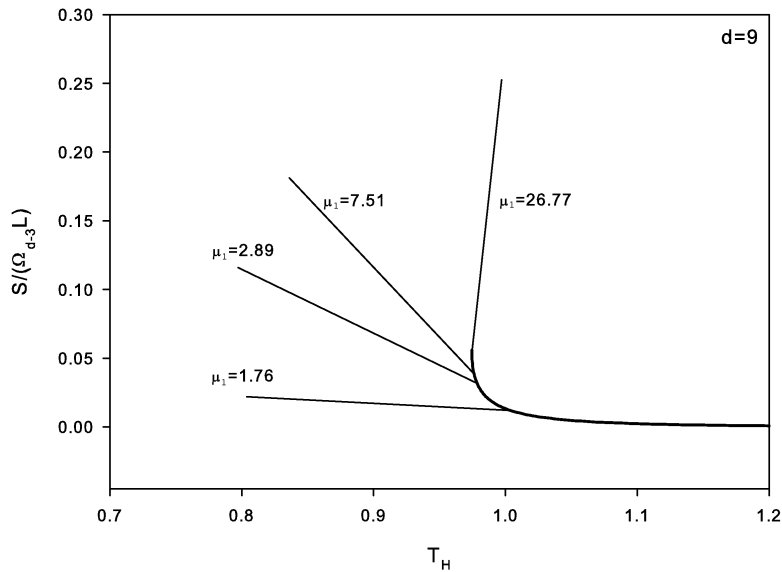}
\caption{The phase diagram in the $S-T_H$ plane for $d=9$. The background and the direction of the corrections for some value of the rescaled length are shown. The same pattern holds for all the number of dimension we considered.}
\label{fig:phase}
\end{center}
\end{figure}

Figure \ref{fig:phase} shows the direction of the phase transition in a $S-T_H$ diagram for $d=9$. In the short black string phase (i.e. with $L\ll\ell$), the entropy increases and the temperature decreases along the non uniform phase while in the long black string phase ($L\approx\ell$) the entropy increases and so does the temperature. Let us emphasize once again that this new phase of non uniform black strings occurs for small black strings ($r_0\ll\ell$) since the length is not well defined for big black strings ($r_0\approx \ell$).

%%%%%%%%%%%%%%%%%%%%%%%%%%%%%%%%%%%%%%%%%%%%%%%%%%%%%%%%%%%%%%%%%%%%%%%%%%%%%%%%%%%%%%%
%%%%%%%%%%%%%%%%%%%%%%%%%%%%%%%%%%%%%%%%%%%%%%%%%%%%%%%%%%%%%%%%%%%%%%%%%%%%%%%%%%%%%%%
%%%%%%%%%%%%%%%%%%%%%%%%%%%%%%%%%%%%%%%%%%%%%%%%%%%%%%%%%%%%%%%%%%%%%%%%%%%%%%%%%%%%%%%
%%%             NUBS
%%%%%%%%%%%%%%%%%%%%%%%%%%%%%%%%%%%%%%%%%%%%%%%%%%%%%%%%%%%%%%%%%%%%%%%%%%%%%%%%%%%%%%%

\section{Non perturbative solutions}
\label{sec:nubs}

In this section, we present the construction of the non perturbative solutions. The problem is given by a system of three coupled partial differential equations and a set of boundary conditions.

\subsection{The coordinate system}
The solutions are a priori independent of the normalisation of the functions $a,b$; this means that the functions $a,b$ determined by \eqref{bcubs} as boundary conditions have to be multiplied by a suitable factor in order to follow the asymptotic behaviour \eqref{ubsas} {\it a posteriori}. This invariance is related to the arbitrary rescaling of $z$ and $t$. In order to avoid this scaling ambiguity, we perform the following change of functions
\be
a(r) = \frac{r^2}{\ell^2}\hat a(r),\ b(r) = \frac{r^2}{\ell^2}\hat b(r),\ f(r) = \frac{r^2}{\ell^2}\hat f(r).
\label{chfun}
\ee
The new functions $\hat a, \hat b, \hat f$ then obey $\hat a(r\rightarrow\infty)=1, \hat b(r\rightarrow\infty)=1$ which are imposed as boundary conditions independently of $\ell$ and $d$. Consequently, the functions reconstructed from $\hat a$ and $\hat b$ are normalized properly.

Note that the $z$ coordinate ranges from $0$ to some length $L$. The value of $L$ is fixed to be the critical length where the linear perturbations are static. This precise value of $L$ is again provided by the stability analysis coming from the first order in perturbation theory \cite{rbd}, see section \ref{sec:stab}.

It must be stressed that the equations resulting from the ansatz \eqref{ansatzbs} require complicated regularity conditions for the functions $A,B,C$ at the horizon. Going in a 'conformal-like' gauge,
\bea
ds^2 = -\frac{g(\tilde r)}{\ell^2}\tilde b(\tilde r)e^{2A(r,z)} dt^2 + e^{2B(r,z)}\left(\frac{\ell^2\tilde r^2d\tilde r^2}{g(\tilde r)^2 \tilde f(\tilde r)}+\frac{g(\tilde r) L^2\tilde a(\tilde r)d\tilde z^2}{\ell^2}  \right) \nonumber\\
+ g(\tilde r)e^{2C(r,z)}d\Omega_{d-3}^2,
\label{confgauge}
\eea
where $g(\tilde r)=\tilde r^2 + r_h^2$, with $r_h$ the horizon radius, leads to much simpler regularity conditions. Note that we rescaled the $z$ coordinate such that $\tilde z\in[0,1],\ \tilde z = z/L$ by factorising the length $L$.

The relation between \eqref{ansatzbs} with the change of functions \eqref{chfun} and \eqref{confgauge} is given by
\bea
\tilde r^2 + r_h^2 = r^2,\ \tilde a(\tilde r) = \hat a(r),\ \tilde b(\tilde r) = \hat b(r), \tilde f(\tilde r) = \hat f(r).
\eea

Note that a subtlety arises when one wants to consider a conformal gauge, starting from the line element \eqref{ansubs}. Naively, one would write 
\bea
&&ds^2 = -\frac{\tilde r^2}{\ell^2}\hat b(\tilde r) dt^2 + \frac{\ell^2}{\tilde r^2 \hat f(\tilde r)}\frac{\tilde r^2}{g(\tilde r)} d\tilde r^2 + g(\tilde r)d\Omega_{d-3}^2 + \frac{\tilde r^2 \hat a(\tilde r)}{\ell^2}dz^2,\nonumber\\
&& g(r)=\tilde r^2+r_h^2,
\eea
following the modification from \eqref{ansubs} to \eqref{confgauge}. This doesn't work at all; the near horizon expansion suffers from divergences of order $(r-r_h)^{-2}$. The correct way to enter the conformal gauge is to replace the diverging $r^2/\ell^2$ by $g(\tilde r)/\ell^2$ since the change of gauge is such that $r^2\rightarrow g(\tilde r)$. To the leading order, it doesn't change anything asymptotically, but it appears that the near horizon expansion doesn't suffer any more from unmanageable divergences. So after some algebra, the conformal gauge with factorized $AdS$ asymptotic is given by \eqref{confgauge}.

From now on, we omit the tilde over the $r$ and $z$ coordinates and the hat over the various functions, except if this spoils clarity.

\subsection{The equations and boundary conditions}
The equations are the Einstein equations with the ansatz \eqref{confgauge}, leading to the following set of coupled partial differential equations:
\bea
&& -\left(\frac{\left(-1+d\right)e^{2B}r^2}{f{g}^2}\right)-\frac{b'}{2rb}+\frac{a'b'}{4ab}-\frac{{b'}^2}{4{b}^2}+\frac{b'f'}{4bf}-
\frac{g'}{2rg}+\frac{a'g'}{4ag}+\frac{\left(2+d\right)b'g'}{4bg}\nonumber\\
&&+\frac{f'g'}{4fg}+\frac{\left(-1+d\right){g'}^2}{4{g}^2}+\frac{b''}{2b}+\frac{g''}{2g}+\frac{{\ell}^4r^2{A^{(0,1)}}^2}{L^2af{g}^3}+ \frac{\left(-3+d\right){\ell}^4r^2A^{(0,1)}C^{(0,1)}}{L^2af{g}^3}\nonumber\\ &&+\frac{{\ell}^4r^2A^{(0,2)}}{L^2af{g}^3}-\frac{A^{(1,0)}}{r}+\frac{a'A^{(1,0)}}{2a}+\frac{b'A^{(1,0)}}{b}+\frac{f'A^{(1,0)}}{2f}+\frac{\left(2+d\right)g'A^{(1,0)}}{2g}\nonumber\\
&&+{A^{(1,0)}}^2+\frac{\left(-3+d\right)b'C^{(1,0)}}{2b}+\frac{\left(-3+d\right)g'C^{(1,0)}}{2g}+\left(-3+d\right)A^{(1,0)}C^{(1,0)}\nonumber\\
&&+A^{(2,0)}=0
\label{eqA}
\eea

\bea
&&\frac{\left(-4+d\right)\left(-3+d\right)e^{2B-2C}{\ell}^2r^2}{2f{g}^3}+\frac{\left(-4+d\right)\left(-1+d\right)e^{2B}r^2}{2f{g}^2}-\frac{a'}{2ra}-\frac{{a'}^2}{4{a}^2}\nonumber\\
&&+\frac{a'f'}{4af}-\frac{g'}{2rg}+\frac{a'g'}{ag}-\frac{\left(-3+d\right)b'g'}{4bg}+\frac{f'g'}{4fg}-\frac{\left(-4+d\right)\left(-1+d\right){g'}^2}{8{g}^2}\nonumber\\
&&+\frac{a''}{2a}+\frac{g''}{2g}-\frac{\left(-3+d\right){\ell}^4r^2A^{(0,1)}C^{(0,1)}}{L^2af{g}^3}-\frac{\left(-4+d\right)\left(-3+d\right){\ell}^4r^2{C^{(0,1)}}^2}{2L^2af{g}^3}\nonumber\\
&&+\frac{{\ell}^4r^2B^{(0,2)}}{L^2af{g}^3}-\frac{\left(-3+d\right) g'A^{(1,0)}}{2g}-\frac{B^{(1,0)}}{r}+\frac{a'B^{(1,0)}}{2a}+\frac{f'B^{(1,0)}}{2f}+\frac{3g'B^{(1,0)}}{2g}\nonumber\\
&&-\frac{\left(-3+d\right)b'C^{(1,0)}}{2b}-\frac{{\left(-3+d\right)}^2g'C^{(1,0)}}{2g}-\left(-3+d\right)A^{(1,0)}C^{(1,0)}\nonumber\\
&&-\frac{\left(-4+d\right)\left(-3+d\right){C^{(1,0)}}^2}{2}+B^{(2,0)}=0
\label{eqB}
\eea

\bea
&\left.\right.&-\left(\frac{\left(-4+d\right)e^{2B-2C}{\ell}^2r^2}{f{g}^3}\right)-\frac{\left(-1+d\right)e^{2B}r^2}{f{g}^2}-\frac{g'}{2rg}+\frac{a'g'}{4ag}+\frac{b'g'}{4bg}+\frac{f'g'}{4fg}\nonumber\\
&&+\frac{\left(-1+d\right){g'}^2}{4{g}^2}+\frac{g''}{2g}+\frac{{\ell}^4r^2A^{(0,1)}C^{(0,1)}}{L^2af{g}^3}+\frac{\left(-3+d\right){\ell}^4r^2{C^{(0,1)}}^2}{L^2af{g}^3} +\frac{{\ell}^4r^2C^{(0,2)}}{L^2af{g}^3}\nonumber\\
&&+\frac{g'A^{(1,0)}}{2g}-\frac{C^{(1,0)}}{r}+\frac{a'C^{(1,0)}}{2a}+\frac{b'C^{(1,0)}}{2b}+\frac{f'C^{(1,0)}}{2f} +\frac{\left(-1+d\right)g'C^{(1,0)}}{g} \nonumber\\
&&+A^{(1,0)}C^{(1,0)} +\left(-3+d\right){C^{(1,0)}}^2+C^{(2,0)}=0
\label{eqC}
\eea
where the exponent $(m,n)$ denotes the $m^{th}$ derivative with respect to $r$ and $n^{th}$ derivative with respect to $z$ and the primes denote the derivative with respect to $r$ for functions depending on $r$ only.

It should be noted that the terms containing second derivatives of the functions $A,B,C$ are all of the form 
\be
\dddf{X}{r} + \frac{\ell^4 r^2}{a f g^3}\frac{1}{L}\dddf{X}{z},
\ee
$X$ denoting $A,B$ and $C$.

The combination of the Einstein equations leading to equations \eqref{eqA}, \eqref{eqB} and \eqref{eqC} are $E_t^t + E_r^r$, $E_z^z$ and $E_\theta^\theta$ where $E_a^b=G_a^b - \Lambda \delta_a^b$ and $\theta$ denotes the angular sector. Note that there are two other linearly independent non vanishing combinations of the Einstein equations; these are constraints: $E_t^t-E_r^r$ and $E_z^t$. It has been argued in \cite{wiseman} that the constraints satisfy the Cauchy-Riemann equations, as a consequence, if they are fulfilled on the boundary of the domain of integration, they will be fulfilled in the interior of the domain as well. Here, although there is a cosmological constant, the argument is the same since the cosmological constant cancels for the particular combination leading to the constraints.

The system of partial differential equations above is supplemented by the following boundary conditions:
\bea
\label{bcnubs}
A(r,z)&=&B(r,z)=C(r,z)=0\mbox{ for } r\rightarrow\infty,\nonumber\\
\partial_z A(r,z)&=&\partial_z B(r,z)=\partial_z C(r,z)=0\mbox{ for }z=0,1,\\
\partial_r A(r,z)&=&\partial_rC(r,z)=0,\ B(r,z)=A(r,z) + d_0\mbox{ for } r=0\nonumber,
\eea
where $d_0$ is a real parameter. The first condition ensures the fact that $A,B,C$ are \emph{corrections} and thus vanish at infinity, the second conditions imposes periodicity along the $z$ coordinate while the last three conditions are the regularity conditions and ensure that the constraints are fulfilled. The parameter $d_0$ is related to the variation of the temperature along the non uniform branch: the variation of temperature along the non uniform branch is a function of to $B-A$ evaluated at the horizon (i.e. of $d_0$); we will come back to the relation between $d_0$ and the temperature later. Let us stress anyway that $d_0$ plays a crucial role in the construction of non uniform solutions: it is the parameter which forces the solution to deviate from the uniform case; if $d_0=0$, the solution to the system of partial differential equations is the trivial solution $A=B=C=0$.

The domain of integration is $z\in[0,1],\ r\in[0,\infty]$ (recall that $r$ is the radial coordinate in the conformal gauge). We used the solver \emph{Fidisol} (see appendix \ref{app:num}) in order to solve this system. Fidisol is sensitive to the mesh and boundary conditions (it needs exact boundary conditions, i.e. imposed at infinity in this case); in order to impose appropriately the boundary conditions \eqref{bcnubs} especially the one at $r=\infty$, we have chosen to work with a compactified variable $x$ ranging from $0$ to $1$:
\be
x = \frac{r-r_h}{r}.
\ee
We have also tried different meshes in the $x$ direction, as well as in the $z$ direction in order to test the robustness of the solution.

Fidisol is based on a Newton-Raphson algorithm and needs a 'good initial profile' (see appendix \ref{app:num})to efficiently provide a proper solution. We tried to start with the uniform string solution, $A=B=C=0$ and to increase the value of $d_0$ but this approach failed to provide convincing solutions. Starting from $A = \epsilon \tilde A_1(\tilde r) \cos 2\pi z,\ B = \epsilon \tilde B_1(\tilde r) \cos 2\pi z, \ C = \epsilon \tilde C_1(\tilde r) \cos 2\pi z$, where $\tilde A_1(\tilde r) = A_1(\sqrt{r^2-r_h^2})$ and similarly for $B_1,C_1$ ( $A_1,B_1,C_1$ being the solution from first order in perturbation theory, section \ref{sec:stab}), for a small real value of $\epsilon$ and a small value of $d_0$ works much better and provides robust solutions. Then, we increased the value of $d_0$ starting with the previous solution as an initial profile.

\subsection{Asymptotic behaviour}
The asymptotic behaviour of the background functions was reminded in the first sections of this chapter; since we factorised the divergent  $\frac{g(r)}{\ell^2}$ term, the asymptotic development in \cite{rms} is of course to be divided by $g(r)/\ell^2$, leading to
\bea
a&=& 1  + \sum_{i=0}^{\left\lfloor \frac{d-4}{2}\right\rfloor} a_i\left(\frac{\ell^2}{g}\right)^{i+1} + c_z\left(\frac{\ell^2}{g}\right)^{\frac{d-1}{2}} + \sum_k\delta_{d,2k+1}\frac{\xi}{2}\log\frac{g}{\ell^2}\left(\frac{\ell^2}{g}\right)^{\frac{d-1}{2}} \nonumber\\
&&+ \mathcal O\left(\frac{\ell^2}{g}\right)^{\frac{d+1}{2}},\nonumber\\
b&=& 1  + \sum_{i=0}^{\left\lfloor \frac{d-4}{2}\right\rfloor} a_i\left(\frac{\ell^2}{g}\right)^{i+1} + c_t\left(\frac{\ell^2}{g}\right)^{\frac{d-1}{2}} + \sum_k\delta_{d,2k+1}\frac{\xi}{2}\log\frac{g}{\ell^2}\left(\frac{\ell^2}{g}\right)^{\frac{d-1}{2}} \nonumber\\
&&+ \mathcal O\left(\frac{\ell^2}{g}\right)^{\frac{d+1}{2}},\nonumber\\
f&=& 1  + \sum_{i=0}^{\left\lfloor \frac{d-4}{2}\right\rfloor} a_i\left(\frac{\ell^2}{g}\right)^{i+1} + (c_t+c_z+c_0)\left(\frac{\ell^2}{g}\right)^{\frac{d-1}{2}} +\nonumber\\ &&\sum_k\delta_{d,2k+1}\frac{\xi}{2}\log\frac{g}{\ell^2}\left(\frac{\ell^2}{g}\right)^{\frac{d-1}{2}} + \mathcal O\left(\frac{\ell^2}{g}\right)^{\frac{d+1}{2}},
\label{asubscg}
\eea
where $g=r^2 + r_h^2$. Note that $1/g$ in the above formula could be further expanded, leading to a much longer expression.

It has been argued in \cite{pnubsads,pnubsd} that the modes of $A,B,C$ in the perturbative approach decay as $\left(\frac{\ell}{r}\right)^{d-1}$ in all order of the perturbations. Since the non perturbative solution is a combination of these modes, the leading order should also decay as
\bea
A(r,z)&=&\alpha_1\left(\frac{\ell}{r}\right)^{d-1} + \mathcal O\left(\frac{\ell}{r}\right)^{d+1},\ B(r,z)=\beta_1\left(\frac{\ell}{r}\right)^{d-1} + \mathcal O\left(\frac{\ell}{r}\right)^{d+1},\nonumber\\ 
C(r,z)&=&\gamma_1\left(\frac{\ell}{r}\right)^{d-1} + \mathcal O\left(\frac{\ell}{r}\right)^{d+1}.
\label{decaynu}
\eea
for some $\alpha_1,\beta_1,\gamma_1\in\mathbb R$. The decay \eqref{decaynu} is confirmed by direct computations on the numerical solution.

\begin{figure}[H]
 \centering
	\includegraphics[scale=.21]{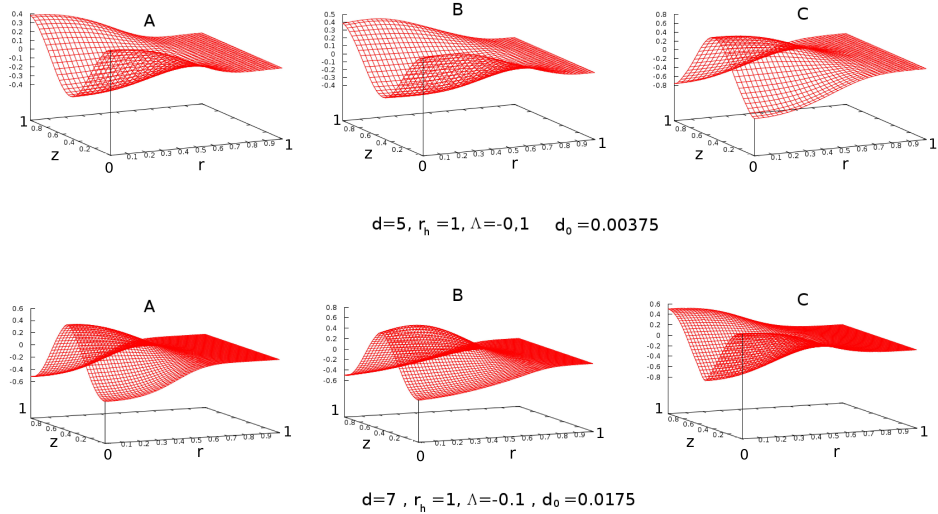}
	\caption{Typical profiles of the correction functions $A,B,C$ for $d=5, r_h=1, \Lambda=-0.1, d_0 = 0.00375$ and $d=7,r_h=1,\Lambda=-0.1, d_0 = 0.0175$.}
	\label{fig:solunu}
\end{figure}

We were able to obtain solutions to the equations \eqref{eqA}, \eqref{eqB}, \eqref{eqC} with the boundary conditions \eqref{bcnubs} for various values of $d$, $r_h/\ell$, $d_0$. For a given value of the number of dimensions, the non uniform solutions are given as families of solutions, labelled by $r_h/\ell$ and $d_0$. The value of $r_h/\ell$ fixes the critical length $L$ of the corresponding uniform string and $d_0$ allows to slide along the non uniform branch for the given value $r_h/\ell$. Before presenting the thermodynamical properties of the non uniform phase, we present generic profiles of the functions $A,B,C$ in figure \ref{fig:solunu}.

%%%%%%%%%%%%%%%%%%%%%%%%%%%%%%%%%%%%%%%%%%%%%%%%%%%%%%%%%%%%%%%%%%%%%%%%%%%%%%%%%%%%%%%%%%%%%%%%%%%%%%%%%%%%%%%%%%%%%%%%

\subsection{Properties of the solution}
\subsubsection{Thermodynamical properties}
The thermodynamical properties of the background have been presented above. The correction functions $A,B,C$ provide corrections on the thermodynamical quantities computed from the background (as the name 'correction function' suggests). The quantities computed on the background solution are denoted with a subscript $U$, while the quantities referring to the full non uniform solution are denoted with a subscript $NU$. 

The corrections on the thermodynamical quantities are computed using the standard procedure (regularity in the euclidean section and one quarter of the horizon area) and are given by
\bea
T_{NU} &=& \frac{1}{4\pi}\sqrt{b'(r_h)f'(r_h)}e^{A(r_h,z)-B(r_h,z)}=  e^{-d_0} T_U\nonumber,\\
S_{NU} &=& \frac{S_U}{L}\int_0^1 L e^{B(r_h,z)+(d-3)C(r_h,z)}dz,
\eea
where $T_{NU}$ and $S_{NU}$ are the Hawking temperature (resp. the entropy) of the non uniform string. Note that the first relation makes the role of $d_0$ clearer: once the background is set, the parameter $d_0$ allows to slide along the non uniform branch in a temperature-entropy phase diagram.

These quantities are defined at the horizon and are relatively easy to compute from a technical point of view. The asymptotic quantities can be parametrized according to
\bea
M_{NU} &=& M_U\left(1+\frac{\delta M}{M_U} \right) ,\nonumber\\
\mathcal T_{NU} &=& \mathcal T_U\left( 1+\frac{\delta \mathcal T}{\mathcal T_U} \right).
\eea
where $\delta M, \delta\mathcal T$ are the corrections to the mass and tension due to the non uniformity and are functions of the parameters $\alpha_1,\beta_1,\gamma_1$ appearing in \eqref{decaynu} and of $c_t,c_z,c_0$ (in \eqref{asubscg}).

Note that at the boundary spacetime, $\partial_z$ and $\partial_t$ are both Killing vectors, so the definition of the mass and tension as functions of the asymptotic parameters \cite{rms} remains the same. Note also that the decay \eqref{decaynu} combined with the asymptotic form of the background solution gives contribution of order $r^{-(d-3)}$ to $g_{tt}, g_{zz}$ and leads to correction on $c_t,c_z$. As a consequence, the mass and tension are also modified; it follows also that another decay of the corrections would then be physically non acceptable.

The quantities $c_t,c_z$ and worse, $\alpha_1,\beta_1,\gamma_1$ are difficult to compute, first because of the sharpness of the decay and second because the determination of these quantities is plagued by numerical noise. 

In order to avoid this problem, we were able to integrate the first law of black holes and then to compute the masses (of the uniform solution \emph{and} of the non uniform solution). Assuming that the Smarr relation \eqref{smarrads} holds for the non uniform solution, we were able to compute the corrections on the tension for the non uniform solution. 

The procedure to integrate the first law is performed in two steps, first at the background level, then for the non uniform solution. 
The procedure for the background level has been described in section \ref{flibg}.

The integration of the first law for the non uniform solution is more straightforward than for the background itself since we the functions $A,B,C$ are corrections to the background and indeed provide \emph{corrections} on the thermodynamical quantities, which naturally appear as functions of the parameter $d_0$. These thermodynamical corrections are related to each other by the first law of black holes. It follows that the procedure to integrate the first law seems more natural in this case; given a uniform black string with horizon radius $r_h$, the non uniform black string with a given value of $d_0$ emanating from this uniform black string has a mass given by
\bea
M_{NU}(r_h) &=& M_U(r_h) + \int_0^{d_0} T_H^{NU}(r_h,d_0)\frac{\delta S_{NU}}{\delta d_0}d\ d_0 \nonumber\\
            &=& M_U(r_h) + T_H^U(r_h)\int_0^{d_0} e^{-d_0}\frac{\delta S_{NU}}{\delta d_0}d\ d_0,
\eea
where $\delta S/\delta d_0$  has to be evaluated numerically. Terms of the form $\mathcal T dL$ don't appear since the length of the non uniform black string is fixed along the non uniform branch.

Once the mass of the non uniform string for a given value of $d_0$ is computed, it is straightforward to obtain the tension for the corresponding value of $d_0$, using the Smarr relation:
\be
\mathcal T_{NU} = \frac{T_H^{NU}S_{NU}}{L} - \frac{M_{NU}}{L}.
\ee

%%%%%%%%%%%%%%%%%%%%%%%%%%%%%%%%%%%%%%%%%%%%%%%%%%%%%%%%%%%%%%%%%%%%%%%%%%%%%%%%%%%%%%%%%%%%%%%%%%%%%%%%%%%%%

\subsubsection{Geometric properties}
The order parameter for the transition from the uniform string phase to the non uniform string is defined by \cite{gubser}
\be
\lambda = \frac{R_{Max}}{R_{Min}}-1,
\label{deflambda}
\ee
where $R_{Min}$ (resp. $R_{Max}$) is the largest (resp. smallest) value of the areal horizon radius, defined as $R = r_h e^{C(r_h,z)}$ in Schwarzschild-like coordinates.

The parameter $\lambda$ measures the deformation of the solution; the uniform string is characterized by $\lambda = 0$, while the hypothetical localized black hole phase has an infinite deformation parameter. In that sense, it measures how 'far' the non uniform solution is from the uniform solution; i.e. it characterizes the non uniformity. This is the reason why it is used as the order parameter in many references dealing with phases of black strings (Wiseman \cite{wiseman}, Wiseman-Kudoh \cite{kuwi}, Gubser \cite{gubser} etc).

\subsubsection{Relation between deformation and thermodynamics}
\label{defthlink}
It has been argued in \cite{flatphase} that using the mass and tension is more relevant in the construction of the phase diagram than another choice of parameters (including $\lambda$); for instance, in the asymptotically locally flat case, the mass and tension remain finite in the three phases (uniform, non uniform and localised black hole/string), while the entropy diverges for small temperature. However, the mass and tension are defined at the boundary spacetime and are difficult to compute from a the numerical data, as already stated. Moreover, in the case of asymptotically locally $AdS$ strings, the fact that the background tension is negative in some region of the parameters makes the picture more difficult to interpret.

An alternative is to study the phase diagram in the $T_H,\ S$ plane; these quantities are defined at the horizon of the black objects, and are easier to compute. Moreover, the entropy does not diverge in the $AdS$ case since there exists a minimum temperature for the uniform strings where the stable branch connects the unstable one.

Two remarks are in order concerning the deformation as a function of the parameter $d_0$, as pointed in \cite{pnubsd}:
\begin{itemize}
 \item $\lambda$ increases with $d_0$,
 \item the slope of $\lambda$ as a function of $d_0$ around $d_0=0$ is larger for larger absolute values of the cosmological constant.
\end{itemize}

These remarks can be formulated by the following statement:
\vskip10pt
{\it The deformation is related to the variation of entropy and $d_0$ is related to the variation of the temperature. Moreover the rate of variation of $\lambda$ with $d_0$ is proportional to the variation of the entropy with respect to the temperature along the non uniform phase.} 
\vskip10pt
The above statement can be argued in three steps:
\begin{enumerate}
 \item For small values of $\lambda$, the non uniform string is well described by the first order of perturbation theory (where $C(r,z)\approx \epsilon C_1(r)\cos 2\pi z$, $\epsilon$ being the small parameter of the perturbation). In this approximation, one can fix the value of $\epsilon$ for a given value of $\lambda$:
\be
\lambda = \frac{e^{\epsilon C_1(r_h)} + \mathcal O(\epsilon)^2}{e^{-\epsilon C_1(r_h))+ \mathcal O(\epsilon)^2}} -1= 2\epsilon C_1(r_h) + \mathcal O(\epsilon)^2.
\ee

From the boundary conditions of the first order in perturbation theory, $C_1(r_h)=1$, leading to $\epsilon = \lambda/2$.

\item The variation of the temperature along the non uniform branch is given by $T_H^{NU} = T_U e^{-d_0}$; a variation $\delta d_0$ in $d_0$ gives rise to a variation of the temperature $\delta T_H^{NU}$ in the non uniform phase according to 
\be
\delta T_H^{NU} = -T_H^Ue^{-d_0}\delta d_0.
\ee

Moreover, in the perturbative approach, the correction to the entropy appears in order $\epsilon^2$; it is then proportional to $\lambda^2$ from the first point:
\be
\delta S^{NU} \approx \lambda^2.
\ee
In every cases we considered, the correction on the entropy was always positive.

\item So, increasing $d_0$ from $0$ to $d_0\ll1$ implies a variation of $\lambda$ of $0$ to $\lambda\ll1$ and
\be
\frac{\delta S^{NU}}{\delta T_H^{NU}} \approx -\frac{1}{T_H^{NU}}\frac{\lambda^2}{d_0} \approx -\frac{1}{T_H^U}\frac{\lambda^2}{d_0}.
\ee
In other words, the value of $\lambda$ corresponding to a (small) value of $d_0$ is a monotonic negative function of  $\frac{\delta S^{NU}}{\delta T_H^{NU}}$, related to $C_p^{NU}$ defined previously:
\be
\lambda^2\approx - C_P^{NU} d_0, \mbox{ for }d_0\ll 1.
\ee
\end{enumerate}

\begin{figure}[H]
\center 
\includegraphics[scale=.3]{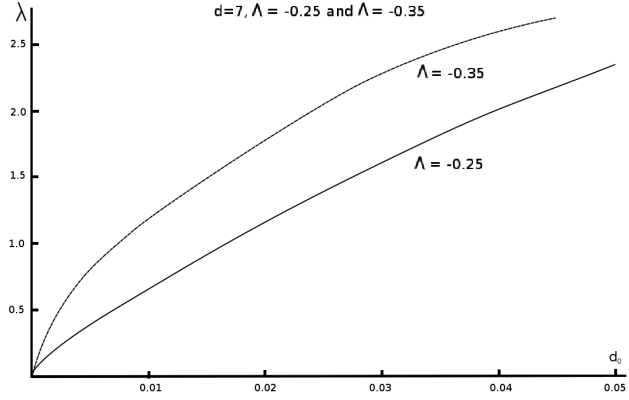}
\caption{For small values of $d_0$, the deformation increases faster for larger absolute values of the cosmological constant, this is to be related to the increase of $\delta S^{NU}/\delta T$ with $\Lambda$ (large absolute values of $\Lambda$ at fixed $r_h$ correspond to large values of $r_h$ at fixed $\Lambda$ from the scaling relations \eqref{ubsscale}, \cite{rms}).}
\label{fig:d7def}
\end{figure}

To sum up, for small values of $d_0$, if $\frac{\delta S^{NU}}{\delta T_H^{NU}}$ is large and negative, $\lambda$ will be large and positive and conversely. This is illustrated in figure \ref{fig:d7def} for $d=7$ and for two values of $\Lambda$.

\subsubsection{Entropy difference at same mass}
\label{dSsameM}
In the case of asymptotically locally flat spacetime, it is possible to evaluate the entropy difference at same mass between the non uniform and uniform string using the relation given in \eqref{dsflat} below (see for example \cite{sorkin, gubser}). To this end, the dimensionless entropy $s$, temperature $\theta$, mass $\mu$ and tension $\tilde\tau$ as functions of the deformation must be introduced:
\be
s := \frac{S}{L^{d-2}},\ \theta:=T_HL,\ \mu:=\frac{M}{L^{d-3}},\ \tilde\tau:=\frac{\tau}{L^{d-3}},
\ee
in units where $G=1$.

Then, we define the variation of the rescaled quantities under a variation of the thermodynamical quantities:
\bea
\frac{\delta s}{s} &=& \frac{\delta S}{S} - (d-2)\frac{\delta L}{L} := s_1\lambda^2 + \Ord{\lambda}{4},\\
\frac{\delta\theta}{\theta} &=& \frac{\delta T_H}{T_H} + \frac{\delta L}{L} := \theta_1\lambda^2 + \Ord{\lambda}{4}\nonumber\\
\frac{\delta\mu}{\mu} &=& \frac{\delta M}{M} - (d-3)\frac{\delta L}{L} := \eta_1\lambda^2 + \Ord{\lambda}{4},\nonumber\\
\frac{\delta\tilde\tau}{\tilde\tau} &=& \frac{\delta \tau}{\tau} - (d-3)\frac{\delta L}{L} := \tau_1\lambda^2 + \Ord{\lambda}{4}.\nonumber
\eea

The entropy difference at same mass (the variation due the non uniformity is compensated in the uniform phase by a variation of the length) is the given by \cite{sorkin,gubser}
\be
\frac{S_{NU}}{S_U} = 1 + \sigma_1 \lambda^2 + \sigma_2 \lambda^4 + ...,
\label{dsflat}
\ee
where $\sigma_1, \sigma_2$ are given by \cite{sorkin}
\be
\sigma_1 = s_1 - \frac{d-2}{d-3}\eta_1,\ \sigma_2 = -\frac{d-2}{2(d-3)}\left( \theta_1 + \frac{1}{d-3}\eta_1 \right)\eta_1.
\ee
In order to derive this \eqref{dsflat}, the relation between the entropy and the mass of the $(d-1)$-Tangherlini black hole, which is a foliation at constant $z$ of the uniform black string, is used at some stage (see \cite{gubser}). The computation of $\sigma_2$ ($\sigma_1$ vanishes \cite{gubser, sorkin}) for various values of the number of dimensions permitted to establish the existence of a critical dimension above which the order of the phase transition between the uniform black string and the non uniform black string changes from one to higher \cite{sorkin}.

However, in the asymptotically locally $AdS$ black string case, the relation between the mass and the entropy for the constant $z$ foliation of the uniform phase is not clear. So one cannot derive a similar relation and numerical computations must be used.

From a technical point of view, the mass and entropy in the uniform phase are functions of $r_h$, say $M(r_h)$ and $S(r_h)$ while the corrections on the mass (resp. on the entropy) along the non uniform string are functions of $r_h$ and $d_0$, say $\delta M(r_h,d_0)$ (resp. $\delta S(r_h,d_0)$), for a fixed value of the $AdS$ length. One can compute the variation of the mass along the non uniform branch and find the value of $r_h$, say $r_h^*$, such that the uniform string has the same mass:
\be
r_h^* \mbox{ is such that } M(r_h) + \delta M(r_h,d_0) = M(r_h^*).
\ee

Once $r_h^*$ is known, it is trivial to compute the difference between the entropy of the non uniform phase emerging from $r_h$ with a given value of $d_0$ and the entropy of the uniform string with horizon radius $r_h^*$, as a function of $d_0$:
\be
\Delta\equiv\frac{S_{NU}-S_U}{S_U} = \frac{(S(r_h) + \delta S(r_h,d_0))-S(r_h^*)}{S(r_h^*)}.
\label{dsnuu}
\ee

We will use the relation \eqref{dsnuu} in order to investigate the order of the phase transition. If the entropy of the uniform string is higher than the entropy of the non uniform string of same mass in the small deformation regime, the system will first choose the uniform phase. At some stage the non uniform string will be more entropic (the non uniform black string somehow interpolate between a localized black hole, which is more entropic than the uniform string as the argument of chapter \ref{chbhbs} suggest), and this phase will be entropically preferred; the phase is discontinuous in $\lambda$ and the order of the phase transition is $1$ \cite{landlif}. Conversely, if the entropy of the non uniform string is larger at same mass, the transition is continuous and the order is higher than one. In other words if $\Delta<0$ for small $d_0$, the order of the phase transition is one, if $\Delta>0$, the order is higher.

We will present the result of this investigation in the next section.

\subsection{Results and discussion}
We first considered the model presented above in 6 dimensions in \cite{pnubsads}, without factorising the asymptotic behaviour of the background functions. Many difficulties were encountered during the procedure, at various levels, but we were able to obtain some preliminary results indicating the existence of the non uniform asymptotically locally $AdS$ black string at least in $6$ dimensions. The results from the non perturbative regime were in agreement with the perturbative approach.

The existence of non uniform asymptotically locally $AdS$ black strings was confirmed in a more careful analysis, using the 'asymptotic factorisation' and for an arbitrary number of dimensions in \cite{nubsd}. However, the numerical procedure is still plagued by many difficulties:

\begin{enumerate}
\item The background functions are numerical and it is not clear how the numerical errors in the background propagates in the corrections. However, the background solution is nevertheless computed quite accurately, with relative errors typically of order $10^{-5}$.

\item If the parameters $\epsilon$ (the small parameter in the perturbative approach) and $d_0$ are not chosen in a suitable manner, the solver produces a uniform solution which seems to be a 'correction' on the uniform string leading to a uniform string with a different temperature. Of course, we are not interested in this kind of solutions. Moreover, it happened that the solution is a kind of combination of this uniform correction and some non uniform correction, leading to a decreasing of the deformation along the non uniform branch.

\item The region of larger horizon radius (but still in the unstable phase of the uniform black string) is even more difficult to investigate: we expect a transition from thermodynamically unstable non uniform black strings to thermodynamically stable non uniform black strings following the arguments presented in section \ref{pertu}.
\end{enumerate}

These thermodynamically stable (resp. unstable) non uniform black string were referred to as small-short (resp. small-long) non uniform black strings since they are characterized by a small horizon radius and a short (resp. large) critical length (compared to the $AdS$ curvature). The boundary separating these two phases, small-short and small-long black string, is characterized by $\delta S/\delta T_H \rightarrow\infty$. As we argued in section \ref{defthlink} the deformation would become extremely high even for very small values of $d_0$ (thus of $\delta T_H$). Of course, this extreme behaviour is not manageable from a technical point of view and sets the limit of our numerical investigation.
However, we were able to find some hints for the existence of non uniform solutions in a neighbourhood of this region, but the corresponding solutions have large numerical errors and are not reliable enough to perform quantitative predictions.

Despite all these difficulties, we were able to obtain accurate solutions in the small-short non uniform black string region, with relative errors of the order of 2-3$\%$. Of course, in the highly deformed region, the solver starts failing giving accurate solutions.

\subsubsection{Hint for a localised black hole phase}
For the reasons described above, we restrict the study of the $AdS$ non uniform black string to the region of the parameter space where $r_h\ll\ell$ and $L\ll\ell$.

We were able to construct several solutions for various values of $d$, $\ell$ (or equivalently $r_h$) and $d_0$. We focused on $d=7$ and constructed a number of solutions for $d=5,8$. We typically considered three or four different values of the cosmological constant, ranging from $-0.05$ to $-0.3$ and $r_h=1$. Then we rescaled the solutions in order to have the results for fixed cosmological constant. For every values of the number of dimensions and cosmological constant, we constructed many solutions with increasing values of $d_0$, typically ranging from $0.0001$ to $0.01$. We were able to construct solutions characterized by $\lambda\approx2.5$ in most cases and higher for $d=7$. It should be mentioned that we did not perform a systematic analysis for every number of dimensions; a deeper analysis, confirming the existence of the solutions and providing a more detailed analysis should be carried in order to complete the picture and to confirm the existence of the solutions.

\begin{figure}[H]
   \center 
    \includegraphics[scale=.23]{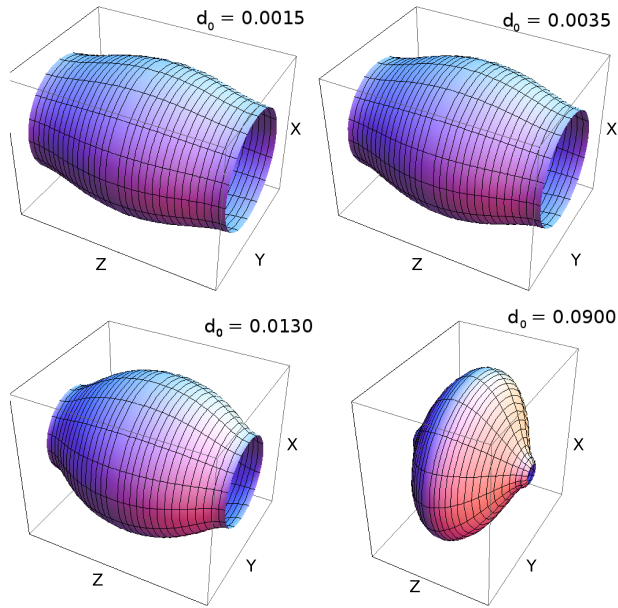}
    \caption{Embedding of the horizon of non uniform black strings with $d=7$, $\Lambda=-0.25$ in a 3 dimensional euclidean space. The equation of the surface is given by $X^2 + Y^2 = r_h^2e^{2C(r_h,z)}$. The horizon looks like a black hole horizon as the deformation increases.}
    \label{fig:d7hor}
  \end{figure}

For all the solutions we found, the deformation is an increasing function of $d_0$ and we systematically found extremely deformed solution, providing strong indications for the existence of an $AdS$ localised black hole phase. In figure \ref{fig:d7hor}, we present embeddings of the event horizon in a 3 dimensional euclidean space i.e. of the areal radius $R(r,z)=re^{C(r,z)}$ evaluated at $r=r_h$; the strings is almost uniform for small values of $d_0$ and looks like a (deformed) sphere touching the borders for larger values of $d_0$.

%%%%%%%%%%%%%%%%%%%%%%%%%%%%%%%%%%%%%%%%%%%%%%%%%%%%%%%%%%%%%%%%%%%%%%%%%%%%%%%%%%%%%%%%%%%%%%%%%%%%%%%%%%%%%
\subsubsection{Horizon radius depending critical dimension}
\begin{figure}[H]
   \center 
    \includegraphics[scale=.3]{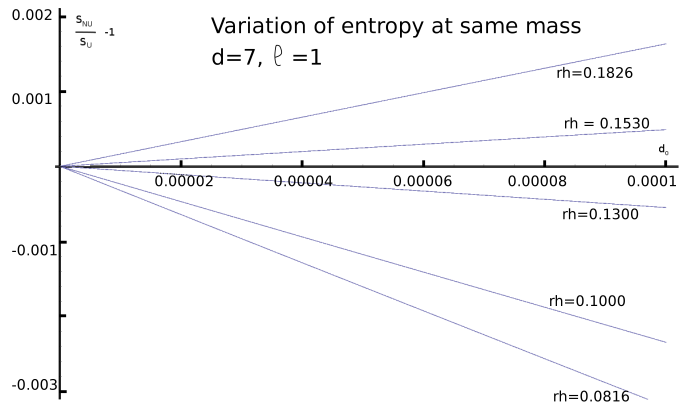}
    \caption{The relative entropy difference between the non uniform phase and uniform phase at same mass as a function of $d_0$ for $d=7,\ell=1$ and various value of $r_h$.}
    \label{fig:criticalrh}
  \end{figure}

We have evaluated the entropy difference between the non uniform black string and the uniform black string with same mass for $d=7$ and various values of $r_h$. Above some critical value of the horizon radius, the uniform black string is more entropic than the non uniform black string, while for larger values of $r_h$, the non uniform black string becomes more entropic. In other words, the phase transition can be of first order or of higher order for the same value of the number of dimension, depending on the value of the horizon radius. We present a plot of the variation of the entropy difference as a function of $d_0$ for $d=7, \ell=1$ in figure \ref{fig:criticalrh} for various values of the horizon radius. Small values have a negative slope while larger value acquire a positive slope, changing the order of the phase transition.

It should be noted that the value of $d_0$ in figure \ref{fig:criticalrh} is very small, and thus the non uniform solutions are weakly deformed. This regime of small deformation is the regime where the solutions to the set of partial non linear differential equations are well approximated by the perturbative solutions. The variation of entropy and mass are computed according to the procedure described in section \ref{dSsameM}, involving a comparison between the entropy of the background and of the non uniform solution. So on the one hand, we have the background uniform solution which is computed very accurately (the relative error is of the order $10^{-5}$) and on the other hand, the non uniform solution is in the best controlled region. Moreover, the quantities involved in the entropy difference computation are extracted from the horizon data, which is the most reliable region.

Figure \ref{fig:criticalrh} thus provides strong indications for the fact that the order of the phase transition can change for the same number of spacetime dimensions. Although we only investigated the case of $d=7$, we expect this property to hold for all values of the dimension. Said differently, the number of dimensions above which the order of the phase transition changes should depend on the value  of $r_h$. 

\section{Concluding remarks}
In this chapter, we studied a part of the phase diagram of the $AdS$ black strings. First, after reviewing the uniform string solution and its thermodynamical properties, we studied the dynamical stability of the solution. We found that the uniform string is unstable for small $AdS$ black strings, i.e. when the horizon radius is small compared to the $AdS$ radius. This is in perfect correspondence with the thermodynamically stable region, along with the Gubser-Mitra conjecture \cite{gmbh}. Inspired by the asymptotically locally flat case, we expected non uniform $AdS$ black string to exist. We computed solutions for such objects, first in a perturbative treatment up to order $2$ then in a non perturbative approach.

The results from the perturbative solutions gave us strong indications for the existence of the $AdS$ non uniform black string phase and permitted to compute the first corrections on the thermodynamical quantities in the non uniform branch. It also allowed to predict a new thermodynamically stable phase of non uniform black string, which we referred to as \emph{small-long} non uniform black string. This is purely an effect of the cosmological constant and is somehow the counterpart of the big $AdS$ uniform black string, except that in this case, $L/\ell\gg1$ instead of $r_h/\ell\gg1$. Moreover, we showed that there is not only \emph{one} non uniform black string (in the sense that in the flat case, the solution depends only on one dimensionless parameter $r_h/\ell$) but a continuity of non uniform black strings emanating from the uniform strings with different critical length. This is also an effect of the cosmological constant, since the $AdS$ radius provides a new length-scale in the theory.

The non perturbative approach confirmed the existence of the non uniform solutions, and gave hints that the counterpart of the critical dimension in the flat case \cite{sorkin} should depend on the value of the $AdS$ radius (or equivalently on the horizon radius). Moreover, from the embeddings of the horizon in euclidean spacetimes, the non uniform phase is likely to connect to an $AdS$ localized black hole phase, still to be constructed.

Note that we revisited some of the thermodynamical properties of the uniform phase, arguing that it has a negative tension for some region of the parameters.

All these results - the new non uniform phase, the negative tension, etc. - should have counterparts in a conformal field theory defined on $S_{d-3}\times S_1$, thanks to the $AdS/CFT$ correspondence, but this aspect hasn't been studied in this thesis.

%% file: ccl.tex
This thesis, entitled 'New features of black strings and branes in higher dimensional gravity due to a cosmological constant' was essentially divided in two parts. The first part dealt with braneworld models while the second part dealt with black objects.

In both cases, we investigated the effect of considering a non vanishing cosmological constant in the model. It turned out that a cosmological constant induces drastic changes on the pattern solutions or on their properties. In this concluding chapter, we will first review our main results, stressing the effect of the cosmological constant; then, we will give some possible perspectives to this thesis.

\section{Braneworld models}
We studied four classes of topological brane models. In these models, the branes consist in a soliton extending in the extradimensions. More precisely, they are topological defects available in the theories under consideration; namely the Einstein abelian Higgs model, the Einstein non abelian Higgs model, the Einstein-Yang-Mills-Higgs model and the baby Skyrme model. The geometry of these models were chosen to be non factorisable. 

We considered a cosmological constant in the entire spacetime (a bulk cosmological constant) and modelled the effect of a cosmological constant on the brane by considering inflating branes. The four dimensional spacetime contributes to the equations associated to these models only through the Ricci scalar associated to the four dimensional subspace, allowing the replacement the four dimensional part by any spacetime with constant curvature. Note that inflating brane models are also relevant in a cosmological context, as braneworld inflationary models.

The consequence of the inclusion of the cosmological constant on the brane and in the entire spacetime is that the solutions admit essentially three type of extradimensional geometry, depending on the sign of the bulk cosmological constant: opened, closed or flat extradimensional geometry. All these geometries have an angular deficit, related to the brane cosmological constant. A common feature of the different models is the occurrence of periodic solutions, providing a natural compactification of the extradimensions. In the case of $6$ dimensions, we have shown that the vacuum system is integrable by quadrature and we were able to express the effective Planck mass in terms of the metric functions. 

%It turned out that it is possible to generate a large effective Planck mass with a fundamental Planck mass of the order of the electroweak mass scale, if the brane is inflating and the bulk cosmological constant is positive. This is an elegant result since the observations points to a positive cosmological constant (in our four dimensional universe), making the model plausible. 

It should be stressed that the branes in these models have an extension, due to the extension of the soliton describing the branes. This is in contrast with the Randall-Sundrum model where the branes are infinitely thin, having the consequence that the metric function are not smooth (the derivative of the metric functions is not continuous). %However, Randall and Sundrum were the first to introduce warped brane models.

We tried to classify the solutions available in the $6$ dimensional Einstein abelian Higgs model according to the values of the bulk cosmological constant, the brane cosmological constant and the self coupling of the Higgs field. The pattern of solutions is very rich but quite involved, due to the number of parameters in the model. We also constructed solutions presenting a 'mirror' symmetry in the framework of non abelian Einstein-Yang-Mills in $7$ dimensions and non abelian Einstein-Goldstone in $d$ dimensions, allowing once again periodic solutions, i.e. naturally compactified spacetimes.

Finally, we turned back to six dimension with the baby Skyrme model; we took advantage of a natural definition of a root mean square radius to characterise the extension of the brane. The baby Skyrme model is a toy model of the Skyrme model, originally designed to describe nucleons. In that sense, the $6$-dimensional baby Skyrme model can be thought as a toy model containing effective hadronic matter, i.e. realistic matter fields. In this case, we also systematically found periodic solutions.

We believe that the presence of periodic solutions (closed geometry) is a generic feature of inflating brane models, independently of the matter content, as long as the matter fields are localized, which is a necessary condition in order to describe localized branes. This idea is comforted by the fact that periodic solutions are indeed vacuum solutions; adding localized matter field should not influence too much the asymptotic form of the spacetime.

\section{Fermion localization on a brane}
We studied a model in $1+1$ dimensions for the localization of fermionic fields on a kink/anti-kink system. Previous works have been made in this direction, for example, the coupling of fermions to kink solution available in the $\phi^4$ model and to a superposition of the latter has been studied in \cite{vachaspati}. However, the superposition is an approximate solution only if the kink and anti-kink are separated enough. 

We took advantage of the existence of a kink/anti-kink solution in the Sine-Gordon model in order to study the bound states of fermions in such a background, without approximation. First, we studied the coupling of fermions to kink or anti-kink solution in the Sine-Gordon model. The kink solution in the Sine-Gordon model is qualitatively the same as in the $\phi^4$ model; as a consequence, the structure of fermionic bound states in both theories is nearly the same.

More interestingly, there is an exact solution in the Sine-Gordon model describing a kink-antikink system. This exact solution is still valid in the limit where the kink and antikink are close. We studied the coupling of fermions to the kink/anti-kink solution available in the Sine-Gordon model and found that when the kink and anti-kink are too close, only a few fermionic bound states exist. When the distance between the kink and anti-kink is increased, new bound state emanate from the continuum. 

Moreover, the eigenvalues degenerate by pair in the limit where the separation between the kink and anti-kink is large enough. This is because the potential seen by the fermion around the kink and around the anti-kink are superpartner potentials. If the kink and antikink are not distant enough, the supersymmetry is somehow broken.

Finally, we studied the stability equation of three particular kink models. The stability equation took the form of a Poschl-Teller equation, which admits exact solutions.

The kink and/or anti-kink can be thought as modelling brane/anti-brane. As a consequence, the $1+1$ dimensional model studied here mimics a five dimensional brane and/or anti-brane.

\section{Higher dimensional black objects}
Black holes in four dimensions are rather well understood. In particular, they are unique, in the sense that they belong to a three parameter family of solutions, the Kerr-Neumann solutions \cite{kn}. The horizon topology of the four dimensional black holes can only be spherical. However, in higher dimensions, there exists black solutions with other horizon topologies. Among the zoo of solutions, there are black hole solutions presenting a spherical horizon topology, but there are also solutions with cylindrical horizon topology (black strings), toroidal horizon topology (black rings \cite{solubr1}), solutions consisting in a black hole surrounded by a black ring (black saturns \cite{blacksaturn}), multi-black rings, etc. 

In this thesis, we focused on spherical horizon topology and cylindrical horizon topology.

\subsection{Charged-rotating de Sitter black hole}
We constructed the charged-rotating de Sitter black hole. Unlike the four dimensional case, such solutions are not known analytically in higher dimensions. The exact black hole solutions available in higher dimensions are the $(A)dS$ rotating higher dimensional black hole (Myers and Perry \cite{solubh} and Gibbons-Lu-Page-Pope \cite{solubh1}), the $(A)dS$ Tangherlini solution, the higher dimensional $(A)dS$ charged black hole solutions. The missing part was the $(A)dS$ charged rotating black hole. Such a solution has been constructed numerically for a negative cosmological constant in \cite{solubh3}, but the case of a positive cosmological constant is more involved, due to the occurrence of a cosmological horizon. 

We were able to integrate the equation of the charged-rotating de Sitter higher dimensional black hole numerically, adding an equation expressing the fact that the cosmological constant is indeed a constant. This 'trick' allowed sufficient freedom to impose regularity conditions on the cosmological horizon. The solution is asymptotically de Sitter, as expected. We focused particularly in the case of five dimensions, hoping that it catches the main features of the generic solutions. 

It must be stressed that the rotating solutions behaves quite differently in $d=5$ and $d>5$, $d$ being the number of dimensions. In five dimensions, the horizon area goes to zero for some maximum angular momentum, whereas in six dimensions and higher, there is no upper bound on the angular momentum. Although there are differences in the thermodynamical properties, the solution behave qualitatively in the same way at least in some region of the parameter space. This is why we expect the de Sitter charged-rotating black holes to exist for any value of $d$. 

Note also the existence of two phases for the $(A)dS$ charged rotating black hole; this is to be put in relation with the existence of two phases in the $AdS$ Tangherlini black hole. This is typically one of the main difference between the solutions with and without a cosmological constant.

\subsection{Black string with positive cosmological constant}

Considering a $(d-1)$ dimensional black hole and adding an extra compact direction leads to a black string, the horizon topology is then cylindrical. The construction of a black string in asymptotically locally flat spacetime is straightforward.

We tried to construct a generalization of black string solutions to asymptotically locally de Sitter spacetimes, but it turned out that such solutions do not exist. There are actually black string solutions to the Einstein equations with a positive cosmological constant, but they are not asymptotically (locally) de Sitter. Instead, the asymptotic spacetime is singular and can be reached by a massive or massless observer in a finite proper time. Moreover, the singularity is located at a finite proper distance from outside the event horizon. This is a naked singularity; there is no second horizon that would 'hide' the latter.

We interpret this non existence as resulting from the fact that the compact extra direction cannot support the negative pressure induced by the positive cosmological constant.

Note that the $AdS$ counterparts of black string solutions indeed exist and have been constructed in \cite{rms}. We revisited the thermodynamical properties of such black strings and as well as their stability and phase diagram. We will review the results in the next section.

\section{Phases of $AdS$ black strings}
Black strings in asymptotically locally flat spacetimes are unstable, thermodynamically as well as dynamically. This was the result of Gregory and Laflamme. The existence of a static perturbation (zero mode) suggested a non uniform black string phase; i.e. a black object with cylindrical horizon topology but which depends non trivially on the compact direction. Note that the existence of a zero mode is the signature of a Gregory-Laflamme instability.

As we said in the previous section, black string solutions have been extended to asymptotically locally $AdS$ spacetimes. There exists two phases of $AdS$ black string solutions; one is thermodynamically stable (big black string), the other is thermodynamically unstable (small black strings).

We studied the dynamical stability of such objects by looking for a static perturbation, following the lines introduced by Gubser \cite{gubser}. Our results suggest that there is a well defined Gregory-Laflamme zero mode for the small black strings, while there is no static perturbation for the big black strings. Moreover, the point where the dynamically stable and unstable branch meet coincides perfectly with the point where the thermodynamical phases meet. This matching between the thermodynamical and dynamical stability was conjectured by Gubser and Mitra \cite{gmbh}.

Let us stress that in the asymptotically locally flat case, there is only one static mode, in the sense that once one uses rescaled variables, the value of the static mode is fixed. In the $AdS$ case, even with rescaled variables, there is a continuity of zero modes, labelled by the horizon radius (or equivalently by the cosmological constant). This is due to the fact that the $AdS$ radius is an additional length scale in the theory.

We constructed the non uniform phase emanating from the unstable black strings. The picture is quite different than in asymptotically flat space: we provided evidences that there should exist a thermodynamically stable phase of non uniform black string in $AdS$, characterized by a long compact direction (relative to the cosmological radius). This new phase seems to be the counterpart of the small/big $AdS$ uniform phase, but where the relevant parameter is the ratio between the length of the compact direction and the $AdS$ radius instead of the ratio between the horizon radius and the $AdS$ radius. We called these solutions short and long black strings for obvious reasons. Note that this phase emanates from the small uniform black strings, since the big uniform black string is stable.

Our results also provide qualitative evidences for the existence of a localized black hole phase in $AdS$. This can be seen by looking at the embeddings of the horizon in an euclidean space; the more we slide along the non uniform branch, the more the deformation increases. We therefore imagine that the deformation can increase until the horizon disconnects and adopts a spherical topology.

Finally, we argued that the order of the phase transition between the uniform and non uniform black string changes depending on the value of the horizon radius (for a given value of the cosmological constant). This is also in contrast with the asymptotically locally flat black string, where the order in the phase transition changes only above a critical value of the number of dimensions ($\approx13$).

\section{Outlook and perspectives}
Many extensions of this thesis are possible... For instance, stability of the baby Skyrme model is currently under investigation \cite{sadelbri}; one can also consider the Skyrme model in $7$ dimensions. Another possibility is to consider other type of matter fields to describe the brane. We may also consider the interaction between two solitonic branes in various models.

Among all possibilities, an interesting outlook for brane models is to couple fermionic fields to the solutions presented in this thesis. We already studied a toy model in $1+1$ dimensions; it would be interesting to study the full $n+4$ dimensional problem.

Concerning the black strings, our results suggest the existence of an $AdS$ localized black hole phase. One obvious continuation of this thesis would be to construct this phase. This is however technically much more involved than the non uniform black string.
Another issue of the black strings in $AdS$ is their connection with the dual conformal field theory in the $AdS/CFT$ context. Note that the uniform $AdS$ black string solution would provide new background for the dual $CFT$.

Finally, the techniques developed to construct non uniform black strings might be applied to black rings, producing an object which is not uniform along the ring.

%% file: reminder.tex
\section{Lagrangian formulation of Einstein equations and boundary terms}
It is possible to derive the Einstein equations from a variational principle. The action giving rise to the Einstein equations (in vacuum) is called the Einstein-Hilbert action:
\be
S_{EH} = \frac{1}{16\pi G}\int_{\mathcal M} \sqrt{-g}R d^dx,
\label{leh}
\ee
where $G$ is the Newton constant, $\mathcal M$ is the $d$-dimensional manifold under consideration, $g$ is the determinant of the metric on the manifold and $R$ is the scalar curvature.

The scalar curvature is constructed from the metric components in the following way: first, the Christoffel symbols are defined as
\be
\Gamma_{ab}^c = \frac{1}{2}g^{cd}\left( -\partial_d g_{ab} + \partial_a g_{bd} + \partial_b g_{ad} \right),
\ee
The indices range from $0$ to $d-1$. Then, the Riemann tensor is constructed:
\be
R^a_{\;bcd} = \partial_c \Gamma_{bd}^a - \partial_d \Gamma_{bc}^a + \Gamma_{cb}^e\Gamma^a_{de} - \Gamma_{bd}^e \Gamma^a_{ce}.
\ee
Finally, the Ricci tensor and Ricci scalar are given by
\be
R_{ab} = R^c_{acb},\ R = g^{ab}R_{ab}.
\ee

If we want to describe matter fields interacting under gravity, the simplest way is to consider an action of the form
% \emph{minimally} couple a matter lagrangian $L_m$:
\be
L = \frac{1}{16\pi G}\int \sqrt{-g}(R + 16\pi G L_m),
\label{lehc}
\ee
where $L_m$ is the lagrangian density describing the matter fields. This way to couple matter fields to gravity is called 'minimal coupling'.

Extremising the action \eqref{lehc} with respect to the variation of the metric components leads to the Einstein equations. 

In order to obtain the equations, we first use the identity
\be
\delta (\sqrt{-g}R + 16\pi G\sqrt{-g}L_m )= \sqrt{-g}\left( R_{ab} - \frac{1}{2}Rg_{ab} - 8\pi GT_{ab} \right)\delta g^{ab},
\ee
where $T_{ab} = -\frac{2}{\sqrt{-g}}\left( \ddf{L_m}{g^{ab}} - \frac{1}{2}g_{ab}L_m \right)$ is the stress-energy tensor.

Let us compute the variation of each terms in order to set this relation:
\bea
\delta g = g g_{ab} \delta g^{ab}\Rightarrow \delta\sqrt{-g} = -\frac{1}{2}\sqrt{-g}g_{ab}\delta g^{ab},\\
\delta R = \delta (R_{ab}g^{ab}) = g^{ab}\delta R_{ab}  + R_{ab}\delta g^{ab}.
\eea
But $\delta R_{ab}$ is divergence and contribute as boundary terms, so it doesn't enter the equations of motions:
\be
\delta R_{ab} = \nabla_c\left( \delta\Gamma^c_{ab} - \delta_b^c\delta\Gamma^e_{ae} \right).
\label{ehdiv}
\ee

Putting everything together, we find
\be
G_{ab} = 8\pi G T_{ab},
\ee
where $G_{ab} = R_{ab} - \frac{1}{2} R g_{ab}$ is the Einstein tensor.

As we have seen, the variational principle gives rise to a divergence term, leading to a boundary term \eqref{ehdiv}. In order to take this divergence term into account and obtain a well defined variational principle, one can add an extra term in the action. This extra term was introduced by Hawking and Gibbons:
\be
S = \frac{1}{16\pi G}\int_{\mathcal M}\sqrt{-g}\left(R + 16\pi G L_m  \right) + \frac{1}{8\pi G}\int_{\partial\mathcal M}\sqrt{-h}K,
\ee
where $\mathcal M$ is the manifold under consideration, $\partial\mathcal M$ is the boundary manifold, $h$ is the determinant of the induced metric on the boundary and $K$ is the trace of the extrinsic curvature of the boundary, $K_{ab}=\nabla_a n_b$, $n^a$ being a vector normal to the boundary.

\section{Symmetries and conserved charges}

\subsection{Komar integrals}
In this subsection, we present a technique to compute conserved quantities, namely the Komar integrals. We follow the approach of reference \cite{wald}. Let $k$ be a killing vector. Consider the quantity
\bea
\label{mf1}
\epsilon^{abm_1\ldots m_{d-2}}\nabla_b \epsilon_{m_1\ldots m_{d-2}cd}\nabla^c k^d
  &=& \epsilon^{m_1\ldots m_{d-2}ab} \epsilon_{m_1\ldots m_{d-2}cd}\nabla_b\nabla^c k^d\nonumber\\
  &=& (d-2)!2!\delta^{[a}_c\delta^{b]}_d\nabla_b\nabla^c k^d\\
  &=& (d-2)!2!\nabla_b\nabla^{[a} k^{b]}\nonumber\\
  &=& (d-2)!2! R^a_b k^b\nonumber  
\eea

Contracting equation \eqref{mf1} with $\epsilon_{an_1\ldots n_{d-1}}$ leads to
\be
\nabla_{[n_1}\epsilon_{n_2\ldots n_{d-1}]cd}\nabla^c\xi^d = \frac{2}{d-1} R^a_bk^b\epsilon_{an_1\ldots n_{d-1}},
\label{mf2}
\ee
since
\bea
&&\epsilon_{an_1\ldots n_{d-1}} \epsilon^{abm_1\ldots m_{d-2}} \epsilon_{m_1\ldots m_{d-2}cd}\\
&&\ \ \ \ \ \ \     = 1!(d-1)! \delta^{[b}_{n_1}\delta^{m_1}_{n_2}\ldots\delta^{m_{d-2}]}_{n_{d-1}}\epsilon_{m_1\ldots m_{d-2}cd}.\nonumber
\eea

In the vacuum region (typically in asymptotic region), the Ricci tensor vanishes. It follows that the left hand side of equation \eqref{mf2} vanishes as well.

In other words, considering the Killing form $\mathcal K = k_m dx^m$, equation \eqref{mf2} states that
\be
d*d\mathcal K = 0,
\label{dkf}
\ee
in regions where $R_{ab}=0$.

Integrating equation \eqref{dkf} over a volume $\mathcal V$ bounded by any two $(d-2)$-spheres $S$ and $S'$ in the exterior vacuum region and applying Stokes theorem results in the fact that the integral
\be
\int_S *d\mathcal K
\ee
is independent on the choice of $S$ and thus is a constant. Expressing the form \eqref{dkf} in components, this integral yields the following expression for a conserved charge associated to a Killing vector $k$, known as the Komar integral:
\be
\mathcal Q_k = -\frac{1}{16\pi G}\int_S \epsilon_{m_1\ldots m_{d-2} ab} \nabla^ak^b d S^{m_1\ldots m_{d-2}},
\ee
where $d S^{m_1\ldots m_{d-2}}$ is the surface element on $S$.

\subsection{Background substraction}
The background substraction method has been introduced by Hawking and Horowitz in \cite{hoha}. The main idea is to first consider the Hamiltonian formulation of general relativity,
\bea
H &=&\frac{1}{16\pi G} \int_{\Sigma_t} (N\mathcal H + N_m \mathcal H^m)\sqrt{\gamma}d^{d-1}x \\
&&- \frac{1}{8\pi G}\int_{S_t^\infty}( N\ ^{(d-2)}K - N^m p_{mn}r^n )\sqrt{\gamma_\infty}d^{d-2}x,\nonumber
\eea
where $\Sigma_t$ is the sliced manifold with respect to some time coordinate, $p_{mn}$ is the momentum associate to the induced metric on $\Sigma_t$ such that $p_{mn} = \ddf{L_{EH}}{\gamma^{mn}}$ ($L_{EH}$ being the Einstein-Hilbert Lagrangian and $\gamma_{mn}$ is defined in the next equation), $N$ and $N^m$ are respectively the lapse function and the shift vector (see \cite{gravitation,wald}) associated to the $t$ coordinate, such that the line element takes the form
\be
ds^2 = -N^2 dt^2 + (dx^a-N^a dt)(dx^b-N^b dt)\gamma_{ab},\mbox{ for some metric }\gamma_{ab}\mbox{ on }\Sigma_t,
\ee
$S_t^\infty = \Sigma_t\cap\Sigma^\infty$, where $\Sigma^\infty$ is the boundary manifold, $n^m$ is the normal unit vector to $\Sigma_t$ while $r^m$ is the normal unit vector to $\Sigma^\infty$. $\ ^{(d-2)}K$ is the trace of the extrinsic curvature of $S_t^\infty$ embedded in $\Sigma_t$ and $\mathcal H, \mathcal H^m$ are constraints associated to the diffeomorphism invariance of the theory \cite{wald}. $\gamma_\infty$ is the determinant of the induced metric on $S_t^\infty$.

A solution to the Einstein equations solves the constraints, so the hamiltonian reduces on-shell to 
\be
H = - \frac{1}{8\pi G}\int_{S_t^\infty}( N\ ^{(d-2)}K - N^m p_{mn}r^n )\sqrt{\gamma_\infty}d^{d-2}x.
\ee
Unfortunately, this Hamiltonian evaluated with a solution to the Einstein equations is sometimes diverging. The simplest example is the Minkowski spacetime in four dimensions expressed in spherical coordinates. The boundary manifold is simply the surface of a sphere of radius $R\rightarrow\infty$ and the extrinsic curvature is $\propto1/R$. The lapse function in this case is just $1$ and the shift vector is the null vector. But the determinant of the induced metric is $R^4\sin^2\theta$, leading to a diverging Hamiltonian $H=\int R \sin\theta d\theta d\phi$ for large values of $R$.

The idea of Hawking and Horowitz is to consider a background reference metric $g^0_{ab}$ and to subtract the contribution to the Hamiltonian of this reference solution. The background has to be static and the solution which we want to compute the Hamiltonian should asymptotically be sufficiently close to the reference background (see \cite{hoha} for the precise meaning of sufficiently close). Then, since the reference background is static, the conjugate momentum to the induced boundary metric vanishes and the only contribution left to the \emph{physical} Hamiltonian is given by
\be
H_{phys} = - \frac{1}{8\pi G}\int_{S_t^\infty}\left( N\ (^{(d-2)}K-^{(d-2)}K_0) - N^m p_{mn}r^n \right)\sqrt{\gamma_\infty}d^{d-2}x,
\ee
where $^{(d-2)}K_0$ is defined in the same way as $^{(d-2)}K$ but for the background reference metric.
For example, the appropriate reference background for Tangherlini solutions is the $d$-dimensional Minkowski space in spherical coordinates.

\section{Linearised gravity}

Let us consider a line element
\be
ds^2 = g_{ab}dx^a dx^b,
\ee
where $g_{ab}$ is a solution of the Einstein equations, i.e. such that $G_{ab}= 8\pi G T_{ab}$ for certain matter fields.

We can consider a perturbation $h_{ab}$ around this solution; the perturbed metric then reads
\be
\tilde g_{ab} = g_{ab} + h_{ab},\ h_{ab}\ll g_{ab}.
\label{gpertu}
\ee
Of course the matter fields are perturbed also, but here, we will focus on the gravitational perturbation since the matter perturbation will depend on the model under consideration.

Taking the trace of the Einstein equation and solving for the scalar curvature yields the following form of the Einstein equation:
\be
R_{ab}=8\pi G\left(  T_{ab} - \frac{1}{d-2}\tilde g_{ab}T \right),
\label{einseq2}
\ee
where $T$ is the trace of the stress tensor and $R_{ab}$ is the Ricci tensor constructed out of \eqref{gpertu}.

To first order in $h_{ab}$, \eqref{einseq2} becomes
\be
-\frac{1}{2}\left( \Box \delta_a^c \delta_b^d + 2 R_{a\ b}^{\ c\ d}\right)h_{cd} = 8\pi G\delta\left(  T_{ab} - \frac{1}{d-2}\tilde g_{ab}T \right).
\label{licheq}
\ee
The right hand side is computed assuming that the matter fields are perturbed also; $\delta$ then represents the variation of $ T_{ab} - \frac{1}{d-2}\tilde g_{ab}T$ limited to first order in $h_{ab}$ and in the matter field perturbation. Equation \eqref{licheq} holds only in the gauge such that $\nabla_a h^{ab} = 0,\ h_{ab}g^{ab}=0$; if we don't impose a gauge choice, there are terms of the form $\nabla_a\nabla_ch^c_b,\ \nabla_a\nabla_b h_{cd} g^{cd}$. Note that the d'Alembertian and the Riemann tensor are evaluated using the background metric $g_{ab}$. The derivation of equation \eqref{licheq} is straightforward.

The left hand side of \eqref{licheq} is the variation of the Ricci tensor and defines the Lichnerowitz operator $(\Delta_L)_{ab}^{\ \ cd}=  \Box \delta_a^c \delta_b^d + 2 R_{a\ b}^{\ c\ d}$ such that $\delta R_{ab} = -\frac{1}{2}(\Delta_L)_{ab}^{\ \ cd}h_{cd}$.

%% file: ads.tex
\section{(A)dS spacetime}

\begin{figure}
\centering
\includegraphics[scale=.5]{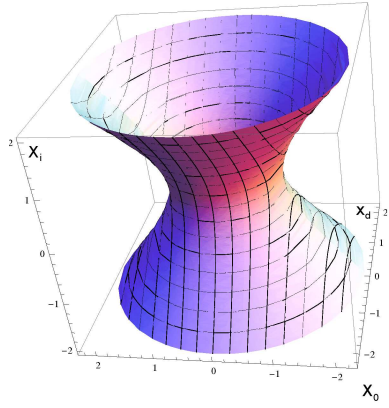}
\caption{Embedding of the $AdS$ hyperboloid in a 3 dimensional Euclidean spacetime. The $AdS$ hyperboloid is embedded in a flat spacetime of signature $(-,-,+\ldots,+)$.}
\label{fig:adshyp}
\end{figure}

From a geometric point of view, the $d$-dimensional $AdS$ spacetime is the surface of a $(d+1)$-dimensional hyperboloid embedded in flat spacetime of signature $(-,-,+,\ldots,+)$. This hyperboloid and the embedding spacetime are invariant under the action of $SO(2,d-2)$. Let us denote the coordinates of the embedding spacetime by $U,V,X_i,\ i=1,\ldots, d-2$; then the equation of the hyperboloid is $H=0$, the metric and $H$ being given by
\bea
ds^2 &=& -dX_0^2 - dX_d^2 + \delta_{ij}dX^idX^j,\ i,j=1,\ldots,d-1,\\
H&=&-X_0^2 - X_d^2 + \sum_{i=1}^{d-1} X_i^2  +R^2=0,\nonumber
\eea
for some $R\in \mathbb R$ and where $\delta_{ij}$ is the Kronecker symbol. This is represented in figure \ref{fig:adshyp} in a 3 dimensional euclidean spacetime.

Defining 
\bea
u &=& \frac{X_0 - X_d}{R^2},\ v = \frac{X_0 + X_d}{R^2},\\
x^i &=& \frac{X_i}{R u},\ t = \frac{X_d}{R u},\nonumber
\eea
the line element on the hyperboloid becomes
\be
ds^2 = \frac{R^2}{z^2}\left( -dt^2 + \delta_{ij}dx^i dx^j + dz^2 \right),\ i,j=1,\ldots,d-2,
\label{adspoinc}
\ee
where $z = 1/u$. These coordinates for $AdS$ spacetime are referred to as the Poincar\'e coordinates.

Note that these coordinates define two patches: $z>0,\ z<0$, corresponding to $X_0-X_d>0$ (resp. $<0$). These two patches are the Poincar\'e patches.
It is clear from \eqref{adspoinc} that the conformal boundary of $AdS_d$ is $\mathcal M_{d-1}$ (located at $z=0$).

$AdS$ spacetimes has a constant negative curvature; the Riemann tensor is given by
\be
R_{abcd} = -\frac{1}{R^2} \left( g_{ac}g_{bd} - g_{ad}g_{bc}\right).
\label{maxsym}
\ee

Note that the Minkowski spacetime obeys the same relation, in the limit where $R\rightarrow\infty$. The counterpart of $AdS$ spacetime with a positive curvature is the de Sitter spacetime, having also such a symmetric Riemann tensor, except that the sign is $+$ instead of $-$.

Such spacetimes are said to be maximally symmetric.

Another choice of parametrisation of the hyperboloid,
\bea
X_0 &=& R \frac{\cos\tau}{\sin\rho},\ X_d = R\frac{\sin\tau}{\sin\rho},\\
X_i &=& R \Omega_i\tan\rho,\ i=1,\ldots,d-1,\nonumber
\eea
where $\sum_{i=1}^{d-1}\Omega_i^2 = 1$, defining the surface of a unit $(d-1)$-ball, gives the following metric form of the $AdS$ line element:
\be
ds^2 =  \frac{R^2}{\cos^2\rho}\left( -d\tau^2 + d\rho^2 + \sin^2\rho d\Omega_{d-2}^2\right),
\ee
where $d\Omega_{d-2}^2$ is the square line element on the $(d-2)$-sphere. These coordinates cover the whole hyperboloid and are referred to as 'global' $AdS$ coordinates.

Setting $r = R \tan\rho$, the line $AdS$ element becomes
\be
ds^2 = -\left(1 + \frac{r^2}{R^2}\right)dt^2 + \frac{dr^2}{1+\frac{r^2}{R^2}} + r^2 d\Omega_{d-2}^2.
\label{adsschc}
\ee
which will be a useful parametrisation for our purpose.

Using \eqref{maxsym}, the $AdS$ spacetime is obviously a solution to 
\be
G_{ab} = -\frac{(d-1)(d-2)}{2R^2}g_{ab},
\ee
$G_{ab}$ being the component of the Einstein tensor, $g_{ab}$ the metric components. In other words, $AdS$ spacetime can be seen as a universe containing a positive energy density $G_0^0$ and a negative pressure.

Note that solutions to Einstein equations such that the Einstein tensor is proportional to the metric tensor are called Einstein spaces.

\section{Asymptotically $AdS$ spacetime}
We will refer to an asymptotically $AdS$ solution as a solution such that the asymptotical form of the solution satisfies globally relation \eqref{maxsym}. However it is not always clear that these asymptotical spacetimes can be recovered as the surface of a hyperboloid embedded in a $d+1$ dimensional flat spacetime. 

For example, the $d$-dimensional $AdS$ black hole:
\be
ds^2 = -f(r)dt^2 + \frac{dr^2}{f(r)} + r^2d\Omega_{d-2}^2,\ f(r) = \frac{r^2}{\ell^2} + 1 - \left(\frac{r_0}{r}\right)^{d-3},
\ee
is clearly asymptotically of the form of \eqref{adsschc}; relation \eqref{maxsym} is immediately satisfied and it is obvious that it describes the surface of the $AdS$ hyperboloid for large values of $r$ (when $r_0/r\ll 1$).

Another example is the $AdS$ black string,
\be
ds^2 = -b(r)dt^2 + \frac{dr^2}{f(r)} + r^2d\Omega_{d-3}^2 + a(r)dz^2,\ z\in[0,L],
\ee
where $a,b,f$ are functions to be computed numerically and where the asymptotic form of these function is
\be
a(r)\propto \frac{r^2}{\ell^2} + a_0 + \mathcal O\left( \frac{\ell}{r} \right)^2,\ b(r)\propto \frac{r^2}{\ell^2} + a_0 + \mathcal O\left( \frac{\ell}{r} \right)^2,\ f(r)\propto \frac{r^2}{\ell^2} + f_0 + \mathcal O\left( \frac{\ell}{r} \right)^2,
\ee
for some values of $a_0,f_0$ (see e.g. \cite{rms}). This spacetime satisfies \eqref{maxsym} asymptotically, but the $z$ coordinate has a finite range. However, it is not clear whether this spacetime describes the surface of the $AdS$ hyperboloid or not, since the asymptotic form of the black string is the $AdS$ space \eqref{adsschc} only to the leading order. We will refer to such spacetimes as asymptotically locally $AdS$ spacetimes.

%% file: susyqm.tex
In this appendix, we review the basic concepts of supersymmetric quantum mechanics, following reference \cite{khare}.

\section{Superpartner potentials and superpotential}
First, consider a Hamiltonian equation for quantum mechanics in $1$ dimensions of the form
\be
H_1\psi_n = -\frac{\hbar^2}{2 m} \dddf{\psi_n}{x} + V_1(x) \psi_n = E_n \psi_n.
\ee

Assuming that the ground state has zero energy, $E_0=0$, it is possible to reconstruct the potential (up to a constant) from the zero energy ground state:
\be
V_1 = \frac{\hbar^2}{2 m}\frac{\psi_0''}{\psi},
\label{defv1psi}
\ee
where the prime denotes the derivative with respect to the only coordinate.

Now, let $H_1 = A^\dagger A$, where 
\be
A = \frac{\hbar}{\sqrt{2m}}\ddx{x} + W,\ A^\dagger = -\frac{\hbar}{\sqrt{2m}}\ddx{x} + W,
\ee
for some function $W$, called the superpotential. Note that if $A\psi_0 = 0$, $H_1\psi_0=0$ is automatically satisfied.

Then, expanding $A^\dagger A$ acting on a test function provides the relation between $V_1$ and $W$:
\be
V_1 = W^2 - \frac{\hbar}{\sqrt{2m}} W'.
\label{defv1}
\ee

Note that this is consistent with $A\psi_0=0$; if this is so, we can write $W$ as a function of the ground state, leading to $W=-\frac{\hbar}{\sqrt{2m}}\frac{\psi_0'}{\psi_0}$. Inserting this expression of $W$ in \eqref{defv1} leads to \eqref{defv1psi}.

Now, let us define another Hamiltonian, $H_2 = A A^\dagger$. This Hamiltonian corresponds to a quantum theory with a new potential $V_2$ given by
\be
V_2 = W^2 + \frac{\hbar}{\sqrt{2m}} W'.
\ee
These two potentials $V_1$ and $V_2$ are said to be superpartner potentials.

\section{Energy eigenvalues}

The striking feature of superpartner potentials is the link between the energy eigenvalues and eigenvectors of each potentials. It turns actually out that the operator $A$ and $A^\dagger$ have the effect of taking an eigenvector of $H_1$ to $H_2$ (resp. $H_2$ to $H_1$). 

The eigenvectors of $H_1$ and $H_2$ obey the following equations
\bea
H_1\psi_n^{(1)} &=& A^\dagger A\psi_n^{(1)} = E_n^{(1)} \psi_n^{(1)}\nonumber\\
H_2\psi_n^{(2)} &=& AA^\dagger\psi_n^{(2)} = E_n^{(2)} \psi_n^{(2)}.
\eea
These two relations imply that 
\bea
H_1A^\dagger\psi_n^{(2)} &=& A^\dagger AA^\dagger\psi_n^{(2)} = E_n^{(2)} A^\dagger\psi_n^{(2)}\nonumber\\
H_2A\psi_n^{(1)} &=& AA^\dagger A\psi_n^{(1)} = E_n^{(1)} A\psi_n^{(1)}.
\eea
which means that $A\psi_n^{(2)}$ is an eigenvector of $H_1$ and $A^\dagger\psi_n^{(1)}$ is an eigenvector of $H_2$. From the fact that that $E_0^{(1)}=0$ and that the energies are positive, we obtain the following relations:
\bea
E_n^{(2)} &=& E_{n+1}^{(1)},\ E_0^{(1)} = 0,\nonumber\\
\psi_n^{(2)} &=& \left(\frac{1}{E_{n+1}^{(1)}}\right)^{\frac{1}{2}} A\psi_{n+1}^{(1)}\nonumber\\
\psi_{n+1}^{(1)} &=& \left(\frac{1}{E_{n}^{(2)}}\right)^{\frac{1}{2}} A^\dagger \psi_{n}^{(2)}\nonumber.
\eea

%% file: num.tex
In this appendix, we introduce the basic ideas used in the two main algorithms used to solve the various differential equations of this thesis, namely Colsys and Fidisol.

\section{Partial differential equations}
As we said in the text, we used the solver Fidisol in order to solve partial differential equations. The Fidisol package has been developed by the Karlsruhe Institut f\"ur Technologie and has been checked on a large number of problems.

The algorithm needs several inputs. First, the effect of introducing a compact coordinate which maps the semi-infinite range $r\in [r_h,\infty]$ to $x\in[0,1]$, such that $x=(r-r_h)/r$ leads to the following substitutions in the differential equations
\begin{eqnarray}
r  \ddf{\hat F}{r}   \rightarrow    (1-x) \ddf{\hat F}{x},\ r^2 \frac{\partial^2\hat F}{\partial r^2}   \rightarrow
(1-x)^2  \frac{\partial^2\hat F}{\partial x^2}  - 2 (1-x) \ddf{\hat F}{x}
\end{eqnarray}
for any function $\hat F_i$.

The equations for $\hat{F}_i$ are then discretized on a non-equidistant grid in $x$ and $z$ (where $z$ is defined in chapter \ref{chadsbs}).
Typical grids used have sizes $60 \times 40$, covering the integration region $0\leq x \leq 1$ and $0\leq z \leq 1$.  

All numerical calculations  are performed by using the program Fidisol, which uses a  Newton-Raphson method. A detailed presentation of the this code is presented in \cite{fidisol}.

This code requests the system of nonlinear partial differential equations to be written in the form
$P(x,z,u,u_{x},u_{z},u_{xx},u_{zz})=0,$ (where $u$ denotes the set of unknown functions and $u_{x}, u_{z}$ are the derivative of $u$ with respect to $x$ and $z$) subject to a set of boundary conditions on a rectangular domain.
The user must deliver to Fidisol the equations, the boundary conditions, the Jacobian matrices for the equations and the boundary conditions, and some initial guess functions.

The numerical procedure works as follows: for an approximate solution $u^{(1)}$, $P(u^{(1)})$ does not vanish. The next step is to consider an improved solution $u^{(2)}=u^{(1)}+\Delta u$, supposing that $P(u^{(1)}+\Delta u)=0$. The expansion in the small parameter $\Delta u$ gives in the first order $0=P(u^{(1)}+\Delta u) \approx P(u^{(1)})+\frac{\partial P}{\partial u }(u^{(1)}) \Delta u +\frac{\partial P}{\partial u_x }(u^{(1)}) \Delta u_x+ \dots\ .$

This equation can be used to determine the correction $\Delta u^{(1)}= \Delta u$. Repeating the calculations iteratively ($u^{(3)}=u^{(3)}+\Delta u^{(2)}$ etc), the approximate solutions will converge, provided the initial guess solution is close enough to the exact solution. The iteration stops after $i$ steps if the Newton residual $P(u^{(i)})$ is smaller than a prescribed tolerance. Therefore it is essential to have a good
first guess, to start the iteration procedure.

In each iteration step a correction to the initial guess configuration is computed. The maximum of the relative defect decreases by a factor of $20$ from one iteration step to another.
However, in some range of the parameters, the convergence is slower. In this case, we can re-iterated the solution until the defect is small enough (about $10^{-4}$). Note, that this defect concerns the discretized equations. The estimates of the relative error of the solution (truncation error) are computed separately. They can be of the order $0.001$. The errors also depend on the order of consistency of the method, $i.e.$ on the order of the discretization of derivatives.

\section{Ordinary differential equations}
The procedure used to integrate systems of ordinary differential equations is very similar to the one used for partial differential equations. One starts with an initial mesh and an initial profile and provides the equations in the following form:
\be
u^{(m_i)}_i = F_i(x,u, u^{(k)}_j), k=1,\ldots,m_j-1,
\ee
where the superscripts denotes the order of the derivative with respect to the variable $x$. The boundary conditions are provided in the following form:
\be
G_i(a_i,u_j, u^{(k)}_j)\ldots) =0,\ k=1,\ldots,m_j-1,
\ee
where $a_i$ is the point where the boundary condition is imposed.

The solver then solves the discretized equations and boundary conditions using a Newton-Raphson method; to this end, it requires the derivatives of the equations and of the boundary conditions with respect to the functions $u_j$ and their derivatives $u_j^{(i)},\ i=1,\ldots,m_j-1$.

Here again, it is important to provide an initial guess sufficiently close to the real solution. It must be noted that Colsys rearranges the mesh in order to minimize the numerical errors. More details can be found in \cite{colsys}.

The typical grid we used consisted in 200 points and the relative errors were of the order of $10^{-5}$ for every cases we considered.